%% file: main.tex
\documentclass[12pt]{iiscthes}
\usepackage{iisclogo}

\usepackage{basicreq}
\usepackage{subfiles}
\graphicspath{{./}{../figures/}{./figures/}}

\def\withcolors{0}

\input{preamble}
\input{tangocolors.sty}

\definecolor{red1}{rgb}{0.4,0,0}
\colorlet{taskyblue}{taskyblueo!0!}
\colorlet{tapink}{pink!0!}
\colorlet{ta2green}{tagreen!0!}
\hypersetup{
colorlinks,
 citecolor=blue,
  linkcolor=red1,
  urlcolor=Blue
}

\begin{document}

\begin{frontmatter}
%
%

\title{Compression for Distributed Optimization and Timely Updates}
\author{Prathamesh Mayekar}
\advisers{Himanshu Tyagi}
\submitdate{April 2022}
\dept{Electrical Communication Engineering}
\enggfaculty
\iisclogotrue 
\tablespagetrue 
 \maketitle

\begin{dedication}
\begin{center}
TO \\[2em]
\large Aai, Baba, and Kshitij
\end{center}
\end{dedication}

%
%
%

\acknowledgements
This dissertation was made possible by a fellowship by the Ministry of Human Resource Development, Gov. of India during my first two academic years at IISc, and a fellowship by Wipro Ltd. during the subsequent three academic years at IISc. I would also like to acknowledge Robert Bosch Centre for Cyber-Physical Systems, IISc and SPCOM, 2018 travel grant for the travel support to the International Symposium on Information Theory (ISIT), 2018.

I consider myself fortunate to have Prof. Himanshu Tyagi as my advisor. He has been extremely generous with his time and ideas and often held my hand through most aspects of academic research, such as reading and writing papers, proving theorems, and preparing research presentations. For instance, before ISIT 2018, he sat through multiple practice talks and gave valuable feedback on each one of them. Moreover, he stayed with me during all conference submissions until we submitted the paper; this was often way past midnight. Without his support and guidance, perhaps this thesis would not have been possible.

 I am thankful to my collaborators Prof. Jayadev Acharya, Prof. Cl\'{e}ment Cannone, Prof. Parimal Parag,  and  Dr. Ananda Theetha Suresh for their guidance throughout our collaborations. I would also like to thank professors at IISc and IITB for their excellent teaching. For instance, at IISc, I thoroughly enjoyed the Information Theory courses by Prof. Himanshu Tyagi, the Optimization reading group led by Prof. Aditya Goplan's research group, the Probability Theory course by Prof. Srikanth Iyer, and the Foundations of Data Science course by Prof. Siddharth Barman. At IITB, I thoroughly enjoyed the Game Theory course by Prof. Ankur Kulkarni,  the Integer Linear Programming course by Ashutosh Mahajan, and the Stochastic Processes course by Prof. Veeraruna Kavitha. The  courses at IITB motivated my decision to pursue a research career, and the courses at IISc became instrumental in my subsequent research work. I would also like to thank Prof. Nandyala Hemachandra and Prof. Veeraruna Kavitha at IEOR, IITB, who encouraged me to pursue a career in academia.

My interactions with fellow graduate students enriched my Ph.D. experience at IISc. I would like to thank  Raghava for delving with me  into the nuances of probability and information theory.  I would like to thank Sanidhay for guiding me through the ups and downs of life at IISc in my early graduate years. I would like to thank Karthik for teaching me how to teach. I want to thank Saumya for sharing with me ``tangocolors.sty'', a customized tex color style file that I still use for all my presentations. Thanks are also due to my labmates Aditya, Mishfad,  Raghava, Shubham, Siddharth, Sahasranand,  Saumya, and Vaishali whose presence in the lab made my work a lot more enjoyable. I am already missing those coffee-break conversations.

My IEOR batchmates made my stay in Bangalore a pleasant one. The weekends spent at Anupam's place made the highs of graduate life even more enjoyable and the lows bearable.

 I would like to thank my parents for their unconditional support, love, and encouragement throughout my graduate studies. Without their reassuring presence and many sacrifices,  both my undergraduate and graduate education would not have been possible. I can never thank them enough! I would like to thank my younger brother Kshitij for patiently listening to all my ramblings and rants about graduate student life, guiding me through most aspects of life outside academia, and, in essence, always being the older, wiser one of the two of us.

\statementoforiginality

I hereby declare that this thesis is my own work and has not been submitted in any form for another degree or diploma at any university or other institute of higher education.

I certify that to the best of my knowledge, the intellectual content of this thesis is the product of my own work and
that all the assistance received in preparing this thesis and sources have been
acknowledged. 

\publications
\textbf{Journal Publications:}
\begin{enumerate}

\item \textbf{P Mayekar} and  H Tyagi. ``RATQ: A Universal Fixed-Length Quantizer for Stochastic Optimization'', IEEE Transactions on Information Theory 67(5) (pp. 3130 - 3154).

\item \textbf{P Mayekar}, P Parag, and  H Tyagi.  ``Optimal source codes for timely updates'', IEEE Transactions on Information Theory 66(6) (pp. 3714-3731).

\end{enumerate}

\vspace{0.5cm}

\textbf{Conference Proceedings:}
\begin{enumerate}
  \item J. Acharya, C. Canonne, \textbf{P. Mayekar}\footnote{Alphabetical Order used for author list.}, and H. Tyagi. ``Information-constrained optimization: Can adaptive processing of gradients help?'', Advances in Neural Information Processing Systems, 2021.
  
 \item  \textbf{P Mayekar},  A. T. Suresh, and H. Tyagi. ``Wyner-Ziv Estimators: Efficient Distributed Mean Estimation with Side Information'', International Conference on
Artificial Intelligence and Statistics (pp. 3502-3510).  PMLR, 2021.

\item  \textbf{P Mayekar} and  H Tyagi. ``Limits on gradient compression for stochastic optimization''.  IEEE International Symposium on Information Theory (pp. 2658-2663). IEEE, 2020.

  \item \textbf{P Mayekar} and  H Tyagi.  ``RATQ: A universal fixed-length quantizer for stochastic optimization''. International Conference on Artificial Intelligence and Statistics (pp. 1399-1409).  PMLR, 2020.
  
  \item \textbf{P Mayekar}, P Parag, and  H Tyagi. 
 ``Optimal lossless source codes for timely updates''. 2018 IEEE International Symposium on Information Theory (ISIT). IEEE, 2018. (Recipient of the Jack Keil Wolf Student Paper Award.)
\end{enumerate}


%
\begin{abstract}
\sl
 The goal of this thesis is to study the compression problems arising in distributed computing systematically.

             In the first part of the thesis, we study gradient compression for distributed first-order optimization. 
             We begin by establishing  information theoretic lower bounds on optimization accuracy when only finite precision gradients are used.
             Also, we develop fast quantizers for gradient compression, which, when used with standard first-order optimization algorithms, match the aforementioned lower bounds.
             
             In the second part of the thesis, we study distributed mean estimation, an important primitive for distributed optimization algorithms. We develop efficient estimators which improve over state of the art by efficiently using the side-information present at the center. We also revisit the Gaussian rate-distortion problem and develop efficient quantizers for this problem in both the side-information and the no side-information setting.

Finally, we study the problem of entropic compression of the symbols transmitted by the edge devices to the center, which commonly arise in cyber-physical systems. Our goal is to design entropic compression schemes that allow the information to be transmitted in a 'timely' manner, which, in turn, enables the center to have access to the latest information for computation. We shed light on the structure of the optimal entropic compression scheme and,  using this structure, we develop efficient algorithms to compute this optimal compression scheme.

\end{abstract}
\newpage
\pdfbookmark[section]{\contentsname}{toc}

\makecontents


\notations
\subfile{notation}

\end{frontmatter}

\chapter{Overview}
 The recent years have witnessed a monumental rise in the data available for machine learning applications. For instance, ImageNet (\cite{deng2009imagenet}), a publicly available image database, has over fourteen million images, all available to train machine learning models. Closely mimicking the data rise is the aspiration to build more capable and accurate machine learning models. This, in turn, has led to the ever-increasing computing power needed to train sophisticated deep learning models. For instance, the ResNet (\cite{he2016deep}) architectures trained on the ImageNet database can have roughly 20-60 million trainable parameters. One approach to ensure fast training of such sophisticated models is to employ distributed optimization methods, where at each iteration, workers {\em quantize} and share their updates, stochastic gradients, with other workers or the center. However, while this distributed optimization approach has enjoyed popularity recently, the precise effect quantization has on convergence rates is not entirely understood. 
 
  A related problem is that of federated learning (\cite{kairouz2019advances}).
 Federated learning is a machine learning paradigm where models are built from decentralized data residing on mobile devices while preserving the privacy of the data. A few of the devices share their stochastic gradient updates with the center in a typical iteration of a federated learning algorithm. In low bandwidth scenarios, efficiently quantizing these updates becomes crucial. Thus, understanding the tradeoff between the precision to which gradients are quantized and the convergence rate becomes crucial in designing efficient federated learning algorithms.
 
 Finally, the problem of timely dissemination of information has become increasingly important in modern cyber-physical systems. For instance, consider the problem of control of the network of autonomous vehicles. In such a setting timely update of the vehicle state becomes exceptionally crucial. In such problems, designing compression specifically tailored to the application of timeliness is crucial.

 Keeping in mind the applications listed above, the thesis considers the following three problems:
 \begin{enumerate}
     \item  { Communication-Constrained First-Order Optimization},
     \item { Efficient Quantization for Federated Learning Primitives},
     \item  { Source Coding Schemes for Timeliness}.
 \end{enumerate}
  The first two parts of the thesis are dedicated to studying the distributed optimization scenarios listed above.   In the first part of the thesis, we build a theory for a distributed optimization setting where the stochastic gradients are quantized to a given precision.  The quantization algorithms we develop in this part improve over the state-of-the-art algorithms in many settings.
 In the second part of the thesis, we revisit primitives often used Federated Learning: 1) Distributed Mean Estimation 2) Gaussian Quantization. We build communication-efficient primitives for both these problems. 
 In the final part of the thesis, we design entropic compression schemes to ensure timely update of information.
 
 \section{Communication-Constrained First-Order Optimization}
  In the first part of the thesis, we study a refinement of the classic query complexity model of Nemirovsky and Yudin (\cite{nemirovsky1983problem}). In our refinement, the gradient estimates supplied by the first-order oracle are not directly available to a first-order optimization algorithm but must pass through a channel, and only the output of the channel is available to the optimization algorithm. While we introduced this refinement to study the effect of communication constraints on convergence rate,  the channel can also be used to model various other information constraints such as local differential privacy constraints and computational constraints.

 In Chapter \ref{Ch:LB}, we derive lower bounds on the optimization error of any first-order optimization algorithm where it only has access to compressed gradients. In Theorem \ref{t:concom_P2} and \ref{t:concom_inf}, we show that for the optimization of convex and $\ell_p$ lipschitz family, any optimization algorithm using gradients compressed to $r$ bits would lead to the following blow-up over the classic convergence rate: for $p \in [1, 2)$, we see a blow-up of $\displaystyle{ \sqrt{\frac{d}{\min\{d, r\}}};}$
 for $p \in [2, \infty]$, we see a  blow-up of
$\displaystyle{
 \left( \sqrt{\frac{d}{ d \wedge 2^r}} \right) \vee \left( \sqrt{\frac{d^{2/p}}{  d \wedge r}} \right),
 }$
 where $d$ is the ambient dimension.
 Our lower bounds also extend to the class of strongly convex and $\ell_2$ lipschitz function, which were missing in the literature of information-constrained optimization. For example, in Theorem \ref{t:sconcom_P2}, we show that for the optimization of strongly convex and $\ell_2$ lipschitz function, any optimization algorithm using gradients compressed to $r$ bits would lead to a  blow-up of at least $ \frac{d}{\min\{d, r\}}$ over the classic convergence rate.
 
 In Chapter \ref{Ch:LB}, we also derive lower bounds for local differential privacy constraints and computational constraints in Theorems \ref{t:conpriv_P2}, \ref{t:conpriv_inf}, and Theorems \ref{t:sconpriv_P2} and \ref{t:concompt_P2} and \ref{t:sconcompt_P2}, respectively. In fact, our lower bounds allows us to establish optimality of the popular random coordinate descent algorithm for convex and $\ell_2$ lipschitz family, when there is a computational constraint of computing just one coordinate. 
 
 Finally, all the lower-bounds derived in Chapter \ref{Ch:LB} allow for adaptive processing of gradients, while the previous literature on information-constrained optimization restricts to non-adaptive protocols. That is, the channel used to process the gradients at a given iteration can be chosen as a function of the information received at the previous iteration. However, as we see in the compressing schemes derived in the next chapters and privacy protocols employed in the literature, the non-adaptive processing of gradients is sufficient to match the lower bounds.
 
 Our proof of lower bounds refines the recipe of \cite{agarwal2012information} and reduces the problem of lower bounding optimization error to that of upper bounding a   mutual information term. The mutual information term then can be bounded by taking recourse to recent strong data processing inequalities in \cite{acharya2020general}. The key observation in our proof is that we only need to bound a coordinate-wise average mutual information term compared to the larger total mutual information considered in \cite{agarwal2012information}. This allows  our recipe to be applicable in settings where the recipe in \cite{agarwal2012information} may not be suitable.
 
 In Chapter \ref{Ch:RATQ}, we focus on developing optimal quantizers to match lower bounds derived in Chapter \ref{Ch:LB} for convex and $\ell_2$ lipschitz functions and strongly convex and $\ell_2$ lipschitz functions. Since we assume that the gradient estimates' have their Euclidean distance almost surely bounded, this problem  essentially reduces to developing efficient quantizers for input vector $Y$ such that
 \[ Y \leq B^2 \quad a.s..\]
 Our main contribution in this chapter is a quantizer RATQ used to quantize such input vectors $Y$. In Corollary \ref{c:PSGD_RATQ_0}, we show that employing RATQ to compress the gradients along with the optimization algorithm projected stochastic gradient descent (PSGD) requires precision of $d \log \log \log \log^*d$ bits to attain the convergence rate of the classic, unrestricted setting for convex, or strongly convex, and $\ell_2$ lipschitz function family. This factor differs by only a minuscule $\log \log \log \log^*d$ from the lower bounds of $\Omega(d)$ on the precision necessary to attain the convergence of classic setting.  Moreover, in Corollary \ref{c:PSGD_RCS_RATQ},  we show that employing a subsampled version of RATQ along with PSGD leads to almost optimal convergence rate, thus matching the lower bounds established in Chapter \ref{Ch:LB} for both convex and strongly convex functions, which are also $\ell_2$ lipschitz.
 
 In fact, our quantizer RATQ is part of a general family of quantizers called adaptive quantizers. An adaptive quantizer uses multiple dynamic ranges, $\{[-M_i, M_i]: i \in [h]\}$, to quantize the input. Once a dynamic range is chosen, the input is quantized uniformly within it using $k$ uniform level. In designing RATQ, we stumble upon the following formula for the mean square error of the adaptive quantizer:
 \[O\left( \frac{\sum_{i \in [h]}M_i^2 . p(M_{i-1})}{(k-1)^2} \right),\]
 where $p(M)$ is the probability of the input vector exceeding the value $M.$ This formula guides the design of all of our subsequent adaptive quantizers.

 Before adaptively quantizing an input, RATQ first preprocesses the input by randomly rotating it. Random rotation allows the input data to be "evenly" distributed across all the coordinates and gives us a handle over the data distribution. This classic idea of Random rotation is also crucial to many of our subsequent quantizers.

Then in Chapter \ref{Ch:RATQ}, we relax the almost sure assumption on gradient estimates noise and study a general noise model for noisy gradient estimate where the expected Euclidean norm square of the estimates' output is bounded, termed the {\em mean square noise model}. We show the theoretical limitations imposed by quantizers for such noise models that do not quantize the norm carefully. We do this by deriving a lower bound on optimization error in Theorems \ref{t:lb_1} and \ref{t:lb_2}, when a popular class of quantizers that quantize the gradient norm uniformly is employed. Our lower bound relies on a novel heavy-tailed construction which may be of independent interest. We then present a fixed-length, {\em gain-shape} variant of our quantizer RATQ, termed A-RATQ. In A-RATQ,  the norm ({\em the gain}) of the input is quantized by employing another adaptive quantizer and the input normalized by the norm ({\em the shape}) is quantized by employing RATQ. In Corollaries \ref{c:PSGD_NRATQ} and \ref{c:PSGD_RCS_ARATQ_fixed}, we show that A-RATQ along with PSGD almost matches the best possible performance for this mean square noise model. Finally, we present a variable-length update to A-RATQ. This variable-length version further improves the performance of A-RATQ but only satisfies the precision constraint in expectation.

In Chapter \ref{Ch:Limits}, we develop quantizers to match the lower bounds for communication-constrained optimization of convex and $\ell_p$ lipschitz family. Our results in this Chapter are primarily restricted to {\em the high-precision regime}. That is, we characterize the minimum precision needed by optimal quantization and optimization algorithms so that optimization with compressed gradient achieves the convergence of the classic, unrestricted setting.
In Theorem \ref{t:main}, we show that for optimization of convex and $\ell_p$ lipschitz family with compressed gradients, if the gradients are compressed to a precision of $d^{2/p} \vee \log d$, for $p \in [2, \infty],$ and $d$, for $p \in [1, 2)$, we can attain the convergence rate of the classic, unrestricted setting. The necessity of this precision follows from the lower bound on optimization error derived in Chapter \ref{Ch:LB}; for sufficiency, we construct new quantizers. For $p \in [2, \infty],$ we propose a Quantizer SimQ$^+$ which along with PSGD exactly matches the lower bounds for $p=2$ and $p=\infty$, and is only a logarithmic factor away from the lower bound for $p \in (2, \infty)$. Interestingly, for $p=\infty$, compressing the gradients to only $\log d$ bits using SimQ$^+$ and then employing PSGD with compressed gradients is sufficient to achieve the convergence of the classic case. Thus, this improves upon the precision required by RATQ and PSGD, $d 
\log \log \log \log^*d,$ in the high-precision regime. For $p \in [1, 2),$ we propose another variant of RATQ, which, combined with appropriate mirror descent algorithms, is almost optimal.

In its simplest form, SimQ$^+$ represents the input in terms of the corner points of the $\ell_1$ ball containing it. To achieve further compression, SimQ$^+$ uses a ``type'' based compression technique. We will use this type-based compression idea in some of our later schemes, too.
 
 \section{Efficient Quantization for Federated Learning Primitives}
 In Chapter \ref{Ch:DME}, we study the primitive of distributed mean estimation. This primitive is a crucial subroutine in distributed learning scenarios when the server uses the average of updates from multiple clients.  \cite{suresh2017distributed} considered a version of this problem where $n$ clients communicate a quantized version of their update to the server, where the total precision of the quantized version can be at the most $r$ bits. The center uses the quantized versions from all the clients to estimate the sample mean of the data. A lower bound of $\displaystyle{\frac{d}{nr}}$ was established on the mean square error (MSE) between the actual sample mean and the estimated sample mean. \cite{suresh2017distributed} also proposed a quantization procedure that matches this lower bound up to $\log \log d$ factor. In Theorem \ref{t:DME_noside},  we derive the best known upper bound on  MSE, which is tight up to a $\log \ln^* d$ factor from the lower bound. Our scheme uses the quantizer RATQ from Chapter \ref{Ch:RATQ}. 
 
 Then, motivated by the fact that in many federated learning scenarios, the server also has access to some side-information, we propose and study distributed mean estimation where the server also has access to side-information. We study this problem in two different settings: 1) the distance between the update and the side-information is known to the clients and the server; 2) the distance between the update and the side-information is unknown to all, the universal setting. For the first setting, we propose a quantizer RMQ and show in Theorem \ref{t:DME_known} that it results in an overall MSE of roughly $\frac{d \Delta^2}{nr}$, where $\Delta^2$ is the distance between the client's update and the respective side-information at the server. Thus, we can break the lower bound of no side-information setting using side-information, as long as $\Delta \leq 1$. Our quantizer RMQ first preprocesses the update using Random rotation like RATQ and then uses a modulo quantizer for each coordinate. Coming to the unknown setting, we propose a quantizer RDAQ and show in Theorem \ref{t:DME_unknown} that results in an overall MSE of roughly $\frac{d \Delta}{ nr}$. Thus, we can again break the lower bound of the no side-information case with accurate side-information. Our quantizer RDAQ  first preprocesses the updates by randomly rotating them and then uses the idea of correlated sampling to provide MSE bounds dependent on the distance without the knowledge of the distance.
 
 In Chapter \ref{Ch:GRD}, we the revisit the Gaussian rate-distortion problem and show that the quantization schemes from earlier Chapters are almost optimal while being efficient for this problem. In Theorem \ref{t:gauss_rd}, we show that a subroutine of RATQ attains a rate very-close to the Gaussian rate-distortion function while being computationally feasible relative to the optimal coding scheme. In Theorem \ref{t:G_WZ}, we show that the simple modulo quantizer achieves a rate close to the rate-distortion function for the version of Gaussian rate-distortion problem with side--information.

 \section{Source Coding for Timeliness}
 In the final part of the thesis, we study a source coding problem to facilitate the timely dissemination of information. This study focuses on communication systems where the time to transmit information is directly proportional to its code length, and the receiver needs to be apprised about only the latest information.  Based on the {\em age of information} metric proposed in \cite{KaulYatesGruteser11}, we measure the performance of our schemes by the average age of information. For information received at time $t$ which was generated at time $U(t) \leq t$, the age of information at the receiver is $t - U(t).$ Our goal is to come up with coding schemes which minimize the average age 
 \[\limsup_{T \to \infty  }\frac{1}{T}\sum_{t=1}^{T}t-u(t).\]
 In Theorem \ref{t:average_age}, we show that the average age equals 
 \[ \E {L}+ \frac{\E{L^2}}{2\E {L}} - \frac 12. \]
 Our proof relies on the modification of the standard renewal reward theorem. We then show in Example \ref{ex:1} that standard prefix-free coding schemes such as Shannon codes can be suboptimal by as far as $O(\log |\X|)$ for these problems, where $|\X|$ is the cardinality of the information. Our main result is Theorem \ref{t:update_optimal}, where we show that the optimal source coding scheme for minimizing average age is Shannon coded corresponding to distribution, which is a tilting of the original distribution. Our proof relies on linearizing the average cost, which, in turn, relies on a variational formula for $L_p$ norm of a random variable.
 
 We then extend our recipe of linearizing the cost and identifying the structure of optimal coding schemes to design source coding schemes to minimize the average delay. In Theorem \ref{t:qmaintheorem}, we show that the optimal source coding scheme here, too, is a Shannon code for the tilting of the original distribution.


\part{Communication-Constrained First-Order Optimization}







\chapter{Lower Bounds for Information-Constrained Optimization}\label{Ch:LB}

\section{Synopsis}
We revisit first-order optimization under local information constraints such as communication, local privacy,  and computational constraints limiting access to a few coordinates of the gradient. In this setting, the optimization algorithm is not allowed to
  directly access the complete output of the gradient oracle, but only gets limited information about it subject to the local information constraints. We consider optimization for both convex and strongly convex functions and obtain tight or nearly tight lower bounds for the convergence rate under all three information constraints.
  
  The results presented in this chapter are from \cite{acharya2021information}.
  \section{Introduction}
  

Distributed optimization has emerged as a central tool in federated learning for building statistical and machine learning models for \newerest{distributed data}.
In addition, large scale optimization is typically implemented in a distributed fashion    
over multiple machines or multiple cores within the same
machine. These distributed implementations fit naturally in the oracle
framework of first-order optimization
(see~\cite{nemirovsky1983problem}) where in each iteration a user or
machine computes the gradient oracle output.  Due to practical local
constraints such as communication bandwidth, privacy concerns, or
computational issues, the entire gradient cannot be made available to
the optimization algorithm. 
Instead, the gradients must be passed
through a mechanism which, respectively, ensures privacy of user data
(local privacy constraints); or compresses them to a small number of
bits (communication constraints); or only computes a few coordinates
of the gradient (computational constraints). 
Motivated by these
applications, in this chapter, we derive lower bounds on first-order optimization under such constraints. While our focus in the rest of the chapters would be to achieve these lower bounds for communication constraints.

When designing a first-order optimization algorithm under local
  information constraints, one not only needs to design the
  optimization algorithm itself, but also the algorithm for local
  processing of the gradient estimates. Many such algorithms have
  been proposed in recent years; see, for
  instance, \cite{duchi2014privacy}, \cite{abadi2016deep}, \cite{agarwal2018cpsgd}, \cite{gandikota2019vqsgd}, \cite{subramani2020enabling}, \cite{girgis2020shuffled},
  and the references therein for privacy constraints; \cite{seide20141}, \cite{alistarh2017qsgd}, \cite{suresh2017distributed}, \cite{konevcny2018randomized}, \cite{faghri2020adaptive}, \cite{ramezani2019nuqsgd}, \cite{lin2020achieving}, \cite{acharya2019distributed}, \cite{chen2020breaking}, \cite{huang2019optimal}, \cite{safaryan2020uncertainty}, \cite{ghosh2020distributed}, \cite{gupta2021localnewton}
  and the references therein for communication
  constraints; \cite{nesterov2013introductory,richtarik2012parallel}
  for computational constraints.  However, these algorithms primarily
  consider \emph{nonadaptive} procedures for gradient processing (with
  the exception of \cite{faghri2020adaptive}): that is, the scheme
  used to process the gradients at any iteration cannot depend
  on the information gleaned from previous
  iterations. In this chapter, we derive lower bounds for optimization under a much larger class of {\em adaptive} gradient processing protocols. 
  As a result, we answer the following open question in this part of the thesis.
\begin{center}
{\em Can adaptively processing gradients improve convergence in
information-constrained optimization?}
\end{center}



For optimization of both
convex and strongly convex function families and under the three
different local constraints mentioned above -- local privacy,
communication, and computational -- we answer this question in the negative. 
\begin{center}
\em
That is, adaptive processing of gradients has no clear advantage over non-adaptive processing  for convex or strongly convex optimization under information-constraints.  
\end{center}

%
\subsection{Main contributions}
We model the information constraints using a family of channels $\W$; see Section~\ref{ssec:info-constraints} for a description of the channel families corresponding to our constraints of interest. We consider first-order optimization where the output of the gradient oracle must be passed through a channel $W$ selected from $\W$.
Specifically, the gradient is sent as input to this channel $W$, and the algorithm receives the output of the channel.
In each iteration of the algorithm, the channel to be used in that iteration can be selected
adaptively based on previously received channel outputs by the algorithm; or channels to
be used throughout can be fixed upfront, nonadaptively. 
The detailed problem setup is given in Section~\ref{ssec:set-up}. We obtain general lower bounds for optimization of convex and strongly convex functions using $\W$,
when adaptivity is allowed. 
These bounds are then applied to the specific constraints of interest to obtain our main results. 

 In terms of overall contribution of this part of this thesis, we show that adaptive gradient processing does {not} help for some of the most typical first-order optimization problems under information constraints. Namely, we prove that for most regimes of local privacy, communication, or computational constraints, adaptive gradient processing has \newest{nearly} the same convergence rate as nonadaptive gradient processing for both convex and strongly convex function families. 
As a consequence, this shows that the nondaptive  LDP algorithms from  \cite{duchi2014privacy} and nonadaptive compression protocols we develop in Chapters \ref{Ch:RATQ} and \ref{Ch:Limits} are (nearly) optimal for private and communication-constrained optimization, respectively, even if adaptive gradient processing is allowed. In another direction, under computational constraints, where we are allowed to compute only one gradient coordinate, we show that standard Random Coordinate Descent ($cf.$~\cite[Section 6.4]{bubeck2015convex}), which employs uniform (nonadaptive) sampling of gradient coordinates, is optimal for both the convex and strongly convex function families. This proves that adaptive sampling of gradient coordinates does not improve over nonadaptive sampling strategies.

\subsection{Remarks on techniques}
 Without information
constraints,~\cite{agarwal2012information} \newerest{provides a general
recipe for proving oracle complexity lower bounds for convex optimization}.  Specifically, it reduces \newerest{optimization problems} with a
first-order \newerest{oracle} to a mean estimation problem whose
probability of error is lower bounded using Fano's method
($cf.$~\cite{yu1997assouad}).
While our work, too, relies on a reduction to mean estimation, we deviate from the
prior
approach, using Assouad's method instead to prove lower bounds for various function families. This different approach, in turn, enables us to derive lower bounds for adaptive processing of gradients.
We then combine our Assouad's type reduction with upper bounds on mutual information derived in   \cite{acharya2020general}, which crucially hold for adaptive protocols.

We note that the prior work in information-constrained optimization -- primarily, locally private optimization -- concerned itself with the family of convex functions, with no lower bounds known for the more restricted family of \emph{strongly convex} functions, even for nonadaptive gradient processing protocols. The key obstacle is the fact that during the reduction from optimization to mean estimation, the known hard instance
for the strongly convex family, even when analyzed for nonadaptive protocols, leads to an estimation problem using adaptive protocols; and thus the lack of known lower bounds for adaptive information-constrained estimation prevented this approach from succeeding.
In more detail, this hard instance has gradients that can depend on the query point which in turn can be chosen
based on previously observed channel outputs, an issue which does not arise in the case of the convex family where the lower bounds are derived using affine functions for which the gradients do not depend on the query point. We manage to circumvent this issue by relying on our different Assouad-type reduction.

\subsection{Prior work}
The \newerest{framework we consider} can be viewed as \newerest{an}
extension of the classical query complexity model
in~\cite{nemirovsky1983problem}.  We refer the reader to textbooks and
 monographs~\cite{nemirovski1995information,nesterov2013introductory,bubeck2015convex}
 for a review of the basic setup.
In the information-constrained setting, motivated by privacy concerns,~\cite{duchi2014privacy} consider \newest{the} problem where the gradient estimates must pass through a locally differentially private (LDP) channel. However, in their setting the LDP channels for \newerest{all} time steps are \newest{selected} at the start of optimization algorithm~--~in other words, the channel selection strategy is nonadaptive. In contrast,  we allow for {\em adaptive} channel selection strategies (as well as other information constraints); as a result, the lower bounds established in these papers do not apply to our setting, and are more restrictive than our bounds.
The results of Duchi and Rogers~\cite{DuchiR19} for Bernoulli product distributions could be combined with our construction to obtain tight lower bounds for optimization in $p \in [1,2]$ under LDP constraints, but would not extend to the entire range of $p$. The work of Braverman, Garg, Ma, Nguyen, and Woodruff~\cite{BGMNW:16} on communication constraints, also for $p \in [1,2]$, is relevant as well; however, their bounds on mutual information cannot be applied directly, as their setting (Gaussian distributions) would not satisfy our almost sure gradient oracle assumption.
\cite{faghri2020adaptive} provide adaptive quantization schemes for convex and $\ell_2$ Lipschitz function family. While the worst-case convergence guarantees for the quantizers in \cite{faghri2020adaptive} are similar to those in \cite{alistarh2017qsgd}, it shows some practical improvements over the state-of-the-art for some specific problem instances. This suggests that while adaptive quantization may not help in the worst case for non-smooth convex and strongly convex optimization, it may be useful for a smaller subclass of convex optimization problems.


\subsection*{Organization}
The rest of the chapter is organized as follows. After formally introducing in Section~\ref{sec:preliminaries} the setting, the function classes considered (convex and strongly convex), and the information constraints we are concerned with, we state and discuss our lower bounds in  Section~\ref{sec:results}.
 Proofs of these lower bounds are given in Section~\ref{sec:proofs}. 

\section{Setup and preliminaries}
\label{sec:preliminaries}
\subsection{Optimization under information constraints}
\label{ssec:set-up}
\begin{figure}
\centering
\begin{subfigure}[t]{\textwidth}
\centering
\input{Figures/Figure_SGD.tex}
\caption{ Classical first-order optimization}
\label{f:CO}
\end{subfigure}
\begin{subfigure}[t]{\textwidth}
\centering
\input{Figures/Figure_SGD3.tex}

\caption{ Information-constrained optimization with adaptive gradient processing.}
\label{f:ICCO}
\end{subfigure}
\begin{subfigure}[t]{\textwidth}
\centering
\input{Figures/Figure_SGD5.tex}
\caption{ Information-constrained optimization with nonadaptive gradient processing.}
\label{f:ICCO_NA}
\end{subfigure}
\end{figure}
We consider the
problem of minimizing an unknown convex function $f\colon\X
\to \R$ over its domain $\X$ 
using \emph{oracle access}
to noisy subgradients of the
function.
That is, the algorithm is not directly given access to the function
but can get subgradients of the function at different points of its choice.
This class of optimization algorithms includes various descent algorithms,
which \newerest{often provide} optimal convergence rate among
all the algorithms in this class ($cf.$~\cite{nemirovsky1983problem}).

In our setup, gradient estimates supplied by the oracle must pass through a channel $W$,\footnote{\new{A channel $W$ with input alphabet $\X$ and output alphabet $\Y$,
  denoted $W\colon \X \to \Y$, represents
  the conditional distribution of the output of
  a randomized function given its input. In particular, $W(\cdot \mid x)$ is
  the conditional distribution of the channel given that the input is $x\in \cX$.}
}
\new{chosen by the algorithm
from a fixed set of channels $\W$,} and the optimization algorithm $\pi$ only has access to the output of this channel.
\new{The \emph{channel family} $\W$ represents information constraints
imposed in our distributed setting.}
In detail, the framework is as follows:
\begin{enumerate}
\item At iteration $t$, the first-order optimization algorithm $\pi$ makes a query for point $x_t$ to the oracle $O$.
\item Upon receiving the point $x_t$, the oracle outputs $\hat{g}(x_t)$, where $\EEC{\hat{g}(x_t)}{ x_t} \in  \partial f (x_t)$ and $\partial f (x_t)$ is the subgradient set of function $f$ at $x_t$.
\item \newerest{The subgradient estimate $\hat{g}(x_t)$ is passed through a channel \new{$W_t\in \W$} and the output $Y_t$ is observed by the first-order optimization algorithm. The algorithm then uses all the messages $\{Y_i\}_{i \in [t]}$ to further update $x_t$ to $x_{t+1}$.}
\end{enumerate}

\newerest{Let $\Pi_T$ be the set of all first-order optimization algorithms that are allowed $T$ queries to the oracle $O$ and after the $t$th query gets back  the output $Y_t$ with distribution $W_{t}(\cdot \mid \hat{g}(x_t))$.} 

Our goal is to select \newerest{gradient processing} channels $W_t$s and an optimization algorithm $\pi$ to \newerest{guarantee a} small worst-case optimization error. Two classes of \emph{channel selection strategies} are of interest: \emph{adaptive} and \emph{nonadaptive}. 

\medskip
\noindent
\begin{defn}
\newerest{Under \textit{adaptive gradient processing}, the channel $W_t$ selected at time $t$ may depend on the previous outputs of channels $\{W_i\}_{i \in [t-1]}$.
  Specifically, denoting by $Y_t$ the output of the channel used at time $t$,
  which takes values in the output alphabet $\Y_t$,
the \emph{adaptive channel selection strategy} $S\eqdef (S_1, \ldots, S_T)$ over $T$ iterations 
consists of mappings $S_t\colon \Y^{t-1} \to \W$ that take $Y_1, \ldots Y_{t-1}$ as input and output a channel $W_t\in\W$ as output.
We write $\mathcal{S}_{\W, T}$ for the collection of all such channel selection strategies.
}
\end{defn}

\medskip
\begin{defn}\label{d:NA}
\noindent
\newerest{Under \textit{nonadaptive gradient processing} all the channels
 $\{W_{t}\}_{t \in [T]}$ through which the gradient estimates must pass are decided at the start of the optimization algorithm. {In other words, conditioned on the shared randomness, the channel $W_t$ is selected independently of all the  gradient observations received by the optimization algorithm until step $t$.} Denote the class of all nonadaptive strategies by $\mathcal{S}_{\W, T}^{\rm{}NA}$.} 
\end{defn}

Figures \ref{f:CO}, \ref{f:ICCO}, and \ref{f:ICCO_NA}, describe the classical optimization framework, information-constrained optimization under adaptive gradient processing, and information-constrained optimization under nonadaptive gradient processing, respectively.

We measure the performance of
an optimization protocol $\pi$ and a channel selection strategy $S$ for a given function $f$ and oracle $O$ using the metric
$\ep(f, O, \pi, S)$ defined as  
\begin{equation}
    \label{eq:def:metric}
  \ep(f, O, \pi, S) = \E{f(x_T)-\min_{x\in \X} f(x)},
\end{equation} 
where the expectation is over the randomness in $x_T$.

For various function and oracle classes, denoted by $\oO$, the channel constraint family $\W$, and the number of \newerest{iterations} $T$, we will characterize the \emph{adaptive minmax optimization error}
\begin{equation}
    \label{eq:def:optimization:error}
  \ep^\ast(\X , \oO, T, \W) =   \inf_{\pi \in \Pi_T}\inf_{S \in \mathcal{S}_{\W, T}} \sup_{(f, O)\in\oO}\ep(f, O, \pi, S)\,,
\end{equation}
and the corresponding \emph{nonadaptive minmax optimization error} 
\begin{equation}
    \label{eq:def:NAoptimization:error}
  \ep^{\rm{}NA \ast}(\X , \oO, T, \W) =   \inf_{\pi \in \Pi_T}\inf_{S \in \mathcal{S}_{\W, T}^{\rm{}NA}} \sup_{(f, O)\in\oO}\ep(f, O, \pi, S)\, .
\end{equation}
\new{Since the adaptive channel selection strategies include the 
  nonadaptive ones, we have $\ep^{\rm{}NA \ast}(\X , \oO, T, \W) \geq \ep^\ast(\X , \oO, T, \W).$ }

\subsection{Function classes}
\newerest{We now define the function classes and the corresponding oracles that we consider.}
  
\paragraph{Convex and $\ell_p$ Lipschitz function family.}
 Our first set of function families are parameterized by \new{a
   number} $p \in [1, \infty]$. Throughout, we restrict
 ourselves to convex functions over \new{a domain} $\X$, i.e.,
functions $f$ satisfying
 \begin{align} \label{e:convexity}
 f(\lambda x +(1-\lambda)y)\leq \lambda f(x) +(1-\lambda)f(y), \quad
 \forall x, y \in \X, \quad \forall \lambda \in [0, 1].
\end{align} 
 Further, for a family parameterized by $p$, we assume that the
 subgradient estimates returned by the first-order oracle for a
 function $f$ satisfy the following two assumptions:
 \begin{align}
\label{e:asmp_unbiasedness}
\E{\hat{g}(x)\mid x} \in \partial f(x), \quad \text{(Unbiased estimates)} \\
\label{e:asmp_as_bound}
\bPr{\norm{\hat{g}(x)}_q^2 \leq B^2\mid x}=1, \quad \text{(Bounded estimates)}
\end{align}
where  $\partial f(x)$ is the set of subgradient for $f$
at $x$ and $q\eqdef p/(p-1)$ is, \newer{as mentioned earlier,} the H\"older conjugate of $p$.

\begin{defn}[Convex and $\ell_p$ Lipschitz function family $\oO_{{\tt c}, p}$]\label{def:c}
We
denote by $\oO_{{\tt c}, p}$ the set of all pairs of functions and oracles satisfying Assumptions \eqref{e:convexity}, \eqref{e:asmp_unbiasedness}, and \eqref{e:asmp_as_bound}.
\end{defn}

We note that \eqref{e:asmp_unbiasedness} is standard in stochastic optimization literature ($cf.$ \cite{nemirovsky1983problem}, \cite{nemirovski1995information}, \cite{bubeck2015convex}, \cite{agarwal2012information}).
To prove convergence guarantees on first-order optimization in the
classic setup (without any information constraints on the oracle), it
is enough to assume $\displaystyle{\E{\norm{\hat{g}(x)}_q^2} \leq B^2}$. We make a
slightly stronger assumption in this case \new{since the more relaxed
  assumption leads to technical difficulties in finding unbiased
  quantizers for gradients.
}

Note that by \eqref{e:asmp_unbiasedness} and \eqref{e:asmp_as_bound}
for every $x\in \cX$
there exists a vector $g\in \partial f(x)$ such that $\norm{g}_q\leq B$.
Further, since $f$ is convex,
$f(x)-f(y)\leq g^T(x-y)$ for every $g\in \partial f(x)$, whereby
$|f(x)-f(y)|\leq B\norm{x-y}_p$.
Namely, $f$ is $B$-Lipschitz continuous in the $\ell_p$ norm.\footnote{The
  same could be said under the weaker assumption
  $\E{\norm{\hat{g}(x)}_q^2 }\leq B^2$.}  

Before proceeding, we recall the optimal convergence results under no information constraints.
No information constraints can be viewed as passing the subgradients estimates through the identity channel. 
\begin{defn}
We denote by $I:\R^d \to \R^d$ the identity channel, where the output always equals the input. Let $\mathcal{I}$ denote the singleton set consisting only of $I$.
\end{defn}

\begin{thm}\label{t:e_infty}
Let
$\mathbb{X}_p(D) \eqdef \{\X \subseteq \R^d: \max_{x,y\in\X}\norm{x-y}_p\leq
D\}.$
There exist absolute constants $c_0$ and $c_1$ where $c_1 \geq c_0 >0$ such that the following hold:
\begin{enumerate}
\item for\footnote{For certain range of $p$ closer to 2 the $\sqrt{\log d}$ factor can be removed; for simplicity, we state the slightly weaker result.} $2 >p \geq 1$,
\[\displaystyle{\frac{c_1  DB \sqrt{\log d}}{\sqrt{T}}  \geq
\sup_{\X\in \mathbb{X}_p(D) }\ep^*(\X , \oO_{{\tt c}, p}, T, \mathcal{I})
\geq   \frac{c_0DB }{\sqrt{T}}.}
\]

\item For $ p \geq 2$,
\[\displaystyle{
  \frac{c_1d^{1/2 -1/p} DB}{\sqrt{T}} \geq \sup_{\X\in \mathbb{X}_p(D)} \ep^*(\X , \oO_{{\tt c}, p}, T, \mathcal{I})
\geq  \frac{c_0 d^{1/2 -1/p} DB}{\sqrt{T}};}\]

\end{enumerate}
\end{thm}
\noindent The lower bounds and the upper bounds can be found, for instance, in \cite[Theorem 1]{agarwal2012information} and \cite[Appendix C]{agarwal2012information}.

\begin{rem}\label{r:Motivationfoalphapin12}
An optimal achievable scheme for $p \in [1, 2)$ is the stochastic
  mirror descent with the mirror maps 
  $\norm{x}_{p^\prime}^2/(p^\prime-1)$, where $p^\prime$ is
  chosen appropriately for a given $p$. When H\"{o}lder conjugate $q$ of $p$ is $o(\log d)$, we choose $p^{\prime}$ to be $p$. When $q$ is $\Omega(\log d)$, we choose $p^{\prime} = \frac{2 \log d}{2 \log d -1 }.$  Further, these algorithms require only
  that the expected squared $\ell_q$ norm of the gradient estimates are
  bounded.
\end{rem}
\begin{rem}\label{r:Motivationfoalphap}
An optimal achievable scheme for $p$ greater than $2$
  is simply projected subgradient descent(PSGD). To see this, note
  that PSGD gives a guarantee of ${D^{\prime}B^{\prime}}/{\sqrt{T}}$
  ($cf.$~\cite{nemirovski1995information}), where $D^{\prime}$ is the
  $\ell_2$ diameter and $B^{\prime}$ is the bound on
  $\E{\norm{\hat{g}}_2^2}$. Using the monotonicity of $\ell_q$ norms
  in $q$,  for $q \geq 2$ we have
  $\E{\norm{\hat{g}}_2^2} \leq \E{\norm{\hat{g}}_q^2} \leq B^2$. Also, the
  $\ell_2$ diameter of a unit $\ell_p$ ball is
  $d^{1/2-1/p}$. It follows that PSGD attains the upper bounds
in Theorem~\ref{t:e_infty}.
\end{rem}

\paragraph{Strongly convex and $\ell_2$ Lipschitz function family.}
We now consider a special subset of the  convex and $\ell_2$ Lipschitz
family described above, where the functions are strongly convex. Recall that for
$\gamma>0$, a function $f$ is \emph{$\gamma$-strongly convex on $\X$} if the following function $h$ is convex:
\begin{align}\label{e:strong_convexity}
h(x)=f(x) - \frac{\gamma}{2}\norm{x}_2^2, \quad \forall x \in \X.
\end{align}
\begin{defn}[Strongly convex and $\ell_2$ Lipschitz function family $\oO_{{\tt sc}}$.] \label{def:sc}
\newerest{We denote by  $\oO_{{\tt sc}}$ the set of all pairs of functions and oracles satisfying \eqref{e:convexity}, \eqref{e:asmp_unbiasedness}, \eqref{e:strong_convexity}, and \eqref{e:asmp_as_bound} for $q=2$.} 
\end{defn}

The strong convexity parameter $\gamma$ is related to the parameter
$B$, the upper bound on the
$\ell_2$ norm of the gradient estimate. We state a relation between
them when the domain $\X$ contains an $\ell_{\infty}$
ball of radius $D$ centered at the origin; this property will be used when we derive lower bounds.  
\begin{lem}
	\label{lemma:B_alpha}
For any $ \X \supseteq \{x: \norm{x}_{\infty} \leq D\}$, we have
$\frac{B}{\gamma}\geq \frac{D d^{1/2}}{4}.$
\end{lem}

\begin{thm}\label{t:sc_infty}
Let
$\mathbb{X}_2(D) \eqdef \{\X \subseteq \R^d: \max_{x,y\in\X}\norm{x-y}_2\leq
D\}.$
There exist absolute constants $c_0$ and $c_1$ where $c_1 \geq c_0 >0$ such that the following hold:

\[ \frac{c_1 B^2}{\gamma T}\geq  \sup_{\X \in \mathbb{X}_2(D)}\ep^*(\X , \oO_{\tt sc}, T, \mathcal{I})
\geq \frac{c_0 B^2}{\gamma T} \]

\end{thm}
\noindent The lower bounds and the upper bounds can be found, for instance, in \cite[Theorem 1]{agarwal2012information} and \cite{nemirovski1995information}.

\begin{rem}\label{r:sc} 
The optimal achievable scheme for strongly convex functions is the stochastic gradient descent algorithm.
\end{rem}  
\subsection{Information constraints}\label{ssec:info-constraints}
We describe  three specific constraints of interest to us: local
privacy, communication, and computation. The first two are
well-studied;
the third \newerest{is new} and arises in procedures such as random coordinate descent.
\paragraph{Local differential privacy.}

\new{To model local privacy,}
we define  the
$\priv$-locally differentially private (LDP) channel family $\W_{{\tt priv},
  \priv}$.
\begin{defn}\label{d:priv_constraints}
A channel $W\colon \R^d \to \R^d$
 is \emph{$\priv$-locally differentially private} ($\priv$-LDP) if  for all $x,x^{\prime}\in\R^d$,
\[	\frac{W (Y \in S \mid X =x)}{ W(Y \in S \mid X =x^{\prime})} \leq e^{\priv}\]
for all Borel measurable subsets $S$ of $\R^d$.
We denote by $\W_{{\tt priv}, \priv}$ the set of all $\priv$-LDP channels.
\end{defn}
\new{When operating under local privacy constraints,} 
 the oracle's subgradient estimates are passed through an $\priv$-LDP channel, and only the output is available to
 the optimization algorithm. \newer{Thus, the resulting process which handles the data of individual users, accessed in each
oracle query, is overall differentially private, a notion of privacy extensively studied and widely used in practice.}   


 \paragraph{Communication constraints.}
\new{To model communication constraints,}
 we define the $\W_{{\tt com}, r}$, the $r$-bit communication-constrained channel family, as follows.
\begin{defn}\label{d:comm_constraints}
A channel $W\colon\R^d \to \{0,1\}^r$ constitutes an \emph{$r$-bit communication-constrained} channel.
We denote by $\W_{{\tt com}, r}$ the set of all $r$-bit communication-constrained channels.
\end{defn}

\paragraph{Computational constraints.}
For high-dimensional optimization, altogether computing the
subgradient estimates \newest{can be computationally expensive}. Often in
such cases, one resorts to computing only a few coordinates of the
gradient estimates and using only them for optimization
(\cite{nesterov2013introductory,richtarik2012parallel}). This
motivates the oblivious sampling channel family $\W_{\tt obl}$, where
the optimization algorithm gets to see only one
\new{randomly chosen}
coordinate of the gradient estimate. 
\begin{defn} 
\label{def:oblv}
An \emph{oblivious sampling} channel $W$ is a channel
$W\colon\R^d\to\R^d$
specified by a probability vector $(p_i)_{i\in[d]}$, $i.e.$,
a vector $p$ 
such that
$p_i\geq 0$ for all $i$ and $\sum_{i \in [d]} p_i=1$. 
For an input $g \in \R^d$, the output distribution of $W$ is given by
$W( g(i) e_i \mid g) = p_i, \forall i \in [d]$.
We denote by  $\W_{\tt obl}$ the set of all oblivious sampling channels.
\end{defn}

Therefore, at most one coordinate of
the oracle's the gradient estimate can be used by the optimization
algorithm. Further, this coordinate is sampled obliviously to the
input gradient estimate itself.
\begin{rem}
\new{We note that the special case of
  $p_i=\frac{1}{d}$ $\forall\, i\in [d]$
  corresponds to sampling employed by standard \emph{Random
  Coordinate Descent} (RCD) ($cf$.~\cite[Section 6.4]{bubeck2015convex}), where at each time step only one
uniformly random coordinate of the gradient is used by the gradient descent algorithm.}
\end{rem}

\section{Main results: lower bounds for information-constrained optimization}
  \label{sec:results}
  For $p\in[1,\infty]$ and $D>0$, let
$\mathbb{X}_p(D) \eqdef \{\X \subseteq \R^d: \max_{x,y\in\X}\norm{x-y}_p\leq
D\}$ be the collection of subsets of $\R^d$ whose $\ell_p$ diameter is
at most $D$. In stating our results, we will fix throughout the
parameter $B>0$, the almost sure bound on the gradient magnitude
defined in~\eqref{e:asmp_as_bound}, as well as the strong convexity
parameter $\gamma>0$ defined in~\eqref{e:strong_convexity} (which,
implicitly, is required to satisfy
Lemma~\ref{lemma:B_alpha}). Throughout this section, our lower bounds
on minmax optimization error focus on tracking the convergence rate
for large $T$, \newerest{a standard regime of interest for the stochastic optimization
setting.}

\subsection{Lower bounds for locally private optimization under adaptive gradient processing}
  \label{sec:results:ldp}
 \newerest{Throughout, we consider $\priv\in[0,1]$, namely the high-privacy regime.}
\paragraph{Convex function family.} \newest{For the convex function family, we prove the following lower bounds.}

\begin{thm}\label{t:conpriv_P2}
Let \new{$p\in[1,2]$}, $\priv \in [0, 1]$, and $D>0$. 
There exist absolute constants \new{$c_0,c_1>0$ such that, for
$T \geq c_0  \frac{d}{\priv^2}$},
\[
 \sup_{\X \in \mathbb{X}_p(D)} \ep^*(\X , \oO_{{\tt c}, p}, T, \W_{{\tt priv}, \priv}) 
\geq 
\frac{c_1DB}{\sqrt{T}} \cdot \sqrt{\frac{d}{\priv^2}}. 
\]
\new{(Moreover, one can take $c_0 \eqdef \frac{1}{2e(e-1)^2}$ and $c_1\eqdef \frac{1}{36(e-1)\sqrt{2e}}$.)}
\end{thm}
\noindent See Section~\ref{s:Proof_conv_P2} for the proof.

\begin{thm}\label{t:conpriv_inf}
\newest{Let $p\in(2,\infty], \priv \in [0, 1]$, and $D>0$. 
There exist absolute constants \new{$c_0,c_1>0$ such that, for}
$T \geq c_0  \frac{d^2}{\priv^2}$},
 \[
 \sup_{\X \in \mathbb{X}_p(D)} \ep^*(\X , \oO_{{\tt c}, p}, T, \W_{{\tt priv}, \priv}) 
\geq  \frac{c_1 DBd^{1/2-1/p}}{\sqrt{T}} \cdot \sqrt{\frac{d}{ \priv^2}}. 
\] 
\new{(Moreover, one can take $c_0$ and $c_1$ as in Theorem~\ref{t:conpriv_P2}.)}
\end{thm}
\noindent See Section~\ref{ssec:con_p_inf} for the proof.

\newerest{
\begin{rem}[Tightness of bounds for convex functions and LDP constraints]
 \cite[Theorem 4 and 5]{duchi2014privacy} 
 provide nonadaptive LDP algorithms which show that Theorem~\ref{t:conpriv_P2} is tight up to logarithmic factors \new{for $p=1$} and Theorem~\ref{t:conpriv_inf} is tight up to constant factors~\new{for all $p\in(2,\infty]$} \new{(to the best of our knowledge, no non-trivial upper bound is known for $p\in(1,2)$.)}.  Therefore, adaptive processing of gradients under LDP cannot significantly improve the convergence rate for convex function families.

Interestingly, for $p=1,$ \cite{duchi2014privacy} also provide a slightly stronger lower bound of $\frac{c_0DB}{\sqrt{T}} \cdot \sqrt{\frac{d \log d}{\priv^2}}$ for nonadaptive protocols, which matches the performance of their nonadaptive protocols up to constant factors. This points to a minor gap in our understanding of adaptive protocols: Can we establish
a stronger lower bound for adaptive protocols to match the performance of the nonadaptive algorithm  of \cite{duchi2014privacy}, or does there exist a better adaptive protocol? 
\end{rem}}

From Theorem~\ref{t:e_infty}, the standard optimization error for $\ell_p$, $p \in [1, \infty]$, convex family blows up by a factor of $\sqrt{d/\priv^2}$ when the gradient estimates are passed through an $\priv$-LDP channel.

\paragraph{Strongly convex family.} \newest{We prove the following result for strongly convex functions.} 
\begin{thm}\label{t:sconpriv_P2} 
Let $\priv\in[0,1]$, and $D>0$. There exist absolute constants \new{$c_0,c_1>0$ such that, for $T \geq c_0\cdot \frac{B^2}{\gamma^2D^2}\cdot\frac{d}{\priv^2}$},
\[
 \sup_{\X \in \mathbb{X}_2(D)} \ep^*(\X , \oO_{{\tt sc}}, T, \W_{{\tt priv}, \priv }) 
\geq 
\frac{c_1 B^2}{\gamma T} \cdot \frac{d}{\priv^2}. 
\]
\end{thm}
\noindent See Section~\ref{ssec:scon} for the proof.

\begin{rem}[Tightness of bounds for strongly convex functions and LDP constraints]
One can use stochastic gradient descent with the nonadaptive protocol from~\cite[Appendix C.2]{duchi2014privacy} to obtain a nonadaptive protocol with
convergence rate
matching the lower bound in Theorem~\ref{t:sconpriv_P2} up to constant factors, establishing that adaptivity does not help for strongly convex functions.
\end{rem}

From Theorem~\ref{t:sc_infty}, the standard optimization error  for strongly convex functions
blows up by a factor of $\frac{d}{\priv^2}$ when the gradient estimates are passed through an $\priv$-LDP channel.

\subsection{Lower bounds on communication-constrained optimization}
  \label{sec:results:communication}
\noindent\textbf{Convex function family.} \newest{For convex functions, we prove the following lower bounds.}
\begin{thm}\label{t:concom_P2}
Let \new{$p\in[1,2]$, and $D>0$}. There exists an absolute constant $c_0>0$ such that, for $r  \in \N$, and $T \geq \frac{d}{6r},$ 
\[
 \sup_{\X \in \mathbb{X}_p(D)} \ep^*(\X , \oO_{{\tt c}, p}, T, \W_{{\tt com}, r  }) 
\geq 
\frac{c_0DB}{\sqrt{T}} \cdot \sqrt{\frac{d}{d \wedge r}} .
\]
\new{(Moreover, one can take $c_0\eqdef \frac{1}{12\sqrt{58}}$.)}
\end{thm}
\noindent See Section~\ref{ssec:scon} for the proof.

\begin{thm}\label{t:concom_inf} 
Let $p \in (2, \infty]$, \new{and $D>0$}. There exists an absolute constant $c_0>0$ such that, for $r \in \N$, and \new{$T \geq \frac{1}{\newer{4}}\cdot\frac{d^2}{2^r\land d}$}, we have
 \[
 \sup_{\X \in \mathbb{X}_p(D)} \ep^*(\X , \oO_{{\tt c}, p}, T, \W_{{\tt com}, r}) 
\geq  \left(\frac{c_0DBd^{1/2-1/p}}{\sqrt{T}} \cdot \sqrt{\frac{d}{ d \wedge 2^r}} \right) \vee \left(  \frac{c_0DB}{\sqrt{T}} \cdot \sqrt{\frac{d}{  d \wedge r}} \right)
\]
\new{(Moreover, one can take $c_0\eqdef \frac{1}{12\sqrt{58}}$.)}  
\end{thm}
\noindent See Section~\ref{ssec:con_p_inf} for the proof.

In Chapters \ref{Ch:RATQ} and \ref{Ch:Limits}, we will derive upper bounds for most regimes of $p$ and $r.$ Specifically, restricting $r \geq r^*(T, p)$ our bounds are tight for all regimes of $p$. Moreover, for $p=1$ and $[2, \infty]$, our bounds are nearly tight for all $r$.

\new{From  Theorem~\ref{t:e_infty}, the standard optimization errors for $\ell_1$ and $\ell_p$, $p \in (2, \infty]$, convex family blow up by a factor of $\sqrt{\frac{d}{d \wedge r}}$ and $\sqrt{\frac{d}{d \wedge 2^r}} \vee \sqrt{\frac{d^{2/p}}{d \wedge r}} $, respectively, when the gradient estimates are  compressed to $r$ bits.}

\paragraph{Strongly convex family.} \newest{We prove the following result for strongly convex functions.} 
\begin{thm}\label{t:sconcom_P2}
\newer{Let $D>0$.} There exist \new{absolute constants $c_0,c_1>0$} such that, for  $ r  \in \N$ and \new{$T \geq c_0 \cdot \frac{B^2}{\gamma^2D^2}\cdot\frac{d}{r}$},
\[
 \sup_{\X \in \mathbb{X}_2(D)} \ep^*(\X , \oO_{{\tt sc}}, T, \W_{{\tt com}, r }) 
\geq 
\frac{c_1 B^2}{\gamma T} \cdot \frac{d}{d \land r} \,.
\]
\end{thm}
\noindent See Section~\ref{ssec:scon} for the proof.


In Chapters \ref{Ch:RATQ}, we will derive upper bounds which are tight upto a factor of $\log \log^* d$ for all $r.$
\new{From Remark~\ref{r:sc}, the standard optimization error  for strongly convex functions
blows up by a factor of $\frac{d}{r}$ when the gradient estimates are compressed to $r$ bits.}

\subsection{Lower bounds on computationally-constrained optimization}
  \label{sec:results:rcd}

We restrict to the case of Euclidean geometry ($p=2$) for the oblivious sampling channel family  $\W_{\tt obl}$. Our motivation for introducing this class was to study the optimality of standard RCD, which is proposed to work in the Euclidean setting alone. Furthermore, if we consider a slightly larger family of channels where the sampling probabilities can depend on the input itself, the resulting family will be similar to the 1-bit communication family, which we have addressed in Section \ref{sec:results:communication}.
\paragraph{Convex family.} For convex functions, we establish the following lower bound, for $p=2$.
\begin{thm}\label{t:concompt_P2}
\newer{Let $D>0$.} There exists an absolute constant $c_0>0$ such that, for \new{$T \geq \frac{d}{4}$}, we have
\[
 \sup_{\X \in \mathbb{X}_2(D)} \ep^*(\X , \oO_{{\tt c}, 2}, T, \W_{{\tt obl} }) 
\geq 
\frac{c_0\sqrt{d}DB}{\sqrt{T}}\,.
\]
\new{(Moreover, one can take $c_0 \eqdef \frac{1}{72}$.)}
\end{thm}
\noindent See Section~\ref{s:Proof_conv_P2} for a proof. 

The standard Random Coordinate Descent (RCD) (see for instance~\cite[Theorem 6.6]{bubeck2015convex}), which employs uniform sampling, matches this lower bound up to constant factors. The optimality of standard RCD \newer{motivates further} the folklore approach of uniformly sampling coordinates for random coordinate descent unless there is an obvious structure to exploit (as
in \cite{nesterov2012efficiency}). This establishes that adaptive sampling
strategies do not improve over nonadaptive sampling strategies for the
family $\W_{{\tt obl}}$.
Also from Theorem~\ref{t:e_infty}, the standard optimization error for $\ell_2$ convex family blows up by a factor of $\sqrt{d}$ when the gradient coordinates are sampled obliviously.
%

\paragraph{Strongly convex family.} For strongly convex functions, we obtain the following lower bound, for $p=2$.
\begin{thm}\label{t:sconcompt_P2}
\newer{Let $D>0$.} \newest{ There exist \new{absolute constants $c_0,c_1>0$} such that, for \new{$T\geq c_0 \cdot d\frac{B^2}{\gamma^2D^2}$}, we have}
\[
 \sup_{\X \in \mathbb{X}_2(D)} \ep^*(\X , \oO_{{\tt sc}}, T, \W_{{\tt obl} }) 
\geq 
\frac{c_1 dB^2}{\gamma T}\,.
\]
\end{thm}
\noindent See Section~\ref{ssec:scon} for the proof.

\new{Once again, the standard RCD algorithm matches this lower bound, which shows that adaptive sampling strategies do not improve over nonadaptive sampling strategies for strongly convex optimization. Further, from Theorem~\ref{t:sc_infty}, the standard optimization error for strongly convex family blows up by a factor of $d$ when the gradient coordinates are sampled obliviously.
}

\begin{figure}[t]
\begin{center}
\renewcommand{\arraystretch}{1}
\begin{tabular}{c|   c| c| c}
\hline
& LDP  &{Communication }&{Computational}\\
&{ constraints }&{-constraints }&{-constraints}\\
\hline

Convex and $\ell_p$ & & & (Only for $p=2$)\\
Lipschitz function & $\displaystyle{\frac{c_1DB}{\sqrt{T}} \cdot \sqrt{\frac{d}{\priv^2}}}$ & $\displaystyle{\frac{c_1DB}{\sqrt{T}} \cdot \sqrt{\frac{d}{d \land r}}}$&  $\displaystyle{\frac{c_1DB}{\sqrt{T}} \cdot \sqrt{d}}$ \\
 family ($p \in [1, 2]$) & (Theorem \ref{t:conpriv_P2}) &  (Theorem \ref{t:concom_P2}) & (Theorem \ref{t:concompt_P2})\\
\hline
Convex and $\ell_p$& & $\left(\frac{c_0DBd^{1/2-1/p}}{\sqrt{T}} \cdot \sqrt{\frac{d}{ d \wedge 2^r}} \right)$   &
\\ Lipschitz function  &  $\displaystyle{ \frac{c_1 DBd^{1/2-1/p}}{\sqrt{T}} \cdot \sqrt{\frac{d}{ \priv^2}}}$ & $\vee   \left(\frac{c_0DB}{\sqrt{T}} \cdot \sqrt{\frac{d}{  d \wedge r}}\right) $& N.A. \\
family ($p \in (2, \infty]$) & (Theorem \ref{t:conpriv_inf}) & (Theorem \ref{t:concom_inf}) \\
\hline
Strongly convex  and $\ell_2$ & & \\ Lipschitz function  & $\displaystyle{\frac{c_1 B^2}{\gamma T} \cdot \frac{d}{\priv^2}}$ &$\displaystyle{\frac{c_1 B^2}{\gamma T} \cdot \frac{d}{r}}$& $\displaystyle{\frac{c_1 dB^2}{\gamma T}}$\\ 
family& (Theorem \ref{t:sconpriv_P2})& (Theorem \ref{t:sconcom_P2})& (Theorem \ref{t:sconcompt_P2})\\ 
\hline
\end{tabular}
\end{center}
\renewcommand{\figurename}{Table}
\caption{Summary of all our lower bounds on gap-to-optimality for information-constrained optimization.}
\label{table:all_results}
\end{figure}\label{t:lb_summary}

  A summary of all our lower bounds is provided in Table \ref{table:all_results}.
\section{Proofs of  lower bounds}\label{sec:proofs}
\subsection{Outline of the proof for our lower bounds}
The proofs of our lower bounds for adaptive protocols follow the same general template,
summarized below.

\medskip
{\bf Step 1. Relating optimality gap to average information:}
We consider a family of functions $\G=\{g_v : v\in\{-1,1\}^d\}$
satisfying suitable conditions and associate with it a ``discrepancy metric''
$\psi(\G)$
that allows us to relate the optimality gap of any algorithm to an average mutual
information quantity. Specifically, for $V$ distributed uniformly over $\{-1,1\}^d$,
we show that the output $\hat x$ of any optimization algorithm satisfies
\[
\E{g_{V}(\hat{x})-\min_{x\in \X}g_{V}(x)} \geq { \frac{d\psi(\G)}{6}}  \left[1 - \sqrt{
\frac{\newer{2}}{d}
    \sum_{i=1}^d  \mutualinfo{V(i)}{Y^T}}\right],
\]
\newer{where} $Y_t$ is the channel output for the gradient in the $t$th iteration and $Y^T:=(Y_1, \dots, Y_T)$.

Heuristically, we have related the gap to optimality to the difficulty of inferring $V$ by observing $Y^T$.
We note that the bound above is similar to that of~\cite{agarwal2012information}, but instead of mutual information $\mutualinfo{V}{Y^T}$ we get the average mutual information per coordinate. This latter quantity is amenable to analysis for adaptive protocols.

\smallskip
{\bf Step 2. Average information bounds:} To bound the average mutual information per coordinate,
$\frac{1}{d}\sum_{i=1}^d  \mutualinfo{V(i)}{Y^T}$, we take recourse to the recently proposed bounds  from~\cite{acharya2020general}. These bounds hold for $Y^T$ which is the output of adaptively selected
channels from a fixed channel family $\W$, with i.i.d.\ input $X^T=(X_1,\dots,X_T)$ generated from
a family of distributions $\{\p_v, v\in \{-1,1\}^d\}$. We view the output of oracle as inputs $X^T$
and derive the required bound.

While results in~\cite{acharya2020general} provided bounds for $\W_{{\tt priv}, \priv}$
and $\W_{{\tt comm}, r}$, we extend the approach to handle $\W_{{\tt obl}}$. Specifically, under a smoothness and symmetry condition on $\{\p_v, v\in \{-1,1\}^d\}$, which has a parameter
$\gamma$ associated with it, we show the following:

For $|\X|<\infty$  and $\X_i:=\{x(i):x\in \X\}$, $i\in [d]$,
we have
\[
\sum_{i=1}^{d}\mutualinfo{V(i)}{Y^T} \leq \frac{C}2\cdot T\gamma^2,
\]
where the constant $C$ depends only on $\{\p_v, v\in \{-1,+1\}^d\}$ and, denoting
by $v^{\oplus i}\in \{-1,1\}^d$ the vector with the sign of the $i$th coordinate
of $v$ flipped, is given by
\[
C=(\max_{i\in[d]}|\X_i|-1)\cdot
\max_{x\in \X}\max_{v\in\{-1,+1\}^d} \max_{i\in[d]}\frac{\p_{v^{\oplus i}}(X(i)=x(i))}
         {\p_v(X(i)=x(i))}.
\]

\smallskip
{\bf Step 3. Use appropriate difficult instances} 
\newer{On the one hand,} to prove lower bounds for the convex family we will use the class of functions $\Gconvex=\{g_v(x) \colon v \in \{-1, 1\}^d  \}$ defined on the domain $\X=\{x \in\R^d: \norm{x}_{\infty} \leq b\}$ comprising functions $g_v$ given below:
 \begin{align*}
g_{v}(x) = a \cdot \sum_{i=1}^{d}|x(i)-v(i)\cdot b|, \quad \forall x \in \X, v \in \{ -1, 1\}^d.
 \end{align*}
 On the other hand, to prove lower bounds for the strongly convex family, we will use the class of functions $\Gstrongconvex=\{g_v(x) \colon v \in \{-1, 1\}^d  \}$
 on $\X=\{x \in\R^d: \norm{x}_{\infty} \leq b\}$ given by
 \begin{align*}
g_{v}(x)= a   \sum_{i=1}^{d} \mleft( \frac{1+2\delta v(i)}{2}  f^+_i(x) + \frac{1-2\delta v(i)}{2} f^-_i(x) \mright), \quad \forall x \in \X, v \in \{ -1, 1\}^d, 
 \end{align*}
 where $f^+_i$ and $f^-_i$, for $i\in[d]$, are given by
\begin{align*}
  f^+_i(x) = \theta b |x(i)+b| + \frac{1-\theta}{4}  (x(i)+b)^2, \qquad
  f^-_i(x) &= \theta b |x(i)-b| + \frac{1-\theta}{4}  (x(i)-b)^2.
\end{align*}

\smallskip
{\bf Step 4. Carefully combine everything:} We obtain our
desired bounds by applying Steps 1 and 2 to difficult instances from Step 3. 
Since the difficult instance for convex family consists of linear functions, the gradient
does not depend on $x$. Thus, we can design oracles which give i.i.d.\ output with distribution independent of the query point $x_t$, whereby the bound in Step 2 can be applied. 
Interestingly, we construct different oracles for $p<2$ and $p\geq 2$.

However, the situation is different for the strongly convex family. The gradients now depend on the query point $x_t$, whereby it is unclear if we can comply with the requirements in Step 2. Interestingly, for communication and local privacy constraints, we construct oracles that allow us to view messages $Y^T$ as \newerest{the} output of adaptively selected channels applied to independent samples from a common distribution $\p_v$. While it is unclear if the same can be done for computational constraints as well, we use an alternative approach and exhibit an oracle for which we can find an intermediate message vector $Z_1, \dots, Z_T$ such that \newer{(i)} $V$ and $Y^T$ are conditionally independent given
$Z^T$ and \newer{(ii)} the message $Z^T$ satisfies the requirements of Step 2. 

\subsection{Relating optimality gap to average information}\label{ssec:gap-information}
\new{In this section, we prove a general lower bound for the expected gap
  to optimality by considering a parameterized family of functions and oracles
which is contained in our oracle family of interest.
We present a bound that relates the expected gap to optimality
to the average mutual information between the channel output and
different coordinates of the unknown parameter. 
This step is \newerest{the} key difference between our approach and that of~\cite{agarwal2012information}, which used Fano's method instead
of our bound below. We remark that the bounds resulting from
Fano's method are typically not amenable to analysis for adaptive protocols.
}

In more
detail, our result can be used to prove bounds for the average
optimization error over any class of functions which satisfies the two
conditions below.  
\begin{assum}\label{e:cond_assouad}
Let $\X\subseteq\R^d$ and $\V=\{-1,1\}^d$. Let $\G=\{g_{v}: v \in \V
\}$ where $g_v:\X\to \R$ are real-valued functions from $\X$ such that
\begin{enumerate}
\item the $g_v$s are \emph{coordinate-wise decomposable}, $i.e.$, there exist
  functions $g_{i,b}\colon \R \to \R$, $i\in[d]$, $b\in\{-1,1\}$,
  such that
		\[
			g_{v}(x)=\sum_{i=1}^{d} g_{i,v(i)}(x(i)).
		\]
\item the minimum of $g_v$ is also a \emph{coordinate-wise minimum}, $i.e.$, if we denote by $x^\ast_{v}$ the minimum of $g_{v}$ over $\X$, then, for all $i\in [d]$, we have
		\[
			x^\ast_{v}(i)=\arg\!\min_{y \in \X_i}g_{i,v(i)}(y),
		\]
		where $\X_i =\{x(i) : x \in \X\}$. \label{listref_2}
\end{enumerate} 
\end{assum}
For $\G$ satisfying Assumptions~\ref{e:cond_assouad} and for $i\in[d]$, we now define the following discrepancy metric: 
\begin{align}
  \label{def:psi}
	\psi_{i}(\G) 
		&\eqdef  \min_{y \in \X_i} \mleft( g_{i, 1}(y) +  g_{i, -1}(y) - \mleft(\min_{y'\in\X_i} g_{i, 1}(y') + \min_{y'\in\X_i}g_{i, -1}(y') \mright) \mright)\\
 	\psi(\G) 
		&\eqdef \min_{i \in [d]} \psi _{i}(\G).
\end{align}
This is a ``coordinate-wise counterpart'' of the metric used in~\cite{agarwal2012information}.
The next lemma follows readily from this definition.
\begin{lem}
	\label{lem:useful:consequence:discrepancy}
Fix  $i\in[d]$. For every $y\in \X_i$, there can be at most one $b\in \{-1, 1\}$ such that 
\[
	g_{i, b}(y) - \min_{y'\in\X_i} g_{i, b}(y') \le \frac{\psi_{i}(\G)}{3}.
\]
\end{lem}
\begin{proof} 
Let $b\in\{-1,1\}$. By definition of $\psi_i(\G)$, for all $y \in \X_i$ we have
 \[
 	\mleft( g_{i, b}(y) - \min_{y'\in\X_i} g_{i, b}(y') \mright) + \mleft( g_{i, -b}(y) - \min_{y'\in\X_i} g_{i, -b}(y') \mright) \geq \psi_{i}(\G).
 \]
For $y$ such that $g_{i, b}(y)- \min_{y'\in\X_i} g_{i, b}(y') \leq \frac{\psi_{i}(\G)}{3}$, we  now must have that 
 \[
 	g_{i, -b}(y)- \min_{y'\in\X_i} g_{i, -b}(y') \geq \frac{2\psi_{i}(\G)}{3}.\qedhere
 \]
\end{proof}
We will use this observation to bound the \new{expected gap to optimality} for any algorithm $\pi$ optimizing an unknown function in $\G$ that has access to only the corresponding first-order oracle.

\begin{lem}\label{l:ImpL}
  Suppose $\G=\{g_v : v\in\{-1,1\}^d\}$ satisfies Assumption~\ref{e:cond_assouad}. Let $\pi$ be any optimization algorithm that adaptively selects the channels $\{W_j\}_{j \in [T]}$.
  For a random variable $V$ distributed uniformly over $\{-1,1\}^d$, the output $\hat{x}$ of $\pi$ when it is applied to a function from $\G$ and any associated (stochastic subgradient) oracle
   satisfies
\[
\E{g_{V}(\hat{x})-g_{V}(x_V^\ast)} \geq { \frac{d\psi(\G)}{6}}  \left[1 - \sqrt{
\frac{1}{d}
    \sum_{i=1}^d  \newer{2}\mutualinfo{V(i)}{Y^T}}\right],
\]
where $\psi(\G)=\min_{j \in [d]}\psi_j(\G)$, $Y_t$ is the channel output for the gradient at time step $t$
and $Y^T:=(Y_1, \dots, Y_T)$.
\end{lem} 
\begin{proof}
  \new{Our proof is based on relating the gap to optimality to the error in estimation of $V$
upon observing $Y^T$.
  }
    Suppose the algorithm $\pi$ along with channels $\{W_j\}_{j \in [T]} $
outputs the point $\hat{x}$ after $T$ iterations. By \newer{linearity of
expectation}, the decomposability of $g_v$, and Markov's inequality,
we have
\begin{align}
  \E{g_{V}(\hat{x}) -g_{V}(x_V^\ast) }
  &= \sum_{i=1}^{d} \E{g_{i,
      V(i)}(\hat{x}(i)) -g_{i,V(i)}(x_V^\ast(i))} \notag\\
  &\geq
\sum_{i=1}^{d} \frac{\psi_i(\G)}{3}\bPr{ g_{i, V(i)}(\hat{x}(i))
  -g_{i, V(i)}(x_V^\ast(i)) \geq \frac{\psi_i(\G)}{3} }
\notag
\\
&\geq \frac{\psi(\G)}{3}
\sum_{i=1}^{d} \bPr{ g_{i, V(i)}(\hat{x}(i))
  -g_{i, V(i)}(x_V^\ast(i)) \geq \frac{\psi_i(\G)}{3} }
\label{l:Impl:postMarkov}.
\end{align}
We proceed to bound each summand separately.

Fix any $i\in[d]$ and
consider the following estimate for $V(i)$: Given $\hat{x}$, we
output a $\hat{V}(i) \in \{-1,1\}$ 
satisfying 
 \[
 g_{i,\hat{V}(i)}(\hat{x}(i)) -
\new{\min_{y^\prime\in \X_i}g_{i,\hat{V}(i)}(y^\prime)} < \frac{\psi_i(G)}{3};
 \]
 if no such $\hat{V}(i)$ exists, we generate $\hat{V}(i)$
 uniformly from $\{-1,1\}$.
 Then, as a consequence of Lemma~\ref{lem:useful:consequence:discrepancy}, we get
 \begin{equation}
  \label{eq:bound:proba:error:test:1}
    \bPr{\hat{V}(i)\neq v(i) } \leq \bPr{ g_{i, v(i)}(\hat{x}(i)) -g_{i, v(i)}(x_v^\ast(i)) \geq \frac{\psi_i(\G)}{3} }.
 \end{equation}

\new{Next, denote by $\p^{Y^T}$ the distribution of $Y^T$ and by
  $\p_{+i}^{Y^T}$  and $\p_{-i}^{Y^T}$, respectively, the distributions of $Y^T$
  given $V(i)=+1$ and $V(i)=-1$. It is easy to verify that
  \[
\p^{Y^T} = \frac 12(\p_{+i}^{Y^T}+ \p_{-i}^{Y^T}), \quad \forall\,i\in[d].
  \]} 
 Noting that $V(i)$ is uniform and the estimate $\hat{V}(i)$ is formed as a function of $Y^T$,
 we get
 \begin{equation}
  \label{eq:bound:proba:error:test:2}
  \bPr{\hat{V}(i)\neq v(i) } \geq \frac{1}{2} - \frac{1}{2}
\totalvardist{\p_{+i}^{Y^T}}{\p_{-i}^{Y^T}}.
 \end{equation}
From this, combining~\eqref{eq:bound:proba:error:test:1} and~\eqref{eq:bound:proba:error:test:2} and plugging the result into~\eqref{l:Impl:postMarkov}, we have 
\begin{align*}
\E{g_{v}(\hat{x}) -g_{v}(x_v^\ast) }  
&\geq \frac{\psi(\G)}{6}\sum_{i=1}^{d}
\left[ 1 - \totalvardist{\p_{+i}^{Y^T}}{\p_{-i}^{Y^T}}\right]\\
&\geq \frac{\psi(\G)}{6}\sum_{i=1}^{d}
\left[ 1 - \totalvardist{\p_{+i}^{Y^T}}{\p^{Y^T}}
- \totalvardist{\p_{-i}^{Y^T}}{\p^{Y^T}}
\right]\\
&\geq \frac{\psi(\G)}{6}\sum_{i=1}^{d}
\left[ 1 - \sqrt{\frac {1}2\kldiv{\p_{+i}^{Y^T}}{\p^{Y^T}}}
- \sqrt{\frac {1}2\kldiv{\p_{-i}^{Y^T}}{\p^{Y^T}}|}
\right]
\\
&\geq \frac{d\psi(\G)}{6}
\left[ 1 - \sqrt{\frac {1}{d}\sum_{i=1}^d
  \kldiv{\p_{+i}^{Y^T}}{\p^{Y^T}}
  +\kldiv{\p_{-i}^{Y^T}}{\p^{Y^T}}}
\right]
\\
&= \frac{d\psi(\G)}{6}
\left[ 1 - \sqrt{\frac {\newer{2}}{d}\sum_{i=1}^d
\mutualinfo{V(i)}{Y^T}}
\right],
\end{align*}
where the second inequality follows from the triangle inequality, the third is Pinsker's
inequality, and the fourth is Jensen's inequality. 
\end{proof}

\subsection{Average information bounds}\label{ssec:avg_info}
The next step in our proof is to bound the average mutual information
that emerged in Section~\ref{ssec:gap-information}.
A general recipe for bounding this average mutual information
has been given recently in~\cite{acharya2020general},
which we recall below. 

Let $\{\p_v, v\in \{-1,1\}^d\}$ be a family of distributions over some domain $\X$
and $\W$ be a fixed channel family.
For $v\in\{-1,1\}^d$ and $i\in[d]$, denote by $v^{\oplus i}$ the element of $\{-1,1\}^d$ obtained by flipping the $i$th coordinate of $v$. For a fixed $v$, we obtain $T$ independent samples $X_1, \dots, X_T$ from $\p_v$.
Let $Y_1, \dots, Y_T$ be the output of channels selected from the channel family $\W$ by
an adaptive channel selection strategy (see Section~\ref{ssec:set-up}) when input to the channel at
time $t$ is $X_t$, $1\leq t\leq T$.\footnote{The bound in~\cite{acharya2020general} allows
  even shared randomness $U$ in its definition of interactive protocols. We have omitted
  $U$ in the description for simplicity.}

For $V$ distributed uniformly on $\{-1,1\}^d$, we are interested in bounding $(1/d)\sum_{i=1}^d\mutualinfo{V(i)}{Y^{T}}$.
In~\cite{acharya2020general}, different bounds were given for this quantity under different assumptions. We state these assumptions below.
\begin{assum}
  \label{assn:decomposition-by-coordinates}
For every $v\in\{-1,1\}^d$ and $i\in[d]$,
there exists $\phi_{v,i}\colon\X\to\R$
such that $\EE{\p_v}{\phi_{v,i}^2}=1$,
$\EE{\p_{v}}{\phi_{v,i}\phi_{v,j}}=\indic{\{i=j\}}$ holds
for all $i,j\in[d]$,
and
\[
 \frac{d\p_{v^{\oplus i}}}{d\p_v}=1+\gamma\phi_{v,i}
,
 \]
 where $\gamma\in \R$ is a fixed constant independent of $v,i$.
\end{assum}
\begin{assum}
  \label{assn:ratio:bounded}
  There exists some $\kappa_{\W} \geq 1$ such that
\[
\max_{v\in \{-1,1\}^d} \max_{y\in \cY} \sup_{W\in \W} \frac{\EE{\p_{v^{\oplus i}}}{W(y\mid X)}}{\EE{\p_{v}}{W(y\mid X)}}
\leq \kappa_{\W}.
\]
\end{assum}
\begin{assum}\label{assn:subgaussianity}
There exists some $\sigma\geq 0$ such
that, for all $v\in\{-1,1\}^d$, the vector $\phi_v(X)\eqdef
(\phi_{v,i}(X))_{i\in[d]}\in\R^d$ is $\sigma^2$-subgaussian
for $X\sim\p_v$.\footnote{Recall that a random variable $Y$ is
$\sigma^2$-subgaussian if $\E{Y}=0$ and $\E{e^{\lambda Y}}\leq
e^{\sigma^2\lambda^2/2}$ for all $\lambda\in\R$; and that a
vector-valued random variable $Y$ is $\sigma^2$-subgaussian if its
projection $\langle{Y,u}\rangle$ is $\sigma^2$-subgaussian for every unit
vector $u$.} Further, for any fixed $z$, the random variables
$\phi_{v,i}(X)$ are independent across $i\in[d]$.
\end{assum}
We then have the following bound local privacy constraints.
\begin{thm}[{\cite[Corollary 6]{acharya2020general}}]
\label{cor:ldp}
Consider $\{\p_v, v\in \{-1,1\}^d\}$ satisfying Assumption~\ref{assn:decomposition-by-coordinates}
and the channel family $\W=\W_{{\tt priv}, \priv}$. Let $V$
be distributed uniformly over $\{-1,1\}^d$
and $Y^T$ be the output of channels selected by
the optimization algorithm
as above.
Then, we have
\[
\sum_{i=1}^{d}\mutualinfo{V(i)}{Y^T} \leq T \cdot \frac{\gamma^2}{2} \cdot e^\priv(e^\priv-1)^2.
\]
\end{thm}
\noindent For the case of communication constraints, we have the analogous statement below:
\begin{thm}[{\cite[Corollary 6]{acharya2020general}}]
  \label{cor:simple-numbits}
  Consider $\{\p_v, v\in \{-1,1\}^d\}$ satisfying Assumptions~\ref{assn:decomposition-by-coordinates}
  and~\ref{assn:ratio:bounded}
  and the channel family $\W=\W_{{\tt com}, r}$.
Let $V$
be distributed uniformly over $\{-1,1\}^d$
and $Y^T$ be the output of channels selected by
the optimization algorithm
as above.
Then, we have
\[
\sum_{i=1}^{d}\mutualinfo{V(i)}{Y^T}  \leq \frac{1}{2}\kappa_{\W_{{\tt com}, r}} \cdot T\gamma^2 (2^r\land d).
\]
Moreover, if Assumption~\ref{assn:subgaussianity} holds as well, we have
\[
\sum_{i=1}^{d}\mutualinfo{V(i)}{Y^T} \leq (\ln 2)\kappa_{\W_{{\tt com}, r}}\,\sigma^2 \cdot T\gamma^2 r.
\]
\end{thm}
Finally, we derive a bound for the oblivious sampling channel family.
\begin{thm}
\label{cor:obl}
Consider $\{\p_v, v\in \{-1,1\}^d\}$ satisfying Assumption~\ref{assn:decomposition-by-coordinates}
and the channel family $\W=\W_{{\tt obl}}$.
Let $V$
be distributed uniformly over $\{-1,1\}^d$
and $Y^T$ be the output of channels selected by
the optimization algorithm
as above.
Further, assume that $|\X|<\infty$. 
Then, we have
\[
\sum_{i=1}^{d}\mutualinfo{V(i)}{Y^T} \leq \frac{C}2\cdot T\gamma^2,
\]
where the constant $C$ depends only on $\{\p_v, v\in \{-1,+1\}^d\}$ and, denoting
$\X_i:=\{x(i):x\in \X\}$, is given by
\[
C=(\max_{i\in[d]}|\X_i|-1)\cdot
\max_{x\in \X}\max_{v\in\{-1,+1\}^d} \max_{i\in[d]}\frac{\p_{v^{\oplus i}}(X(i)=x(i))}
         {\p_v(X(i)=x(i))}.
\]
\end{thm}
\begin{proof}
We recall another result from~\cite[Theorem 5]{acharya2020general}:
Under Assumptions~\ref{assn:decomposition-by-coordinates}
  and~\ref{assn:ratio:bounded}, we have\footnote{This is the general bound underlying Theorem~\ref{cor:ldp}.}
\[
\sum_{i=1}^{d} \mutualinfo{V(i)}{Y^T} \leq \frac{1}{2}\kappa_{\W_{\tt obl}}  \cdot T\gamma^2 \max_{v 
\in \{-1, 1\}^d}\max_{W\in\W_{\tt obl}} \sum_{y\in\Y} \frac{\operatorname{Var}_{\p_{v}}[W(y\mid X)]}{\EE{\p_{v}}{W(y\mid X)}}.
\]
We now evaluate various parameters involved in this bound.
  Let $W$ be a oblivious sampling channel specified by the probability vector $(p_i)_{i \in [d]}.$ 
Note that a channel $W\in \W_{{\tt obl}}$ can be equivalently viewed as having output alphabet
$\Y=\{(i, z)\colon z\in \X_i, i\in [d] \}$.
Recall that for an input $x$, the channel output is $x(i)$ with probability $p_i$, $i\in[d]$,
$i.e.$, for $y=(i,z)$, $W(y\mid x) =p_i\indic{\{x(i)=z\}}$.
Thus, we have
\begin{align*}
  \sum_{y\in\Y} \frac{\operatorname{Var}_{\p_{v}}[W(y\mid X)]}{\EE{\p_{v}}{W(y\mid X)}}
  &= \sum_{i=1}^d\sum_{z\in\X_i}\frac{p_i^2\bPr{X(i)=z}-p_i^2\bPr{X(i)=z}^2}{p_i\bPr{X(i)=z}}
  \\
  &= \sum_{i=1}^dp_i(|\X_i|-1)
  \\
  &\leq \max_{i\in[d]}|\X_i|-1.
\end{align*}
Furthermore, proceeding similarly, we get that Assumption~\ref{assn:ratio:bounded} holds as well
with
\[
\kappa_{\W_{\tt obl}}=\max_{x\in \X}\max_{v\in\{-1,+1\}^d} \max_{i\in[d]}\frac{\p_{v^{\oplus i}}(X(i)=x(i))}
         {\p_v(X(i)=x(i))}.
\]
The proof is completed by combining the bounds above.
\end{proof}

\subsection{The difficult instances for our lower bounds}
With our general tools ready, we now describe the precise constructions of function families
we use to get our lower bounds. 
We first provide the details of a family $\Gconvex(a,b)$ of convex functions, before turning to $\Gstrongconvex(a,b,\delta,\theta)$, our family of hard instances for the strongly convex setting. In both cases, our families of hard instances are parameterized (by $a,b$ and $a,b,\delta,\theta$, respectively), and setting those parameters carefully will enable us to prove our various results.

\paragraph{Difficult functions for the convex family.}
To prove lower bounds for the convex family, we will use the class of functions $\Gconvex(a,b)$ below, parameterized by $a,b>0$ and defined on the domain $\X$ as follows:
 \begin{align}\label{e:conv_bottl}
 \nonumber & \X=\{x \in\R^d: \norm{x}_{\infty} \leq b\},\\
\nonumber &g_{v}(x) = a \cdot \sum_{i=1}^{d}|x(i)-v(i)\cdot b|, \quad \forall x \in \X, v \in \{ -1, 1\}^d, \text{ and} \\
&\Gconvex=\{g_v(x) \colon v \in \{-1, 1\}^d  \}.
 \end{align}
 
 Observe that the class $\Gconvex$ satisfies the conditions in Assumption~\ref{e:cond_assouad} with $g_{i,1}(x) = a|x(i)- b|$ and $g_{i,-1}(x)=a|x(i)+b|$ and $\X_i = [-b, b]$ for all $i \in [d].$
Further, we can bound the discreprency metric for this class as follows.
\begin{lem}\label{l:psiGc}
For the class of functions $\Gconvex$ defined in \eqref{e:conv_bottl}, we have
$\psi(\Gconvex) \geq 2 a b$.
\end{lem}
\begin{proof}
  Note that $\min_{x \in [-b, b]}g_{i, 1}(x)=\min_{x \in [-b, b]}g_{i, -1}(x)=0$. Therefore, for all $i\in[d]$, 
\[
  \psi_i(\Gconvex) = \min_{x \in [-b, b]}  \left( a |x(i)- b| +a|x(i)+ b| \right) \geq 2 a b,
\]
where the inequality follows from the triangle inequality.
\end{proof}

\paragraph{Difficult functions for the strongly convex family.}
To prove lower bounds for the strongly convex family, we will use the class of functions $\Gstrongconvex(a,b,\delta,\theta)$, parameterized by $a,b>0$, $\delta>0$, and $\theta\in[0,1]$, and defined on the domain $\X$ as follows:
 \begin{align}\label{e:sconv_bottl}
 \nonumber & \X=\{x \in\R^d: \norm{x}_{\infty} \leq b\},\\
\nonumber &g_{v}(x)= a   \sum_{i=1}^{d} \mleft( \frac{1+2\delta v(i)}{2}  f^+_i(x) + \frac{1-2\delta v(i)}{2} f^-_i(x) \mright), \quad \forall x \in \X, v \in \{ -1, 1\}^d, \text{ and} \\
&\Gstrongconvex=\{g_v(x) \colon v \in \{-1, 1\}^d  \},
 \end{align}
 where $f^+_i$ and $f^-_i$, for $i\in[d]$, are given by
\begin{align}\label{e:f+f_}
  f^+_i(x) &= \theta b |x(i)+b| + \frac{1-\theta}{4}  (x(i)+b)^2,  \\
  f^-_i(x) &= \theta b |x(i)-b| + \frac{1-\theta}{4}  (x(i)-b)^2,
\end{align}
for all $x\in\X$. We can check that, for every $v\in\{-1,1\}^d$, the function $g_v$ is then $\gamma$-strongly convex for $\gamma \eqdef a\cdot \frac{1-\theta}{4}$. Moreover, we have the following bound for the discrepancy metric.
\begin{lem}\label{l:psiGsc}
For the class of functions $\Gstrongconvex$ defined in \eqref{e:sconv_bottl}, if $\frac{1-\theta}{1+\theta}\geq 2 \delta$ then
$
\psi(\Gstrongconvex) \geq \frac{2 a b^2 \delta^2  }{1-\theta}\,.
$
\end{lem}
\begin{proof}
This follows from similar calculations as in \cite[Appendix A]{agarwal2012information}; \newer{we provide the proof here for completeness. Fixing any $v\in\{-1,1\}^d$, we first note that by definition of $\Gstrongconvex$, the function $g_v$ can be indeed be decomposed as
$
  g_v(x) = \sum_{i=1}^d g_{i,v(i)}(x_i)
$ 
for $x\in\cX$ (i.e., $\norm{x}_\infty\leq b$), where, for $i\in[d]$, $\nu\in\{-1,1\}$ and $y \in \cX_i \eqdef [-b,b]$,
\begin{align*}
    g_{i,\nu}(y) 
    &= a\mleft(  \frac{1+2\delta \nu}{2} \mleft( \theta b |y+b| + \frac{1-\theta}{4}(y+b)^2 \mright) + \frac{1-2\delta \nu}{2} \mleft( \theta b |y-b| + \frac{1-\theta}{4}(y-b)^2 \mright) \mright) \\
    &= a\mleft(  \frac{1-\theta}{4}y^2 + \frac{1+3\theta}{4}b^2 + \delta\nu (1+\theta)by \mright)
\end{align*}
where the second line relies on the fact that $|y+b|=y+b$ and $|y-b|=b-y$ for $|y|\leq b$. One can easily see, e.g., by differentiation, that $g_{i,\nu}$ is minimized at $y^\ast \eqdef -2\delta\nu\frac{1+\theta}{1-\theta}b$ which does satisfy $|y^\ast|\leq b$ given our assumption $\frac{1-\theta}{1+\theta}\geq 2 \delta$. It follows that $\min_{y\in\cX_i} g_{i,1}(y) = \min_{y\in\cX_i} g_{i,-1}(y) = ab^2 \mleft( \frac{1+3\theta}{4} - \delta^2\frac{(1+\theta)^2}{1-\theta} \mright)$.
Similarly, we have, for $y\in\cX_i$,
\begin{align*}
    g_{i,1}(y) + g_{i,-1}(y) 
    &= a\mleft(  \frac{1-\theta}{2}y^2 + \frac{1+3\theta}{2}b^2 \mright)
\end{align*}
which is minimized at $y^\ast=0$, where it takes value $ab^2\frac{1+3\theta}{2}$. Putting it together,
\[
  \psi_i(\Gstrongconvex) = \min_{y\in\cX_i} ( g_{i,1}(y) + g_{i,-1}(y) ) - (\min_{y\in\cX_i} g_{i,1}(y) + \min_{y\in\cX_i} g_{i,-1}(y) ) = 2ab^2\delta^2\frac{(1+\theta)^2}{1-\theta}\,.
\]
Finally, $\psi(\Gstrongconvex) = \min_{i\in[d]} \psi_i(\Gstrongconvex) = 2ab^2\delta^2\frac{(1+\theta)^2}{1-\theta} \geq \frac{2ab^2\delta^2}{1-\theta}$, as claimed.
}
\end{proof}

\subsection{Convex Lipschitz functions for $p\in[1,2]$: Proof of Theorems \ref{t:conpriv_P2},   \ref{t:concom_P2}, and \ref{t:concompt_P2}}\label{s:Proof_conv_P2}
We first prove Theorems \ref{t:conpriv_P2} and \ref{t:concom_P2}, our lower bounds on optimization of convex functions for $p\in[1,2]$ under privacy and communication constraints, respectively. We consider the class of functions $\G_{\tt c}$ defined in \eqref{e:conv_bottl} with parameters $a\eqdef 2B \delta/d^{1/q}$ and $b \eqdef  D/(2d^{1/p})$. That is, $\X=\{x \in \R^d:\norm{x}_\infty\leq D/(2d^{1/p}) \}$ and 
\begin{equation}
  \label{eq:def:gv:convex:p2}
g_{v}(x) \eqdef \frac {2B\delta}{d^{1/q}} \sum_{i=1}^{d} \mleft|x(i) - \frac{v(i) D}{2 d^{1/p}}\mright|
\qquad x \in \X, v \in \{-1, 1 \}^d.
\end{equation}
Note that the gradient of $g_v$ is equal to $-2B\delta v /d^{1/q}$ at every $x\in \X$.

For each $g_v$, consider the corresponding gradient oracle $O_v$ which outputs 
independent values for each coordinate,
with the $i$th coordinate taking values $-B/d^{1/q}$ and $B/d^{1/q}$ with probabilities $(1+2\delta v(i))/2$ and $(1-2\delta v(i))/2$, respectively, for some parameter $\delta>0$ to be suitably chosen later.

Clearly, $\X \in \mathbb{X}_p(D)$ and all  the functions $g_v$ and the corresponding  oracles $O_v$ belong to the convex function family $\oO_{{\tt c}, p}$.
We begin by noting that for $V$ distributed uniformly over $\{-1,1\}^d$, we have
\[ \sup_{\X \in \mathbb{X}_p(D)} \ep^*(\X , \oO_{{\tt c}, p}, T, \W_{{\tt priv}, \priv}) 
\geq \E{g_{V}(x_T)-g_{V}(x_V^\ast)},  \]
where the expectation is over $v$ as well as the randomness in $x_T$.

From Lemma~\ref{l:ImpL} and \ref{l:psiGc}, we have
\begin{equation}\label{e:C_P2}
  \E{g_{V}(x_T)-g_{V}(x_V^\ast)} \geq {\frac{d\cdot a b}{3} \cdot}  \left[1 -  \sqrt{
\frac{\newer{2}}{d}
      \sum_{i=1}^d \mutualinfo{V(i)}{Y^T}}\right],
\end{equation}
where $Y^T=(Y_1, ..., Y_T) $ are the channel outputs for the gradient estimates supplied by the oracle for the $T$ queries. 

Next, we apply the average information bound from Section~\ref{ssec:avg_info}.
To do so, observe that by the definition of our oracle,
the oracle output at each time step is an independent draw from the 
product distribution $\p_v$ on $\Omega \eqdef \mleft\{-\frac{B}{d^{1/q}},\frac{B}{d^{1/q}}\mright\}^d$ (in particular, $\p_v$ is the same at each time step, as it does not depend on the query $x_t$ at time step $t$ to the oracle). We treat the output of the independent outputs of the oracle as i.i.d. samples $X_1, ...,X_T$
in Section~\ref{ssec:avg_info} and the corresponding channel outputs as $Y^T$.
We can check that, for every $i\in[d]$, we have 
\begin{equation}
\label{eq:convex:p12:ratio}
    \frac{\p_{v^{\oplus i}}(x)}{\p_v(x)} = \frac{1+2\delta v(i) \sign(x(i))}{1-2\delta v(i) \sign(x(i))}
\end{equation}
for all $x\in\Omega$, and that Assumption~\ref{assn:decomposition-by-coordinates} is satisfied with
\begin{equation}
\label{eq:convex:p12:gamma:phi}
  \gamma \eqdef \frac{4\delta}{\sqrt{1-4\delta^2}}, \qquad \phi_{i,v}(x) \eqdef \frac{v(i) \sign(x(i))+2\delta}{\sqrt{1-4\delta^2}}\,.
\end{equation}
Furthermore, noting that Assumption~\ref{assn:ratio:bounded} always holds with
\[
\kappa_\W= \max_{v\in \{-1,1\}^d}\max_{x\in \Omega}\max_{i\in [d]}
    \frac{\p_{v^{\oplus i}}(x)}{\p_v(x)},
\]
it is satisfied with $\kappa_{\W} = 2$  (regardless of $\W$), as long as $\delta \leq 1/6$,
since the right-side above is bounded by $2$ for such a $\delta$. Finally, Assumption~\ref{assn:subgaussianity}, is also satisfied as $(\phi_{i,v}(X))_{i\in[d]}$ for $X\sim\p_v$ is $\sigma^2$-subgaussian for $\sigma^2 \eqdef \frac{1}{1-4\delta^2}$. 

\paragraph{Completing the proof of Theorem \ref{t:conpriv_P2} (LDP constraints).}
From Theorem~\ref{cor:ldp} and the bounds derived above, we have 
\[
  \sum_{i=1}^d \mutualinfo{V(i)}{Y^T} \leq T\cdot  \frac{8\delta^2}{1-4\delta^2}\cdot e^\priv(e^\priv-1)^2,
\]
and therefore,
\[
  \sum_{i=1}^d \mutualinfo{V(i)}{Y^T} 
\leq c\cdot T \delta^2 \priv^2,
\]
where $c\eqdef 9e(e-1)^2$ (recalling that $\priv\in(0,1]$ and $\delta \leq 1/6$). 
Substituting this bound on the average mutual information in \eqref{e:C_P2} along with the values of $a$ and $b$, we have
\[
\E{g_{V}(x_T)-g_{V}(x_V^\ast)} \geq {\frac{DB\delta}{3} \cdot}  \left[1 -  \sqrt{
    \frac{\newer{2}c T\delta^2 \priv^2}d}\right].
\]
Upon setting $\delta \eqdef \sqrt{\frac{d}{\newer{8}cT\priv^2}}$, we get
\[
\E{g_{V}(x_T) -g_{V}(x_V^\ast) } 
\geq \frac{1}{12\sqrt{\newer{2}c}}\cdot\frac{DB}{\sqrt{T}}\cdot \sqrt{\frac{d}{\priv^2}},
\]
where we require $T \geq \frac{9}{\newer{2}c}\cdot\frac{d}{\priv^2}$ in order to enforce $\delta \leq 1/6$.
\qed 

\paragraph{Completing the proof of Theorem \ref{t:concom_P2} (Communication constraints).}
From Theorem~\ref{cor:simple-numbits} and $\gamma$, $\sigma$, and $\kappa_\W$ set as discussed above, we have 
\[
  \sum_{i=1}^d\mutualinfo{V(i)}{Y^T} \leq \frac{32(\ln 2)}{(1-4\delta^2)^2}\cdot T \delta^2 r,
\]
whereby, using $\delta\leq 1/6$,
\[
\sum_{i=1}^d\mutualinfo{V(i)}{Y^T}
\leq 29 T \delta^2 r.
\]
Substituting this bound on mutual information in \eqref{e:C_P2} along with the values of $a$ and $b$, we have
\[
\E{g_{V}(x_T)-g_{V}(x_V^\ast)} \geq \frac{DB\delta}{3} \cdot  \left[1 -  \frac{1}{\sqrt{d}}\cdot\sqrt{\newer{58} T\delta^2 r}\right].
\]
Setting $\delta\eqdef \sqrt{\frac{d}{\newer{232} r T}}$, we finally get
\[
\E{g_{V}(x_T) -g_{V}(x_V^\ast) } \geq \frac{1}{12\sqrt{\newer{58}}}\cdot \frac{DB}{\sqrt{T}}\cdot \sqrt{\frac{d}{r}},
\]
where we require $T\geq \frac{9}{ \newer{58}}\cdot \frac{d}{r}$ in order to enforce $\delta \leq 1/6$.
\qed
\begin{rem}\label{rem:varaible_length_opt}
 Finally, we remark that the lower bound as in Theorem \ref{t:concom_P2} also holds when the communication constraint of $r$ bits is satisfied in expectation and not in the strict worst-case sense.  The modification to the proof above is minimal. The only change  is that the mutual information term is now bounded by using a strong data processing inequality from \cite{duchi2014optimality}. This bound holds for variable-length quantizers that satisfy the communication constraints in expectation, as opposed to the current bound from \cite{acharya2020general}, which only holds for fixed-length quantizers. Specifically, from~\cite[Proposition
  2]{duchi2014optimality}, we have that 
 \[\sum_{i=1}^d\mutualinfo{V(i)}{Y^T} \leq  c T\delta^2 r,\]
 when the communication channel restricts the length of the outputs $Y_t$ to $r$ bits in expectation.  A caveat here is that the strong data processing inequality from \cite{duchi2014optimality} holds only for nonadaptive channel selection strategies. Thus we have the same lower bound on optimization error of family $\oO_{{\tt c}, p}$, $p \in [1, 2],$ as in   Theorem \ref{t:concom_P2} even when the communication constraints are satisfied in expectation as long as the gradient processing is done in a nonadaptive manner.
 
\end{rem}



\paragraph{Completing the proof of Theorem \ref{t:concompt_P2} (Computational constraints).}
Note that the sets $\X_i$s in Theorem~\ref{cor:obl} have $|\X_i|=2$ for our oracle.
Further,
\[
\frac{\p_{v^{\oplus i}}(X(i)=x(i))}{\p_v(X(i)=x(i))} =
\frac{\p_{v^{\oplus i}}(x)}{\p_v(x)}=
\frac{1+2\delta v(i) \sign(x(i))}{1-2\delta v(i) \sign(x(i))}\leq 2,
\]
when $\delta\leq 1/6$. Thus, the constant $C$
 in Theorem~\ref{cor:obl} is less than $2$, whereby
\[
\sum_{i=1}^d\mutualinfo{V(i)}{Y^T} \leq  \frac{16 \delta^2}{1-4\delta^2}\cdot T ,
\]
whereby, using $\delta\leq 1/6$,
\[
\sum_{i=1}^d\mutualinfo{V(i)}{Y^T}  \leq 18T\delta^2.
\]
Substituting this bound on mutual information in \eqref{e:C_P2} along with the values of $a$ and $b$, we have
\[
\E{g_{V}(x_T)-g_{V}(x_V^\ast)} \geq \frac{DB\delta}{3} \cdot  \left[1 -  \frac{1}{\sqrt{d}}\cdot\sqrt{\newer{36} T\delta^2 }\right].
\]
Setting $\delta\eqdef \sqrt{\frac{d}{\newer{144}T}}$, we finally get
\[
\E{g_{V}(x_T) -g_{V}(x_V^\ast) } \geq \frac{1}{\newer{72}}\cdot \frac{DB\sqrt{d}}{\sqrt{T}},
\]
where we require $T\geq \frac{d}{\newer{4}}$ in order to enforce $\delta \leq 1/6$.
\subsection{Convex Lipschitz functions for $p\in (2,\infty]$: Proof of Theorems \ref{t:conpriv_inf} and \ref{t:concom_inf}}\label{ssec:con_p_inf}
Next, we establish Theorems \ref{t:conpriv_inf} and \ref{t:concom_inf}, the analogous lower bounds on optimization of convex functions when $p\in[2,\infty)$. We again consider the class of functions $\G_{\tt c}$ defined in \eqref{e:conv_bottl}, this time with parameters $a\eqdef 2B \delta/d$ and $b \eqdef  D/(2d^{1/p})$ That is, here $\X=\{x:\norm{x}_\infty\leq D/(2d^{1/p}) \}$ and 
 \[
g_{v}(x) \eqdef \frac {2B\delta}{d} \sum_{i=1}^{d} \left|x(i) - \frac{v(i) D}{2 d^{1/p}}\right|.
\quad \forall x \in \X, v \in \{-1, 1 \}^d.
\]
It follows that the gradient of $g_v$ is equal to $-2B\delta v /d$ at every $x\in \X$.

For each $g_v$, consider then the gradient oracle $O_v$ which
outputs $0$ in all but a randomly chosen coordinate; if that coordinate is $i$,
it takes values $-B$ and $B$ with probabilities $\frac{1+2\delta v(i))}{2d}$ and $\frac{1-2\delta v(i)}{2d}$, respectively, for some parameter $\delta\in(0,1/6]$ to be suitably chosen later. Thus, the oracle is no longer
a product distribution.

Clearly, $\X \in \mathbb{X}_p(D)$ and all  the functions $g_v$ and the corresponding  oracles $O_v$ belong to the convex function family $\oO_{{\tt c}, p}$. Proceeding as in Section \ref{s:Proof_conv_P2}, we get
for a uniformly distributed $V$ that
\begin{equation}
	\label{e:C_Pinf}
	\E{g_{V}(x_T)-g_{V}(x_V^\ast)} \geq { \frac{DB \delta}{3d^{1/p}} \cdot}  \left[1 -  \sqrt{
\frac 1d
            \sum_{i=1}^d \newer{2}\mutualinfo{V(i)}{Y^T}}\right] .
\end{equation}
Further, proceeding as in the previous section to bound the average information,
we note that the oracle outputs independent samples from the \newer{distribution $\p_v$ on $\Omega \eqdef \mleft\{-B,0,B\mright\}^d$}  
 at each time. It can be checked  easily that, for every $i\in[d]$, the expression of the ratio $\frac{\p_{v^{\oplus i}}}{p_v}$ given in~\eqref{eq:convex:p12:ratio} still holds (as only the denominators of the Bernoulli parameters have changed, and they cancel out in the ratio), and that Assumption~\ref{assn:decomposition-by-coordinates} is satisfied with the following $\gamma$, $\phi_{i,v}$s:
\begin{equation}
	\label{eq:convex:pinf:gamma:phi}
  \gamma \eqdef \frac{1}{\sqrt{d}}\cdot\frac{4\delta}{\sqrt{1-4\delta^2}}, \qquad \phi_{i,v}(x) \eqdef \sqrt{d}\cdot\frac{v(i) \sign(x(i))+2\delta}{\sqrt{1-4\delta^2}}\,.
\end{equation}
Observe the difference with the expressions from the previous section (specifically,~\eqref{eq:convex:p12:gamma:phi}), as the orthonormality assumption now crucially introduces a factor $1/\sqrt{d}$ in the value of $\gamma$. Finally, because we will enforce $\delta \leq 1/6$ we also can take $\kappa_{\W_{{\tt com}, r}} = 2$ for the communication constraints, as before. We remark that $\phi_{i,v}(X)$ is no longer subgaussian.

\paragraph{Completing the proof of Theorem \ref{t:conpriv_inf} (LDP constraints).}
From Theorem~\ref{cor:ldp} and the value of $\gamma$ above, we get, analogously to the previous section,  
\[
\sum_{i=1}^d \mutualinfo{V(i)}{Y^T} \leq
c\cdot \frac{T \delta^2 \priv^2}{d},
\]
where $c\eqdef 9e(e-1)^2$ (recalling that $\priv\in(0,1]$ and $\delta \leq 1/6$). 
Substituting this bound on mutual information in \eqref{e:C_Pinf}, we obtain
\[
  \E{g_{V}(x_T)-g_{V}(x_V^\ast)} \geq \frac{DB\delta}{3d^{1/p}} \,  \left[1 -  \sqrt{\frac{\newer{2}c T\delta^2 \priv^2}{d^2}}\right].
\]
Optimizing over $\delta$, we set $\delta \eqdef \sqrt{\frac{d^2}{8cT\priv^2}}$ and get
\[
\E{g_{V}(x_T) -g_{V}(x_V^\ast) }
\geq \frac{1}{12\sqrt{\newer{2}c}}\cdot\frac{DBd^{1/2-1/p}}{\sqrt{T}}\cdot \sqrt{\frac{d}{\priv^2}},
\]
where we require $T \geq \frac{9}{\newer{2}c}\cdot\frac{d^2}{\priv^2}$ in order to guarantee $\delta \leq 1/6$. This concludes the proof.
\qed 

\paragraph{Completing the proof of Theorem \ref{t:concom_inf} (Communication constraints).}
We prove the two parts of the lower bounds separately, starting with the first. From Theorem~\ref{cor:simple-numbits} and the setting of $\gamma$ and $\kappa_\W$ as above, we have 
\[
\sum_{i=1}^d \mutualinfo{V(i)}{Y^T}
  \leq \frac{16}{1-4\delta^2}\cdot T \delta^2 \frac{2^r\land d}{d},
\]
whereby, using $\delta\leq 1/6$,
\[
\sum_{i=1}^d \mutualinfo{V(i)}{Y^T}\leq 18 T \delta^2 \frac{2^r\land d}{d}.
\]
Substituting this bound on mutual information in \eqref{e:C_Pinf}, we have
\[
\E{g_{V}(x_T)-g_{V}(x_V^\ast)}
\geq \frac{DB\delta}{3d^{1/p}} \cdot  \left[1 -  \frac{1}{\sqrt{d}}\cdot\sqrt{\newer{36} T\delta^2\frac{2^r\land d}{d}}\right].
\]
Setting $\delta\eqdef \sqrt{\frac{d^2}{\newer{144}(2^r\land d)T}}$, we finally get
\[
\E{g_{V}(x_T) -g_{V}(x_V^\ast) } \geq \frac{1}{\newer{72}}\cdot \frac{DB d^{1/2-1/p}}{\sqrt{T}}\cdot \sqrt{\frac{d}{2^r\land d}},
\]
where we require $T\geq \frac{1}{\newer{4}}\cdot \frac{d^2}{2^r\land d}$ in order to guarantee $\delta \leq 1/6$.

The second bound follows by noting that the lower bound in Theorem \ref{t:concom_P2} is still valid. Finally, since $\frac{d^2}{2^r\land d} \geq \frac{d}{r}$ for all $1\leq r\leq d$, both bounds apply whenever $T = \Omega\mleft(\frac{d^2}{2^r\land d}\mright)$, as claimed.
\qed
 
\subsection{Strongly convex functions: Proof of Theorem \ref{t:sconpriv_P2}, \ref{t:sconcom_P2}, and \ref{t:sconcompt_P2}}\label{ssec:scon}
Next, we establish our lower bounds on strongly convex optimization. We consider the class of functions $\G_{\tt sc}$ defined in \eqref{e:sconv_bottl} with parameters $a\eqdef B /(\sqrt{d} b)$ and $b \eqdef D/(2\sqrt{d}).$ That is, $\X=\{x:\norm{x}_\infty\leq D/(2 \sqrt{d}) \}$, and, for every $x \in \X$ and $v \in \{ -1, 1\}^d$,
 \[
	g_{v}(x) \eqdef \frac {B}{b\cdot \sqrt{d}}  \sum_{i=1}^{d} \frac{1+2\delta v(i)}{2} f^+_i(x) + \frac{1-2\delta v(i)}{2}f^-_i(x),
\]
and
\[
	f^+_i(x)= \theta b |x(i)+b| + \frac{1-\theta}{4}  (x(i)+b)^2 ~ \text{and} ~ f^-_i(x)= \theta b |x(i)-b| + \frac{1-\theta}{4}  (x(i)-b)^2.
\]
Moreover, in order to ensure that the every $g_v$ is $\gamma$-strongly convex, we choose $\theta\eqdef 1-\frac{4\gamma}{a}$ (so that $a\frac{1-\theta}{4} =\gamma$). It remains to specify $\delta$, which we will choose such that $0< \delta \leq \frac{1}{2}\cdot \frac{1-\theta}{1+\theta}$ in the course of the proof.

For each $g_v$, consider the gradient oracle $O_v$ which on query $x$ outputs independent values for each coordinate,
with the $i$th coordinate taking values
$\frac{B}{b \sqrt{d}}\cdot \frac{\partial f^+_i(x)}{\partial x_i}$ 
and 
$\frac{B}{b \sqrt{d}}\cdot \frac{\partial f^-_i(x)}{\partial x_i}$ 
with probabilities 
$\frac{1+2\delta v(i))}{2}$
 and 
 $\frac{1-2\delta v(i)}{2}$, 
respectively. 

Note that we have $\left| \frac{\partial f^+_i(x)}{\partial x_i} \right|, \left| \frac{\partial f^-_i(x)}{\partial x_i} \right| \leq b$ for all $x$ and $i$, and therefore the gradient estimate $\hat{g}(x)$ supplied by the oracle $O_v$ at $x$ satisfies $\norm{\hat{g}(x)}_2^2\leq B^2$ with probability one, for every query $x\in\X$. 
Further, it is clear that $\X \in \mathbb{X}_2(D)$ and all the functions $g_v$ and the corresponding  oracles $O_v$ belong to the strongly convex function family $\oO_{{\tt sc}}$.

Using our assumption that $\delta \leq \frac{1}{2}\cdot \frac{1-\theta}{1+\theta}$, we obtain by Lemma \ref{l:psiGsc}
\begin{equation}
	\label{eq:inequality:psiGsc}
	\psi(\G_{\tt sc})\geq \frac{2 a b^2 \delta^2  }{1-\theta} =  \frac{2 a^2 b^2 \delta^2  }{4 \gamma}=  \frac{ B^2 \delta^2  }{2 d \gamma},
\end{equation}
where we first plug in $a(1-\theta)=4\gamma$ and then substitute for $a$ and $b$.

\paragraph{Completing the proof of Theorem \ref{t:sconcom_P2} (Communication constraints).}
By proceeding as in Section \ref{s:Proof_conv_P2}, from Lemma~\ref{l:ImpL} and using the inequality~\eqref{eq:inequality:psiGsc} above, we have
\begin{equation}
	\label{eq:lb:assouadtype:sc}
	\sup_{\X \in \mathbb{X}_2(D)} \ep^*(\X , \oO_{{\tt sc}}, T,
        \W_{{\tt com}, r })  
	\geq  \frac{B^2\delta^2}{12\gamma}  \left[1 -
          \sqrt{\frac {\newer{2}}{d}\sum_{i=1}^d\mutualinfo{V(i)}{Y^T}}
\right]. 
\end{equation} 
It remains to bound $\sum_{i=1}^d\mutualinfo{V(i)}{Y^T}$
 to complete the proof. Note that unlike the proof in
 Section~\ref{s:Proof_conv_P2}, the gradient estimates have different
 distributions for different $x$. However, for a point $x$ we can
 still express the gradient estimate $\hat{z}(x) $ of $g_v(x)$ given by $O_v$ as follows:
 abbreviating $f^{\prime+}_i(x):=\frac{\partial f^+_i(x)}{\partial x_i}$ and
 $f^{\prime -}_i(x):=\frac{\partial f^-_i(x)}{\partial x_i}$, we have
\begin{align}\label{e:preprocessing}
 \hat{z}(x)(i)= a  Z_i f^{\prime+}_i(x) + a (1-Z_i) f^{\prime-}_i(x),
\end{align}
where $Z_i\sim\operatorname{Ber}(1/2+\delta v(i))$ and the $Z_i$'s are mutually independent. 
Thus, for a fixed $x$, $\hat{z}(x)$ can be viewed as a function of $\{
Z_i\}_{i \in [d]}$. Furthermore, for a channel $W \in \W_{{\tt com},
  r}$ consider the channel $W^{\prime}_x$ which first passes the
Bernoulli vector $\{Z_i\}_{i \in [d]}$  through the function
$\hat{z}(x)(i)$ and the resulting output is passed through the channel
$W$. This composed channel $W_x$ belongs to $\W_{{\tt com}, r}$, too.

Therefore, we can treat the independent copies of $Z\sim \p_v$
revealed by the oracle as i.i.d. random variables $X_1, ..., X_n$
in Section~\ref{ssec:avg_info}.
Further, note that at time $t$, the query is for a point $x_t$ which is a random
function of $Y^{t-1}$, and so,
$Y^T$ can be viewed as the channel
outputs with adaptively selected channels from $\W_{{\tt com},r}$. Thus, we can apply
the bounds in Theorem~\ref{cor:simple-numbits}. 

Doing so, analogously to the computations in Section~\ref{s:Proof_conv_P2},\footnote{As we have, in both cases, unknown Bernoulli product distribution over $\{-1,1\}^d$ with bias vector $\frac{1}{2}+\delta v$.} we get
\[
\sum_{i \in [d]}I(v(i) \wedge \{Y_i\}_{i \in [T]} ) \leq 
\sum_{i=1}^d\mutualinfo{V(i)}{Y^T}
\leq c \delta^2 r T,
\] 
for an appropriate constant $c$, which in view
of~\eqref{eq:lb:assouadtype:sc} leads to
\[
	\sup_{\X \in \mathbb{X}_2(D)} \ep^*(\X , \oO_{{\tt sc}}, T,
        \W_{{\tt com}, r }) \geq { \frac{ B^2\delta^2}{12\gamma}
          \cdot} \left[1 - \frac{1}{\sqrt{d}}\cdot\sqrt{\newer{2}cT\delta^2
            r}\right] = \frac{1}{\newer{192}c}\cdot \frac{ B^2}{\gamma T}\cdot
        \frac{d}{r}
\]
the last equality by setting $\delta \eqdef \sqrt{\frac{d}{\newer{8}c
    Tr}}$. Finally, observe that this choice of $\delta$ indeed
satisfies $\delta < \frac{1}{2}\cdot\frac{1-\theta}{1+\theta}$, as
long as $T \geq \newer{2}c\cdot\frac{ B^2}{D^2}\cdot\frac{d }{\gamma^2 r}$. This
completes the proof. \qed

\paragraph{Completing the proof of Theorem \ref{t:sconpriv_P2} (Privacy constraints).}
Proceeding as in the proof of Theorem~\ref{t:sconcom_P2} above, we
have the analogue of~\eqref{eq:lb:assouadtype:sc},
\[
	\sup_{\X \in \mathbb{X}_2(D)} \ep^*(\X , \oO_{{\tt sc}}, T,
        \W_{{\tt priv}, \priv })
	\geq
	 \frac{B^2\delta^2}{12\gamma}  \left[1 -
          \sqrt{\frac {\newer{2}}{d}\sum_{i=1}^d\mutualinfo{V(i)}{Y^T}}\right].
\]
As stated in the proof of Theorem \ref{t:sconcom_P2}, the privatization of the gradient $\hat{z}(x)$ can be viewed as first preprocessing $\{Z_i\}_{i \in [d]}$ and the passing the preprocessed output through the LDP channel. Such a composed channel also belongs to $\W_{{\tt priv}, p}$.
Thus, we can apply the bound in Theorem~\ref{cor:ldp} and proceed as in the proof of Theorem~\ref{t:conpriv_P2} to obtain
\[
	\sum_{i=1}^d\mutualinfo{V(i)}{Y^T} \leq c T \delta^2 \priv^2
\]
where $c>0$ is an absolute constant. Choosing $\delta \eqdef \sqrt{\frac{d}{\newer{8}c T \priv^2}}$, which makes $2\delta$ less than $\frac{1-\theta}{1+\theta}$ for   $T \geq \newer{2}c\cdot \frac{B^2}{D^2}\cdot\frac{d }{\gamma^2 \priv^2}$, for some universal positive constant $c$, then yields
\[
	\sup_{\X \in \mathbb{X}_2(D)} \ep^*(\X , \oO_{{\tt sc}}, T, \W_{{\tt priv}, \priv }) 
	\geq c_0\cdot\frac{B^2}{\gamma T}\cdot \frac{d}{\priv^2}
\]
for some absolute constant $c_0>0$, 
concluding the proof. \qed

\paragraph{Completing the proof of Theorem \ref{t:sconcompt_P2} (Computational constraints).}
As before, we can get
\[
	\sup_{\X \in \mathbb{X}_2(D)} \ep^*(\X , \oO_{{\tt sc}}, T,\W_{\tt obl})
	\geq
	  \frac{B^2\delta^2}{12\gamma}  \left[1 -
          \sqrt{\frac {\newer{2}}{d}\sum_{i=1}^d\mutualinfo{V(i)}{Y^T}}\right].
\]
Recall that we can express the subgradient estimate as in \eqref{e:preprocessing}. Note that for an oblivious sampling channel $W_t$  used at time $t$, specified by a probability vector $(p_j)_{j \in [d]}$, the output
is given by
\eq{\newest{
Y_i = (a Z_{J_t} f^{\prime+}_{J_t}(x) + a(1-Z_{J_t})f^{\prime-}_{J_t}(x))e_{J_t}, }
}
where $J_t=j$ with probability $p_j$.
To proceed, we observe that the Markov relation
\newest{$V \text{---} \{Z_{J_t}, J_t\}_{t\in [T]} \text{---} {Y^T}$} holds.
Indeed, we can confirm this by noting that
\newest{$\{Z_{J_t}\}_{t \in [T]}$} are generated i.i.d. from $\p_V$
and, for each $t\in[T]$, $Y_t$ is a function of \newest{$(Y^{t-1}, Z_{J_t}, J_t)$} and a local randomness $U$
available only to the optimization algorithm which
is independent jointly of $V$ and \newest{$\{Z_{J_t}, J_t\}_{t \in [T]}$.}
It follows that $Y^{T}$ itself is a function of $U$ and \newest{$\{Z_{J_t}, J_t\}_{t \in [T]}$},
which gives 
\begin{align}\label{eq:sc_nice_markovchain}
\newest{\condmutualinfo{V}{Y^T}{\{Z_{J_t}, J_t\}_{t\in [T]}}
\leq \condmutualinfo{V}{U}{\{Z_{J_t}, J_t\}_{t\in [T]}}
=0.}
\end{align}

From the previous observation, we also get
that
the Markov relation
\newest{$V(i) \text{---} \{Z_{J_t}, J_t\}_{t\in [T]} \text{---} {Y^T}$} holds
for every $i\in[d]$.
Thus,
by the data processing inequality for mutual information, we have
\[\newest{
\sum_{i=1}^d\mutualinfo{V(i)}{Y^T}
\leq
\sum_{i=1}^d\mutualinfo{V(i)}{\{Z_{J_t}, J_t\}_{t \in [T]}}.}
\]
Now since vector $(Z_j)_{j \in [d]}$ is a Bernoulli vector, the mutual information on the right-side can be bounded by the same computation as in the proof of Theorem \ref{t:concompt_P2} using Theorem \ref{cor:obl}. \newest{ This follows by observing that for all $t \in [T]$, $(Z_{J_t}, J_t)$ is a function of $Z_{J_t} e_{J_t}$, which  in turn can be seen as a output of the oblivious sampling channel for an input vector $(Z_j)_{j \in [d]}$.} Therefore, we have
\[
\sum_{i=1}^d\mutualinfo{V(i)}{Y^T}
\leq cT\delta^2
\]
for an appropriate constant $c$ and $\delta \leq \frac{1}{6},$
which in view of~\eqref{eq:lb:assouadtype:sc} leads to
\[
	\sup_{\X \in \mathbb{X}_2(D)} \ep^*(\X , \oO_{{\tt sc}}, T, \W_{{\tt com}, r }) 
	\geq  \frac{ B^2\delta^2}{12\gamma}  \left[1 -  \frac{1}{\sqrt{d}}\cdot\sqrt{\newer{2}cT\delta^2 }\right]
	= \frac{1}{c_0}\cdot \frac{d B^2}{\gamma T},
\]
where the last identity is obtained by setting $\delta \eqdef c_1\sqrt{\frac{d}{ T}}$, where $c_0$ and $c_1$ are universal positive constants.
Finally, observe that this choice of $\delta$ indeed satisfies $\delta < \frac{1}{2}\cdot\frac{1-\theta}{1+\theta}$, as long as $T \geq c_2\cdot \frac{B^2}{D^2}\cdot\frac{d^{2}}{\gamma }$, for some universal positive constant $c_2$. This completes the proof. \qed

\section{Concluding Remarks}
In this chapter, we derived lower bounds on the optimization error incurred by any first-order algorithm when the stochastic gradients used by the optimization algorithm need to be further processed to satisfy information constraints. We also saw that the gradient processing schemes proposed in the literature and appropriate first-order algorithms almost match our derived lower bounds in the case of privacy and computational constraints. Therefore, in the rest of the first part, we will develop gradient compression algorithms to match the lower bounds for communication-constrained optimization.



\chapter{Communication-Constrained  Optimization over Euclidean Space}\label{Ch:RATQ}
\section{Synopsis}
For communication-constrained optimization over the Euclidean Space, where the subgradient estimate's norm is almost surely bounded, we present Rotated Adaptive Tetra-iterated Quantizer (RATQ), a
fixed-length quantizer for subgradient estimates.  RATQ is easy to implement and involves only a Hadamard transform computation and adaptive uniform quantization with appropriately chosen dynamic ranges. We show that RATQ along with PSGD achieves the lower bound for communication-constrained optimization over Euclidean Space.

 We further extend our results for communication-Constrained optimization over Euclidean space when the subgradient estimates are mean square bounded. In this setting, we use a gain-shape subgradient quantizer which separately quantizes the Euclidean norm and uses RATQ to quantize the normalized unit norm vector. We
 establish lower bounds for performance of any optimization procedure
 and shape quantizer, when used with a uniform gain
 quantizer. Finally, we propose an adaptive quantizer for gain which
 when used with RATQ for shape quantizer outperforms uniform gain
 quantization and is, in fact, close to optimal.
 
 The results presented in this chapter are from \cite{mayekar2020ratqa} and  \cite{mayekar2020ratqj}.

\section{Introduction}
In this chapter, we develop new algorithms to match the lower bounds for communication-constrained optimization over Euclidean Space. More precisely, we study communication-constrained optimization for convex  and $\ell_2$ lipschitz function family as well as strongly convex and $\ell_2$ lipschitz function family.    We consider two {\em oracle models}: the first
where the subgradient estimate's Euclidean norm is {\em almost surely bounded} and the second where it is {\em mean square bounded}.   While our lower bounds in Chapter \ref{Ch:LB} were derived for the simpler, almost surely bounded oracles, where the subgradient estimates have their noise almost surely bounded. In this setting, we also develop algorithms for the more general mean square bounded oracle. Our main contributions include new quantizers for the two oracle models
and theoretical insights into the limitations imposed by heavy-tailed
gradient distributions admitted under the mean square bounded
oracles. A more specific description of our results and their relation
to prior work is provided below.


 \subsection{Main contributions}\label{ss:contributions}

 We start with almost surely bounded oracles and consider communication-constrained optimization for convex  and $\ell_2$ lipschitz function family. Our 
   precision-dependent lower bound  in Theorem \ref{t:concom_P2}  shows that no optimization
   protocol using a first order oracle and gradient updates of
   precision $r<d$ bits can have optimization error smaller than
   roughly $\sqrt{d}/\sqrt{rT}$. In particular, we need precision
   exceeding $\Omega(d)$ bits to get the classic convergence rate of
   $1/\sqrt{T}$ for convex functions.
 As our main contribution, we propose a new fixed-length
   quantizer we term {\em{Rotated Adaptive Tetra-iterated Quantizer }}
   (RATQ) that along with projected subgradient descent (PSGD) is
   merely a factor of $O(\log \log \log \log^* d)$ far from this
   minimum precision required to attain the $O(1/\sqrt{T})$
   convergence rate.  In a different setting, when the precision is
   fixed upfront to $r$, we modify RATQ by roughly quantizing and sending only
a subset of coordinates of the rotated vector. 
We show that this modified version of RATQ is only a factor $O(\log
   \log^*d)$ far from the optimal convergence rate.  For almost sure  bounded oracles, all our results for convex and $\ell_2$ lipschitz family are then extended to strongly convex and $\ell_2$ lipschitz family.


 \newest{For mean square bounded oracles,  we state our results for convex and $\ell_2$ lipschitz family. However, most of our results in this setting can be extended to strongly convex family {\em{mutatis mutandis}}. In this setting, most of the prior work makes an additional assumption that the gradient norm can be expressed using only a finite number of bits without accruing any quantization error. One of our main contributions for mean square bounded oracles is to analyse the quantization error without any such additional assumptions.}
For such oracles, we establish an
information theoretic lower bound in Section \ref{s:ug} which shows
(using a heavy-tailed oracle) that the precision used for gain\footnote{\newest{In the vector quantization literature, the norm of the vector to be quantized is called the {\em {gain}} and vector normalized by this norm is called the {\em{shape}}.}}
quantizer must exceed $\log T$ when the gain is quantized uniformly
for $T$ iterations and we seek $O(1/\sqrt{T})$ optimization
accuracy. Thus, \newest{if $32$ bits are used
  to describe the gain using a uniform quantizer,}
they will suffice for roughly a billion iterations.
Interestingly, we
present a new, adaptive gain quantizer which can attain the same
performance using only $\log \log T$ bits for quantizing gain.
\newest{In particular, using our scheme, only $5$ bits assigned for describing
gain will suffice for a billion iterations; these many bits will work for less than 100 iterations using uniform gain quantizers.}
\newest{ In a different setting, when the precision is fixed upfront we propose a quantizer which along with PSGD achieves the almost optimal convergence rate.}


 \subsection{Remarks on techniques}\label{ss:Prior_work}

 In this work we use adaptive quantizers with multiple
  dynamic-ranges $\{[-M_i, M_i]: i \in [h] \}$, with possibly a
  different dynamic range chosen for each coordinate. Once a
  dynamic-range $[-M_i, M_i]$ is chosen for a coordinate, the
  coordinate is  quantized uniformly within this
  dynamic-range using $k$ levels. Using a different dynamic-range for
  each coordinate allows us to reduce error per coordinate, but costs
  us in communication since we need to communicate which $M_i$ is used
  for each coordinate. In devising our scheme, we need to carefully
  balance this tradeoff. We do this by taking recourse to the
  following observation: when the same dynamic range is chosen for all
  coordinates, the mean square error per coordinate roughly grows as
\begin{align}\label{e:genprincipal}
O\left(\frac{\sum_{ i \in [h]}M_{i}^2 \cdot p(M_{i-1})}
{(k-1)^2}\right),
\end{align}
where $p(M)$ is the probability of the $\ell_\infty$ norm of the input
vector exceeding $M$ and $k$ denotes the number of levels of the
uniform quantizer. This observation allows us to relate the mean
square error to the tail-probabilities of the $\ell_\infty$ norm of
the input vector. In particular, we exploit it to decide on the
subvectors which we quantize using the same dynamic range. 
 
\newest{
As an aside, we believe that this approach  and equation \eqref{e:genprincipal}, in particular, will yield
very efficient rate-distortion codes for various sources with different tail probabilities, answering questions of fundamental interest and having many applications. We point out an application to the classic Gaussian rate-distortion problem in Chapter \ref{Ch:GRD}.}

We use another classic trick (see~\cite{gersho2012vector}): we
  transform the input vector before we apply our adaptive
  quantizer. In particular, we use a randomized transform
that expresses the input vector over a random basis.
    The specific choice of our random transform is determined by our
  assumption for the gradients, namely that their $\ell_2$ norms are
  almost surely bounded by $B$.

Drawing from these ideas, we propose the quantizer RATQ for quantizing random vectors with $\ell_2$ norm almost surely bounded by $B$.

We remark that using an adaptively chosen dynamic-range can
  alternatively be implemented by transforming the input using a
  monotone function. This, too, is a classic technique in quantization
  known as {\it companding}
  ($cf.$~\cite{gersho2012vector}). Companding is known as a popular
  alternative to entropic coding for fixed-length codes. However, to
  the best of our knowledge, this work is the first to combine it with
  other techniques and rigorously analyze it for the $\ell_2$ norm
  bounded vector quantization problem. Perhaps it is a bit surprising
  that this combination of classic technique was not analysed for
  constructing an efficient covering of the unit Euclidean ball, the
  problem underlying our quantization problem.

Moving to oracles with mean square
  bounded $\ell_2$ norms, we take recourse to gain-shape quantizers
  and quantize the (normalized) shape vector using RATQ. However,
  unlike prior work, we rigorously treat gain quantization. Our
  proposed quantizer for gain is once again an adaptive 
  quantizer.

Our lower bounds derived in Chapter \ref{Ch:LB} use almost surely bounded oracles as the difficult oracle. However, this only allows us to obtain lower bounds for the almost surely bounded setting. For the mean square bounded setting, we need a new construction with “heavy tails”. In particular, our proposed heavy-tailed construction shows a bottleneck for uniform gain quantizers which can be circumvented by our proposed quantizer, thereby establishing a strict improvement over uniform gain quantizers. 
 \subsection{Prior work}\label{ss:Prior_work}
Our work is motivated by the results
 in~\cite{suresh2017distributed,alistarh2017qsgd}, and we elaborate on
 the connection.
Specifically,~\cite{alistarh2017qsgd} considers a problem very similar
to ours. The paper~\cite{suresh2017distributed} considers the related
problem of distributed mean estimation -- we elaborate on the distributed mean estimation results in Chapter \ref{Ch:DME}  -- but the quantizer and its
analysis is directly applicable to distributed optimization. The two
papers present different quantizers that encode each input using a
variable number of bits. \newest{ Both these quantizers require the optimal expected
precision to achieve the $1/\sqrt{T}$ convergence rate for almost surely bounded oracles. However, their worst-case
(fixed-length) performance maybe suboptimal. Our proposed quantizer RATQ requires a precision only slightly more than the optimal precision to achieve the $1/\sqrt{T}$ convergence rate for almost surely bounded oracles, while still being  fixed-length. Moreover, in the slightly different setting of operating for any precision constraint $r$ less than the dimension, we significantly improve upon the current state-of-the-art.}

In fact, the problem of designing fixed-length quantizers for almost
surely bounded oracles is closely related to designing small-size
covering for the Euclidean unit ball. There has been a longstanding
interest in this problem in the vector quantization and information
theory literature ($cf.$~\cite{Wyner1967,gersho2012vector, hughes1987gaussian, csiszar1991capacity, Lapidoth1997TIT, Dumer2007}).


In a slightly different direction, a seminal, but perhaps not so
  widely known, result of~\cite{ziv1985universal} provides a very simple
  universal quantizer for random vectors with independent and
  identically distributed (\iid) coordinates, with each coordinate
  almost surely bounded. In this scheme, we first quantize each
  coordinate uniformly, separately using a ``scalar-quantizer,'' and
  then apply a universal entropic compression scheme to the quantized
  vector. We note that the variable-length schemes proposed
  in~\cite{alistarh2017qsgd,suresh2017distributed} are very similar,
  albeit with a specific choice of the entropic compression scheme.

All these schemes \newest{(the ones in~\cite{ziv1985universal, alistarh2017qsgd, suresh2017distributed})} are variable-length schemes, while it is
  desirable to get a fixed-length scheme for the ease of both protocol
  and hardware implementation.  We remark that
  indeed~\cite{suresh2017distributed} presents an interesting
  randomly-rotate and quantize fixed-length scheme, but it still
  requires communicating $O(\log \log d)$ times more than the optimal
  fixed-length quantizer for the unit Euclidean ball given
  in~\cite{Wyner1967}.  To the best of our knowledge, prior to our
  work, the quantizer in~\cite{suresh2017distributed} is the best
  known efficient fixed-length quantizer for the unit Euclidean ball.

In fact, a randomized orthogonal transform scheme similar to that in~\cite{suresh2017distributed} appeared
almost concurrently in~\cite{HadadErez16} as well, where an  analysis for Gaussian source is presented.
  However, a rate-distortion analysis has not been done in~\cite{HadadErez16}.
Remarkably, an early instance of the ``rotated dithering'' scheme for distributing energy equally appears in
the image compression literature in~\cite{OstromoukhovHA94}, albeit without formal error or performance analysis.
Another interesting scheme was proposed in~\cite{AkyolRose13} where 
nonuniform quantization (using {\em companding}) was combined with dithering. 
Our adaptive choice of dynamic range for uniform quantizers is similar, in essence, to companding.
But our scheme differs from the one in~\cite{AkyolRose13} in several ways: 
First,~\cite{AkyolRose13} uses the knowledge of input distribution to design 
their companding function, whereas we only need knowledge of the tail behaviour of
the input distribution in our setting; second, we apply a random rotation to our input leading to a universal quantizer, which is not needed in \cite{AkyolRose13}; and finally, the specific structure of our quantizer with adaptive dynamic ranges makes it amenable to mean square error analysis for a large variety of sources. 


\newest{Another scheme, similar, in spirit, to \cite{suresh2017distributed}, appears earlier in~\cite{lyubarskii2010uncertainty}. In this scheme, the input vector is preprocessed using a  redundant system of vectors  (the
  resulting representation is called Kashin's representation) instead
  of random rotation as in \cite{suresh2017distributed}. 
  In theory,  if the underlying system of vectors satisfies certain desirable properties, then preprocessing the vector in this manner and then uniformly quantizing each coordinate in the representation will lead to an orderwise optimal fixed-length quantizer for the unit ball. Unfortunately, \cite{lyubarskii2010uncertainty} provides only a randomized constructions\footnote{Note that randomly producing the optimal quantizer with high probability is different from constructing the optimal random quantizer, as we do. The former only gives high probability guarantees for bounds on loss, but need not yield a bound for expected loss.}
  for the system of vectors that satisfy the aforementioned properties with high probability. Thus, the scheme in~\cite{lyubarskii2010uncertainty} is not explicit.
Further, as a side remark, we note that the preprocessing step in
\cite{lyubarskii2010uncertainty} requires $O(d^2 \log d)$ real
operations, much worse than the $O(d\log d)$ real operations required
by RATQ.}

\newest{ Another recent independent work~\cite{gandikota2019vqsgd}
  presents a different scheme where a different random transform is
  used instead of random rotation. The optimal scheme in
  \cite{gandikota2019vqsgd} is similar to the one in
  \cite{lyubarskii2010uncertainty} and in essence to classical
  information-theoretic schemes ($cf.$~\cite{Wyner1967},
  \cite{Lapidoth1997TIT}). Specifically, \cite{gandikota2019vqsgd}
  provides a randomized algorithm that outputs the orderwise optimal
  quantizer for the unit ball without an explicit
  construction. Moreover, the time complexity of the overall
  quantization procedure in \cite{gandikota2019vqsgd} is much worse
  than RATQ. }


Returning to the literature on quantizers for first order stochastic
optimization, prior works including~\cite{alistarh2017qsgd} remain
vague about the analysis for mean square bounded oracles.  Most of the
works use gain-shape quantizers that separately quantize the Euclidean
norm ({\em gain}) and the normalized vector ({\em shape}). But they
operate under an engineering assumption: ``the standard $32$ bit
precision suffices for describing the gain.''
\newest{One of our goals in this
work is to carefully examine this heuristic. For
instance, can we use a simple uniform quantizer for gain with $32$ bits, or even say $8$ bits?}


\newest{Independent of our work, a non-uniform quantizer
  similar to the one we use for gain-quantization with geometrically
  increasing dynamic-ranges appears in~\cite{ramezani2019nuqsgd}.
  However, there are some key differences between the two quantizers.
  First, note that we use this quantizer for
  gain-quantization, while \cite{ramezani2019nuqsgd} uses it to
  quantize the shape. Second, in the case of our quantizer the
  geometrically growing dynamic ranges are further quantized uniformly
  whereas \cite{ramezani2019nuqsgd} chooses to use geometrically
  growing points as final quantization points.
  Third, the analysis of
  mean square quantization error in this work and
  \cite{ramezani2019nuqsgd} differ significantly. In particular, our
  mean square analysis follows from the general principle stated in
  \eqref{e:genprincipal}, whereas \cite{ramezani2019nuqsgd} builds
  upon the analysis of QSGD in
  ~\cite{alistarh2017qsgd}. \new{Finally, the setting considered in \cite{ramezani2019nuqsgd} is similar to
that of~\cite{alistarh2017qsgd} where quantization is followed by entropic compression.}  In particular, the fixed-length performance may be 
  suboptimal for almost surely bounded oracles
  and mean square bounded oracles are not handled.}



 \subsection*{Organization}
 We formalize our problem in the next section and describe our results
   for almost surely and mean
 square bounded oracles in Sections~\ref{s:as} and~\ref{s:ms},
 respectively, along with some of the shorter proofs. The more
 elaborate proofs are provided in 
 Section~\ref{s:Ch_l2_opt_proof}, with additional details relegated to Section~\ref{s:norm_proofs}.

\section{Setup and preliminaries}
\subsection{Setup}\label{s:problemsetup}
In the next two chapters, we develop efficient subgradient compression algorithms for the setting of communication-constrained first-order optimization described in the previous Chapter. Our domain $\X$ throughout this chapter has Euclidean diameter less than $D$. That is,
\begin{align}\label{eq:domain}
    \X \in \mathbb{X}_2(D)=\{ \X^{\prime}: \sup_{x, y\in \X^{\prime}}\norm{x-y}_2 \leq D.\}
\end{align}
For the domain of optimization $\X$, we develop subgradient compression schemes for function and oracle families given by $\oO_{{\tt c}, 2}$ and $\oO_{{\tt sc}}$, which are defined in  Definitions \ref{def:c} and \ref{def:sc}, respectively. 

We remark that the typical assumption made on the optimization literature on the oracle noise is that it is \emph{mean square bounded}. That is, for a query point $x \in \X,$ the oracle random 
estimates of the subgradient $\hat{g}(x)$ which 
 for all  $x \in \X$ satisfy
\begin{equation}\label{e:asmp_L2_bound}
\E{\norm{\hat{g}(x)}_2^2|x} \leq B^2,
\end{equation}
where $\partial f(x)$ denotes the set of subgradients of $f$ at $x$. In this chapter, we also want to develop schemes for mean square bounded oracles. Towards that end, we define generalization of class $\oO_{{\tt c}, 2}^{\tt m}$ below.


\begin{defn}[Convex and $\ell_2$ Lipschitz function family for a mean square bounded oracle $\oO_{{\tt c}, 2}^{\tt m}$]\label{def:c_m}
We
denote by $\oO_{{\tt c}, 2}^{\tt m}$ the set of all pairs of functions and oracles satisfying Assumptions \eqref{e:convexity}, \eqref{e:asmp_unbiasedness}, and \eqref{e:asmp_L2_bound}.
\end{defn}


Thus the assumption \eqref{e:asmp_as_bound} is replaced by \eqref{e:asmp_L2_bound} for mean square bounded families.
Clearly, 
\eq{&\ep^*(\X , \oO_{{\tt c}, 2}^{\tt m}, T, \W_{{\tt com}, r)  }  \geq  \ep^*(\X , \oO_{{\tt c}, 2}, T, \W_{{\tt com}, r  }) .}

Therefore, the lower bounds derived in Chapter \ref{Ch:LB} derived for almost surely bounded oracles still hold for mean square bounded oracles. Although, we still derive another lower for mean square bounded oracle to point out the limitations of specific quantization procedures in Section \ref{s:ms}.

\subsection{Structure of our protocols}\label{s:stucture_quant_l2_lp}
It will be instructive to recall the definition of the quantizer, since the the outputs of the oracle are passed through a
quantizer. An {\em $r$-bit quantizer} consists of randomized mappings\footnote{We can use public randomness $U$ for randomizing.}
$(\Qenc, \Qdec)$ with the encoder mapping $\Qenc:\R^d\to\{0,1\}^{r}$
and the decoder mapping $\Qdec: \{0,1\}^r\to \R^d$. The 
overall quantizer is given by the composition mapping $Q=\Qdec\circ
\Qenc$.

In both this Chapter and the next Chapter, 
 we restrict to {\em memoryless} quantization schemes where the same quantizer will be applied to each new gradient vector, without using any information from the previous updates. 
Specifically, at each instant $t$ and for any precision $r$, the quantizers in  $\W_r$ do not use any information from the previous time instants to quantize the subgradient outputted by $O$ at $t$. Our primary motivation for restriction to memoryless quantization schemes is ease of implementation and their application to other problems, as we see in Chapter \ref{Ch:DME} and \ref{Ch:GRD}.

Thus our channel selection strategy $S$ is nonadaptive (recall definition \ref{d:NA})  and is simply denoted by the quantizer $Q$ we choose to use. Thus the optimization error for a function $f$ and oracle $O$ when employing a first order optimization $\pi$ and quantizer $Q$ is given  by  \[\ep(f, O, \pi, Q)=\E{f(x_T)}-\E{f(x^\ast}.\]

\subsection{Quantizer performance for finite precision optimization}
Our overall optimization protocol throughout is the {\em projected SGD} (PSGD)
(see~\cite{bubeck2015convex}).  In fact, we establish lower bound
showing roughly the optimality of PSGD with our quantizers. 

In PSGD the  standard SGD updates are projected back 
 to the domain using the projection map $\Gamma_\X$ given by 
$\Gamma_{\X}(y) := \min_{x \in \X} \norm{x-y}_2.$
We use the {\em quantized PSGD} algorithm described in Algorithm~\ref{a:SGD_Q}. 
\begin{figure}[h]
\centering
\begin{tikzpicture}[scale=1, every node/.style={scale=1}]
\node[draw,text width= 8 cm , text height= ,] {%
\begin{varwidth}{\linewidth}       
            \algrenewcommand\algorithmicindent{0.7em}
\begin{algorithmic}[1]
   \Statex \textbf{Require:} $x_0\in \X, \eta \in \R^+$, $T$ and
   access to composed oracle $QO$ \For{$t=0$ to $T-1$}

$x_{t+1}=\Gamma_{\X} \left(x_{t}-\eta_t Q(\hat{g}(x_{t}))\right)$
   \EndFor \State \textbf{Output:} $\frac 1 T \cdot {\sum_{t=1}^T x_t}$
\end{algorithmic}
\end{varwidth}};
 \end{tikzpicture}

 \renewcommand{\figurename}{Algorithm}
\caption{Quantized PSGD with quantizer $Q$}
\label{a:SGD_Q}

\end{figure}

The quantized output $Q(\hat{g}(x_t))$, too, constitutes a noisy
oracle, but it can be biased for mean square bounded oracles. Though biased first-order
oracles were considered in~\cite{hu2016bandit}, the effect of
quantizer-bias has not been studied in the past.
The performance of a quantizer $Q$, 
when it is used with PSGD  for mean square  bounded oracles, is controlled by 
the worst-case $L_2$ norm $\alpha_2^{\tt m}(Q)$ of its output and the worst-case
bias $\beta_2^{\tt m}(Q)$ defined as\footnote{We omit the
dependence on $B$ and $d$ from our notation.}
\begin{align}
    \alpha_2^{\tt m}(Q)&\eqdef \sup_{Y \in \R^d: \E{\norm{Y}_2^2}\leq B^2}
    \sqrt{\E{\norm{Q(Y)}_2^2}},
    \nonumber
\\
\beta_2^{\tt m}(Q)&\eqdef \sup_{Y \in \R^d:
  \E{\norm{Y}_2^2}\leq B^2} \norm{\E{Y-Q(Y)}}_2.
\label{e:alpha,beta}
\end{align}
The corresponding quantities for almost surely bounded oracles are
    \begin{align}
      \alpha_2(Q)&\eqdef \sup_{Y \in \R^d: \norm{Y}_2\leq B \text{ a.s.}}
      \sqrt{\E{\norm{Q(Y)}_2^2}},
      \nonumber
  \\
  \beta_2(Q)&\eqdef \sup_{Y \in \R^d:
    \norm{Y}_2\leq B \text{ a.s.}} \norm{\E{Y-Q(Y)}}_2.
  \label{e:alpha0,beta0}
    \end{align}
    Using a slight modification
of the standard proof of convergence for PSGD, we get the following result for convex functions.
\begin{thm}\label{t:basic_convergence}
Let the domain $\X$ satisfy \ref{eq:domain}. 
For any quantizer $Q$, the output $x_T$ of optimization protocol $\pi$
given in Algorithm \ref{a:SGD_Q} satisfies
\begin{align*}
\sup_{(f, O) \in \oO_{{\tt c}, 2}}\ep(f, O, \pi, Q)&\leq
D\left(\frac{\alpha_2(Q)}{\sqrt{T}}+ \beta_2(Q)\right),
\\
  \sup_{(f, O) \in \oO_{{\tt c}, 2}^{\tt m}}\ep(f, O, \pi, Q)&\leq
D\left(\frac{\alpha_2^{\tt m}(Q)}{\sqrt{T}}+ \beta_2^{\tt m}(Q)\right),
\end{align*}
when the parameter $\eta_t=\eta$, for all $t$, is set to $D/(\alpha_2(Q) \sqrt{T})$ and
$D/(\alpha_2^{\tt m}(Q) \sqrt{T})$, respectively.
\end{thm}
\noindent See Section \ref{ap:QPSGD} for the proof.

\begin{rem}[Knowledge of time horizon in setting the learning rate.]
Note that the choice of learning rate $\eta_t$ in \ref{t:basic_convergence} requires the knowledge of the time horizon $T$. In fact, all the convergence results in this thesis require setting $\eta_t$ based on the time horizon. One could employ the doubling trick --see, for instance, \cite[Pg. 129]{nemirovski1995information} -- to remove this restriction. However, this would add a multiplicative $\sqrt{\log T} $ factor to the convergence rate.
\end{rem}
We have also have the following counterpart of the previous result for strongly convex functions.

\begin{thm}\label{t:basic_convergence_sc}
Let the domain $\X$ satisfy \ref{eq:domain}. For any quantizer $Q$, the output $x_T$ of optimization protocol $\pi$
given in Algorithm \ref{a:SGD_Q} satisfies
\begin{align*}
\sup_{(f, O) \in \oO_{{\tt sc}}}\ep(f, O, \pi, Q)&\leq
D\left(\frac{\alpha_2(Q)^2}{D \gamma T}+ \beta_2(Q)\right).
\end{align*}
when the parameter $\eta_t$ is set to $2/\gamma(t+1)$.
\end{thm}
\noindent See Section \ref{ap:QPSGD_SC} for the proof.
\begin{rem}[Choice of learning rate]  For the class of convex functions, we fix  the parameter $\eta_t$ of Algorithm \ref{a:SGD_Q} to a constant value $\eta$, for all $t$. $\eta$ is set to  $D/(\alpha_2(Q) \sqrt{T})$  and $D/(\alpha_2^{\tt m}(Q) \sqrt{T})$ for all the results in Section \ref{s:as} and Section \ref{s:ms}, respectively. For the class of strongly convex functions, we fix the parameter $\eta_t = 2/\gamma(t+1)$ all the results in Section \ref{s:as}.
  \end{rem}

 
 \section{Main results for almost surely bounded oracles}\label{s:as}
Our main results will be organized along two regimes: the high-precision
and the low-precision regime. For the 
high-precision regime, we seek to attain the optimal, classic convergence rate of
$1/\sqrt{T}$, for convex functions, and $1/T$, for strongly convex functions, using the minimum precision possible. For the low-precision regime, we 
seek to attain the fastest convergence rate possible for a given, fixed precision $r$.

From  our lower bounds on minmax optimization error of  families $\oO_{{\tt c}, 2}$ and $\oO_{{\tt sc}}$ under communication constraints, which are derived in Theorem \ref{t:concom_P2} and \ref{t:sconcom_P2}, we have the following corollaries. Our corollaries show that there is no hope of
getting the desired convergence rate of $1/\sqrt{T}$ for convex function and $\ell_2$ lipschitz function families  ($\oO_{{\tt c}, 2}$, $\oO_{{\tt c}, 2}^{\tt m}$) and $1/T$ for strongly convex function families $\oO_{{\tt sc}}$  by using a
precision of less than $d$.  
\begin{cor}\label{c:OmegaD}
Let $\mathbb{X}_2(D)=\{\X^{\prime}: \sup_{x, y \in \X^{\prime}}\norm{x-y}_2\leq D\}$. Then, the
precision $r$ must be at least $\Omega(d),$
for either one of the following to hold: 
\eq{
&\displaystyle{\sup_{\X \in \mathbb{X}_2(D)}\ep^*(\X , \oO_{{\tt c}, 2}, T, \W_{{\tt com}, r)  }}\leq \frac{DB}{\sqrt{T}},   \quad \displaystyle{\sup_{\X \in \mathbb{X}_2(D)}\ep^*(\X , \oO_{{\tt c}, 2}^{\tt m}, T, \W_{{\tt com}, r  })} \leq \frac{DB}{\sqrt{T}},\\
&\text{and} 
\displaystyle{\sup_{\X \in \mathbb{X}_2(D)}\ep^*(\X , \oO_{{\tt sc}}, T, \W_{{\tt com}, r)  }}\leq \frac{B^2}{\gamma T}.
} 
\end{cor}

Thus, from  Corollary \ref{c:OmegaD}, our quantization schemes in high-precision regime will use a precision of atleast $d$ bits. 
\subsection{RATQ: Our quantizer for the  $\ell_2$ ball}\label{s:alg_RATQ}
We propose {\em{Rotated Adaptive
    Tetra-iterated Quantizer }}(RATQ) to quantize any random vector
$Y$ with $\norm{Y}_2^2\leq B^2$, which is what we need for almost surely bounded oracles.
RATQ first rotates the input vector, then divides the coordinates of the rotated vectors into smaller groups,
and finally quantizes each subgroup-vector using a {\em Coordinate-wise Uniform Quantizer} (CUQ). 
However, the dynamic-range used for each subvector is chosen adaptively from a set of tetra-iterated levels. 
We call this adaptive quantizer {\em Adaptive Tetra-iterated Uniform Quantizer} (ATUQ), and it is  the main workhorse of our construction. The encoder and decoder for RATQ are given in Algorithm~\ref{a:E_RATQ} and Algorithm~\ref{a:D_RATQ}, respectively. 
The details of all the components involved are described below.
 
 \begin{figure}[ht]
\centering
\begin{tikzpicture}[scale=1, every node/.style={scale=1}]
\node[draw, text width= 11 cm, text height=,] {%
\begin{varwidth}{\linewidth}
            
            \algrenewcommand\algorithmicindent{0.7em}
 \renewcommand{\thealgorithm}{}
\begin{algorithmic}[1]
\Require  Input $Y\in \R^d$, rotation matrix R
  
 \State Compute $\tilde{Y}=RY$ 

\For {$i \in
    [d/s]$}
  
  $\displaystyle{\tilde{Y}_i ^{T} =[\tilde{Y}((i-1)s+1), \cdots \tilde{Y}(\min\{is,d\})]^{T}}$ 
 
  \EndFor

   \State \textbf{Output:} 
   $\displaystyle{\Qenc_{{\tt at}, R}(Y) =\{\Qenc_{{\tt at}}(\tilde{Y_1})\cdots\Qenc_{{\tt
      at}}(\tilde{Y}_{\ceil{d/s}})\}}$
\end{algorithmic}
\end{varwidth}};
 \end{tikzpicture}
 \renewcommand{\figurename}{Algorithm}
 \caption{Encoder $\Qenc_{{\tt at}, R}(Y)$ for RATQ}\label{a:E_RATQ}
 \end{figure}

 \begin{figure}[ht]
\centering
\begin{tikzpicture}[scale=1, every node/.style={scale=1}]
\node[draw, text width= 11 cm, text height=,] {%
\begin{varwidth}{\linewidth}
            \algrenewcommand\algorithmicindent{0.7em}
 \renewcommand{\thealgorithm}{}
\renewcommand{\thealgorithm}{}
\begin{algorithmic}[1]
    \Require 
  Input $\{Z_i, j_i \}$ for $i \in [\ceil{d/s}]$, rotation matrix R

\State $\hat{Y}^{T}= [\Qdec_{\tt at}( Z_1, j_1 ), \cdots$,
  $\Qdec_{{\tt at}}( Z_{\ceil{d/s}}, j_{\ceil{d/s}} ) ]^T$
\State \textbf{Output:} $\Qdec_{{\tt at},R}( \{Z_i, j_i\}_{i=1}^{\ceil{d/s}})=R^{-1}\hat{ Y}$
\end{algorithmic}
\end{varwidth}};
 \end{tikzpicture}
 \renewcommand{\figurename}{Algorithm}
 \caption{Decoder $\Qdec_{{\tt at}, R}(Z, j)$ for RATQ}\label{a:D_RATQ}
 \end{figure}

\paragraph{Rotation and division into subvectors.} 
RATQ first rotates the input vector by multiplying it with a random Hadamard matrix.
Specifically, denoting by $H$ the $d\times d$
Walsh-Hadamard Matrix (see \cite{horadam2012hadamard})\footnote{We
  assume that $d$ is a power of $2$.}, define
\begin{equation}\label{e:R}
R\eqdef \frac{1}{\sqrt{d}}\cdot HD,
\end{equation}
where $D$ is a diagonal matrix with each diagonal entry generated uniformly from $\{-1, +1\}$. 
The input vector $y$ is multiplied by $R$ in the rotation step. 
The matrix $D$ can be generated using shared randomness
between the encoder and decoder.

Next, the rotated vector of dimension $d$ is partitioned into
 $\ceil{d/s}$ smaller subvectors. The $i^{th}$ subvector comprises
the coordinates $\{(i-1)s+1,\cdots, \min\{is,d\}\}$,
 for all $i \in [d/s].$ Note that the dimension of all the sub
 vectors except the last one is $s$, with the last one having a
 dimension of $d-s\floor{d/s}$.

\begin{rem}
As an aside, we remark that preprocessing the data by such a random transform $R$ was  used by \cite{ailon2006approximate} for Fast Johnson Lindestrauss transform. 
\end{rem}

We now describe the advantage of random rotation. The advantage of subvector division will be clear once we describe the rest of the scheme.
\begin{rem}[Advantage of  random rotation]\label{r:Rotation_Gain}
While by almost sure assumption the input vector to the quantizer is inside the Euclidean ball of radius $B$, to set the dynamic range\footnote{We mean the interval $[-M,M].$}, we need upper bounds for each coordinate of the vector. After random rotation, each coordinate of the input vector is a centered subgaussian random variable with a variance of $O(B^2/d)$, as opposed to a variance factor of $O(B^2)$, which is all that can be said for the original input  vector.
\end{rem}

\paragraph{Coordinate-wise Uniform Quantizer (CUQ).} RATQ uses CUQ as a subroutine;  we describe the latter for $d$ dimensional inputs, but it will only be applied to subvectors of lower dimension in RATQ. CUQ has a dynamic range $[-M, M]$ associated with it, and it uniformly quantizes each coordinate of the input to $k$-levels as long as the component is
within the dynamic-range $[-M, M]$. 
Specifically, it 
partitions the interval $[-M, M]$ into parts $I_\ell\eqdef
(B_{M,k}(\ell), B_{M,k}(\ell+1)]$, $\ell\in\{0,\ldots, k-1\} $, where
  $B_{M,k}(\ell)$ are given by \eq{ &B_{M,k}(\ell) := -M+\ell \cdot
    \frac{2M}{k-1}, \quad \forall\,  \ell \in \{0,\ldots,
    k-1\}.} 
    Then, for a coordinate $y \in (B_{M,k}(\ell), B_{M,k}(\ell+1)],$ CUQ randomly outputs either $B_{M,k}(\ell)$ or $ B_{M,k}(\ell+1)$ with probabilities such that the output value equals the input $y$ in expectation.
Note that  each output coordinate of the CUQ encoder takes $k+1$ values -- $k$ of these symbols correspond to the $k$ uniform levels and the additional   symbol corresponds to the overflow symbol $\emptyset$.  Thus we need a total precision of $d\ceil{\log(k+1)}$ bits to represent the output of the CUQ encoder. The encoder and decoders used in CUQ are given in Algorithms~\ref{a:E_CUQ} and~\ref{a:D_CUQ}, respectively. 
In the decoder, we have set $B_{M,k}(\emptyset)$ to $0$.

\begin{figure}[ht]
\centering
\begin{tikzpicture}[scale=1, every node/.style={scale=1}]
\node[draw,text width= 11 cm, text height=,] {%
\begin{varwidth}{\linewidth}
            
            \algrenewcommand\algorithmicindent{0.7em}
\begin{algorithmic}[1]
\vspace{-0.5cm}
\Require Parameter $M \in \R^+$ and input $Y \in \R^d$
  \For {$i \in [d]$} \If{$|Y(i)| > M$}
   
   $Z(i)=\emptyset$
 
 \Else \For {$\ell \in \{0,\ldots, k-1\}$}\label{step:UQ} \If {${Y}(i)
   \in (B_{M,k}(\ell), B_{M,k+1}(\ell+1)]$}
 \Statex  \hspace{1cm}
     $
      {Z}(i) =
\begin{cases}
\ell+1, \quad w.p. ~ \frac{{Y}(i) - B_{M,k}(\ell)}{B_{M,k}(\ell+1)-B_{M,k}(\ell)}
 \\ \ell, \hspace{0.6cm} \quad w.p. ~  
  \frac{B_{M,k}(\ell+1) - {Y}(i)}{B_{M,k}(\ell+1)-B_{M,k}(\ell)} 
\end{cases}
$
   
\EndIf \EndFor \EndIf \EndFor \State \textbf{Output:} $\Qenc_{\tt u}(Y; M)=Z$
\end{algorithmic}  
\end{varwidth}};
 \end{tikzpicture}
 \renewcommand{\figurename}{Algorithm}
 \caption{Encoder  $\Qenc_{\tt u}(Y; M)$ of CUQ}\label{a:E_CUQ}
 \end{figure}

 \begin{figure}[ht]
\centering
\begin{tikzpicture}[scale=1, every node/.style={scale=1}]
\node[draw, text width= 11 cm, text height=,] {%
\begin{varwidth}{\linewidth}
            
            \algrenewcommand\algorithmicindent{0.7em}
 \renewcommand{\thealgorithm}{}
\begin{algorithmic}[1]
  \Require  Parameter  $M \in \R^+$, input $\displaystyle{Z \in \{0,
  \dots, k-1, \emptyset\}^d}$ \State Set $\hat{Y}(i)= B_{M,k}(Z(i))$, for all $i\in
       [d]$ \State \textbf{Output:} $\Qdec_{\tt u}(Z;
       M)=\hat{Y}$\label{step:output_coordinate}
\end{algorithmic}
\end{varwidth}};
 \end{tikzpicture}
 \renewcommand{\figurename}{Algorithm}
 \caption{Decoder $\Qdec_{\tt u}(Z; M)$ of CUQ}\label{a:D_CUQ}
 \end{figure}


\paragraph{Adaptive Tetra-iterated Uniform Quantizer (ATUQ).}
The quantizer ATUQ is CUQ with its dynamic-range chosen in an adaptive manner. 
In order to a quantize a particular input vector, it first chooses a 
 dynamic range from $[-M_i, M_i]$, $1\leq i\leq h$.
 To describe these $M_i$s, we first define the $i^{th}$
tetra-iteration for $e$, denoted by {$e^{*i}$, recursively as follows:
\[
\newest{e^{*0}:=1}, \quad e^{*1}:=e, \quad e^{*i}:=e^{e^{*(i-1)}}, \quad i\in \N.
 \]
Also, for any non negative number $b$, we define 
$\ln^*b :=\inf\{i \in \N :  e^{*i} \geq b \}.$}
With this notation, the values $M_i$s are defined in terms of 
$m$ and $m_0$ as follows: \eq{  
M_{i}^2= m \cdot e^{*i}+m_0,\quad \forall\, i \in
  \{0, \ldots, h-1\},}
\newest{  where the parameters $m$ and $m_0$ will be set later.}
ATUQ finds the smallest level $M_i$ which
 bounds the infinity norm of the input vector; if no such $M_i$ exists, it simply uses $M_{h-1}$.  
It then uses CUQ with dynamic range $[-M_i, M_i]$ to quantize the input vector. In RATQ, we apply ATUQ to each subvector.
The decoder of ATUQ is simply the decoder of CUQ using the dynamic range outputted by the ATUQ encoder.

   Note that in order to represent the output of ATUQ for $d$ dimensional inputs, we need a precision of  at the most $\ceil{\log h}+d\ceil{\log (k+1)}$ bits: $\ceil{\log h}$ bits to represent the dynamic range and at the most $d\ceil{\log (k+1)}$ bits to represent the output of CUQ. 
The encoder and decoder for ATUQ are given in Algorithms~\ref{a:E_ATQ} and~\ref{a:D_ATQ}, respectively.

\begin{figure}[ht]
\centering
\begin{tikzpicture}[scale=1, every node/.style={scale=1}]
\node[draw, text width= 11 cm, text height=,] {%
\begin{varwidth}{\linewidth}
            
            \algrenewcommand\algorithmicindent{0.7em}
 \renewcommand{\thealgorithm}{}
\begin{algorithmic}[1]
\Require Input $Y \in \R^d$ \If { $\norm{Y}_\infty > M_{h-1}$}
 
 Set $M^*=M_{h-1}$

\Else

Set $j^*
 =\min \{
 j: \norm{Y}_{\infty} \leq M_j\},$
  $M^*=M_{j^*}$ \EndIf \State Set $Z=\Qenc_{\tt u}(Y; M^*)$

\State \textbf{Output:} $\Qenc_{\tt at}(Y)=\{Z, j^*\}$
\end{algorithmic}
\end{varwidth}};
 \end{tikzpicture}
 \renewcommand{\figurename}{Algorithm}
 \caption{Encoder $\Qenc_{{\tt at}}(Y)$ for ATUQ}\label{a:E_ATQ}
 \end{figure}

\begin{figure}[ht]
\centering
\begin{tikzpicture}[scale=1, every node/.style={scale=1}]
\node[draw, text width= 11 cm, text height=,] {%
\begin{varwidth}{\linewidth}
            \algrenewcommand\algorithmicindent{0.7em}
 \renewcommand{\thealgorithm}{}
\renewcommand{\thealgorithm}{}
\begin{algorithmic}[1]
  \Require
  Input $\{Z,j\}$ with $Z\in \{0,\dots, k-1, \emptyset \}^d$ and $j\in \{0, \dots
  h-1\}$
 
  \State \textbf{Output:} $\Qdec_{\tt at}(Z,j)= \Qdec_{\tt u}(Z;
  M_{j})$
 
\end{algorithmic}
\end{varwidth}};
 \end{tikzpicture}
 \renewcommand{\figurename}{Algorithm}
 \caption{Decoder $\Qdec_{\tt at}(Z, j)$ for ATUQ}\label{a:D_ATQ}
 \end{figure}

When ATUQ is applied to each subvector in RATQ, each of the  $\ceil{d/s}$ subvectors are represented using less than $\ceil{\log h} + s\ceil{\log (k+1)}$ bits. Thus, the overall precision for RATQ is less than\footnote{$\log$ denotes the logarithm to the base $2$,  $\ln$ denotes logarithm to the base $e$.} 
\[\ceil{d/s}\cdot \ceil{\log h}+ d \ceil{\log (k+1)}\] bits.
The decoder of RATQ is simply formed  by collecting the output of the ATUQ decoders for all the subvectors to form a $d$-dimensional vector, and rotating it back using the matrix $R^{-1}$  (the inverse of the rotation matrix used at the encoder).

\begin{rem}[Advantage of  division  into  subvectors]\label{r:subsvectors_Gain}
The overall precision of RATQ allows us to understand the advantage of clubbing multiple coordinates into subvectors. Since we use the same dynamic range for all  coordinates of a subvector, we save on coordinate-wise communication of the dynamic range.
 \end{rem}
 
 \begin{rem}[Mean square error of ATUQ]
The per coordinate mean square error between the input to ATUQ and its output   roughly grows as
\begin{align}\label{e:DR_MSE}
O\left(\frac{\sum_{ i \in [h]}M_{i}^2 \cdot p(M_{i-1})}
{(k-1)^2}\right),
\end{align}
where $p(M)$ is the probability of the $\ell_\infty$ norm of the input
vector exceeding $M$ and $k$ denotes the number of levels of the
uniform quantizer. This observation allows us to relate the mean
square error to the tail-probabilities of the $\ell_\infty$ norm of
the input vector. In particular, we exploit it to decide on the \newest{dimension} $s$ of 
subvectors as well as the growth rate of $M_i$s.  
\end{rem}

\begin{rem}[Growth rate of Tetration] \label{r:RATQ}
A key distinguishing feature of RATQ is choosing the set of $M_i$s to grow as a tetration, roughly as $M_{i+1}=e^{M_{i}}$. The large
  growth rate of a tetration allows us to cover the complete range of
  each coordinate using only a
  small number of dynamic ranges, which leads to an unbiased quantizer and reduces the
  communication. Also, after random rotation, each
  coordinate of the vector is a centered subgaussian random
  variable with a variance-parameter of $O(B^2/d)$ (see Remark \ref{r:Rotation_Gain}), which, despite the large
  growth rate of a tetration, ensures that the per coordinate mean
  square error between the quantized output and the input is almost a
  constant, as can be seen from \eqref{e:DR_MSE}.
\end{rem}
 
                             {
\paragraph{Choice of parameters.}
Throughout the remainder of this section, we set our parameters $m$, $m_0$, and $h$ as follows
\begin{align}
m=\frac{3B^2}{d}, \quad m_0=\frac{2B^2 }{d} \cdot  \ln s, \quad \log h=\ceil{\log(1+\ln^\ast(d/3))}.
\label{e:RATQ_levels}
\end{align}
In particular, this results in $M_{h-1} \geq B$ whereby,
for an input $Y$ with $\norm{Y}_2^2 \leq B^2$, RATQ outputs an unbiased estimate of $Y$. 
}


                             \newest{We close with a remark on the computational complexity of RATQ.
        \begin{rem}[Computational complexity of RATQ]
                                 Since $R$ is a Hadamard matrix,
                                 the matrix multiplication at the encoder
                                 and the decoder requires $O(d \log d)$ real operations\footnote{Each addition, subtraction, multiplication, or division operation on the real field will be referred to as a real operation}. Further, it takes 
                                 $O(\log h)$ real operations to find the dynamic-range for each subvector, whereby the overall complexity for finding dynamic-ranges for $\ceil{d/s}$ subvectors is $O(\ceil{d/s} \log h)$ real operations; we can represent
each of these dynamic-ranges as a $\log h$-bit binary string in
another $O(\ceil{d/s} \log h)$ real operations, too.
Note that the encoding complexity of CUQ for an input subvector of dimension  $s$ is  $s$ real operations, and thus, the overall complexity of $\ceil{d/s}$ CUQ operations is $O(d)$ real operations. Finally, we need $O(d \log k)$ real operations to represent
the quantized values of $d$ coordinates to $k$ levels 
$\log k$-bit binary strings.
Putting it all together, the overall complexity of the encoding procedure is $O(d \log d + d \log h/s + d \log k)$. By similar arguments, the complexity of real operations at the decoder would also be $O(d \log d + d \log h/s + d \log k)$. 
 
 Throughout the chapter,  our choice of parameters $s$, $h$, and $k$ for RATQ, which is roughly the optimal choice of these parameters for quantizing the $\ell_2$ ball, would result in quantities $d \log h/s $ and $d 
\log k$ to be much lesser than $d 
\log d$. Thus, for parameters as chosen in this chapter or other reasonable choices of $s$, $h$, $k$, the encoding and decoding complexity of RATQ is $O(d \log d)$. Note that the random rotation based quantizer from \cite{suresh2017distributed} also has encoding and decoing complexity of $O(d \log d)$.
\end{rem}}

\subsection{RATQ in the high-precision regime}\label{s:RATQh}
The following result shows that RATQ is unbiased for almost surely bounded inputs and provides a bound for its worst-case second order moment; this constitutes a key technical tool for characterizing the performance of RATQ.
\begin{thm}[Performance of RATQ]\label{t:RATQ_alpha0_beta0}
Let $Q_{{\tt at}, R}$ be
the quantizer RATQ with $M_j$s set by~\eqref{e:RATQ_levels}. Then, 
for all $s,k\in \N$,
\begin{align} \label{e:RATQ_alpha_bound}
\alpha_2(Q_{{\tt at}, R})&\leq B\sqrt{
\frac{9+3\ln s}{(k-1)^2}+1}, \quad 
 \beta_2(Q_{{\tt at}, R})= 0. 
\end{align}
\end{thm}
\noindent The proof is deferred to Section~\ref{s:ProofRATQ}.

Thus, $\alpha_2$ is lower when $s$ is small, but the overall precision needed
grows since the number of subvectors increases. The following choice of
parameters yields almost optimal performance:
\begin{align}
  s=\log h,\quad \log(k+1) = \ceil{\log (2  + \sqrt{9  + 3\ln s })}. 
\label{e:RATQ_bits}
\end{align}
For these choices, we obtain the following.
\begin{cor}\label{c:PSGD_RATQ_0} The overall precision $r$ used by the quantizer $Q=Q_{{\tt at},R}$
  with parameters set as in~\eqref{e:RATQ_levels},~\eqref{e:RATQ_bits} satisfies
\[
r\leq d(1+\Delta_1)+\Delta_2,
\] 
where
$\Delta_1=\ceil{\log \left(2+  \sqrt{  9 + 3 \ln \Delta_2 }\right)}$ and 
$\Delta_2=\ceil{\log (1+\ln^*({d}/{3})) }$.\\
Furthermore, the optimization protocol $\pi$
  given in Algorithm~\ref{a:SGD_Q} satisfies
  \[\sup_{(f,O)\in \oO_{{\tt c}, 2}}\ep(f, O, \pi, Q) \leq \frac{\sqrt{2}DB}{\sqrt{T}} \text{and}\]
   \[\sup_{(f,O)\in \oO_{{\tt sc}}}\ep(f, O, \pi, Q) \leq \frac{2B^2}{\gamma T}.\]
\end{cor}
\begin{proof}
By the description RATQ, it encodes the subgradients using a fixed-length code of at the most 
$ \ceil{d/s}\cdot \ceil{\log h}+ d \ceil{\log (k+1)}$ bits. Upon substituting $s$, $\log h$, and $\log(k+1)$
as in \eqref{e:RATQ_bits} and \eqref{e:RATQ_levels}, we obtain that
the total precision is bounded above by $d(1+\Delta_1)+\Delta_2$. 

For the second statement of the corollary, we have
\eq{
\sup_{(f, O) \in \oO_{{\tt c}, 2}}\ep(f, O, \pi, Q) &
\leq
D\left(\frac{\alpha_2(Q_{{\tt at}, R})}{\sqrt{ T}}+ \beta_2(Q_{{\tt at}, R})\right)\\
&\leq \frac{DB}{\sqrt{ T}}\cdot\sqrt{ \frac{9+3 \ln s}{(k-1)^2}+1 }\hspace{0.1cm}\\
&\leq \frac{\sqrt{2}DB }{\sqrt{T}},
}
where the first inequality follows by Theorem \ref{t:basic_convergence}, the second inequality follows by upper bounding $\alpha_2(Q_{{\tt at}, R})$ and $\beta_2(Q_{{\tt at}, R})$ using Theorem \ref{t:RATQ_alpha0_beta0}, and the third follows by substituting the parameters in the corollary statement. The upper bound for the strongly convex family follows  in precisely the same manner. In particular, by combining Theorem \ref{t:basic_convergence_sc} with  Theorem \ref{t:RATQ_alpha0_beta0}, we have
\eq{
\sup_{(f, O) \in \oO_{{\tt sc}}}\ep(f, O, \pi, Q) &
\leq \frac{2B^2}{\gamma T}.
}

\end{proof}

{
\begin{rem}
The precision requirement in Corollary \ref{c:PSGD_RATQ_0} matches the $d$-bit lower bound of Corollary \ref{c:OmegaD} upto a multiplicative factor of $O\left(\log \log \log \ln^*(d/3) \right)$.
\end{rem}
}

\subsection{RATQ in the low-precision regime}\label{s:LowPrec}
We present a general method for reducing precision to much below
$d$. This scheme is applicable when the output of the quantizer's encoder is a $d$ length vector, where each coordinate is a separate fixed-length code. We simply reduce the length of the output message vector from the
quantizer's encoder by sub-sampling a subset of coordinates using shared
randomness.  The decoder obtains the values of these coordinates using
the decoder for the original quantizer and sets the rest of the
coordinate-values to zero.  This subsampling layer, which we call the
{\em Random Coordinate Sampler} (RCS), can be added to 
RATQ after applying random rotation. In particular, 
 $RATQ$ we need the parameter $s$ of these quantizers to be set to 1. This requirement of setting $s=1$ ensures that the subsampled coordinates of the rotated vector can be decoded separately. This is a randomized scheme and
requires the encoder and the decoder to share a random set
$S\subset[d]$ distributed uniformly over all subsets of $[d]$ of
cardinality $\mu d$.
 
The encoder $\Qenc_{S}$ of RCS simply outputs the vector
\[\Qenc_{S}(Y)\eqdef\{Y(i), i\in S\},\] and the decoder
$\Qdec_{S}(\tilde{Y})$, when applied to a vector $\tilde{Y}\in \R^{\mu
  d}$, outputs\[\Qdec_S(\tilde{Y})\eqdef\mu^{-1}\sum_{i\in
  S}\tilde{Y}(i)e_i,\] where $e_i$ denotes the $i$th element of
standard basis for $\R^d$.

We can compose RCS with  RATQ with parameter $s=1$  by setting the encoder to $\Qenc_{S}\circ \Qenc$, and setting the
decoder to $\Qdec\circ \Qdec_{S}$. Here we follow the convention that
all $0$-coordinates outputted by $\Qdec_{S}$ are decoded as $0$ by
$\Qdec$.    
 Note that 
since we need to retain RATQ encoder output for only $\mu d$ coordinates, the overall precision of the quantizer is reduced by a factor of $\mu$. We analyze the performance of this combined quantizer in the following theorem.
\begin{thm}\label{t:RCS_RAQ_alpha_beta}
Let $Q_{{\tt at}, R}$ be RATQ  with $s=1$ and $\tilde{Q}$ be the combination of RCS
and $Q_{{\tt at}, R}$ as described above. Then,

\[
\E{\tilde{Q}(Y) | Y }   = \E{
  Q_{{\tt at}, R}(RY)| Y}   \quad \text{and} \quad  \E{\norm{\tilde{Q}(Y) }_2^2| Y }   = \frac{1}{\mu}\E{\norm{Q_{{\tt at}, R}(RY)}_2^2 |Y},
\] 
which further leads to
\[ \alpha_2(\tilde{Q})\leq
\frac{\alpha_2(Q_{{\tt at}, R})}{\sqrt{\mu}} \quad \text{and}
\quad \beta_2(\tilde{Q})= \beta_2(Q_{{\tt at}, R}).\] 
\end{thm}

\begin{proof}
By the description of $Q_{{\tt at},R}$, we have  
\[
\tilde{Q}(Y) = \frac{1}{\mu}
      R^{-1} \sum_{i \in S} Q_{{\tt at},I}(RY) (i)e_i,
\]
where $Q_{{\tt at},I}$ is the output vector formed by combining the $d$
  quantized values outputted by ATUQ ($Q_{\tt at}$) when input is the 
rotated vector. Namely,   
\[
Q_{{\tt at}, I}(RY)=[Q_{{\tt at}}(RY(1)), \cdots,  Q_{{\tt
      at}}(RY(d))]^T.
\]

For the mean of $\tilde{Q}(Y)$, 
it holds that
\begin{align}\label{e:mean_RCS}
 \nonumber
\E{\tilde{Q}(Y) | Y } & = \E{
      R^{-1} \sum_{i \in d} Q_{{\tt at}, I}(RY)(i) e_i \frac{1}{\mu}\mathbbm{1}_{i \in S} |  Y}\\ \nonumber
  & = \sum_{i \in d} \E{
      R^{-1}  Q_{{\tt at}, I}(RY) (i)e_i|  Y}\cdot   \frac{1}{\mu} \E{\mathbbm{1}_{i \in S}|  Y }\\ \nonumber
      & = \sum_{i \in d} \E{
      R^{-1}  Q_{{\tt at}, I}(RY) (i)e_i| Y} \\ \nonumber
       & =  \E{
  R^{-1} \sum_{i \in d} Q_{{\tt at}, I}(RY) (i)e_i| Y}  \\ \nonumber
       & =  \E{
  R^{-1} Q_{{\tt at}, I}(RY)| Y}  \\
     & =  \E{
Q_{{\tt at}, R}(RY)| Y},
\end{align}
where the second identity follows from the fact that randomness used
to generate a set $S$ is independent of the randomness
used in the quantizer and the randomness of $Y$;  the third identity holds since
$P(i \in S)= \mu$. 

Next, moving to the computation of the second moment of the output of $\tilde{Q}$,
we have
\begin{align} \nonumber
\E{\norm{\tilde{Q}(Y)}_2^2|Y} &=\E{\norm{ \frac{1}{\mu}
      R^{-1} \sum_{i \in S} Q_{{\tt at}, I}(RY)(i) e_i}_2^2|Y} \\ \nonumber
   &=   
      \frac{1}{\mu^2}\E{\norm{
     \sum_{i \in S} Q_{{\tt at}, I}(RY)(i) e_i}_2^2|Y}\\ \nonumber
&=\frac{1}{\mu^2}\sum_{i \in [d]}\E{Q_{{\tt at}, I}(RY)(i)^2 |Y}\E{\mathbbm{1}_{i \in S}|Y}\\ \nonumber
&=\frac{1}{\mu}\E{\norm{Q_{\tt at}(RY)}_2^2 |Y} \\ 
&=\frac{1}{\mu}\E{\norm{Q_{{\tt at}, R}(RY)}_2^2 |Y},
\end{align}
where the second identity follows from the fact that $R$ is a unitary matrix and
the remaining steps follow simply by the description of the quantizers used.
It follows that
\eq{
\alpha(\tilde{Q})= \frac{1}{\sqrt{\mu}}\alpha(Q_{{\tt at}, R}),\quad  \beta(\tilde{Q})=\beta(Q_{{\tt at}, R}).
}
\end{proof}


{We now set  the parameter  $k$ to be a constant and sample roughly $r$ coordinates. Specifically,
we set
\begin{align}
  s&=1,\quad \log( k+1)=3,
\nonumber
\\ 
\mu d&=\min\{d, \floor{r/({3+\ceil{\log (1+\ln^*({d}/{3}) )
  }})}\}.
\label{e:RATQ_RCS_params}
\end{align}

For these choices, we have the following corollary.
 \begin{cor}\label{c:PSGD_RCS_RATQ}
For $r \geq 3+\ceil{\log (1+\ln^*({d}/{3}) ) }$, let $Q$
be the composition of RCS and RATQ with parameters set as in~\eqref{e:RATQ_levels},~\eqref{e:RATQ_RCS_params}.
 Then, the optimization protocol $\pi$ in
Algorithm~\ref{a:SGD_Q} satisfies
 \[\sup_{(f, O)\in \oO_{{\tt c}, 2}} \ep(f, O, \pi, Q) \leq 
    \frac{\sqrt{2}DB}{\sqrt{\mu T}},\]
    \[\sup_{(f, O)\in \oO_{{\tt sc}}} \ep(f, O, \pi, Q) \leq 
    \frac{2B^2}{\mu \gamma T}   .\]
\end{cor}
\begin{proof}

When $Q$ is a composition of 
 RCS and RATQ, from Theorem \ref{t:RCS_RAQ_alpha_beta}
 $
 \alpha_2^{\tt m}(Q) \leq \frac{1}{\sqrt{\mu}}\alpha(Q_{{\tt at}, R}), \quad \beta_2^{\tt m}(Q) \leq  \beta(Q_{{\tt at}, R}),
$
which by Theorem \ref{t:basic_convergence} yields
\eq{
\sup_{(f, O) \in \oO_{{\tt c}, 2}}\ep(f, O, \pi, Q) &
\leq
D\left(\frac{ \alpha_2(Q_{{\tt at}, R})}{\sqrt{\mu T}}+ \beta_2(Q_{{\tt at}, R})\right)\\
&\leq \frac{DB}{\sqrt{\mu T}}\cdot\sqrt{ \frac{9}{(k-1)^2}+1 }\hspace{0.1cm}\\
&\leq \frac{\sqrt{2}DB }{\sqrt{T}}\cdot\frac{\sqrt{d}}{\sqrt{\min{\{d, \floor{r/(3+\log \ln^*(d/3))}\}} }},
}
where the second  inequality follows from Theorem \ref{t:RATQ_alpha0_beta0} with $s=1$, and the final inequality is obtained upon substituting the parameters as in the statement of the result. Similarly, for strongly convex function family, the result follows by combining Theorem \ref{t:sconcom_P2} and \ref{t:RCS_RAQ_alpha_beta}.

\end{proof}
\begin{rem}
Note that the convergence rate slows down by a $\mu$ specified in~\eqref{e:RATQ_RCS_params}, which matches the lower bounds in Theorem \ref{t:concom_P2} (for $p=2$) and \ref{t:sconcom_P2} upto a multiplicative factor of $O(\log \ln ^* (d/3))$
\end{rem}
}

\section{Main results for mean square bounded oracles}\label{s:ms} 
In this section, we present results only for the convex family. We do this to avoid repetition since we can derive upper bounds for strongly convex family precisely in the same manner as convex family, as seen from the previous section.

{
With the mean square  bounded assumption, 
we now need to quantize 
random vectors $Y$ such that $\E{\norm{Y}_2^2} \leq B^2$. 
We take recourse to the standard {\em gain-shape} quantization
paradigm in vector quantization ($cf.$\cite{gersho2012vector}). 
\begin{defn}[Gain-shape quantizer]\label{Gain-shape quantizer}
A Quantizer Q is defined to be a gain-shape quantizer if it has the following form
\[Q(Y)=Q_g(\norm{Y}_2)\cdot Q_s(Y/\norm{Y}_2),\]
where $Q_g$ is any $\R \rightarrow \R$ quantizer and $Q_s$ is any $\R^d \rightarrow \R^d$ quantizer.
\end{defn}
Specifically, we separately quantize the {\em gain}
$\norm{Y}_2$ and the {\em shape}{\footnote{\newest{For the event $\norm{Y}_2=0$, we  follow the convention that $Y/\norm{Y}_2=e_1$.}} $Y/\norm{Y}_2$ of $Y$, and form the
estimate of $Y$ by simply multiplying the estimates for the gain and
the shape.
 Note that we already have a good shape quantizer: RATQ. 
We only need to modify the parameters in~\eqref{e:RATQ_levels} to make
it work for the unit sphere; we set
\begin{align}
m=\frac{3}{d}, \quad m_0=\frac{2}{d} \cdot \ln s,  \quad \log h=\ceil{\log(1+\ln^\ast(d/3))}.
\label{e:RATQ_unit_levels}
\end{align}

}

  We now proceed to derive the worst-case $\alpha$ and $\beta$ for  a general gain-shape. In order to make clear the dependence on $B$ and $d$, we refine our notations for  $\{\alpha_2^{\tt m}(Q), \beta_2^{\tt m}(Q)\}$ and $\{\alpha_2(Q), \beta_2(Q)\}$, defined in \eqref{e:alpha,beta} and \eqref{e:alpha0,beta0}, respectively, to   $\{\alpha_2^{\tt m}(Q; B, d), \beta_2^{\tt m}(Q; B, d)\}$  and  $\{\alpha_2(Q; B, d), \beta_2(Q; B, d)\}$.


\begin{thm}\label{t:gain_RATQ}
Let $Q(Y)= Q_{1}(\norm{Y}_2)\cdot Q_{2}(Y/\norm{Y}_2) ,$ where $Q_{1}$ is any gain quantizer and   $Q_{2}$ is any shape quantizer.
Also, suppose $Q_{1}(\norm{Y}_2)$ and  $Q_{2}(Y/\norm{Y}_2)$ are conditionally independent given $Y$. Then,
\begin{align*}  
\alpha_2^{\tt m}(Q; B, d)&\leq   \alpha_2^{\tt m}(Q_{1}; B, 1) \cdot \alpha_2(Q_{ 2};  1, d).
\end{align*}
Furthermore, suppose that $Q_2$ satisfies
\[
\E{Q_{2}(y_s) } = y_s, \quad \forall y_s \quad  s.t. \quad \norm{y_s}_2^2 \leq  1 .
\]
Then, we have\footnote{\newest{The quantity on the right-side of this bound exceeds the bias $\beta_2^{\tt m}(Q_1; B,1)$. Nonetheless, in all our bounds for bias, this is the quantity we have been handling.}}
\eq{\beta_2^{\tt m}(Q; B, d)&\leq \sup_{Y \in \R^d: \E{\norm{Y}_2^2} \leq B^2 }\E{\bigg|\E{Q_{1}(\norm{Y}_2)- \norm{Y}_{2} \mid Y} \bigg|}.}

\end{thm}

\begin{proof}
    Denote by $Y_s$ the shape of the vector $Y$ given by
    \[
    Y_s := \frac{Y}{\norm{Y}_2}.
    \]
\paragraph{The worst-case second moment:}
Towards evaluating $\alpha(Q; B, d)$, we have 
\eq{
\E{\norm{Q(Y)}_2^2} 
& =\E{Q_{1}(\norm{Y}_2)^2 \norm{Q_{2}(Y_s)}_2^2} \\
&=\E{\E{Q_{1}(\norm{Y}_2)^2 \norm{Q_{2}(Y_s)}_2^2|Y}}\\
&=\E{\E{Q_{1}(\norm{Y}_2)^2 |Y} \E{ \norm{Q_{2}(Y_s)}_2^2|Y}}\\
&=\E{\E{Q_{1}(\norm{Y}_2)^2 |Y} \E{ \norm{Q_{2}(Y_s)}_2^2|Y_s}},
}
where the third identity follows by conditional independence of $Q_{1}(\norm{Y}_2)^2$ and $ \norm{Q_{2}(Y_s)}_2^2$
given $Y$ and the fourth follows from the  law of iterated expectations. 

Consider the random variable $\E{ \norm{Q_{2}(Y_s)}_2^2|Y_s}$. We claim that this is less than $\alpha_{0}(Q_{2}; 1, d)$ almost surely. Towards this end, note that
\[
\E{ \norm{Q_{2}(Y_s)}_2^2|Y_s=y} =\E{\norm{Q_{2}(y)}_2^2},
\]
since the randomness used in implementation of $Q_2$ is independent of the input random variable $Y$.
Moreover, for any $y$ with $\norm{y}_2^2 \leq 1$, we have from the definition of $ \alpha_2(Q_{2}; 1, d)$ that
$\E{\norm{Q_{{2}}(y)}_2^2 }  \leq  \alpha_2(Q_{2}; 1, d)^2$.
Therefore, for any $Y$ with $\E{\norm{Y}_2^2} \leq B^2$, we have
\begin{align}\label{e:cond_secmom_gain-RATQ}
\nonumber
\E{\norm{Q(Y)}_2^2} &=\E{\E{Q_{1}(\norm{Y}_2)^2 |Y} \E{ \norm{Q_{2}(Y_s)}_2^2|Y_s}}\\
&\leq \E{\E{Q_{1}(\norm{Y}_2)^2 |Y} } \cdot \alpha_2(Q_{2}; 1, d)^2
\nonumber
\\
&= \E{Q_{1}(\norm{Y}_2)^2 } \cdot \alpha_2(Q_{2}; 1, d)^2
\nonumber
\\
&\leq \alpha_2^{\tt m}(Q_{1}; B, 1)^2 \cdot \alpha_2(Q_{2}; 1, d)^2
 .
 \end{align}
Taking the supremum  of the left-side over all random vectors $Y$ with $\E{\norm{Y}_2^2}\leq B^2$ gives us the desired bound for $\alpha_2^{\tt m}(Q; B, d)$.

\paragraph{The worst-case bias:}  Towards evaluating $\beta_2^{\tt m}(Q; B, d)$, we note from  our hypothesis that
 $\E{ Q_{2}(Y_s)| Y} = \E{ Q_{2}(Y_s)| Y_s} = Y_s$, which further yields 
\begin{align} \label{e:RCSbias-gain-ratq}
\E{Q(Y)-Y} &=\E{\E{Q_{1}(\norm{Y}_2) Q_{ 2}(Y_s)-Y|Y}} \nonumber
\\ \nonumber
&=\E{\E{Q_{1}(\norm{Y}_2)|Y} \E{Q_{2}(Y_s)|Y}-Y}\\
&=\E{\E{Q_{1}(\norm{Y}_2)|Y}  Y_s -\norm{Y}_2Y_s} \nonumber
\\
&=\E{\E{Q_{1}(\norm{Y}_2)- \norm{Y}_{2} |Y} Y_s},
\end{align}
where the second identity uses conditional independence of 
$Q_{1}(\norm{Y}_2)$ and $Q_{ 2}(Y_s)$.
By using the conditional Jensen's inequality, we get
 \eq{ 
\norm{ \E{Q(Y)-Y}}_2 &=
\norm{\E{\E{Q_{1}(\norm{Y}_2)- \norm{Y}_{2} |Y} Y_s}}_2\\
&\leq \E{\norm{\E{Q_{1}(\norm{Y}_2)- \norm{Y}_{2} |Y} Y_s}_2}\\
&= \E{\bigg|\E{Q_{1}(\norm{Y}_2)- \norm{Y}_{2} |Y} \bigg|}.}
 \end{proof}


}

We remark that
quantizers proposed in most of the prior work can be cast in this
gain-shape framework. Most works simply state that gain is a
single parameter which can be quantized using a fixed number of bits;
for instance, a single double precision number is prescribed for
storing the gain. However, the quantizer is not specified. We carefully
analyze this problem and establish lower bounds when a uniform quantizer
with a fixed dynamic range is used for quantizing the gain. Further, we
present our own quantizer which significantly outperforms  uniform gain
quantization.

\subsection{Limitation of uniform gain quantization}\label{s:ug}
We establish lower bounds for a general class of gain-shape quantizers 
$Q(y)=Q_g(\norm{y}_2)Q_s(y/\norm{y}_2)$ of precision $r$ that satisfy the following {\em structural assumptions}: 
\begin{enumerate}\label{en:Assumptions}
\item {\bf (Independent gain-shape quantization)} For any given $y\in \R^d$, the output of the gain and the shape quantizers are independent. 

\item {\bf (Bounded dynamic-range)} \newest{For any $y \in \R^d$, there exists a  $M > 0$ such that whenever $\norm{y}_2> M$, $Q(y)$ has a fixed distribution $P_\emptyset$. }
\item {\bf (Uniformity)} There exists $m \in [M/{2^r}, M]$ such that for every  $t$ in $[0,  m ]$,
\begin{enumerate}
\item ${\tt supp}(Q_g(t))\subseteq \{0,m\}$;
\item If $P(Q_g(t)=m)>0$, then 
\[
\frac{P(Q_g(t_2)=m)}{ P(Q_g(t_1)=m)}\leq~ \frac{t_2}{t_1}, \quad \forall\, 0\leq t_1 \leq t_2 \leq m.
\]
\end{enumerate}
\end{enumerate}

The first two assumptions are perhaps clear and hold for a large class of quantizers. The third one is the true limitation and is satisfied by different forms of uniform gain quantizers. For instance, for the one-dimensional version of CUQ  with dynamic range $[0,M]$, which is an unbiased, uniform gain quantizer with $k_g$ levels, it holds with $m=M/(k_g-1)$ (corresponding to the innermost level $[0,M/(k_g-1)]$). It can also be shown to include a deterministic uniform quantizer that rounds-off at the mid-point. The third condition, in essence, captures the unbiasedness requirement that the probability of declaring higher level is proportional to the value. Note that $(t_2/t_1)$ on the right-side can be replaced with any constant multiple of $(t_2/t_1)$. \newest{For easy reference, we will refer to these assumptions  as Structural Assumptions 1-3.}

Below we present lower bounds for performance of any optimization
protocol using a gain-shape quantizer that satisfies the assumptions
above. We present separate results for high-precision and
low-precision regimes, but both are obtained using a 
general construction that 
exploits the admissibility of heavy-tail  
 distributions for mean square  bounded oracles. This  construction
  is new and may be of independent interest.

\begin{thm}\label{t:lb_1}
  Consider a gain-shape quantizer $Q$ satisfying 
  \newest{Structural Assumptions 1-3.} 
 Suppose that for $\X=\{x:\norm{x}_2\leq D/2\}$ we can find an optimization protocol
$\pi$ which, using at most $T$ iterations, achieves
$
\sup_{f, O \in \oO}\ep(f, O, \pi, Q) \leq
\frac{3DB}{\sqrt{T}}.
$
Then, we can find a universal constant $c$ such that the overall precision $r$ of the quantizer must satisfy 
\[
r \geq c(d+\log T).
\] 
 \end{thm}

 \begin{thm}\label{t:lb_2}
Consider a gain-shape quantizer $Q$ satisfying \newest{Structural Assumptions 1-3.} Suppose that the number of bits \newest{$r_g$} used by the gain quantizer are fixed
independently of $T$. Then, for
$\X=\{x: \norm{x}_2\leq D/2\}$, there exists $(f,O)\in \oO$ such that
for any optimization protocol $\pi$ using at most $T$ iterations, we
must have
\[
\ep(f, O, \pi, Q) \geq \frac{\newest{c(r_g)}DB}{T^{1/3}},
\]
where \newest{$c({r_g})$} is a constant depending only on the number of bits used by
the gain quantizer (but not on $T$).
 \end{thm}
 \noindent The proofs of Theorems~\ref{t:lb_1} and~\ref{t:lb_2} are technical and long; we defer
them to Section \ref{s:lbproof}.
 
\newest{ \begin{rem}
    Thus, from Theorem \ref{t:lb_1}, for any optimization algorithm to achieve the error of $O(1/\sqrt{T})$ after $T$ iterations, when used a quantizer  satisfying the Structural Assumptions 1-3, the precision of the quantizer at every iteration must scale at least
    as roughly $\log T$. Conversely, from Theorem \ref{t:lb_2}, if we use a quantizer
    with precision fixed independently of $T$, then any optimization algorithm
used with this quantizer must have error at least $\Omega(1/T^{1/3})$.
 \end{rem}}
 
 \subsection{A-RATQ in the high-precision regime}\label{s:ARATQh}
Instead of quantizing the gain uniformly, we propose to use an
adaptive quantizer termed {\em Adaptive Geometric Uniform Quantizer} (AGUQ) for
gain. AGUQ operates similar to the one-dimensional
ATUQ, except the possible dynamic-ranges $M_{g,0}, \ldots, M_{g,h}$ grow
geometrically (and not using tetra-iterations of ATUQ) 
as follows:
\begin{align}
M_{g,j}^2= B^2 \cdot a_g^{j}, \quad 0\leq j \leq h_g-1.
\label{e:AGUQ_levels}
\end{align}
Specifically, for a given gain $G\geq 0$, AGUQ first identifies the
smallest $j$ such that $G\leq M_{g,j}$ and then represents $G$ using the one-dimensional version of CUQ with a dynamic range $[0, M_{g,j}]$ and
$k_g$ uniform levels
\eq{ &B_{M_{g,j},k}(\ell) := \ell \cdot
    \frac{M_{g, j}}{k_g-1}, \quad \forall\,  \ell \in \{0,\ldots,
    k-1\}.} 
 As in ATUQ, if $G>M_{h_g -1}$,
the overflow $\emptyset$ symbol is used and the decoder simply outputs
$0$. The overall procedure is the similar to
Algorithms~\eqref{a:E_ATQ} and~\eqref{a:D_ATQ} for $s=1, h=h_g$,  and
$M_j=M_{g,j}$, $0\leq j\leq h_g-1$; the only changes is that now we
restrict to nonnegative interval $[0,M_{g,j}]$ for the one-dimensional version of CUQ with uniform levels $k_g$.

\newest{The following result characterizes the performance of one-dimensional quantizer AGUQ; it is
the only component missing in the analysis of A-RATQ.}
\begin{lem}\label{l:AGUQ_alpha_beta}
Let $Q_{{\tt a}}$ be
the quantizer AGUQ described above, with $h_g \geq 2$. Then,
\begin{align*}
  &\alpha_2^{\tt m}(Q_{{\tt a}}; B, 1)\leq B\sqrt{ \frac{1 }{4(k_g-1)^2} + 
    {\frac{a_g(h_g-1)}{4(k_g-1)^2}} + 1 },\\
  &\beta_2^{\tt m}(Q_{{\tt a}}; B, 1) \leq \sup_{Y \geq 0 \text{ a.s.  } : \E{Y^2} \leq B^2 }\E{\bigg|\E{Q_{a}(Y)- Y |Y} \bigg|}
\leq \frac{B^2}{M_{g, h_g-1}}.
\end{align*}
\end{lem}

\noindent Proof of this result, too, is deferred to Section~\ref{s:AGUQ}. Note that we have derived a bound for a quantity that is slightly larger than the bias of $Q_{\tt a}$, since we want to use this result along with Theorem~\ref{t:gain_RATQ}.  

Thus, our overall quantizer termed the {\em adaptive-RATQ} (A-RATQ) is given by 
\[Q(Y) := Q_{a}(\norm{Y}_2) \cdot Q_{{\tt at}, R}(Y/\norm{Y}_2),\]
where $Q_{a}$ denotes the one dimensional AGUQ and $Q_{{\tt at}, R}$
denotes the $d\text{-dimensional}$ RATQ. Note that we use independent
randomness for  $Q_{a}(\norm{Y}_2)$ and $Q_{{\tt at},
  R}(Y/\norm{Y}_2)$, rendering them  conditionally independent given $Y$. 

The parameters $s,k$ for RATQ and $a_g, k_g$ for AGUQ are yet
to be set. We first present a result which holds for all choices of
these parameters.


\begin{thm}[Performance of A-RATQ]\label{t:NRATQ_alpha_beta}
For $Q$ set to A-RATQ with parameters set as
in~\eqref{e:RATQ_unit_levels},~\eqref{e:AGUQ_levels}, we have 
\begin{align*}  
&\alpha_2^{\tt m}(Q; B, d)\leq  B\sqrt{ \frac{1}{4(k_g-1)^2} + 
    {\frac{a_g(h_g-1)}{4(k_g-1)^2}} + 1 } \cdot \sqrt{
\frac{9+3\ln s}{(k-1)^2}+1} ,\\
  &\beta_2^{\tt m}(Q; B, d)\leq \frac{B^2}{M_{g, h_g-1}}.
\end{align*}
\end{thm}
\begin{proof}~\paragraph{The worst-case second moment  of A-RATQ:}
By Theorem~\ref{t:gain_RATQ} we have
 \eq{
 \alpha_2^{\tt m}(Q; B, d) 
&\leq  \alpha_2^{\tt m}(Q_{\tt a}; B, 1) \cdot   \alpha_2(Q_{{\tt at}, R}; 1, d)\\
& \leq \alpha_2^{\tt m}(Q_{\tt a}; B, 1) \cdot \sqrt{
\frac{9+3\ln s}{(k-1)^2}+1}\\
& \leq B\sqrt{ \frac{1}{4(k_g-1)^2} + 
    {\frac{a_g(h_g-1)}{4(k_g-1)^2}} + 1 }  \cdot \sqrt{
\frac{9+3\ln s}{(k-1)^2}+1} ,
}
where the second inequality used Theorem~\ref{t:RATQ_alpha0_beta0} with $B=1$,
and the third follows by Lemma~\ref{l:AGUQ_alpha_beta}. 
\paragraph{The worst-case bias of A-RATQ:}
With parameters of RATQ set as in~\eqref{e:RATQ_unit_levels}, we have that 
\eq{ 
\E{Q_{{\tt at}, R}(y)}=y, \quad \forall y \quad s.t \quad \norm{y}_2^2 \leq 1. }
 Therefore, by Theorem~\ref{t:gain_RATQ} it follows that
 \eq{ 
\beta_2^{\tt m}(Q; B, d)&
\leq
\sup_{Y: \E{\norm{Y}_2^2} \leq B^2 }\E{\bigg|\E{Q_{\tt a}(\norm{Y}_2)- \norm{Y}_{2} |Y} \bigg|}
\leq \frac{B^2}{M_{g,h_g-1}},}
 where the second  inequality follows from Lemma \ref{l:AGUQ_alpha_beta}.
\end{proof}

Note that RATQ yields an unbiased estimator; the bias in A-RATQ arises
from AGUQ since the gain is not bounded. Further, AGUQ uses a precision
of $\ceil{\log h_g} + \ceil{\log (k_g+1)}$ bits, and therefore, the
overall precision of A-RATQ is $\ceil{\log h_g} + \ceil{\log (k_g+1)}+
\ceil{d/s}\ceil{\log h}+ d \ceil{\log (k+1)}$ bits.

In the high-precision regime, we set 
\begin{align}
a_g&=2,\quad \log h_g =
\ceil{\log(1+\frac{1}{2}\log T)}, 
\nonumber
\\
\log(k_g+1)&=\ceil{\log \left( 2+\frac{1}{2}\sqrt{\log T +1} \right)}.
\label{e:ARATQ_gain_bits}
\end{align}

\begin{cor}\label{c:PSGD_NRATQ} Denote by $Q$ the quantizer A-RATQ
  with parameters set as
  in~\eqref{e:RATQ_unit_levels},~\eqref{e:RATQ_bits},
  and~\eqref{e:ARATQ_gain_bits}. Then,  
the overall precision $r$ used by $Q$
 is less than 
\[
d(1+\Delta_1)+\Delta_2+  \ceil{\log \bigg( 2+\sqrt{\log T +1} \bigg)},
\] 
where $\Delta_1=\ceil{\log \left(2+  \sqrt{  9 + 3 \ln \Delta_2 }\right)}$ and 
$\Delta_2=\ceil{\log (1+\ln^*({d}/{3})) }$, the same as
Corollary~\ref{c:PSGD_RATQ_0}. 
Furthermore, the optimization protocol $\pi$
  given in algorithm \ref{a:SGD_Q} satisfies
 $\sup_{(f,O)\in \oO}  \ep(f, O, \pi, Q) \leq {3DB}/{\sqrt{T}}$.
\end{cor}
\begin{proof}
The proof is similar to the proof of Corollary \ref{c:PSGD_RATQ_0}. The first statement follows by simply upper bounding the precision of the fixed-length code for A-RATQ with parameters as in the statement. The second statement follows by bounding  
$\sup_{(f,O)\in \oO}  \ep(f, O, \pi, Q)$ using Theorem \ref{t:basic_convergence}, using the upper bounds for $\alpha$ and $\beta$ given in Theorem~\ref{t:NRATQ_alpha_beta}, and finally substituting the parameters.
\end{proof}

\begin{rem}
The  precision used in Corollary \ref{c:PSGD_NRATQ} 
  matches the lower bound in Corollary \ref{c:OmegaD} upto an additive
  factor  of $\log \log T$ (ignoring the mild factor of  $\log \log \log
  \ln^* (d/3)$), which is much lower than the $\log T$ lower bound we established for uniform
  gain quantizers. \newest{Hence, the precision requirement of A-RATQ in the high-precision regime is considerably smaller than the precision requirement of uniform gain quantizers established in Theorem~\ref{t:lb_1} for  $ \log T \gg d(1+\Delta_1)$,  while remaining roughly the same for $\log  T =O(d(1+\Delta_1))$.}
\end{rem}

 \subsection{A-RATQ in the low-precision regime}\label{s:ARATQl}
  In order to operate with a fixed precision $r$, we combine
 A-RATQ with RCS. 
  We simply combine RCS with RATQ as 
 in Section \ref{s:LowPrec} to limit the precision and use AGUQ as the gain quantizer. Note that we use independent
randomness in our gain quantizer  $Q_{a}(\norm{Y}_2)$ and our shape quantizer $\tilde{Q}(Y/\norm{Y}_2)$, rendering them  conditionally independent given $Y$. We have the following theorem characterizing $\alpha$ and $\beta$ for this  quantizer. 

 \begin{thm}\label{t:RCS_Gain-RATQ}
Let $Q(Y)=Q_a(\norm{Y})\cdot \tilde{Q}(Y/\norm{Y}_2)$, where $\tilde{Q}$ is the composition of RCS and RATQ described in Theorem \ref{t:RCS_RAQ_alpha_beta} with parameters $m$, $m_0$, and $h$ of RATQ as in \eqref{e:RATQ_unit_levels}  and $Q_a$ is AGUQ.  Then, 
\eq{ \alpha_2^{\tt m}(Q; B, d)&\leq  B\sqrt{ \frac{1}{4(k_g-1)^2} + 
    {\frac{a_g(h_g-1)}{4(k_g-1)^2}} + 1 }  \cdot \frac{1}{\sqrt{\mu}}  \sqrt{
\frac{9+3\ln s}{(k-1)^2}+1 },\\
  \beta_2^{\tt m}(Q; B, d)&\leq \frac{B^2}{M_{g, h_g-1}}.}

\end{thm}
\begin{proof}~\paragraph{The worst-case second moment:} 
Starting by applying Theorem~\ref{t:gain_RATQ}, we have
 \eq{
 \alpha_2^{\tt m}(Q; B, d) 
 &\leq  \alpha_2^{\tt m}(Q_{\tt a}; B, 1) \cdot   \alpha_2(\tilde{Q}; 1, d)\\
&\leq  \alpha_2^{\tt m}(Q_{\tt a}; B, 1) \cdot   \frac{1}{\sqrt{\mu}}\alpha_2(Q_{{\tt at}, R}; 1, d)\\
& \leq B\sqrt{ \frac{1}{4(k_g-1)^2} + 
    {\frac{a_g(h_g-1)}{4(k_g-1)^2}} + 1 }  \cdot \frac{1}{\sqrt{\mu}}\sqrt{
\frac{9+3\ln s}{(k-1)^2}+1} ,
}
where the second inequality follows by Theorem~\ref{t:RCS_RAQ_alpha_beta} and the third follows by Theorem~\ref{t:RATQ_alpha0_beta0} and Lemma~\ref{l:AGUQ_alpha_beta}.
\paragraph{The worst-case bias:} With parameters of RATQ set as in~\eqref{e:RATQ_unit_levels}, we have that 
\eq{ 
\E{\tilde{Q}(y)}=y, \quad \forall y \quad s.t \quad \norm{y}_2^2 \leq 1. }
 Therefore, by Theorem \ref{t:gain_RATQ} we get
 \eq{ 
\beta_2^{\tt m}(Q; B, d)&
\leq
\sup_{Y: \E{\norm{Y}_2^2} \leq B^2 }\E{\bigg|\E{Q_{\tt a}(\norm{Y}_2)- \norm{Y}_{2} |Y} \bigg|}\\
&\leq \frac{B^2}{M_{g,h_g-1}},}
 where the second  inequality follows from Lemma \ref{l:AGUQ_alpha_beta}.

\end{proof}
 
 We divide the total precision $r$ into $r_g$ and $r_s$ bits: $r_g$ to quantize the gain, $r_s$ to quantize the subsampled shape vector.
We set  
\begin{align}
 & s, k, \text{~and~} \mu d \text{~as in \eqref{e:RATQ_RCS_params}, with $r_s$ replacing $r$,}
\nonumber 
 \\ 
&{\log h_g}=\log (k_g+1)={\frac{r_g}{2}}, 
~a_g=\left(\mu T\right)^{\frac{1}{h_g+1}}
\label{e:ARATQ_RCS_params}
\end{align}
That is, our shape quantizer simply quantizes $\mu d$ randomly chosen
coordinates of the rotated vector using ATUQ with $r_s$ bits, and the
remaining bits are used by the gain quantizer AGUQ. 
The result below shows the performance of this quantizer.

 \begin{cor}\label{c:PSGD_RCS_ARATQ_fixed}
For any $r$ with gain quantizer being assigned  $r_g \geq 4$ bits and shape quantizer being assigned $r_s \geq 3+\ceil{\log (1+\ln^*({d}/{3}) ) }$, let $Q$
be the combination of RCS and A-RATQ with parameters set as
in~\eqref{e:RATQ_unit_levels},~\eqref{e:AGUQ_levels},~\eqref{e:ARATQ_RCS_params}. Then
for $\mu T\geq 1$, the optimization protocol $\pi$ in
Algorithm~\ref{a:SGD_Q} 
can obtain  
\[\sup_{(f, O)\in \oO} \ep(f, O, \pi, Q)   \leq
O\left(DB \left(\frac{d}{T \min\{d, \frac{r_s}{\log \ln^*(d/3)}\}}\right)^{\frac 12\cdot\frac {2^{r_{g}/2}-1}{2^{r_{g}/2}+1}} \right).
\]
 \end{cor}
 \begin{proof} By using Theorem~\ref{t:basic_convergence} to upper bound $\sup_{(f, O) \in \oO_{{\tt c}, 2}^{\tt m}}\ep(f, O, \pi, Q)$ and then Theorem \ref{t:RCS_Gain-RATQ} to upper-bound $\alpha$ and $\beta$, we get
\eq{
\sup_{(f, O) \in \oO_{{\tt c}, 2}^{\tt m}}\ep(f, O, \pi, Q) &
\leq  D\left(\frac{1}{\sqrt{\mu T}}\sqrt{ \frac{B^2}{4(k_g-1)^2}+\frac{a_g(h_g-1)B^2}{4(k_g-1)^2}
+B^2}\sqrt{
\frac{9+3\ln s}{(k-1)^2}+1}+ \frac{B^2}{M_{g, h-1}}\right).
}
 By substituting the parameters as in the statement and using the fact that $\mu T \geq 1$ completes the proof.
\end{proof}
 
 \begin{rem}\label{r:rem_MS}
Our fixed precision quantizer in
  Corollary~\ref{c:PSGD_RCS_ARATQ_fixed} establishes that using only a constant number of bits for gain-quantization, we get very close to the lower  bound in Theorem~\ref{t:concom_P2}. For instance, given access to a large enough precision $r$,  if we set $r_g$ to be $16$ bits, we get
  \[\sup_{(f, O)\in \oO} \ep(f, O, \pi, Q)   \leq
O\left(DB \left(\frac{d}{T \min\{d, \frac{r -16}{\log \ln^*(d/3)}\}}\right)^{\frac{1}{2}\cdot \frac{255}{257}}\right).
\]
Here, the ratio of $d/(\min\{d, \frac{r -16}{\log \ln^*(d/3)}\})$ is very close to the optimal ratio of $d/(\min\{d, r\})$, and the exponent $255/(2\cdot 257)$ is close to the optimal exponent $1/2$.
 
  \end{rem}

\begin{rem}
  We remark that A-RATQ satisfies Assumptions (1) and (2) in Section \ref{s:ug} but not (3), and breaches the lower bound  for uniform gain quantizers established in Section~\ref{s:ug}.
  \end{rem}


\subsection{A variable-length quantizer}
So far in this thesis, we have restricted our quantizers to be fixed-length. We now present a variable-length quantizer for mean square bounded oracles which improves over the convergence rate of Corollary \ref{c:PSGD_RCS_ARATQ_fixed}. The quantizer we present is a gain-shape quantizer that uses RATQ as the shape quantizer but uses a variable-length version of the gain quantizer called AGUQ$^+$, an update on AGUQ presented in the previous section.

AGUQ$^+$ differs from AGUQ in two crucial aspects: 1)
The number of uniform levels of the uniform quantizer for different dynamic ranges is different. Denote by $k_{g, j}$ the number of uniform levels corresponding to the range $[0, M_{g, j}]$. We choose  $k_{g, j}$ to grow geometrically.
2)  AGUQ$^+$ employs entropic compression further to reduce the expected code-length of the quantized representation. 
Besides the differences, 
the quantization in AGUQ$^+$ is the same as that of AGUQ. 

Specifically, AGUQ$^+$ is once again an adaptive quantizer like AGUQ with dynamic ranges growing in a goemetric manner, precisely in the same way as in \eqref{e:AGUQ_levels}.  
For a given gain $G\geq 0$, AGUQ$^+$ first identifies the
smallest $j$ such that $G\leq M_{g,j}$ and then represents $G$ using the one-dimensional version of CUQ with a dynamic range $[0, M_{g,j}]$ and
$k_{g, j}$ uniform levels
\eq{ &B_{M_{g,j}}(\ell) := \ell \cdot
    \frac{M_{g, j}}{k_{g, j}-1}, \quad \forall\,  \ell \in \{0,\ldots,
    k_{g, j}-1\}.} 
 As in AGUQ, if $G>M_{g, h_g -1}$,
the overflow $\emptyset$ symbol is used and the decoder simply outputs
$0$. 

Note that since we are quantizing unbounded random variables, it is difficult to avoid bias. Nevertheless, we will make the effect of the bias negligible. Specifically, for the application of communication-constrained optimization of convex functions, it will be desirable to have a bias of at the most $1/\sqrt{T}$. To achieve this, we set
\begin{align}\label{e:AGUQparam}
a_g=2, \quad h_g=1+\frac{1}{2}\log T, \quad k_{g, j}+1= 2^{j+1}.
\end{align}
We now describe the variable length coding procedure.  The variable-length bit string itself is a concatenation of two separate bit strings. The first string represents the non-uniform level $j$ in $\{0, \cdots, h_g-1\}$ using the first $h_g$ symbols of the  Huffman code for geometric distribution with parameter $1/2$. The second string uses a fixed-length code of $k_{g, j}$ bits. We will show that the total number of bits used for both the strings would be $O(1)$ in expectation.

\begin{rem}[Entropic compression in $AGUQ^+$]
$AGUQ^+$ improves over vanilla $AGUQ$ by exploiting that the probability of a larger dynamic range being chosen decays exponentially with the dynamic range level $j$. For concreteness, since $\E{Y^2}\leq B^2,$ we have by Markov's inequality
$P(|Y| > M_{g, j-1})\leq B^2/M^2_{g, j-1}= a_g^{-j}=2^{-j+1}.$ This probability bound allows us to use a code similar to that of Huffman code for geometric distribution and only have a constant code-length.

\end{rem}
The following result characterizes the performance of one-dimensional quantizer AGUQ$^+$.
\begin{lem}\label{l:AGUQ+_alpha_beta}
Let $Q_{{\tt a^+}}$ be
the quantizer  AGUQ$^+$ described above with parameters set as in \eqref{e:AGUQ_levels} and \eqref{e:AGUQparam}. Then,  for 
  $Y$ such that $\E{Y^2}\leq B^2,$  $Q_{\tt a^+}(Y)$ can be represented using at the most $O(1)$ bits of precision in expectation,
\begin{align*}
  &\alpha_2^{\tt m}(Q_{\tt a^+}; B, 1) \leq O\left(1\right), \text{ and } \\& \beta_2^{\tt m}(Q_{{\tt a^+}}; B, 1) \leq \sup_{Y \geq 0 \text{ a.s.  } : \E{Y^2} \leq B^2 }\E{\bigg|\E{Q_{a}(Y)- Y |Y} \bigg|} \leq   O\left(\frac{B}{\sqrt{T}}\right).
\end{align*}
\end{lem}
\noindent The proof is deferred to Section \ref{s:proof_Aguq^+}

\begin{rem}
Thus employing a gain-shape quantizer where the gain is quantized by AGUQ$^+$
 and the shape is quantized by the subsampled version of RATQ along with PSGD improves over the convergence guarantees of Corollary \ref{c:PSGD_RCS_ARATQ_fixed}, and essentially has the same order of convergence guarantees as that in Corollary \ref{c:PSGD_RCS_RATQ}. That is, this particular gain-shape quantizer along with PSGD achieves roughly the convergence rate of $O\left( \frac{DB}{\sqrt{T}}\cdot \frac{d}{r} \right)$, which is the same as that in lower-bound for communication constrained optimization of convex and $\ell_2$ lipschitz functions, $\oO_{{\tt c}, 2}^{m}$, as stated in Theorem \ref{t:concom_P2}. From Remark \ref{rem:varaible_length_opt}, the lower bound in Theorem \ref{t:concom_P2} holds for variable length quantizers, too, provided the gradient processing is done in a nonadaptive manner. Thus employing a gain-shape quantizer where the gain is quantized by AGUQ$^+$
 and the shape is quantized by the subsampled version of RATQ along with PSGD is optimal for communication-constrained optimization of $\oO_{{\tt c}, 2}^{m}$ when nonadaptive, variable-length quantizers are allowed.
 \end{rem}
 


\section{Main proofs}\label{s:Ch_l2_opt_proof}

\subsection{ Proof of Theorem \ref{t:basic_convergence}}\label{ap:QPSGD}

We proceed as in the standard proof of convergence (see, for
instance,~\cite{bubeck2015convex}): Denoting by $\Gamma_\X(x)$ the
projection of $x$ on the set $\X$ (in the Euclidean norm), the error
at time $t$ can be bounded as
\eq{
  \norm{x_t-x^*}_{2}^{2} &= \norm{\Gamma_{\X}\big(x_{t-1}-\eta
    Q(\hat{g}(x_{t-1}))\big)-x^*}_{2}^{2} \\
    & \leq \norm{\big(x_{t-1}-\eta
    Q(\hat{g}(x_{t-1}))\big)-x^*}_{2}^{2} \\ &
    = \norm{x_{t-1}-x^*}_{2}^{2} + \norm{\eta
    Q(\hat{g}(x_{t-1}))}_{2}^{2}- 2 \eta
    (x_{t-1}-x^*)^{T}Q(\hat{g}(x_{t-1}))\\
    &=\norm{x_{t-1}-x^*}_{2}^{2} + \norm{\eta
    Q(\hat{g}(x_{t-1}))}_{2}^{2}- 2 \eta
    (x_{t-1}-x^*)^{T}\big(Q(\hat{g}(x_{t-1}))
    - \hat{g}(x_{t-1})\big) \\& \quad - 2 \eta
    (x_{t-1}-x^*)^{T}\hat{g}(x_{t-1}), } where the first inequality is
    a well known property of the projection operator $\Gamma$ (see,
    for instance, Lemma 3.1,
\cite{bubeck2015convex}).  By rearranging the terms, we have
\begin{align}\label{eq:gd_me}
  2\eta (x_{t-1}-x^*)^{T}\hat{g}(x_{t-1})
  &\leq \norm{x_{t-1}-x^*}_{2}^{2} - \norm{x_{t}-x^*}_{2}^{2}
  + \norm{\eta Q(\hat{g}(x_{t-1}))}_{2}^{2} \\ & \hspace{1cm} -2 \eta
  (x_{t-1}-x^*)^{T}\left(Q(\hat{g}(x_{t-1}))
  - \hat{g}(x_{t-1})\right).  
 \end{align}
 Also, since $\E{\hat{g}(x_{t-1})|
  x_{t-1}}$ is a subgradient at $x_{t-1}$ for the convex function $f$,
  upon taking expectation \newest{over the randomness in the subgradient estimates as well as the quantizer output} we get
\begin{align}\label{eq:gd_con}
\E{f(x_{t-1})-f(x^*)}\leq
\E{(x_{t-1}-x^*)^{T}\E{\hat{g}(x_{t-1})| x_{t-1}}}, 
 \end{align}
which with the previous bound yields
\eq{
  2\eta\E{f(x_{t-1})-f(x^*)} & \leq \E{\norm{x_{t-1}-x^*}_{2}^{2} }
  - \E{\norm{x_{t}-x^*}_{2}^{2} }+ \eta^2 \E{ \norm{
  Q(\hat{g}(x_{t-1}))}_{2}^{2}} \\ & \hspace{1cm} -
  2 \eta \E{(x_{t-1}-x^*)^{T}\big(Q(\hat{g}(x_{t-1}))
  - \hat{g}(x_{t-1})\big)}.  } Next, by the Cauchy-Schwarz inequality
  and the assumption in (1), the third term on the right-side above
  can be bounded further to obtain
\eq{2\eta\E{f(x_{t-1})-f(x^*)} & \leq \E{\norm{x_{t-1}-x^*}_{2}^{2} }
  - \E{\norm{x_{t}-x^*}_{2}^{2} }+ \eta^2 \E{ \norm{
      Q(\hat{g}(x_{t-1}))}_{2}^{2}} \\& \hspace{2 cm} + 2 \eta\cdot
      D\cdot \E{\norm{\E{Q(\hat{g}(x_{t-1}))
      - \hat{g}(x_{t-1})|x_{t-1} }}_2} .  } Finally, we note that, by
      the definition of $\alpha$ and $\beta$, for $L_2$-bounded
      oracles we have
\begin{align*}
\E{ \norm{
    Q(\hat{g}(x_{t-1}))}_{2}^{2}} &\leq \alpha_2^{\tt m}(Q)^2,
\\
\norm{\E{Q(\hat{g}(x_{t-1})) - \hat{g}(x_{t-1})|x_{t-1} }}_2&\leq \beta_2^{\tt m}(Q),
\end{align*}

which gives
\eq{2\eta\E{f(x_{t-1})-f(x^*)} & \leq \E{\norm{x_{t-1}-x^*}_{2}^{2} }
  - \E{\norm{x_{t}-x^*}_{2}^{2} }+ \eta^2 \alpha_2^{\tt m}(Q)^2+ 2 \eta
    D \beta_2^{\tt m}(Q). }
 
Therefore, by summing from $t=2$ to $T+1$, dividing by $T$, and using
assumption that the domain $\X$ has diameter at the most $D$, we have
\eq{
  2\eta\E{f(\bar{x}_{T})-f(x^*)}
  & \leq \frac{D^2}{T}+ \eta^2 \alpha_2^{\tt m}(Q)^2+ 2 \eta D \beta_2^{\tt m}(Q). }
 
The first statement of Theorem~\ref{t:basic_convergence} follows upon
dividing by $\eta$ and setting the value of $\eta$ as in the
statement. The second statement holds in a similar manner by replacing
$\alpha$ and $\beta$ with $ \alpha_2$ and $ \beta_2$, respectively.
\qed

\subsection{Proof of Theorem \ref{t:basic_convergence_sc}}\label{ap:QPSGD_SC}

Note that since the gradient descent step remains  the same, \eqref{eq:gd_me} still holds. Except now,   instead of \eqref{eq:gd_con}, we have a stronger inequality
\begin{align}\label{eq:gd_sc}
  \E{f(x_{t-1})-f(x^*)}\leq
\E{(x_{t-1}-x^*)^{T}\E{\hat{g}(x_{t-1})| x_{t-1}}} -\frac{\gamma}{2}\norm{x_{t-1}-x^*}_2^2,
\end{align}
Combining \eqref{eq:gd_me} and \eqref{eq:gd_sc}, and then using the definition of $\alpha$, $\beta$ as in the previous proof, we get 
\eq{
2\eta_{t-1}\E{f(x_{t-1})-f(x^*)} & \leq \E{\norm{x_{t-1}-x^*}_{2}^{2} }
  - \E{\norm{x_{t}-x^*}_{2}^{2} }+ \eta^2_{t-1} \alpha_2(Q)^2+ 2 \eta_{t-1}
    D \beta_2(Q) \\& \hspace{1cm} -\frac{\gamma}{2}\norm{x_{t-1}-x^*}_2^2. 
    }
    Then multiplying by $(t-1)/2\eta_{t-1}$, we get
    \eq{
    (t-1)\E{f(x_{t-1})-f(x^*)} & \leq \frac{(t-1)\eta_{t-1}}{2} \alpha_2(Q)^2+\left(\frac{t-1}{2\eta_{t-1}}-\frac{(t-1)\gamma}{2} \right)\E{\norm{x_{t-1}-x^*}_{2}^{2} }
  \\& \hspace{1cm} -\frac{t-1}{2\eta_{t-1}}\E{\norm{x_{t}-x^*}_{2}^{2} }+  (t-1)
    D \beta_2(Q). 
    }
    
    Substituting for $\eta_{t-1}$ as in the theorem statement, summing the resultant inequality from $t=1$ to $T$, and then applying Jensen's inequality completes the proof.



\subsection{Proof of Theorem~\ref{t:RATQ_alpha0_beta0}}\label{s:ProofRATQ}
\paragraph{Step 1: Analysis of CUQ.}
We first prove a result for CUQ (with a dynamic range of $[-M, M]$)
which will bound the expected value of
\[
\sum_{i\in[d]}\big(Q_{{\tt
    u}}(Y)(i)-{Y(i)}\big)^2\mathbbm{1}_{\{|{Y(i)|}\leq M\}},
\]
namely the mean square error when there is no overflow. This will be
useful in the analysis of RATQ, too.     
\begin{lem}\label{l:sup}
  For an $\R^d$-valued random variable $Y$ and $Q_{{\tt u}}$ denoting the quantizer CUQ  
  with parameters $M$ (with dynamic range $[-M,M]$)
 and $k$, let $Q_{{\tt u}}(Y)$ be the quantized value of $Y$.  Then, 
 \eq{ \E{ \sum_{i\in[d]}\big(Q_{{\tt
        u}}(Y)(i)-{Y(i)}\big)^2\mathbbm{1}_{\{{|Y(i)|}\leq M\}} \mid Y
   } \leq \frac{dM^2}{(k-1)^2} \left( \frac{1}{d}\sum_{j\in[d]}
   \mathbbm{1}_{\{{|Y(j)|}\leq M\}}\right).  }
\end{lem}
\noindent The proof is relatively straightforward with the
calculations similar to \cite[Theorem 2]{suresh2017distributed}; it is
deferred to Section \ref{ap:CUQ}.

Also, the quantizer AGUQ in Section~\ref{s:ARATQh} uses the
one-dimensional CUQ with dynamic range $[0, M]$ as a subroutine. The
uniform levels for this variant of CUQ are given by
\[
B_{M,k}(\ell)=\ell\cdot\frac{M}{k-1}, \forall  \ell \in [k-1].
\]
We have the following lemma for this variant of CUQ.
\begin{lem}\label{l:sup0m}
  For an $\R$-valued random variable $Y$ which is almost surely nonnegative
  and the quantizer $Q_{\tt u}$ with dynamic range $[0,M]$ and parameter $k$, let $Q_{{\tt u}}(Y)$
denote the quantized value of $Y$. Then, 
\eq{
  \E{\big(Q_{{\tt
        u}}(Y)-{Y}\big)^2\mathbbm{1}_{\{{|Y|}\leq M\}} \mid Y } 
  \leq \frac{M^2}{4(k-1)^2}  \left( 
  \mathbbm{1}_{\{{|Y|}\leq M\}}\right).
} 
\end{lem}
\noindent The proof is very similar to the proof of Lemma \ref{l:sup} and is deferred to Section \ref{ap:CUQ}.

\paragraph{Step 2: Mean square error for adaptive quantizers.}
The quantizers RATQ and A-RATQ use ATUQ as subroutine; in addition, A-RATQ uses AGUQ for gain quantization. Thus, in order to analyze RATQ and A-RATQ,  we need to analyze
 ATUQ  and AGUQ first. 

In this step we provide a general bound on the mean square error of adaptive quantizers. We capture the performances of ATUQ and AGUQ in two separate results below. 
\begin{lem}\label{l:sup2}
For an $\R^d$-valued random variable  $Y$ and $Q$ denoting the quantizer ATUQ with dynamic-range
parameters $M_j$s, we have
\eq{
\E{ \sum_{i\in[d]}\big(Q(Y)(i)-{Y(i)}\big)^2\mathbbm{1}_{\{{|Y(i)|}\leq M_{h-1}\}}}  \leq \frac{d}{(k-1)^2}\left(m+m_0+ \sum_{j=1}^{h-1}
M_j^2P\left(\norm{Y}_{\infty} > M_{j-1}\right)\right). 
}
\end{lem}
\begin{proof}
  Consider the events $A_j$s corresponding to different levels used by
  the adaptive quantizer of the norm, defined as follows:
  \begin{align*}
    A_0 &:=   \{\norm{Y}_{\infty} \leq m \},
    \\
    A_j &:= \{M_{j-1}  <\norm{Y}_{\infty} \leq M_{j} \}, \quad \forall
    j \in [h-2],\\
     A_{h-1} &:= \{M_{h-2}  <\norm{Y}_{\infty} \}
.  \end{align*}
By construction, $\sum_{j =0}^{h-1}\indic{A_j}=1 \text{ a.s.}$. Therefore, we have
 \eq{ 
\E{\sum_{i\in[d]}\big(Q(Y)(i)-{Y(i)}\big)^2\mathbbm{1}_{\{{|Y(i)|}\leq M_{h-1}\}}}
&= \E{\norm{Q_{{\tt }}(Y)-{Y}}_2^2\indic{A_0}} + \sum_{j=1}^{h-2}\E{\norm{Q_{{\tt }}(Y)-{Y}}_2^2\indic{A_j}} \\
&\hspace{1cm} 
+\E{\sum_{i\in[d]}\big(Q_{{\tt
        u}}(Y)(i)-{Y(i)}\big)^2\mathbbm{1}_{\{{|Y(i)|}\leq M_{h-1}\}}\indic{A_{h-1}}}.
  }
Note that $ \mathbbm{1}_{A_0}$ implies
that we are using a $k$-level uniform quantization with a dynamic
range of $[-m, m]$.  Therefore, this term can be bounded by
Lemma~\ref{l:sup} as follows:
\eq{
\E{\norm{Q_{{\tt }}(Y)-{Y}}_2^2\indic{A_0}}\ \leq \frac{dm}{(k-1)^2}.
}  
Under the event $A_j$ with $j \in [h-1]$, we use   a $k$-level uniform
quantization with a dynamic range of $[-M_{j}, M_{j} ].$ Therefore, by
Lemma~\ref{l:sup}, we have 
\eq{
 \E{\norm{Q_{{\tt }}(Y)-{Y}}_2^2\indic{A_j}}
\leq & \frac{dM_j^2}{(k-1)^2}  \E{\mathbbm{1}_{A_j}}\\
\leq & \frac{dM_j^2}{(k-1)^2}  P\left(\norm{{Y}}_{\infty} > M_{j-1}\right).
}
\end{proof}
Note that the proof above does note use specific form of $M_j$'s and therefore applies as it is for the one-dimensional AGUQ gain quantizer used in A-RATQ; the only change is the fact that instead of using Lemma \ref{l:sup} for uniform quantization we use Lemma \ref{l:sup0m}. This leads to the following lemma, which will be useful later in the analysis of A-RATQ.

\begin{lem}\label{l:sup3}
For an $\R$-valued random variable $Y$ which is almost surely nonnegative
and $Q$ denoting the quantizer \newest{AGUQ} with dynamic-range
parameters $M_{g,j}$s, we have
\eq{
\E{(Q_{{\tt }}(Y)-{Y})^2\mathbbm{1}_{\{|Y| \leq
    M_{g, h-1}\}}}  \leq \frac{1}{4(k-1)^2}\left(B^2+ \sum_{j=1}^{h-1}
M_{g,j}^2P\left(|Y| > M_{g,j-1}\right)\right). 
}
\end{lem}
\noindent The proof is similar to that of Lemma~\ref{l:sup2} and is omitted.

\paragraph{Step 3: Mean square error of ATUQ for a subgaussian input vector.}
In our analysis, we need to evaluate the performance of ATUQ for {\em subgaussian} input vectors.
\begin{defn}[$cf.$~\cite{boucheron2013concentration}]\label{d:subg}
 A centered random variable $X$ is said to be {subgaussian} with {variance factor $v$} if for all $\lambda$ in $\R$, we have
 \[
 \ln \E{e^{\lambda X}}\leq \frac{\lambda^2 v}{2}.
 \]
\end{defn}
The following well-known fact ($cf.$~\cite[Chapter 2]{boucheron2013concentration}) will be used throughout. 
 \begin{lem}\label{l:standar_subg}
For a centered subgaussian random variable $X$ with variance factor $v$ the 
\begin{align*}
&P(|X| > x) \leq 2 e^{-x^2/2v}, \quad\forall\, x>0,
\\
&\E{X^2} \leq 4 v, \quad \E{X^4} \leq 32 v^2.
\end{align*}
\end{lem} 
 

 Next, consider the quantizer $Q_{{\tt at}, I}$ which is similar to RATQ but skips the rotation step. Specifically, $Q_{{\tt at}, I}$ is obtained by replacing the random matrix $R$ in the encoder and decoder of RATQ (given in Algorithms \ref{a:E_RATQ} and \ref{a:D_RATQ}, respectively) by the identity matrix $I$. Symbolically, the quantizer $Q_{{\tt at}, I}$ can be described as follows for the $d$-dimensional input vector $Y$
 \begin{equation}\label{e:Q_at_I}
 Q_{{\tt at}, I}(Y)=[Q_{\tt at}(Y_1)^T, \cdots, Q_{\tt at}(Y_{\ceil{d/s}})^T],
 \end{equation}
where $Q_{\tt at}$ is the quantizer ATUQ and $Y_i$ is the $i^{th}$  subvector of $Y$. Recall that the $i^{th}$ subvector $Y_i$ comprises
the coordinates $\{(i-1)s+1,\cdots, \min\{is,d\}\}$,
 for all $i \in [d/s].$ Also, recall that the dimension of all the sub
 vectors except the last one is $s$, with the last one having 
 dimension $d-s\floor{d/s}$.

 Notice that like RATQ,  $Q_{at, I}$ has parameters $k$, $h$, $s$, $m$, and $m_0$ which need to be set.
 We set the parameters $m$ and $m_0$ to be $3v$ and $2v \ln s$,
 respectively, and prove a general lemma in terms of the other
 parameters of $Q_{at, I}$ for a subgaussian input vector.
\begin{lem}\label{l:subg_mse}
Consider $Y=[Y(1),\ldots, Y(d))]^T,$ where for all $i$ in $[d]$,
$Y(i)$ is a centered subgaussian random variable with variance factor
$v$. Let $Q$ denote the quantizer $Q_{{\tt at}, I}$ with parameters
$m$ and $m_0$ set to $3v$ and $2v \ln s$, respectively.
Then, for every $s,k, h \in \N$, we have 
\[
\frac{1}{d}\cdot \E{\sum_{i \in [d]}(Y(i)-Q(Y)(i))^2  \indic{\{|Y(i)|\leq M_{h-1}}\}} \leq  v \cdot \frac{9+3\ln s}{(k-1)^2}.
\] 
\end{lem} 
\begin{proof} Since 
  \eq{
    \E{\sum_{i \in [d]}(Y(i)-Q(Y)(i))^2  \indic{\{|Y(i)|\leq M_{h-1}}\}}
    &= \sum_{i=1}^{ \ceil{\frac{d}{s}}}\sum_{j =  
    (i-1)s+1}^{\min\{is,d\}}
    \E{\left(Q_{{\tt at}}(Y)(j)-Y(j)\right)^2 \indic{\{|Y(j)| \leq M_{h-1}\}}},
  }
by using Lemma $\ref{l:sup2}$ for each of the $\ceil{d/s}$
subvectors, we get
\eq{
  \nonumber
  &\lefteqn{\E{\sum_{i \in [d]}(Y(i)-Q(Y)(i))^2  \indic{\{|Y(i)|\leq M_{h-1}}\}}}
\\
  &\leq
\frac{s}{(k-1)^2}
  \sum_{i \in
  1}^{\floor{\frac{d}{s}}} \left(  m + m_0 +\sum_{j
  \in [h-1]}  M_{ j}^2 P\left(\norm{Y_{i,
    s}}_{\infty} > M_{j-1}\right) \right)
\\ \nonumber
& \hspace{0.3cm}
+\frac{(d-s\floor{\frac{d}{s}})
  }{(k-1)^2}\left(m
+  m_0+\sum_{j \in [h-1]}
  M_{ j}^2P\left(\norm{Y_{\ceil{d/s}, s}}_{\infty} >
M_{j-1}\right) \right)  .
}
For all $i \in \floor{d/s}$, it follows from the union bound that
\eq{
P\left(\norm{Y_{1,
    s}}_{\infty} > M_{j-1}\right)
\leq 2s e^{\frac{-M_{j-1}^2}{2v}}. 
}
Also, since $d-s\floor{d/s}\leq s$, we have
\eq{
P\left(\norm{Y_{\ceil{d/s},
    s}}_{\infty} > M_{j-1}\right)
\leq  2s e^{\frac{-M_{j-1}^2}{2v}}.
} 
Using these tail bounds in the previous inequality, we get
\eq{
  \nonumber \E{\sum_{i \in [d]}(Y(i)-Q(Y)(i))^2  \indic{\{|Y(i)|\leq M_{h-1}}\}} \leq \frac{d}{(k-1)^2}\left(m+m_0+2s\sum_{j \in [h-1]} M_{ j}^2 e^{\frac{-M_{j-1}^2}{2v}}\right).
} 
Setting  $m=3v$ and $m_0=2v$, the summation
on the right-side is bounded further as
\eq{
  &\lefteqn{2s\left(\frac{3v}{s }\sum_{j=1}^{h-1}(e^{*j})\cdot e^{- 1.5{e^{*(j-1)}}}\right)+2s\left(\frac{2v}{s }\sum_{j=1}^{h-1}e^{- 1.5{e^{*(j-1)}}}\right)}
  \\
&= 6v\sum_{j=1}^{h-1}e^{-0.5{e^{*(j-1)}}} +4v\ln s \sum_{j=1}^{h-1}e^{-1.5{e^{*(j-1)}}}\\
& \leq 6 v \sum_{j=1}^{\infty}e^{-0.5{e^{*(j-1)}}}+ 4v\ln s  \sum_{j=1}^{h-1}e^{-1.5{e^{*(j-1)}}}\\
& \leq 6v+v\ln s ,
}
where we use a bound of $1$ for $\sum_{j=1}^{\infty}e^{-0.5{e^{*(j-1)}}}$, whose validity
can be seen as follows\footnote{In fact, these bounds motivate the use of tetration
as our choice for $M_j$s.}
\begin{align*}
\sum_{j=1}^{\infty}e^{-0.5{e^{*(j-1)}}}
&={e^{-0.5}}+  {e^{-0.5e}}+ {e^{-0.5e^e}}
+\sum_{j=3}^{\infty}e^{-0.5{e^{*(j)}}}
\\
&\leq {e^{-0.5}}+  {e^{-0.5e}}+ {e^{-0.5e^e}}
+\sum_{j=3}^{\infty}e^{-0.5{je^e}}
\\
&\leq {e^{-0.5}}+  {e^{-0.5e}}+ {e^{-0.5e^e}}+\frac1{e^{e^e}-1}
\\
&\leq 1,
\end{align*}
and $1/4$ for $\sum_{j=1}^{h-1}e^{-1.5{e^{*(j-1)}}}$, whose validity can be seen as follows
\eq{
\sum_{j=1}^{\infty}e^{-1.5{e^{*(j-1)}}}
&={e^{-1.5}}+  {e^{-1.5e}}+ {e^{-1.5e^e}}
+\sum_{j=3}^{\infty}e^{-1.5{e^{*(j)}}}
\\
&\leq {e^{-1.5}}+  {e^{-1.5e}}+ {e^{-1.5e^e}}
+\sum_{j=3}^{\infty}e^{-1.5{je^e}}
\\
&\leq {e^{-1.5}}+  {e^{-1.5e}}+ {e^{-1.5e^e}}+\frac1{e^{3e^e}-1/e^{1.5e^e}}
\\
&\leq 0.2401.
}
Therefore, we obtain 
\[\frac{1}{d}\cdot \E{\sum_{i \in [d]}(Y(i)-Q(Y)(i))^2  \indic{|Y(i)|\leq M_{h-1}}} \leq v \cdot  \frac{9+3\ln s}{(k-1)^2}.\]

\end{proof} 
We remark that calculations present in this lemma are at the heart of the analysis of RATQ.
Also, this lemma will be useful for other applications discussed in Chapters \ref{Ch:DME} and \ref{Ch:GRD}. 

\paragraph{Step 4: Completing the proof.}
Recall that the random matrix $R$ defined in~\eqref{e:R} is  used at the encoder of RATQ to randomly rotate the input vector. We observe that the rotated vector has subgaussian entries. 
\begin{lem}\label{l:concentration_a.s.}
For an $\R^d$-valued random variable $Y$ such that $\norm{Y}_2^2 \leq B^2 \text{ a.s.}$, all
coordinates of the rotated vector $RY$ are centered subgaussian random
variables with a variance factor of ${B^2}/{d}$, whereby
\[
P(|{RY}(j)|\geq M) \leq 2  e^{-dM^2/2B^2}, \quad \forall\,  j \in [d],
\]
where $RY(j)$ is the $j^{th}$ coordinate of the rotated vector.
\end{lem}
\noindent The proof uses similar calculations as~\cite{ailon2006approximate} and~\cite{suresh2017distributed}; it is deferred to Section \ref{ap:R_subg}. 

Intuitively, the Lemma~\ref{l:concentration_a.s.} highlights the fact that overall energy $\norm{Y}_2^2$ in the input vector $Y$ is divided equally among all the coordinates after random rotation.

\paragraph{The worst-case second moment of RATQ.}
Note that by the description of RATQ which will be denoted by $Q_{{\tt at}, R}(RY),$ we have that
\[Q_{{\tt at}, R}(Y) =R^{-1} Q_{{\tt at}, I}(RY), \]
where $Q_{{\tt at}, I}$ is as defined in \eqref{e:Q_at_I}. Thus,
\begin{align}\label{e:Q_at}
Q_{{\tt at}, I}(RY)=[Q_{{\tt at}}(RY_{1,s})^{T}, \cdots,  Q_{{\tt
      at}}(RY_{\ceil{d/s},s})^{T}]^T,
      \end{align} where
the subvector $RY_{i,s}$ is given by
\begin{align*}\label{e:ithsubvector}
RY_{i, s}=[RY((i-1)s+1), \cdots, RY(\min\{is,d\})]^T. 
\end{align*}

To compute $\alpha(Q_{{\tt at}, R}(Y))$, we will first compute the
second moment for the output of RATQ. Specifically, using the fact $R$
is a unitary transform, we obtain
\eq{
  \E{\norm{Q_{{\tt at}, R}(Y)}_2^2}&=\E{\norm{R^{-1}Q_{{\tt at}, I}(RY)}_2^2}
  \\
  &=\E{\norm{Q_{{\tt at}, I}(RY)}_2^2}
  \\ 
  &=\sum_{j \in [d] } \E{(Q_{{\tt at}, I}(RY)(j))^2}
  \\
&=\sum_{i=1}^{\ceil{\frac{d}{s}}}\sum_{j = (i-1)s+1}^{\min\{is,d\}} \E{(Q_{{\tt at}, I}(RY)(j))^2}.
}

We now observe that for our choice of $m$ and $h$ for RATQ given by \eqref{e:RATQ_levels},  we have 
\[
M_{h-1}^2\geq m(e^{*{\log_{e}^*(d/3)}})=(3B^2/d).(d/3)=B^2.
\]
Using this observation and noting that $R$ is a unitary matrix, 
we have that 
\[
\indic{\{\norm{RY}_{2} \leq M_{h-1}\}}=1 \text{ a.s.}.
\]
Also, noting that $|RY(j)| \leq \norm{RY}_{2}=\norm{Y}_2=B$  a.s., for all $j\in [d]$,
we get
\begin{equation}\label{e:RATQ_Unbiased}
\indic{\{|RY(j)| \leq M_{h-1}\}} =1 \text{ a.s.}, \forall j \in [d].
\end{equation}
Proceeding with these observations, we get
\eq{
  \E{\norm{Q_{{\tt at}, R}(Y)}_2^2}
& \leq \sum_{i=1}^{ \ceil{\frac{d}{s}}}\sum_{j =
    (i-1)s+1}^{\min\{is,d\}}
  \E{(Q_{{\tt at}, I}(RY)(j))^2 \indic{\{|RY(j)| \leq M_{h-1}\}}}\\
  & = \sum_{i=1}^{ \ceil{\frac{d}{s}}}\sum_{j =
    (i-1)s+1}^{\min\{is,d\}}
  \E{(Q_{{\tt at}, I}(RY)(j)-RY(j)+RY(j))^2 \indic{\{|RY(j)| \leq M_{h-1}\}}}\\
  &\leq  \sum_{i=1}^{ \ceil{\frac{d}{s}}}\sum_{j =
    (i-1)s+1}^{\min\{is,d\}}
  \E{\left((Q_{{\tt at}, I}(RY)(j)-RY(j))^2+ RY(j)^2\right) \indic{\{|RY(j)| \leq M_{h-1}\}}} ,
}
where the previous inequality uses the fact that, under the event $\{|RY(j)|\leq M_{h-1}\}$,
$Q_{{\tt at}, I}(RY)(j)$ is an unbiased estimate of $RY(j)$. Namely,
\[
\E{Q_{{\tt at}, I}(RY)(j)\indic{\{|RY(j)|
\leq M_{h-1}\}}\mid R, Y} = \E{RY(j)\indic{\{|RY(j)|
\leq M_{h-1}\}}\mid R, Y}.
\] 
Therefore, noting that $R$ is a unitary matrix, we have  
\begin{align*}
\E{\norm{Q_{{\tt at}, R}(Y)}_2^2} \leq  \E{\sum_{j \in [d]}(RY(i)-Q_{{\tt at}, I}(RY)(j))^2  \indic{\{|RY(j)|\leq M_{h-1}\}}} +\E{\norm{Y}_2^2}.
\end{align*}

To bound the first term on the right-side we have the following lemma.
\begin{lem}\label{l:useful}
For an $\R^d$-valued random variable $Y$ such that $\norm{Y}_2^2 \leq B^2$ {a.s.}. Then, for $m$ and $m_0$ set to be $3B^2/d$ and $(2B^2/d) \ln s$, respectively, we have that 
\eq{
  \E{\sum_{j \in [d]}(RY(i)-Q_{{\tt at}, I}(RY)(j))^2  \indic{\{|RY(j)|\leq M_{h-1}\}}} \leq B^2 \cdot  \frac{9+3\ln s}{(k-1)^2}.
}
\end{lem}
\begin{proof}
  By Lemma~\ref{l:concentration_a.s.} we have that all coordinates
  $RY(j)$ are centered subgaussian random variable with variance
  factor $B^2/d$.  Thus, the parameters $m$ and $m_0$ of RATQ set as
  in~\eqref{e:RATQ_levels}, and equal $3v$ and $2v \ln s$, respectively,
  where $v$ is the variance factor of each subgaussian coordinate. The result follows by invoking  Lemma~\ref{l:subg_mse}.
\end{proof}

Therefore, for any $Y$ such that $\norm{Y}_2^2 \leq B^2$, we have
\eq{
\E{\norm{Q_{{\tt at}, R}(Y)}_2^2} \leq  B^2 \cdot  \frac{9+3\ln s}{(k-1)^2} + B^2
,}
whereby
\[
 \alpha_2(Q_{{\tt at}, R})\leq B\sqrt{ \frac{9 +
3 \ln s}{(k-1)^2} + 1}.
\]

\paragraph{The worst-case bias of RATQ.} By \eqref{e:RATQ_Unbiased} we have that the input always remains
in the dynamic-range of the quantizer, resulting in unbiased quantized
values. In other words,  $ \beta_2(Q_{{\tt at, R}})=0$.

\subsection{Proof of Lemma \ref{l:AGUQ_alpha_beta}}
\label{s:AGUQ}
We first note AGUQ is used to quantize a scalar $Y$.
It follows from the description of the quantizer that
\begin{equation}\label{e:MeanAGUQ}
\mathbbm{1}_{\{|{Y}| \leq M_{g, h_g-1}\}} \E{Q_{\tt a}(Y) |Y} = \mathbbm{1}_{\{|{Y}| \leq M_{g, h_g-1}\}}   Y,
\end{equation}
and that\footnote{Once again, this follows from our convention that the outflow symbol is evaluated to $0$.}
 \begin{equation}\label{e:MeanAGUQSec}
 \mathbbm{1}_{\{|{Y}| > M_{g, h_g-1}\}} Q_{\tt a}(Y)  = 0.
 \end{equation} 
 
\paragraph{The worst-case second moment of AGUQ.}
Towards evaluating $\alpha(Q_{{\tt a}})$ for AGUQ,  for any $Y \in \R$ we have
\begin{align}
\nonumber
\E{Q_{\tt a}(Y)^2} &= \E{Q_{\tt a}(Y)^2\mathbbm{1}_{\{|Y| \leq M_{g, h_g-1}\}}}+\E{Q_{\tt a}(Y)^2\mathbbm{1}_{\{|Y| > M_{g, h-1}\}}}\\
\nonumber
&= \E{(Q_{\tt a}(Y)-Y+Y)^2\mathbbm{1}_{\{|Y| \leq M_{g, h_g-1}\}}}+\E{Q_{\tt a}(Y)^2\mathbbm{1}_{\{|Y| > M_{g, h_g-1}\}}}
\\ \nonumber
&= \E{(Q_{\tt a}(Y)-Y)^2\mathbbm{1}_{\{|Y| \leq M_{g, h_g-1}\}}}+\E{Y^2\mathbbm{1}_{\{|Y| \leq M_{g, h_g-1}\}}},
\nonumber
\end{align}
where the last identity uses
\eqref{e:MeanAGUQSec}, and the fact that $\E{(Q_{\tt
    a}(Y)-Y)Y\mathbbm{1}_{\{|Y| \leq M_{g, h-1}\}}|Y}=0,$ which
follows from \eqref{e:MeanAGUQ}.  From Lemma \ref{l:sup3} it follows
that
\eq{ \E{(Q_{{\tt a}}(Y)-Y)^2\mathbbm{1}_{\{|Y| \leq M_{g,
        h-1}\}}} \leq \frac{1}{4(k_g-1)^2}\left(B^2+ \sum_{j=1}^{h-1}
  M_j^2P\left(|Y| > M_{g, j-1}\right)\right).
}
By Markov's
inequality we get that for any random variable $Y$ with $\E{Y^2}\leq B^2$, we have
$P(|Y| > M_{g, j-1})\leq B^2/M^2_{g, j-1},$ which further leads to
\eq{
  \E{(Q_{{\tt a}}(Y)-Y)_2^2\mathbbm{1}_{\{|Y| \leq
    M_{g, h-1}\}}}  &\leq \frac{B^2}{4(k_g-1)^2}+ \sum_{j=1}^{h_g-1}
 \frac{M_{g,j}^2 }{4(k_g-1)^2} \frac{B^2}{M_{g,j-1}^2}\\
 &=  \frac{B^2}{4(k_g-1)^2}+ \frac{a_g(h_g-1)B^2}{4(k_g-1)^2}.
}
Therefore, we have
\eq{
\E{Q_{\tt a}(Y)^2} \leq \frac{B^2}{4(k_g-1)^2}+ \frac{a_g(h_g-1)B^2}{4(k_g-1)^2} +\E{Y^2\mathbbm{1}_{\{|Y| \leq M_{g, h-1}\}}}.
}
The result follows upon taking the supremum of the left-side over all
random variables $Y$ with $\E{Y^2} \leq B^2$.

\paragraph{The worst-case bias of AGUQ.}
Towards evaluating $\beta(Q_{\tt a})$, we note first using Jensen's inequality that
\eq{
  \big|\E{Q_{\tt a}(Y)-Y}\big| 
  &\leq \E{\big|\E{Q_{\tt a}(Y)-Y|Y}\big|}.
}
Then, for $Y$ with $\E{Y^2}\leq B^2$, using \eqref{e:MeanAGUQ} and Markov's inequality, we get
\begin{align}\label{e:conditionBeta_AGUQ}
\nonumber \E{|\E{Q_{\tt a}(Y)-Y|Y}|}
&= \E{|Y|\mathbbm{1}_{\{|Y| \geq M_{g, h-1}\}}}\\
\nonumber &\leq \sqrt{\E{Y^2}P(|Y| \geq M_{g, h-1})}\\
&\leq \frac{B^2}{M_{g, h-1}}.
\end{align}
Therefore, for any $Y$ with $\E{Y^2}\leq B^2$, we have 
\eq{
  \big|\E{Q_{\tt a}(Y)-Y}\big|  \leq
\sup_{Y\geq 0 \text{a.s.}: \E{Y^2}\leq B^2}\E{\big|\E{Q_{\tt a}(Y)-Y|Y}\big|}\leq
  \frac{B^2}{M_{g, h-1}}.
}
The result follows upon taking the supremum of left-side over all
random variables $Y$ with $\E{Y^2} \leq B^2$.  \qed

\subsection{Proof of Lemma \ref{l:AGUQ+_alpha_beta}}\label{s:proof_Aguq^+}

\paragraph{$O(1)$ expected precision.}
Recall that the variable-length bit string is a concatenation of two bit strings: The first bit string represents the dynamic range $M_{g, j}$, $j \in \{0,  \ldots h-1\}$; the second-bit string represents the uniform level within that dynamic range.

The first string uses the first $h$ symbols of the Huffman codes corresponding to the geometric distribution with parameter  $1/2.$ Its code length can be bounded as follows.
 By Markov's
inequality, we have
$P(|Y| > M_{g, j-1})\leq B^2/M^2_{g, j-1}= a_g^{-j}=2^{-j+1}.$ 
For a symbol $j \in \{0, \cdots, h-1\}$ representing the chosen dynamic range, let $\ell(j)$ denote the length of the codeword of that symbol.
Therefore, the expected codelength $\E{L} $ can be upper bounded as follows.
\eq{
\E{L}& \leq \sum_{j=0}^{h-1}P(|Y| > M_{g, j-1}) \ell(j)\\
&\leq 2\sum_{j=0}^{h-1} 2^{-j} \ell(j).
}
Since we will assign code-lengths $\ell(j)$ as that assigned to the first $h$ symbols of the Huffman code corresponding to the geometric distribtution with parameter $1/2$, we have
 the following.
\eq{
\E{L}& \leq 2\sum_{j=0}^{h-1} 2^{-j} \ell(j)\\
& = 4\sum_{j=0}^{h-1} 2^{-j-1} \ell(j)\\
&\leq 4 (H(Z) +1),
}
where $Z$ is the geometric distribution with parameter $1/2$. Above, the final inequality can be seen by the fact that expected code-length for Huffman codes for a particular pmf is upper bounded by one plus entropy of that pmf. This bounds the code-length of the first bit string by a constant.

Coming to the bounding the code-length of the second bit string, the expected code length of the second string is upper bounded by
\eq{\sum_{j=0}^{h-1}P(|Y| > M_{g, j-1}) \log (k_{g, j}+1)
\leq  2\sum_{j=0}^{h-1} 2^{-j} (j+1)
\leq 12.
}

\paragraph{Worst-case second moment  of AGUQ$^+$.}
The only change from the worst-case second moment upper bound calculations in the proof of Lemma \ref{l:AGUQ_alpha_beta} is that the number of uniform levels for different dynamic ranges is different. The rest of proof remains precisely the same, and is skipped.

\paragraph{Bias of AGUQ$^+$.}
The proof is identical to one in Lemma \ref{l:AGUQ_alpha_beta}, and is skipped.
\qed

\subsection{Proof of Theorems \ref{t:lb_1} and \ref{t:lb_2}}\label{s:lbproof}
Before we proceed with our lower bounds, we will set up some
notation. We consider quantizers of the form
\[
Q(Y)=Q_g(\norm{Y}_2)Q_s(Y/\norm{Y}_2).
\] 
Let $W(\cdot |y)$, $W_g(\cdot| y)$, and $W_s(\cdot |y)$, respectively,
denote the distribution of the output of quantizers $Q(y)$, $Q_g(y),$ and $Q_s(y)$.
We prove a general lower bound for a quantizer satisfying Structural Assumptions
1-3 in Section~\ref{s:ug} in terms of the precision $r$; Theorems~\ref{t:lb_1} and~\ref{t:lb_2}
are obtained as corollaries of this
general lower bound. 
\begin{thm}\label{t:lb}
  Suppose that $\X$ contains the set $\{x\in \R^d: \norm{x}_2
  \leq D/2\}$.  
  Consider a gain-shape quantizer $Q$ of precision $r$ satisfying the  Assumptions 1-3 in Section~\ref{s:ug}. Then, there exists an oracle $(f, O)\in \oO$ such that for any optimization protocol $\pi$ using $T$
  iterations, we have 
  \[
  \ep(f, O, \pi, Q) \geq 
\frac{DB}{2\sqrt{2}}\min\bigg\{ \frac{1 }{2^r} ,\frac 1{4\cdot2^{r/3}T^{1/3}},\frac{1}{2(2T)^{1/3}}  \bigg\}.
\]
\end{thm}
\begin{proof}
Consider the function $f_\alpha: \R^d \to \R$, $\alpha\in\{-1,1\}$ given by
  \[
f_\alpha(x) \eqdef \delta \frac{B}{\sqrt{2}}|x(1)- \alpha D/2|, \quad \alpha \in \{-1, 1\}.
\]
Note that the functions $f_1$ and $f_{-1}$ are convex and depend only on the first
coordinate of $x$. Further, for $x\in\X$, the \newest{subgradient} of $f_\alpha$
is \newest{$-\delta \alpha B e_1/\sqrt{2}$}, \newest{since $sign(x(1) - \alpha D/ 2) =-sign(\alpha)$}, where $e_1$ is the vector $[1, 0, 0, \dots, 0]^T$.  We consider oracles $O_\alpha$, $\alpha\in\{-1,1\}$, that produce noisy \newest{subgradient} updates with distribution 
\eq{
 P_{\alpha}\left(\frac{B}{\sqrt{2}}e_1\right) =
   \frac{1-\delta^2}{2} , \quad  P_{\alpha}\left(\frac{-B}{\sqrt{2}}e_1\right) =
   \frac{1-\delta^2}{2} ,\quad
P_{\alpha}\big(-\frac{\alpha B}{\sqrt{2}\delta}e_1\big) =
   \delta^2.
}  
 It is easy to check that the oracle outputs satisfy ~\eqref{e:asmp_unbiasedness}
and~\eqref{e:asmp_L2_bound} described in Section \ref{s:problemsetup}. That is, the output of $O_{\alpha}$ is an unbiased estimate of the subgradient of $f_{\alpha}$, and the expected Euclidean norm square of the oracle output is bounded by $B^2$. 

We now take recourse to the standard reduction of optimization to
hypothesis testing: To estimate the optimal value of $f_{1}$ and
$f_{-1}$ to an accuracy $\delta$, the optimization protocol must
determine if the oracle outputs are generated by $P_{1}$ or
$P_{-1}$.  However in order to distinguish between $P_{1}$ or
$P_{-1}$, the optimization protocol only has access to the quantized oracle outputs.  Specifically, the protocol sees the samples from $Q(Y)$ at every time step, where $Y$ has distribution either $P_1$ or $P_{-1}$.

Denoting by $P_\alpha W$ the distribution
of the output  $ Q(Y)$ when the input $Y$ is
generated from $P_\alpha$,  
 we have from the standard reduction (see, for instance, \cite[Theorem 5.2.4]{duchi2017introductory}) that
 \[
\max_{\alpha\in\{-1,1\}}\ep(f, O, \pi, Q) \geq 
\frac{DB}{2\sqrt{2}}\delta\bigg( 1- \sqrt{\frac{T}{2}\chi^2(P_1W,P_{-1}W}\bigg), 
\]
where $\displaystyle{\chi^2(P, Q)=\sum_{x}(P(x)-Q(x))^2/Q(x)}$ denotes the chi-squared divergence between $P$ and $Q$.

Note that Assumption 2 on the structure of the quantizer implies that when $M< B/{\delta\sqrt{2}}$, the distributions $P_1W$ and $P_{-1}W$ are the same. It follows that for every $\delta<\min\{\sqrt{B^2/2M^2}, 1\}$, the
left-side of the previous inequality exceeds $(DB/2\sqrt{2})\delta$, whereby
\begin{align}
\max_{\alpha\in\{-1,1\}}\ep(f, O, \pi, Q)
\geq \frac{DB}{2\sqrt{2}}\min \bigg\{\frac{B}{\sqrt{2}M}, 1\bigg\}.
\label{e:bound1}
\end{align}

Next, we consider the following modification of the previous
construction in the case when ${B}/\sqrt{2}< m$:  
 \eq{ 
  P_{\alpha}\left(\frac{B}{\sqrt{2}}e_1\right) =
   \frac{1-\delta^{1+y}}{2} , \quad  P_{\alpha}\left(\frac{-B}{\sqrt{2}}e_1\right) =
   \frac{1-\delta^{1+y}}{2} ,\quad
P_{\alpha}\big(-\frac{\alpha B}{\sqrt{2}\delta^y}e_1\big) =
   \delta^{1+y}.
}
for $y\in [0,1]$. Once again, the oracle outputs satisfy ~\eqref{e:asmp_unbiasedness}
and~\eqref{e:asmp_L2_bound} described in Section \ref{s:problemsetup}. In this case, the  vector $Y \sim P_{\alpha}$ has entries with $\ell_2$ norm at the most $ B/(\sqrt{2}\delta^y)$. We set $y$ such that this value is less than $m$ and $\chi^2(P_1W,P_{-1}W)$ is minimized.  Note that if $B/(\delta^y\sqrt{2d})< m$, then ${\tt supp}(Q_{g}(\norm{a})) \subseteq \{0, m\}$ for all the $a$'s in the support of $P_1$ or $P_{-1}$. 
      
      For all $z \neq 0$, $z \in  {\tt supp}(Q(a))$, when $a$ is in the support of $P_1$ or $P_{-1}$, we have
\eq{W\big(z \mid a \big)=W_g\big(m\mid \norm{a}_2\big) W_s\big(\frac{z}{m}\mid\frac{a}{\norm{a}_2}\big).}
Therefore,
 \eq{
 P_{1}W(z)-P_{-1}W(z) &=\delta^{1+y}  W_g\big(m\mid\frac{B}{\sqrt{2}\delta^y}\big) \left( W_s\big(\frac{z}{m}\mid -e_1\big)   - W_s\big(\frac{z}{m}\mid e_1\big)\right)
 }
\eq{
P_{-1}W(z) &\geq \frac{1-\delta^{1+y}}{2} W_g\big(m\mid \frac{ B}{\sqrt{2}}\big) W_s\big(\frac{z}{m}\mid e_1\big) + \frac{1-\delta^{1+y}}{2}  W_g\big(m\mid \frac{ B}{\sqrt{2}}\big) W_s\big(\frac{z}{m}\mid -e_1\big). }
Using the preceding two inequalities 
\eq{
\frac{(P_{1}W(z)-P_{-1}W(z))^2}{P_{-1}W(z)} &\leq  \frac{\delta^{2+2y}  W_g\big(m\mid \frac{B}{\sqrt{2}\delta^y}\big)^2 \left( W_s\big(\frac{z}{m}\mid e_1\big)   - W_s\big(\frac{z}{m}\mid -e_1\big)\right)^2}{\frac{1-\delta^{1+y}}{2} W_g\big(m\mid \frac{ B}{\sqrt{2}}\big) W_s\big(\frac{z}{m}\mid e_1\big) + \frac{1-\delta^{1+y}}{2}  W_g\big(m\mid \frac{ B}{\sqrt{2}}\big) W_s\big(\frac{z}{m}\mid -e_1\big)}\\
&\leq \frac{2\delta^{2+2y}}{1-\delta^{1+y}} \cdot \frac{ W_g\big(m\mid \frac{B}{\sqrt{2}\delta^y}\big)^2 \left(W_s\big(\frac{z}{m}\mid e_1\big)   + W_s\big(\frac{z}{m}\mid -e_1\big)\right)}{W_g\big(m\mid \frac{ B}{\sqrt{2}}\big)} \\
&\leq  \frac{2\delta^{2+y}}{1-\delta^{1+y}} \cdot  W_g\big(m\mid \frac{B}{\sqrt{2}\delta^y}\big) \left(W_s\big(\frac{z}{m}\mid e_1\big)   + W_s\big(\frac{z}{m}\mid -e_1\big)\right)\\
&\leq  \frac{2\delta^{2+y}}{1-\delta^{1+y}} \cdot \left(W_s\big(\frac{z}{m}\mid e_1\big)   + W_s\big(\frac{z}{m}\mid -e_1\big)\right),
}
\noindent where the third inequality uses Assumption 3b for the quantizer in Section \ref{s:ug}, i.e., it uses
\[
\frac{ W_g\big(m\mid \frac{B}{\sqrt{2}\delta^y}\big)}{W_g\big(m\mid \frac{ B}{\sqrt{2}}\big)} \leq \delta^{-y}
.
\]
For $z = 0$, $z \in  {\tt supp}(Q(a))$, when $a$ is in the support of $P_1$ or $P_{-1}$, we have
\eq{W\big(0\mid a \big)=W_g\big(0\mid \norm{a}_2\big) + W_g\big(m\mid \norm{a}_2\big) W_s\big(0\mid a/\norm{a}_2\big).}

Therefore, by similar calculations for $z\neq 0$, we have
\eq{
\frac{(P_{1}W(0)-P_{-1}W(0))^2}{P_{-1}W(0)} &\leq \frac{\delta^{2+2y} W_g\big(m\mid\frac{B}{\sqrt{2}\delta^y}\big)^2 \left( W_s\big(0\mid e_1\big)   + W_s\big(0\mid -e_1\big)\right)^2}{(\frac{1-\delta^{1+y}}{2}) W_g\big(m\mid \frac{ B}{\sqrt{2}}\big) W_s\big(0\mid e_1\big) + (\frac{1-\delta^{1+y}}{2})  W_g\big(m\mid \frac{ B}{\sqrt{2}}\big) W_s\big(0\mid -e_1\big)}\\
&\leq  \frac{2\delta^{2+y}}{1-\delta^{1+y}} \left( W_s\big(0\mid e_1\big)   + W_s\big(0\mid -e_1\big)\right)
.}

In conclusion,
\eq{  
\chi^2(P_1W, P_{-1}W) \leq \frac{4\delta^{2+y}}{1-\delta^{1+y}}.
}
Now, if $\delta < 1/2,$ we have
\eq{  
\chi^2(P_1W, P_{-1}W) \leq 8\delta^{2+y}.
}
 Upon setting $\delta=(16 T)^{-1/(2+y)}$, which satisfies $\delta<1/2$ for all $T$, 
we get
\begin{align}
\max_{\alpha\in\{-1,1\}}\ep(f, O, \pi, Q)\geq \frac{DB}{2\sqrt{2}}\delta( 1- \sqrt{4T\delta^{2+y}})
= \frac{DB}{4\sqrt{2}}\left(\frac{1}{16T}\right)^{\frac 1 {2+y}}.
\label{e:bound2}
\end{align}
But we can only set $\delta$ to this value if
\begin{align}
\frac{B}{\sqrt{2}}\cdot \left({16 T}\right)^{\frac y {2+y}}< m .
\label{e:y_constraint}
\end{align}
Thus, for each $y$ such that~\eqref{e:y_constraint} holds, we get~\eqref{e:bound2}.
Taking the the supremum of RHS in \eqref{e:bound2} over all  $y\in[0,1]$ such that \eqref{e:y_constraint} holds, we obtain whenever $B/\sqrt{2} \leq m,$
\begin{align*}
\max_{\alpha\in\{-1,1\}}\ep(f, O, \pi, Q)\geq 
\frac{DB}{2\sqrt{2}} \cdot \min\left\{ \frac{1}{8}\sqrt{\frac{m\sqrt{2}}{BT}} , \frac{1}{2(2T)^{1/3}} \right\},
\end{align*}
where we use the following lemma proved in Section~\ref{app:elemineq}.
\begin{lem}\label{l:elemineq}
For $a, c > 0,$ and $b > 1$. 
\[\sup_{y \in [0,1]: a(b)^{y/(2+y)}< c.} a\left(\frac{1}{b}\right)^{\frac{1}{2+y}}= \min\left\{ \sqrt{\frac{ca}{b}}, \frac{a}{b^\frac{1}{3}}\right\}\]
\end{lem}
Upon combining this bound with~\eqref{e:bound1}, we obtain 
\[
\sup_{(f,O)\in \oO}\varepsilon( f_\alpha, \pi^{QO})\geq 
\frac{DB}{2\sqrt{2}}\max\left\{\min\left\{\frac{c m}{M}, 1\right\}, \min\left\{\frac{1}{8}\sqrt{\frac{1}{cT}}, \frac{1}{2(2T)^{1/3}}\right \}\mathbbm{1}_{\{c<1\}}\right\},
\]
where $c=B/(m\sqrt{2})$. By making cases $1\leq c$, $ \frac{1}{8(2T)^{1/3}} \leq c<1$, and $c < \frac{1}{8(2T)^{1/3}}$, and using the fact that for $a,b \geq 0$, $\max\{a,b\}\geq  a^{1/3}b^{2/3}$ in the second case, we get
 \[
\sup_{(f,O)\in \oO}\varepsilon( f_\alpha, \pi^{QO})\geq 
\frac{DB}{2\sqrt{2}}\min\bigg\{1, \frac{1 }{(M/m)} ,\frac{1}{4 (M/m)^{1/3}T^{1/3}},\frac{1}{2(2T)^{1/3}}  \bigg\}.
\]
By Assumption 3 in Section \ref{s:ug}, we know that $\frac{M}{m}\leq 2^r$. Therefore,
 \[
\sup_{(f,O)\in \oO}\varepsilon( f_\alpha, \pi^{QO})\geq 
\frac{DB}{2\sqrt{2}}\min\bigg\{ \frac{1 }{2^r} ,\frac{1}{4 (2)^{r/3}T^{1/3}},\frac{1}{2(2T)^{1/3}}  \bigg\}.
\]
 \end{proof}

Theorem~\ref{t:lb_2} follows as an immediate corollary;
Theorem~\ref{t:lb_1}, too, is obtained by noting that
\[
\sup_{(f,O)\in
  \oO}\varepsilon( f_\alpha, \pi^{QO})< \frac{3DB}{\sqrt{T}}
\]
holds only if  $\sqrt{T} < 2^r$.

\section{Remaining proofs for the main results}\label{s:norm_proofs}

\subsection{Analysis of CUQ: Proof of Lemmas \ref{l:sup} and \ref{l:sup0m} }\label{ap:CUQ}

\paragraph{Proof of Lemma \ref{l:sup}:}

Denoting by $\B_{j,\ell}$ the event $\big\{ {Y}(j) \in [B_{M,k}(\ell),
  B_{M,k}(\ell+1))\big\}$, we get
\eq{
&\lefteqn{\E{\sum_{j\in[d]}\big(Q_{{\tt
u}}(Y)(j)-{Y(j)}\big)^2\mathbbm{1}_{\{{|Y(j)|}\leq M\}}\mid Y}}
\\
&= \sum_{j \in [d]} \sum_{\ell =0}^{k-1}\E{\big(Q_{{\tt
       u}}(Y)(j)-{Y}(j)\big)^2 \mathbbm{1}_{\B_{j,\ell}}\mid
       {Y}} \mathbbm{1}_{\{{|Y(j)|}\leq M\}}.  }  For the term inside the summation on
       the right-side, we obtain
\begin{align}\label{e:residual}
 \nonumber &\lefteqn{\E{\big(Q_{{\tt
u}}(Y)(j)-{Y}(j)\big)^2 \mathbbm{1}_{\B_{j,\ell}}\mid
{Y}}} \\ \nonumber &= \left(( B_{M,k}(\ell+1)-{Y}(j))^2 \frac{{Y}(j)
-B_{M,k}(\ell)
}{B_{M,k}(\ell+1)-B_{M,k}(\ell)}\right) \mathbbm{1}_{\B_{j,\ell}}
\\ \nonumber
&\hspace{2cm} + \left(( B_{M,k}(\ell)-{Y}(j))^2\frac{ B_{M,k}(\ell+1)-
{Y}(j)}{B_{M,k}(\ell+1)-B_{M,k}(\ell)} \right) \mathbbm{1}_{\B_{j,\ell}}
\\ \nonumber
&= \left( B_{M,k}(\ell+1)-{Y}(j)) ({Y}(j) -
B_{M,k}(\ell)\right)\mathbbm{1}_{\B_{j,\ell}}
\\ \nonumber
&\leq \frac{1}{4}\left(B_{M,k}(\ell+1)- B_{M,k}(\ell) \right)^2
\\ 
&=\frac{M^2}{ (k-1)^2},
\end{align}
where the inequality uses the GM-AM inequality and the final identity
is simply by the definition of $B_{M,k}(\ell)$.  Upon combining the
bounds above, we obtain \eq{
\E{\sum_{j\in[d]}\big(Q_{{\tt u}}(Y)(j)-{Y(j)}\big)^2\mathbbm{1}_{\{{|Y(j)|}\leq M\}}\mid Y} \leq
            \frac{dM^2}{
            (k-1)^2} \cdot \frac{1}{d}\sum_{j\in[d]} \mathbbm{1}_{\{{|Y(j)|}\leq
            M\}}.}  \qed

\paragraph{Proof of Lemma \ref{l:sup0m}}
The proof is the same as the proof of Lemma~\ref{l:sup}, except that
we need to set $d=1$ and replace the identity used
in\eqref{e:residual} with
\[
B_{M,k}(\ell+1)- B_{M,k}(\ell)=\frac{M}{k-1}.
\]

\subsection{Proof of Lemma \ref{l:concentration_a.s.}}\label{ap:R_subg}

For the rotation matrix $R=(1/\sqrt{d})HD$, each entry of $RY(j)$ of
the rotated matrix has the same distribution as $(1/\sqrt{d})V^T Y$,
where $V=[V(1), ..., V(d)]^{T}$ has independent Rademacher entries. We
will use this observation to bound the moment generating function of
$RY(i)$ conditioned on $Y$. Towards that end, we have
\eq{
\E{ e^{\lambda RY(i)}\mid Y}
& = \prod_{i=1}^{d} \E{ e^{\lambda V(i)Y(i)/\sqrt{d}} \mid Y}\\ &
= \prod_{i=1}^{d} \frac{e^{\lambda Y(i)/\sqrt{d}}+e^{- \lambda
Y(i)/\sqrt{d}}}{2}
\\
&\leq \prod_{i=1}^{d} e^{\lambda^2 Y(i)^2/2d}\\ &=
e^{\lambda^2 \norm{Y}_{2}^2/2d}, } where the first identity follows
from independence of $V(i)$s and the first inequality follows by the
fact that $(e^{x}+e^{-x})/2$ is less than $e^{x^2/2}$, which in turn
can be seen from the Taylor series expansion of these terms. Thus, we
have proved the following:
\begin{align}
\E{ e^{\lambda RY(i)}\mid Y} \leq e^{\lambda^2 \norm{Y}_2^2/2d}, \quad \forall \lambda\in \R, \forall i \in [d]. 
\label{e:subg}
\end{align}
 Note that $\norm{Y}_2^2$ can be further bounded by $B^2$, which along
 with \eqref{e:subg} leads to \[\E{ e^{\lambda RY(i)}} \leq
 e^{\lambda^2 B^2/2d} \quad \forall \lambda\in \R, \forall i \in
 [d]. \] Using this inequality and the observation that $\E{RY(i)}=0$,
 we note that $RY(i)$ is a centered subgaussian with a variance
 parameter $B^2/d$.  The second statement of the lemma trivially
 follows from Lemma \ref{l:standar_subg}.
\qed.

\subsection{Proof of Lemma~\ref{l:elemineq}}\label{app:elemineq}
For any $y \in [0,1]$ such that $ab^{y/(2+y)}< c$, we have
$ab^{y/(2+y)} < \min\left\{c, ab^{1/3} \right\}$.  By multiplying by
$a/b$ on both sides and taking square root, we get
\[
\frac{a}{b^{\frac{1}{2+y}}} < \min\left\{\sqrt{\frac{ca}{b}},\frac{a}{b^{1/3}}\right\},
\]
which gives
\[
\sup_{y \in [0,1]: a(b)^{y/(2+y)}< c.}\frac{a}{b^{\frac{1}{2+y}}} \leq \min\left\{\sqrt{\frac{ca}{b}},\frac{a}{b^{1/3}}\right\}.
\]
Making cases $ab^{1/3} \geq c$ and $ab^{1/3} < c$, we note that the
supremum on the left-side equals the right-side in both the cases.
\qed

\section{Concluding Remarks}
 In this chapter,  we developed quantizers for communication-constrained optimization over Euclidean spaces. The problem here essentially reduces to minimizing the worst-case $\ell_2$ norm, $\alpha_2(Q)$ or $\alpha_2^{\tt m}(Q)$, of the quantized gradient under the constraint that worst-case $\ell_2$ bias between the quantized gradient and input gradient, $\beta_2(Q)$ or $\beta_2^{\tt m}(Q)$, is small and the output of the quantizer can be represented in $r$ bits. In fact, in the next chapter, we will see that for designing gradient quantizers for communication-constrained optimization over $\ell_p$ spaces, we need to solve a similar problem where the $\ell_q$ norms are considered instead, where  $q$ is the H\"{o}lder conjugate of $p$.
Since the only knowledge we have of the input gradient to be quantized is either an almost sure or mean square-bound on the $\ell_2$ norm of the gradient, we first preprocess the gradient to have some handle over the distribution of the gradient coordinates. We do this by randomly rotating the gradient input since it divides the overall gradient value  equally across coordinates. We believe that without any more information on gradient distribution, this is a reasonable preprocessing. We will resort to this idea of random rotation once again in Chapter \ref{Ch:DME}. Another idea we will pick up from the quantizer design in this chapter is that of adaptive quantization. As we saw in our achievability proofs, adaptive gradient quantization became crucial to come up with tight upper bounds. We will see in Chapter \ref{Ch:GRD} that this idea can also be used for classic information theory problems such as the  Gaussian rate-distortion problem.

\chapter{Communication-Constrained  Optimization over $\ell_p$ Spaces}\label{Ch:Limits}
\section{Synopsis}
 In this chapter, we study communication-constrained optimization for $\ell_p$ lipschitz and convex function family.  For this class of functions, we characterize the  minimum
    precision to which the oracle output must be quantized to retain the unrestricted
    convergence rates?  We characterize this precision for every
  $p\geq 1$ by accessing the information theoretic lower bounds derived in Chapter \ref{Ch:LB} and by
  providing quantizers that (almost) achieve these lower bounds.  Our
  quantizers are new and easy to implement.  In particular, our
  results are exact for $p=2$ and $p=\infty$, showing the minimum
  precision needed in these settings are $\Theta(d)$ and $\Theta(\log
  d)$, respectively. The latter result is surprising since recovering
  the gradient vector will require $\Omega(d)$ bits.

The results presented in this Chapter are from \cite{mayekar2020limits}.

\section{Introduction}

In this chapter, we develop new algorithms to match the lower-bounds for communication-constrained optimization over $\ell_p$ spaces. Specifically, we study communication-constrained optimization for convex and $\ell_p$ lipschitz function family. We study this problem in the \emph{high-precision regime} introduced in Chapter \ref{Ch:RATQ}. That is, we ask what is the minimum precision to which the subgradient estimates' supplied by the oracle must be quantized so that we can attain the convergence of the classic, unrestricted setting. We derive a lower bound on this precision using the lower bounds on optimization error derived in Chapter \ref{Ch:LB}. As our main contribution, we propose simple, efficient subgradient quantization algorithms which along with appropriate mirror decent algorithms match these lower bounds.


\subsection{Main Contributions}
We show that for $p\in [1,2]$ and $p\geq 2$, respectively, roughly $d$
  and $d^{2/p}\log (d^{1-2/p}+1)$ bits are necessary and sufficient
  for retaining the standard convergence rates. These bounds are tight
  upto an $O(\log d)$ factor, in general, but are exact for $p=2$ and
  $p=\infty$. Prior work has only considered the problem for the
  Euclidean case, and not for general $\ell_p$ geometry. Note that in the previous Chapter, we show that RATQ along with PSGD requires a precision of $O(d \log \log \log \ln^*d)$ per iteration to attain the classic convergence rate. In this chapter we get rid of the nagging $ \log \log \log \ln^*d$ factor and establish tight
  bounds.

We use different quantizers for $p\geq 2$ and $p\in [1,2]$.  In the
  $p\geq 2$ range, we use a quantizer we call SimQ$^+$.  SimQ$^+$, in
  turn, uses multiple repetitions of another quantizer we call SimQ
  which expresses a vector as a convex combination of corner points of
  an $\ell_1$ ball. It is SimQ that yields an $O(\log d)$ bit
  quantizer for optimization over $\ell_\infty$.  Also, SimQ$^+$
  yields the exact upper bound in the $\ell_2$ case.  In the $[1,2]$
    range, we divide the vector into two parts with small and large
    coordinates. We use a uniform quantizer for the first part and
    RATQ of Chapter~\ref{Ch:RATQ} for the second part.

The main observation in our analysis for upper bound is that the role
of quantizer in optimization is not to express the gradient with small
error. It suffices to have an unbiased estimate with appropriately
bounded norms.


 \subsection{Prior Work}
 A detailed literature review on the quantizer used in communication-constrained optimization is presented in Chapter \ref{Ch:RATQ}. Most of the literature in this area looks at the Euclidean setting. In the general setting of Information-constrained optimization, convex and $\ell_p$ Lipschitz family was considered in~\cite{feldman2017statistical}, in a statistical query setup, and \cite{duchi2014privacy}, in a local differential privacy setup.
 
 Finally, we remark that while our quantizers are related to the ones used in prior works, our main contribution is to show that our specific design choices yield optimal precision. For instance, the quantizers in~\cite{gandikota2019vqsgd} express the input as a
  convex combination of a set of points, similar to SimQ.  
  One of the quantizers in \cite{gandikota2019vqsgd} uses a similar set of points as that of SimQ with a different scaling. However, the quantizers in \cite{gandikota2019vqsgd} are designed keeping in mind other objectives, and they fall short of attaining the optimal precision guarantees of SimQ and SimQ$^+$. SimQ is also closely related Maurey's empirical method  (see  \cite{pisier1981remarques}, or \cite{wu2017lecture} for a recent reference), however, the use in gradient quantization is new.



\subsection*{Orgainization}
In the next section, we describe the setup and the structure of the schemes we will be employing. In Section \ref{e:highprecision_rp}, we describe the main result of the paper --  a characterization of the minimum precision required to attain classic convergence rate for all $p$. In Section \ref{s:SimQ} and \ref{s:infty}, we describe our quantizers used to achieve our upper bounds.  Finally, we close with comments on achieving tight upper bounds for the lower bounds in Theorems \ref{t:concom_P2} and \ref{t:concom_P2} for all $p$ and for generalizing our results to mean square bounded oracles in Section \ref{s:comment_general}.


\section{Setup and preliminaries}
\subsection{Setup}
We consider optimization domains $\X_p$ such that $\ell_p$ diameter is less than $D$. That is,
\begin{align}\label{eq:domain_X_p}
    \X_p \in \mathbb{X}_p(D) \eqdef \{ \X^{\prime}: \sup_{x, y\in \X^{\prime}}\norm{x-y}_p \leq D.\}
\end{align}
For the domain of optimization $\X_p$, we develop subgradient compression schemes for function and oracle families given by $\oO_{{\tt c}, p}$, which are defined in Definition \ref{def:c}.

We want to study communication-constrained optimization for $\oO_{{\tt c}, p}$ in the high precision regime. 
Thus,
 the fundamental quantity of interest in this work is the
minimum precision to achieve the optimization accuracy of the classic case, denoted by  $r^*(T, p)$. Symbolically,
\begin{align}\label{e:r*} 
r^*(T, p)
&\eqdef \inf \{r \in \N:  \sup_{\X \in \mathbb{X}_p(D)} \ep^*(\X , \oO_{{\tt c}, p}, T, \W_{{\tt com}, r  })  \leq
\mathcal{U}(T, p) \},    
 \end{align}
where
  \vspace{-0.15cm}
 \begin{align}\label{e:e_r}
  \mathcal{U}(T, p) &:= \frac{4c_1d^{1/2 -1/p} DB}{\sqrt{T}}, \quad \forall p \in (2, \infty],\\ \nonumber
  \mathcal{U}(T, p) &:= \frac{4c_1 \sqrt{\log d} DB }{\sqrt{T}}, \quad \forall p \in [1, 2),
 \end{align}
 and  $\sup_{\X \in \mathbb{X}_p(D)} \ep^*(\X , \oO_{{\tt c}, p}, T, \W_{{\tt com}, r  })$ is as defined in Chapter \ref{Ch:LB}. 
Recall that $\mathcal{U}(T, p)$ denotes the classic convergence rate for the family $\oO_{{\tt c}, p}$ given in Theorem \ref{t:e_infty}, where the oracle output is available as it is to the optimization algorithm, without any quantization.

\subsection{Quantizer performance for finite precision optimization}
As described in Section \ref{s:stucture_quant_l2_lp}, we restrict to memoryless quantization schemes, where the same quantizer will be used for each new subgradient vector, without any information about the previous updates.   Also recall from Section  \ref{s:stucture_quant_l2_lp}, our nonadaptive channel selection strategy is simply denote by quantizer $Q$. Further, the optimization error for a function $f$ and oracle $O$ when employing a first order optimization $\pi$ and quantizer $Q$ is given  by 
\[\mathcal{E}(f, O, \pi, Q)=\E{f(x_T)}-
\E{f(x^\ast)}.\]

We now define $\alpha_p(Q)$, which generalizes the definition of $\alpha_2(Q)$ in Chapter \ref{Ch:RATQ}, to charactize the performance of a quantizer $Q$ for optimization of convex and $\ell_p$ lipschitz funciton class.  Since we restrict to unbiased quantizers, we don't need to define $\beta$.

\[\displaystyle{\alpha_{p}(Q) := \sup_{Y \in \R^d: \norm{Y}_q^2\leq B^2
  \text{ a.s.}} 
\sqrt{\E{\norm{Q(Y)}_2^2}}, \quad  p \in (2, \infty]}, \]

\[\displaystyle{\alpha_{p}(Q) := \sup_{Y \in \R^d: \norm{Y}_q^2\leq B^2 \text{ a.s.}}
      \sqrt{\E{\norm{Q(Y)}_q^2}}, \quad p\in [1,2].}\]
      
Note that for all $p\geq 1$, the composed oracle $QO$ 
satisfies assumption \eqref{e:asmp_unbiasedness}. We employ the stochastic mirror descent (SMD) algorithm with mirror maps given by Remarks~\ref{r:Motivationfoalphapin12}
and~\ref{r:Motivationfoalphap} to use the output from the composed oracle. The algorithm's description is given in Algorithm 
\ref{a:SMD_Q}. 
Recall that for a mirror map $\Phi$, the Bregman divergence associated with $\Phi$ is defined as \[D_{\Phi}(x,y) \colon= \Phi(x)-\Phi(y)- \langle \nabla \Phi(y), x-y \rangle.
\]

\begin{figure}[h]
\centering
\begin{tikzpicture}[scale=1, every node/.style={scale=1}]
\node[draw,text width= 14 cm , text height= ,] {%
\begin{varwidth}{\linewidth}       
            \algrenewcommand\algorithmicindent{0.7em}
\begin{algorithmic}[1]
   \Statex \textbf{Require:} $x_0\in \X, \eta \in \R^+$, $T$ and
   access to composed oracle $QO$ \For{$t=0$ to $T-1$}

$x_{t+1}= \arg\min_{x \in \X} (\eta_t \langle x, Q(\hat{g}(x_t)) \rangle) +D_{\Phi_a}(x, x_t))$
   \EndFor \State \textbf{Output:} $\frac 1 T \cdot {\sum_{t=1}^T x_t}$
\end{algorithmic}
\end{varwidth}};
 \end{tikzpicture}

 \renewcommand{\figurename}{Algorithm}
\caption{Quantized SMD with quantizer $Q$}
\label{a:SMD_Q}

\end{figure}

Moreover, in view of Remarks~\ref{r:Motivationfoalphapin12}
and~\ref{r:Motivationfoalphap} , we have the following convergence 
guarantees for first-order stochastic optimization using gradients
quantized by $Q$. 

\begin{thm}\label{t:e_uq}
Consider a quantizer $Q$ for the gradients. Then the algorithm \ref{a:SMD_Q} with mirror maps as in Remarks \ref{r:Motivationfoalphapin12} and \ref{r:Motivationfoalphap} and an unbiased quantizer $Q$
performs as follows.
\begin{enumerate}

\item For $2 \geq p \geq 1$,
\[
\displaystyle{\frac{c_1\sqrt{\log d}D \alpha_{p}(Q)}{\sqrt{T}} \geq \sup_{(f, O)     \in \oO_{{\tt c}, p}}\ep(f, O, \pi, Q)
;}
\]

\item For $ p > 2$,
\[\displaystyle{
  \frac{c_1d^{1/2 -1/p} D\alpha_{p}(Q)}{\sqrt{T}}\geq  \sup_{(f, O)     \in \oO_{{\tt c}, p}}\ep(f, O, \pi, Q).}
  \]
\end{enumerate}
\end{thm}
\begin{proof}
The proof straightaway follows from Theorem \ref{t:e_infty} and Remarks \ref{r:Motivationfoalphapin12} and \ref{r:Motivationfoalphap}. For completeness, we provide the details below.

 First statement simply follows by noting that the  bounds in Theorem \ref{t:basic_convergence} hold when instead of $\norm{\hat{g}(x)}_q \leq B,$ we have $\E{\norm{\hat{g}(x)}_q^2} \leq B^2$ and then using definition of $\alpha_p.$ The second statement simply follows by noting that  the bounds in Theorem \ref{t:basic_convergence}  are obtained by employing PSGD. Thus it suffices to only have a bound on $\E{\norm{\hat{g}(x)}_2^2}$ and then using the definition of $\alpha_p.$

\end{proof}

An interesting insight offered by the result above, which is perhaps simple in
hindsight, is that even when dealing with $\ell_p$ oracles for $p> 2$, we only need to be concerned about the expected $\ell_2$ norm of
the quantizers output. This  follows from the fact that PSGD is the optimal  optimization algorithm for $p>2$ and it's  convergence rate is only concerned with the $\ell_2$ norm of the quantizers output. It is this insight that leads to the realization that SimQ$^+$ is optimal for these settings.

In the rest of the Chapter, we design unbiased, fixed length
quantizers which have $\alpha_p( \cdot)$ of the same order as
$B$. Then, using Theorem \ref{t:e_uq}  the quantized updates give the
same convergence guarantees as that of the classical case, which 
leads to upper bounds for $r^*(T, p)$. Further, we using Theorems \ref{t:concom_P2} and \ref{t:concom_inf}, we derive lower bounds for
$r^*(T, p)$ to prove optimality of our quantizers.



\section{Main Result: Characterization of $r^*(T, p)$}\label{e:highprecision_rp}

   The main result of this Chapter is the almost complete
   characterization of $r^*(T,p)$. We divide the result into cases
   $p\in [1,2]$ and $p\geq 2$; as mentioned earlier, we use different
     quantizers for these two cases.
  \begin{thm}~\label{t:main}
    For stochastic optimization using $T$ accesses to a first-order
    oracle, the following bounds for $r^*(T,p)$ hold.
\begin{enumerate}
\item For $p > 2$, we have \eq{ d^{2/p}\log &\,(2e\cdot d^{1-2/p}+2e) 
  \geq r^*(T, p) \geq \left(\frac{c_0}{4c_1}\cdot d^{1/p}\right)^2 \vee 
2 \log  \left(\frac{c_0}{4c_1}\cdot d^{1/2}\right) .} 
  \item For $2 \geq p \geq 1,$ we have \eq{d\left(
\ceil{\log(2\sqrt{2} {\Delta_1}^{1/q}+2)}+3\right) +\Delta_2   \geq ~ r^*(T,
    p) \geq \left(\frac{c_0}{4c_1 \sqrt{\log d}}\right)^2 \cdot d,}
    where
$\Delta_1 =\ceil{\log\left(2+\sqrt{  18 + 6\ln \Delta_2} \cdot d^{1/2-1/q}\right)}$ and $\Delta_2 =\ceil{\log (1+\ln^*({d}/{3})) }.$
\end{enumerate}
\end{thm} 
\noindent Note that for $p > 2$ the upper bounds and lower bounds for  $r^*(T,p)$ are off by nominal factor of $\log (d^{1-2/p}+1)$. Also, for $p \in [1,  2]$ the bounds are roughly off by $O(\log d \cdot\log (\log d^{1/2-1/q})^{1/q})$ (ignoring the $\log^*d$ terms).

  We present the quantizers achieving these upper bounds, and the
  proof of the upper bounds, in the next two sections. For $p > 2$,
  we use a quantizer SimQ and its extension SimQ$^+$, presented in
  Section~\ref{s:SimQ}. For $p\in [1,2]$, we use a combination of
    uniform quantization and the quantizer RATQ
    from previous chapter, presented in
   Section~\ref{s:p-less-than-two}.
 The lower bounds on $r^*(T,
    p)$ follow immediately by the lower bounds in Theorem \ref{t:concom_inf} and \ref{t:concom_P2}.


We highlight the most interesting features of the result above in
separate remarks below.  
\begin{rem}[$r^*(T, p) $ is independent of $T$]
Theorem \ref{t:main} shows that $ r^*(T, p)$ is a function only of $p$
and $d$, and is
independent of $T$.  The number of queries $T$ is a proxy for the desired
optimization accuracy. Therefore, the fact that $r^*(T, p)$ is
independent of such a parameter is interesting.  We note, however,
that for oracle models with milder assumptions, such as mean square
bounded oracles, this may not hold. In fact, the results of previous chapter suggest that for mean square bounded oracles $r^*(T,2)$ is dependent on $T$.
\end{rem}

\begin{rem}[Optimality for $p=\infty$]
  Our bounds match for $p=\infty$, namely our quantizer SimQ offers optimal
  convergence rate with gradient updates at the least precision. A
  surprising observation is that this precision is merely $O(\log d)$,
  much smaller than $O(d)$ bits needed to recover the gradient vector
  under any reasonable loss function.
 \end{rem}

\begin{rem}[Optimality for $p=2$]
The high-precision regime for $p=2$ was already considered in the previous chapter. Both \cite{alistarh2017qsgd} and
\cite{suresh2017distributed} give variable-length quantization schemes
to exactly achieve the lower bound on $r^*(T, p) $, but the worst-case precision can be
order-wise greater than $d$. The quanitzer RATQ from the previous Chapter was within a small factor of
$O(\log\log \log \ln ^*d)$ of this lower bound. In this chapter, we  remove 
this nagging factor using a different fixed-length quantizer SimQ$^+$.
\end{rem}

\begin{rem}[Fixed precision]
The quantizer RATQ  remains
optimal upto a factor of $O(\sqrt{\log \ln^* d})$ for the more general problem of  characterizing $\displaystyle{\sup_{\X \in \mathbb{X}_p(D)}\mathcal{E}^*(\X, \oO_{{\tt c}, p}, T, \W_{{\tt com}, r})}$
for any precision $r$ less than $d$ bits. In this
setting of small precision, the performance of SimQ$^+$ is much worse.
%
\end{rem}

\section{Our quantizers for $p > 2$}\label{s:SimQ}

We present our quantizer SimQ and its extension SimQ$^+$. The former
is seen to be optimal for $p=\infty$ while the latter for $p=2$.

\subsection{An optimal quantizer for $p=\infty$}\label{s:infty}

\begin{figure}[ht]
\centering
\begin{tikzpicture}[scale=1, every node/.style={scale=1}]
\node[draw, text width= 11 cm, text height=,] {%
\begin{varwidth}{\linewidth}
            
            \algrenewcommand\algorithmicindent{0.7em}
            \renewcommand{\thealgorithm}{}
\begin{algorithmic}[1]
\Require Input $Y\in \R^d$, Parameter $B$ \State $i^*=\begin{cases}
i\quad w.p. \quad |Y(i)|/B\\ 0 \quad w.p. \quad 1-\norm{Y}_1/B
 \end{cases} $

\If { $i^* \in [d]$} \Statex \hspace{1cm} $j^*=sign(Y(i^*))$

\Else \Statex \hspace{1cm} ~$j^*=1$

  \EndIf \State \textbf{Output:} $\Qenc_{{\tt SimQ}}(Y; B) =i^*\cdot
  j^* $ \
\end{algorithmic}
\end{varwidth}};
 \end{tikzpicture}
 \caption{Encoder $\Qenc_{{\tt SimQ}}(Y; B)$ for
   SimQ}\label{a:E_SIMQ}
 \end{figure}

 \begin{figure}[ht]
\centering
\begin{tikzpicture}[scale=1, every node/.style={scale=1}]
\node[draw, text width= 11 cm, text height=,] {%
\begin{varwidth}{\linewidth}
            \algrenewcommand\algorithmicindent{0.7em}
            \renewcommand{\thealgorithm}{}
            \renewcommand{\thealgorithm}{}
\begin{algorithmic}[1]
    \Require Input $i^\prime \in \{-d, -(d-1), \cdots 0,\cdots, d\} $
    \If { $i^\prime \neq 0$} \Statex \hspace{1cm} $Z=B
    sign(i^{\prime}) e_{|i^{\prime}|} $

\Else \Statex \hspace{1cm} ~$Z=0$

  \EndIf

\State \textbf{Output:} $\Qdec_{{\tt SimQ}}( i^\prime; B)=Z$
\end{algorithmic}
\end{varwidth}};
 \end{tikzpicture}
 \renewcommand{\figurename}{Algorithm}
 \caption{Decoder $\Qdec_{{\tt SimQ}}(i^\prime; B)$ for
   SimQ}\label{a:D_SIMQ} 
 \end{figure}
 \paragraph*{\textbf{ Simplex Quantizer ($\textbf{SimQ}$)}}
Our first quantizer SimQ is described in Algorithms~\ref{a:E_SIMQ}
and~\ref{a:D_SIMQ}.  For $p=\infty$, our quantizer's input vector $Y$
is an unbiased estimate of the subgradient of the function at the
point queried and satisfies $\norm{Y}_1 \leq B$. SimQ takes such a
$Y$ as an input and produces an output vector which, too, satisfies
both these properties. The main idea behind SimQ is the fact that
any point inside the unit $\ell_1$ ball can be represented as a convex
combination of at the most $2d$ points: $\{e_i,-e_i: i \in [d]\}$. With this observation, we can create an
unbiased estimate of the input vector using only these $2d$
corner points along with the zero vector. Since all of these
$ 2d + 1$ points have a $\ell_2$ norm of at the most $B$, the output
vector, too, has a $\ell_2$ norm of at the most $B$.


  \begin{thm}\label{t:SimQ}
 Let $Q$ be the quantizer SimQ described in Algorithms \ref{a:E_SIMQ}, \ref{a:D_SIMQ}. Then, for 
  $Y$ such that $\norm{Y}_1 \leq B~a.s.$,
   $Q(Y)$ can be represented in $\log (2d+1)$ bits, $\E{Q(Y)|Y}=Y$, and
   $\alpha_{\infty}(Q)\leq B$.
 \end{thm}
 \begin{proof}
Since $i^*\in[d]$ and $j^*\in\{-1,1\}$, we can represent the output of
the encoder of SimQ using $\log (2d+1)$ bits. Next, denoting
the quantizer SimQ by $Q$, note that
\[
\E{Q(Y)|Y }= \sum_{i=1}^d B \cdot sign(Y(i))\cdot e_i \cdot
\frac{|Y(i)|}{B} = Y,
\]
namely SimQ is unbiased. To complete the proof, note that
$\norm{Q(Y)}_2^2 \leq B^2~a.s.$.
\end{proof}
 Theorem \ref{t:SimQ} along with Theorem \ref{t:e_uq} establishes 
Theorem~\ref{t:main} for $p=\infty$.

\subsection{Our Quantizer for $p \in [2, \infty)$}
For this case,
    we need to quantize inputs that are bounded in
    $\ell_q$ norm with $q\in (1,2]$ so that the quantized output is unbiased and has small expected $\ell_2$ norm square; we will use SimQ$^+$ to do this.

\paragraph*{$\textbf{SimQ}^+$}
The quantizer SimQ$^+$ outputs the average of $k$ independent repetitions of the SimQ
quantizer for a given input vector. The input vectors $Y$ satisfy
 $\norm{Y}_1\leq B d^{1/p}$. Therefore, we use SimQ with
parameter $B d^{1/p}$ instead of $B$. The repetitions help reduce the
error to compensate for the extra loss factor. Specifically, the output of
SimQ$^+$ denoted by  $Q(Y)$ is given by
 \begin{align}\label{e:SIMQ+}
 Q(Y) =\frac{1}{k}\cdot\sum_{i =1}^{k} Q^i_{\tt SimQ}(Y; Bd^{1/p}),
  \end{align}
 where $Q^i_{\tt SimQ}$ are independent iterations of SimQ.


The next component of SimQ$^+$ is how the encoder of SimQ$^+$ expresses the output of these $k$ copies of SimQ to attain compression. If represented naively, this
will require $O(d^{2/p}\log d)$. But we can do much better since we only need the average value of these entries. For that, we can simply
report the {\em type} of this vector -- the frequency of each
 index in the $k$ length sequence. The signs of the
 input coordinates for the non-zero entries can be sent separately.

 Note that there are $d+1$ indices
 overall, as SimQ can pick any index from $\{0, \ldots d\}$.
 Therefore, the total number of types is ${{d+k}\choose{k}}$, which
 can at the most be $(\frac{ed+ek}{k})^k$ bits. Hence, the precision
 needed to represent the type is at the most $k \log e +
 k\log(\frac{d}{k}+1)$.

 The type of the input can be used to determine a set $\mathcal{I}_0$ of non-zero indices
 that appear at least once. There are at most $k$ such entries. Therefore, we can use
 a binary vector of length $k$ to store the signs for these entries. We use this representation in SimQ$^+$, with the indices in $\mathcal{I}_0$ represented in the vector in increasing order.
%
 

\begin{thm}\label{t:p2infty}
 For a $p \in [2, \infty)$, let $Q$ be the quantizer SimQ$^+$ described in \eqref{e:SIMQ+}. Then, for 
  $Y$ such that $\norm{Y}_q \leq B~a.s.,$
   $Q(Y)$ can be represented in $k \log e + k\log(\frac{d}{k}+1)+k$ bits, $\E{Q(Y)|Y}=Y$, and
   $\displaystyle{\alpha_{p}(Q) \leq \sqrt{\frac{B^2 d^{2/p}}{k}+B^2}}$.
 \end{thm}
\begin{proof}
  We already saw how to represent the output of the $k$ copies of SimQ
using $k \log e + k\log(\frac{d}{k}+1)+k$ bits. 
For bounding $\alpha_p(Q)$, note from~\eqref{e:SIMQ+}
that SimQ$^+$ is an unbiased quantizer since SimQ is unbiased.
Further, denoting by $Q_i(Y)$ the output $Q^i_{\tt SimQ}(Y;
Bd^{1/p})$, we get
\eq{
  \E{\norm{Q(Y)}_2^2}&=\E{\norm{Q(Y)-Y}_2^2}+\E{\norm{Y}_2^2}\\ &=\frac{1}{k^2} \sum_{i=1}^{k}\E{\E{\norm{Q_i(Y)-Y}_2^2|Y}}+\E{\norm{Y}_2^2}\\ &= \frac{\E{\norm{Q_1(Y)-Y}_2^2}}{k}+\E{\norm{Y}_2^2}
\\
  &\leq
  \frac{d^{2/p}B^2}{k}+B^2,
} where the first identity uses the fact
that $Q(Y)$ is an unbiased estimate of $Y$; the second uses the fact
that $Q_i(Y)-Y$ are zero-mean, independent random variables when conditioned on $Y$;
the third uses the fact that $Q_i(Y)-Y$ are identically distributed;
and the final inequality is by the performance of SimQ.
\end{proof}

The proof of upper bound for $p\in [2,\infty)$ in Theorem~\ref{t:main} is completed by setting $k=d^{2/p}$ and using Theorems~\ref{t:p2infty} and~\ref{t:e_uq}.

{

\section{Our Quantizers for $p \in [1,2]$}\label{s:p-less-than-two}
For p in $[1,2]$, the oracle yields unbiased subgradient estimates
  $Y$ such that $\norm{Y}_q \leq B$ almost surely. Our goal is to
  quantize such $Y$s in an unbiased manner and ensure that
  $\E{\norm{Q(Y)}_q^2}$ is $O(B^2)$. It can be seen that a simple unbiased
  uniform quantizer will achieve this using $d (\log d^{1/q}+1)$.}
  However, our goal here is to get a result that is stronger than this
  baseline performance. To that end, we split the input vector $Y$ in
  two parts $Y_1$ and $Y_2$ with the first part having $\ell_\infty$
  norm less than $c$ and the second part having less than $d/\Delta_1$
  nonzero coordinates. We use an ``$\ell_\infty$ ball quantizer'' (a
  uniform quantizer) for $Y_1$ and an ``$\ell_2$ ball quantizer'' for
  $Y_2$.

Specifically, set $\displaystyle{c :=\frac{B
    \Delta_1^{1/q}}{d^{1/q}}},$ where $\Delta_1$ is that in Theorem
\ref{t:main}.  Then, define
\begin{align}\label{e:Y1}
Y_1 :=\sum _{i=1}^{d} Y(i)\indic{\{|Y(i)| \leq c\}} e_i,\text{~~} Y_2
:=\sum _{i=1}^{d} Y(i)\indic{\{|Y(i)| > c\}} e_i.
\end{align}
Clearly, $\norm{Y_1}_\infty\leq c$. Further, since $\norm{Y}_q \leq
B$, the number of nonzero coordinates in $Y_2$ can be at the most
$B^q/c^q=d/\Delta_1$. For quantizing $Y_1$, we use the coordinate-wise
 uniform quantizer (CUQ) described in Section \ref{s:alg_RATQ}. In order to quantize $Y_1$ in \eqref{e:Y1},
 we set the parameters of CUQ to
 \begin{align}\label{e:CUQparam_lp}
 M = c, \quad \log(k+1)=\ceil{\log(2\sqrt{2}\Delta_1^{1/q}+2)}.
 \end{align}


\begin{lem}\label{l1}
Let $Q_{\tt u}$ be the quantizer CUQ with parameters $M$ and $k$ set
as in \eqref{e:CUQparam_lp}. Then, for $Y$ such that $\norm{Y}_q \leq
B~\text{a.s.}$ and $Y_1$ as that in \eqref{e:Y1}, $Q_{\tt u}(Y_1)$ can
be represented in $ d \ceil{\log(2\sqrt{2}\Delta_1^{1/q}+2)} $ bits,
$\E{Q_{\tt u}(Y_1)\mid|Y}=Y_1$, and $\E{\norm{Q_{\tt u}(Y_1)}_q^2}\leq
3B^2$.
\end{lem}
\begin{proof}
 CUQ requires a precision of $d\log(k+1)$, which coincides with the
 statement above for our choice of $k$.  To see unbiasedness, note
 that CUQ is an unbiased quantizer as long as all the coordinates of
 the input do not exceed $M$. Since we have set $M=c$ and
 $\norm{Y_1}_\infty=c$, this property holds.  Finally, to show that
 $\E{\norm{Q_{\tt u}(Y_1)}_q^2}\leq 3B^2$, note that $\E{\norm{Q_{\tt
       u}(Y_1)}_q^2} \leq 2\E{\norm{Q_{\tt u}(Y_1)
     -Y_1}_q^2}+2\E{\norm{Y_1}_q^2}.$ Also,
$\E{\norm{Q_{\tt u}(Y_1) -Y_1}_q^2} \leq B^2,$
where we use the fact that for
$M$ set as in \eqref{e:CUQparam_lp} we have that $|Q_{\tt
  u}(Y_1)(i)-Y_1(i)| \leq \frac{2M}{(k-1)}$ a.s., $\forall i \in [d]$,
by the description of CUQ.
\end{proof}
In order to quantize $Y_2$, we indicate the coordinates with non-zero
entries. This takes less than $d$ bits.  Then, we quantize the
restriction $Y_2^\prime$ of $Y_2$ to these nonzero entries. Recall
that the dimension of $Y_2^{\prime}$ is less than $d^{\prime}
:={d}/{\Delta_1}$. Also, the $\ell_2$ norm of $Y_2^{\prime}$ is less
than $\norm{Y}_2\leq\norm{Y}_q d^{1/2-1/q} \leq B d^{1/2-1/q} =:
B^{\prime}$.

We need a quantizer $Q$ such that $\E{\norm{Q(Y_2^{\prime})}_q^2}$ is
$O(B^2)$.  As seen in the proof of Lemma~\ref{l1}, one way to do this
is to ensure $\E{\norm{Q(Y_2^{\prime})-Y^{\prime}}_q^2}$ is $O(B^2), $
which, in turn, can be ensured if
$\E{\norm{Q(Y_2^{\prime})-Y_2^{\prime}}_2^2}$ is $O(B^2)$.  To achieve
this, we can use an unbiased quantizer for the unit $\ell_2$ ball in
$\R^d$, which can quantize it to an MSE of $O( 1/d^{1-2/q})$ using
$O(d \log (d^{1/2-1/q})$ bits.
We note that SimQ$^+$, while optimal for the stochastic optimization
use-case, does not yield the required scaling of bits in MSE.  A natural
candidate quantizer is RATQ, which is, in fact,
close to information theoretically optimal.
 We 
  set the parameters of RATQ in terms of
 $B^{\prime}$ and $d^{\prime}$. 
 We set 
\begin{align}\label{e:param_RATQ_lp}
&\nonumber
m=\frac{3{B^{\prime}}^2}{d^{\prime}}, \quad m_0=\frac{2{B^{\prime}}^2 }{d^{\prime}} \cdot  \ln s, \quad \log h=\ceil{\log(1+\ln^\ast(d^{\prime}/3))},
\\
& s=\log h, \quad \log(k+1) = \Delta_1.
\end{align}
%


\begin{lem}\label{l2}
Let $Q_{{\tt at}, R}$ be the quantizer RATQ with parameters set as \eqref{e:param_RATQ_lp}.
Then, for $Y$ such that $\norm{Y}_q \leq B ~{a.s.}$  and $Y_2^{\prime}$ the restriction of $Y_2$ in \eqref{e:Y1},    $Q_{{\tt at}, R}(Y_2^{\prime})$ can be represented in $2d+\Delta_2$ bits, $\E{Q_{{\tt at}, R}(Y_2^{\prime})\mid|Y}=Y_2^{\prime}$, and $\E{\norm{Q_{{\tt at}, R}(Y_2^{\prime})}_q^2}\leq 3B^2.$
\end{lem}
\begin{proof}

First, we note that the output of RATQ can be represented in $\ceil{d^{\prime}/s}(\log h) +d \log (k+1)$ bits, which, in this case, is less than
\eq{
\frac{d}{\Delta_1 \log h} \cdot \left(\log h \right)+ \log h+\left(\frac{d}{\Delta_1 } \log(k+1)\right) \leq 2d+\Delta_2.
}
For unbiasedness, note that for our choice of $m, m_0, h$, RATQ is always an unbiased quantizer of the input.
Finally, for showing $\E{\norm{Q_{{\tt at}, R}(Y_2^{\prime})}_q^2}\leq 3B^2$, 
we note that
\eq{
  \E{\norm{Q_{{\tt at}, R}(Y_2^{\prime})}_q^2}
  &\leq 2\E{\norm{Q_{{\tt at}, R}(Y_2^{\prime})-Y_2^{\prime}}_q^2}
  +2\E{\norm{Y_2^{\prime}}_q^2}
\\ 
  &\leq 2\E{\norm{Q_{{\tt at}, R}(Y_2^{\prime})-Y_2^{\prime}}_q^2}
  +2B^2
\\ 
  &\leq 2\E{\norm{Q_{{\tt at}, R}(Y_2^{\prime})-Y_2^{\prime}}_2^2}
  +2B^2.
}
The proof will be complete upon showing that $\E{\norm{Q_{{\tt at},R}(Y_2^{\prime})-Y_2^{\prime}}_2^2} \leq B^2/2$, towards
which we apply Lemma  \ref{l:useful} to get 
\[
\E{\norm{Q_{{\tt at}, R}(Y_2^{\prime})-Y_2^{\prime}}_2^2}
\leq B^2 d^{1-2/q} \cdot  \frac{9+3\ln s}{(k-1)^2},
\]
and substituting our choice of $k$.
\end{proof}

The overall quantizer $Q$ of input vector $Y$ is the sum of quantized
outputs of $Y_1$ and $Y_2$. By Lemmas \ref{l1} and \ref{l2}, the
quantized output of $Y$ can be represented in $d\left(
\ceil{\log(2\sqrt{2} {\Delta_1}^{1/q}+2)}+3\right) +\Delta_2$ bits\footnote{This
accounts for the communication needed to send the nonzero indices of
$Y_2$, too.}. Furthermore, $\alpha_p(Q) \leq \sqrt{12}B$. These facts
along with Theorem \ref{t:e_uq} prove the upper bound in
Theorem~\ref{t:main} for $p\in [1,2]$.

\section{Characterization of general tradeoff and mean square bounded oracles}\label{s:comment_general}
We close with the remark that an almost complete characterization of optimization error $\mathcal{E}^*(\X, \oO_{{\tt c}, p}, T, \W_{{\tt com}, r})$
 for any $r, p$ (namely, the low-precision regime) can be obtained using our
quantizers and the ideas developed in this Chapter. We describe in detail algorithms to achieve these bounds below.

\subsection{Upper Bounds on $\mathcal{E}^*(\X, \oO_{{\tt c}, 1}, T, \W_{{\tt com}, r})$ for $p \in (2, \infty]$}
For upper bounds when $p \in [2, \infty)$,
  note that the parameter $k$ of SimQ$^+$ gives us a nice lever to operate
  under any precision constraint $r \geq \log d$. It turns out that
  such a quantizer along with PSGD leads to upper bounds which are off by at the most
$\sqrt{\log d}$ factor from the lower bounds in Theorem \ref{t:concom_inf}.
    
    \subsection{Upper Bounds on $\mathcal{E}^*(\X, \oO_{{\tt c}, 1}, T, \W_{{\tt com}, r})$}\label{app:scheme}
    We now describe a new scheme to match the lower bound for $p=1$.
Our scheme divides the entire horizon of $T$ iterations into $Tr/d$ different phases. For any phase $t \in [Tr/d]$, the same point $x_t$ in the domain is queried $d/r$ times. For each of the $d/r$ queries in a phase, we use $r$-bit quantizers to quantize  different coordinates of the subgradient output. At a high level, we want to use these $r$ bits to send $1$ bit each for $r$ different coordinates, sending $1$ bit for each coordinate across the phases. However,
there is one technical difficulty. We have not assumed that making queries for the same point
gives identically distributed random variables. We circumvent this difficulty using random
permutations to create unbiased estimates for the subgradients. 

Specifically, for a permutation $\sigma\colon [d] \to [d]$ chosen uniformly at random using public randomness, we select the coordinates $\sigma(1+(i-1)\cdot r)$ to $\sigma(i\cdot r)$  of the subgradient estimate $\hat{g}_i$ supplied by the oracle for the $i$th query in the $t$th phase (i.e., $i$th time we query the point $x_t$) and quantize all of these coordinates using an $1$-bit unbiased quantizer for the interval $[-B, B]$. Note that such a quantizer can be formed
since $\norm{\hat{g}_i}_\infty\leq B$. 

Using this procedure, the quantized gradient for every query in each phase can be stored in $r$ bits. Furthermore, using all the $d/r$ quantized estimates received in a phase, we can create an estimate of the subgradient by simply adding all the estimates. Denote by $\bar{Q}_t$ our subgradient estimate in the $t$th phase. Then, 
\[
\bar{Q}_t=\sum_{i=1}^{d} Q_{\pi(i)}(\hat{g}_i)e_{\sigma(i)},
\]
where
$\hat{g}_i$ is the subgradient estimate returned by the oracle when we query $x_t$ for the $i$th time and $Q_i$ is a $1$-bit unbiased estimator of the $i$th coordinate of gradient estimate given below: For all vectors $g$, such that $\norm{g}_{\infty}\leq B$, we have
\[
Q_i(g)
 = 
\begin{cases} 
B &\text{ w.p. }  \quad\frac{g(i)+B}{2B}\\
-B   &\text{ w.p. }  \quad\frac{B-g(i)}{2B}
\end{cases}. 
\] 
Then, we use $\bar{Q}_t$ to update $x_t$ to $x_{t+1}$ using stochastic mirror descent with mirror map 
\[
\phi_a(x) \colon= \frac{\norm{x}_a^2}{a-1},
\]
where
$a =\frac{2\log d}{2\log d-1}$.


\begin{figure}[ht]
\centering
\begin{tikzpicture}[scale=1, every node/.style={scale=1}]
\node[draw, text width= 15cm, text height=,] {%
\begin{varwidth}{\linewidth}
            
            \algrenewcommand\algorithmicindent{1em}
            \renewcommand{\thealgorithm}{}

\begin{algorithmic}[1]
 \For{$t\in [Tr/d]$}
 
  \For{$i\in [d/r]$}
   
 \State {\color{red1} At Center:}\;
      Query the oracle for  $x_t$\;
      
      \vspace{0.2cm}
 \State      {\color{blue} At Oracle:}\;
        Output the $r$-bit vector of $1$-bit unbiased estimates of the 
  \Statex      $r$ coordinates $\{1+(i-1)\cdot r, \ldots ,i\cdot r \}$ of $\hat{g}_i(x_t)$ given by 
        \[
          \bar{Q}_t = \sum_{j=1+(i-1)\cdot r}^{i \cdot r} Q_{\pi(j)}(\hat{g}_i(x_t))e_{\sigma(j)}
        \]
        \;
\State         {\color{red1} At Center:}\;
  $x_{t+1} = \arg\min_{x \in \X} (\eta_t \langle x, \bar{Q}_t \rangle) +D_{\Phi_a}(x, x_t))$ \;
    \EndFor
    \EndFor
\State \textbf{Output:} $\frac{\sum_{i=1}^{T}x_t}{T}$
\end{algorithmic}
\end{varwidth}};
\end{tikzpicture}

\caption{$\pi^*$ Almost optimal Scheme for Communication constrained optimization for convex and $\ell_1$ lipschitz family}\label{alg:optimal_l1}
\end{figure}
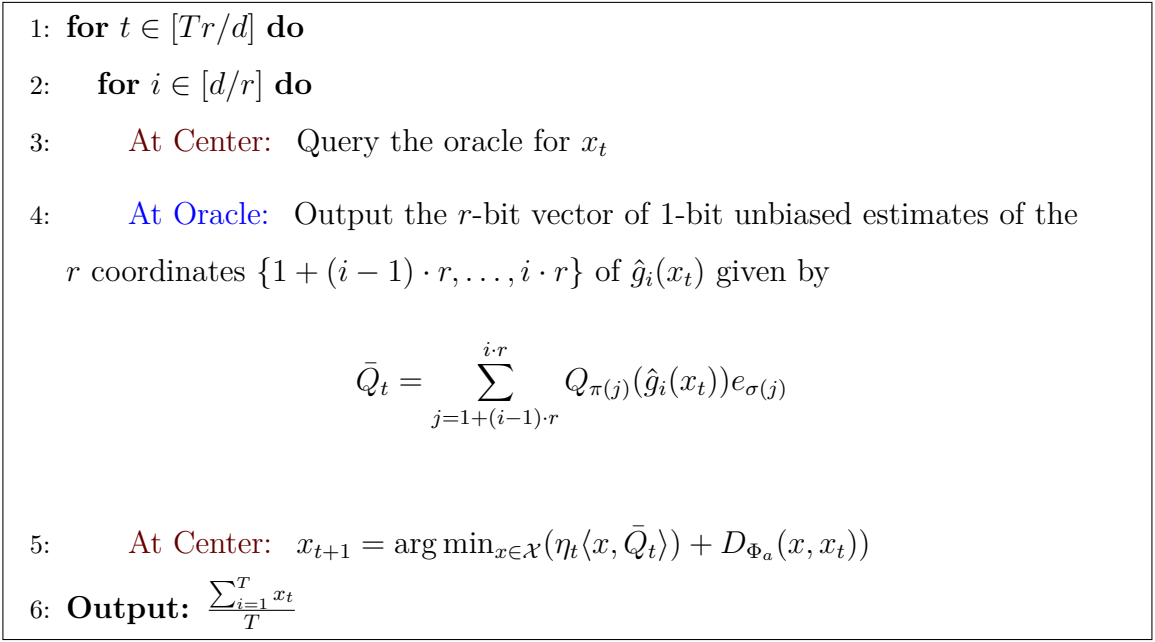
\begin{thm}
 For $ r  \in \N$, we have
\[
 \sup_{\X \in \mathbb{X}_1(D)} \ep^*(\X , \oO_{{\tt c}, 1}, T, \W_{{\tt com}, r  }) 
\leq 
\frac{c_0DB \sqrt{\log d}}{\sqrt{T}} \cdot \sqrt{\frac{d}{d \wedge r}} 
\]
for every $D>0$.
\end{thm}
\begin{proof}
  Note that our first order optimization algorithm $\pi^*$ uses $Tr/d$ iterations. Moreover, the subgradient estimates $\bar{Q}_t$ are unbiased and have their infinity norm bounded by $B$.
  Namely, we have obtained an unbiased subgradient oracle which produces estimates
  with infinity norm bounded by $B$. Thus, using the standard analysis of
  mirror descent using noisy subgradient oracle for optimization over an $\ell_1$ ball
  with mirror map
  $\phi_a(x) \colon= \frac{\norm{x}_a^2}{a-1}$  (see Theorem \ref{t:e_uq}), the proof is complete.
\end{proof}
 
\subsection{Upper Bounds on $\mathcal{E}^*(\X, \oO_{{\tt c}, p}, T, \W_{{\tt com}, r})$ for $p \in (1, 2)$.}
    For $p \in (1, 2)$, the scheme described in Algorithm \ref{alg:optimal_l1} can still be used. However, the upper bounds would be off by a factor of $d^{1/q}$. This factor  increases with increase in $p$ and and as we get closer to $2$, employing optimal quantizers for the Euclidean setting is a better option. For instance, employing RATQ  along with the appropriate mirror descent algorithm would lead to optimization error off by $d^{1/2-q}$ from the lower bound. Thus, inteprolationg between the schemes for $p=1$ and $p=2$, we match the lower bound in Theorem \ref{t:concom_P2} upto a factor of $\min{d^{1/q},  d^{1/2-q}}$ for any precision $r$ and $p \in (1, 2).$
    
    Finally, we believe that  removing these
      remaining factors can lead to new quantizers, and is of research
      interest.

\subsection{Mean square bounded oracles}
For mean square bounded oracles mentioned in Chapter \ref{Ch:RATQ}, the bias in the quantized oracle
        output is nearly inevitable. In our previous chapter, we proposed appropriate {\em{gain-shape}}
        quantizers for quantizing the oracle output in the Euclidean
        setup, which resulted in lesser bias over standard
        quantizers. This idea is valid for the general $\ell_p$ setup;
        in particular, we can use a gain quantizer to quantize the
        $\ell_q$ norm of the oracle output and a
        {{shape}} quantizer to quantize the oracle output vector normalized by the
        $\ell_q$ norm, the shape of the oracle output vector.
        Note that the shape vector has
        has $\ell_q$ norm $1$, which
        allows us to use the quantizers developed in this chapter to
        quantize the shape. The gain is a scalar random variable which
        has its second moment bounded by $B^2$. To quantize such a
        random variable, we can use AGUQ from previous chapter.
      Clearly, the lower bounds
      for almost surely bounded oracles remain valid for mean square bounded oracles as well. Additionally, we can also derive lower bounds for a specific class
        of quantizers, such as those derived in previous chapter, which help in capturing the reduction in the convergence rate due to mean square bounded noise. 




\part{Efficient Quantization for Federated Learning Primitives}
\chapter{Communication-Efficient Distributed Mean Estimation }\label{Ch:DME}
\section{Synopsis}
Communication efficient distributed mean estimation is an important primitive that arises in many distributed learning and optimization scenarios such as federated learning. Without any probabilistic assumptions on the underlying data, we study the problem of distributed mean estimation in two different settings: 1) where the server does not have access to side information and 2) where the server has access to side-information. In the first setting, we use RATQ proposed in Chapter \ref{Ch:RATQ} and improve over the state of the art. 

In the second setting, we propose \emph{Wyner-Ziv estimators}, which are communication and computationally efficient and near-optimal when an upper bound for the distance between the side information and the data is known. 
  In a different direction, when there is no knowledge assumed about the distance between side information and the data, we present an alternative
  Wyner-Ziv estimator that uses correlated sampling. This latter setting offers {\em universal recovery guarantees}, and perhaps will be of interest in practice when the number of users is large and keeping track of the distances between the data and the side information may not be possible.
  
  The results presented in this Chapter are from \cite{mayekar2020ratqj} and \cite{mayekarwyner}.
\section{Introduction}
Consider the problem of distributed mean estimation for $n$ vectors $\{x_i\}_{i=1}^{n}$ in $\R^d$, where $x_i$ is available to client $i$. Each client communicates to a server using a few bits to enable the server to compute the  empirical mean
\begin{align}\label{e:sample_mean}
\bar{x}=\frac{1}{n}\sum_{i=1}^n x_i.
\end{align}

This estimation problem has become a crucial primitive for distributed optimization scenarios such as federated learning, where the data is distributed across multiple clients.  One of the main bottlenecks in such distributed scenarios is the significant communication cost incurred due to client communication at each iteration of the distributed algorithm. This has spurred a recent line of work which seeks  to design quantizers  to express $x_i$s using  a low precision and, yet, enable the server to compute a high accuracy estimate of $\bar{x}$ (see \cite{suresh2017distributed}, \cite{konevcny2018randomized}, \cite{chen2020breaking}, \cite{huang2019optimal},    and the references therein).

Most of the recent works on distributed mean estimation focus on the setting where the server  must estimate the sample mean based on the client vectors, and nothing else. However, in practice, the server may also have access to some side information. For example, consider the task of training a machine learning model based on remote client data as well as some publicly accessible data. At each iteration, the server communicates its global model to the client, based on which the clients compute their updates (the gradient estimates based on their local data), compress them, and then send them to the server. The server may choose to compute its own update using the publicly available dataset to complement the updates from the client.
In a related setting, the server can use the previously received gradients as side information for the next gradients expected from the clients. Similarly, distributed mean estimation with side information can be used for variance reduction in other problems such as power iteration or parallel SGD ($cf.$~\cite{davies2020distributed}). 

Motivated by these observations, for the distributed mean estimation problem described at the start of the section, we study both the settings of distributed mean estimation:
\begin{enumerate}
\item The {\em{no side information}} setting, where the server does not have access to any side information.
\item The {\em{side information}} setting, where the server has access to some side information $\{y_i\}_{i=1}^{n}$ in $\R^d$, in addition to the communication from clients. Here, $y_i$ can be viewed as server's initial 
estimate (guess)
of $x_i$. We emphasize that the side information $y_i$ is available only to the sever and can, therefore, be used for estimating the mean at the server, but is not available to the clients while quantizing the updates $\{x_i\}_{i=1}^n$. 
\end{enumerate}

We close this section with the remark  that distributed mean estimation in the no side information setting can be viewed as a special case of distributed mean estimation in the side information setting, where the side-information $\{y_i\}_{i=1}^{n}$ is set to $0.$ We will, therefore, describe our model for the side-information setting.

\subsection{The model} 
\begin{figure}[ht]
\begin{center}

\begin{tikzpicture}[scale=1, transform shape,
    pre/.style={=stealth',semithick},
    post/.style={->,shorten >=1pt,>=stealth',semithick},
dimarrow/.style={->, >=latex, line width=1pt},
dimmarrow/.style={<->, >=latex, line width=1pt},
    ]
\clip (2,17.35) rectangle  (11,13.65) ;

\draw[ fill = brown!30! ] (3.5, 17) rectangle(7.5, 16);
\node[align=center] at (5.5,16.5) {$(y_1,\ldots, y_n)$} ;
\node[align=center] at (5.5,17.2) {Server} ;

\draw[ fill = orange!50!] (3, 15) rectangle(4, 14);
\node[align=center] at (3.5,14.5) {$x_1$};
\node[align=center] at (3.5,13.8) {Client 1} ;

\draw[ fill = orange!50!] (4.5, 15) rectangle(5.5, 14);
\node[align=center] at (5,14.5) {$x_2$};
\node[align=center] at (5,13.8) {Client 2} ;

\draw[ fill = orange!50!] (7, 15) rectangle(8, 14);
\node[align=center] at (7.5, 14.5) {$x_n$}; 
\node[align=center] at (7.5,13.8) {Client n} ;

\draw[dashed, line width=1pt, ta3aluminium] (5.7,14.5) to (6.8,14.5);

\draw[dimarrow, ta3aluminium] (3.5,15) to (8-3,15.9); 
\draw[dimarrow, ta3aluminium]  (5,15) to(9.5-3.5,15.9) ;
\draw[dimarrow, ta3aluminium] (7.5,15) to(9.5-3,15.9)  ;

\end{tikzpicture}

\end{center}
\caption{Problem setting of mean estimation with side information}
\label{fig:setup}
\end{figure}
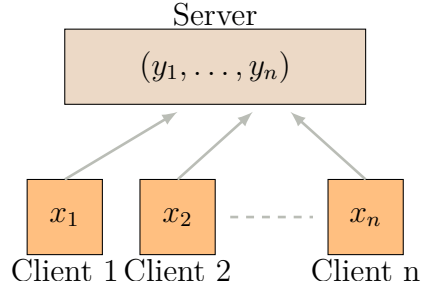

Consider the input $\mathbf{x} :=(x_1,\ldots, x_n)$ and the side information $\mathbf{y} :=(y_1,\ldots, y_n)$.
The clients use a communication protocol to send $r$ bits each about their observed vector
to the server. For the ease of implementation, we restrict to non-interactive
protocols. Specifically,
we allow {\em simultaneous message passing} (SMP) protocols $\pi=(\pi_1, ...,\pi_n)$ where the communication
$C_i=\pi_i(x_i, U)\in \{0,1\}^r$ of client $i$, $i\in [n]$,   can only depend on its local observation $x_i$ and public randomness $U$.
Note that the clients are not aware of side information $\mathbf{y}$, which is available only to the server.
In effect, the message $C_i$ is obtained by {\em quantizing} $x_i$ using an appropriately chosen
randomized quantizer. 
Denoting the overall communication by $C^n:=(C_1, C_2, ..., C_n)$,
the server uses the transcript $(C^n, U)$ of the protocol
and the side information $\mathbf{y}$ to form the estimate 
of the sample mean\footnote{While side information $y_i$ is associated with client $i$, we do
not enforce this association in our general formulation at this point.} $\hat{\bar{x}}=\hat{\bar{x}}(C^n, U, \mathbf{y})$;
see Figure~\ref{fig:setup} for a depiction of our setting.
We call such a $\pi$ an {\em $r$-bit SMP protocol} 
with input $(\mathbf{x}, \mathbf{y})$
and output $\hat{\bar{x}}$.

We measure the performance of protocol $\pi$ for inputs $\mathbf{x}$ and $\mathbf{y}$ 
and output $\hat{\bar{x}}$
using mean squared error (MSE)
given by
 \[
 \mathcal{E}(\pi, \mathbf{x}, \mathbf{y}):=\E{\norm{\hat{\bar{x}}-\bar{x}}_2^2},
 \]
where the expectation is over the public randomness $U$ and $\bar{x}$ is given in \eqref{e:sample_mean}.
We study the MSE of protocols for $\mathbf{x}$
and $\mathbf{y}$ such that 
the Euclidean distance between $x_i$ and $y_i$ is at most $\Delta_i$, i.e.,
\begin{align}
\norm{x_i-y_i}_2 \leq \Delta_i, \quad \forall \,i \in[n].
\label{eq:delta_cond}
\end{align}
Denoting $\mathbf{\Delta} :=(\Delta_1, \ldots, \Delta_n)$, we are interested in the performance
of our protocols for the following settings: 
\begin{enumerate}
\item {\bf The {\em no side information} setting}, where $\Delta_i =1$ and $y_i=0,$ for all $i \in [n].$ That is, the server does not have access to side information and the input vectors lie in the unit Euclidean ball.
\item {\bf The  {\em side information} setting}, where the server has access to some side-information. We study two different cases for this setting, which are described as follow:
\begin{enumerate}
\item {\bf  The {\em \known} setting}, where $\Delta_i$ is known to client $i$ and the server; 

\item {\bf The {\em \unknown}setting}, where $\Delta_i$s are unknown to everyone.
\end{enumerate}
\end{enumerate}

In all these settings, we seek to find efficient $r$-bit quantizers for $x_i$ that will allow accurate sample mean estimation. We now point out the difference between the two different settings of distributed mean estimation in presence of side information.
In the \known setting, the quantizers of different clients can be chosen using the knowledge of $\mathbf{\Delta}$;
in the \unknown setting, they must be fixed irrespective of $\mathbf{\Delta}$.

In another direction, we distinguish the {\em small-precision} setting of $r\leq d$ from the {\em large-precision}\footnote{The definition of large-precision setting is different from the definition of high-precision setting described in the first part of the thesis. Observe that the large precision setting here simply refers to the setting of  $r >d$, whereas in the high-precision setting from earlier chapters we looked to characterize the minimum precision required to attain the classical convergence rate.} setting of $r>d$.
The former is perhaps of more relevance for federated learning and high-dimensional distributed optimization, while the latter
has received a lot of attention in the information theory literature on rate-distortion theory.

As a benchmark, we recall the result for distributed mean estimation with no side-information from \cite{suresh2017distributed}.
 \cite{suresh2017distributed} showed that  the minmax MSE in the no side-information setting is 
\begin{equation}
    \label{eq:opt}
{\Omega}\left(\frac{d}{nr} \right).
\end{equation}
Further, \cite{suresh2017distributed} derive an upper bound which matches the lower bound upto a factor of $\log \log d.$
\subsection{Our contributions}\label{s:OurC}
In the no side-information setting, we improve over the upper bound of \cite{suresh2017distributed} and match the lower bound upto a miniscule $\log \log ^*d $ factor by using RATQ from Chapter \ref{Ch:RATQ}.

In the side-information setting, drawing on ideas from distributed quantization problems 
in information theory ($cf.$~\cite{wyner1976rate}), 
specifically the Wyner-Ziv problem, we present {\em Wyner-Ziv estimators}.
In the \known setting, for a fixed $\mathbf{\Delta}$,
and the small-precision setting of $r\leq d$,
we propose an {\em $r$-bit SMP protocol} $\pi^*_{\tt k}$ which satisfies
\[
\mathcal{E}(\pi^*_{\tt k},  \mathbf{x}, \mathbf{y})={O}\left(\sum_{i=1}^{n}\frac{\Delta_i^2}{n} \cdot \frac{d \log \log n}{nr} \right),
\]
for all $\mathbf{x}$ and $\mathbf{y}$ satisfying~\eqref{eq:delta_cond}. 
 Thus, in the case where all $x_i$s lie in the Euclidean ball of radius $1$, we improve upon the optimal estimator for distributed mean estimation in the no side information setting \theerthaedit{}{\eqref{eq:opt}} in the regime $\sum_{i=1}^{n}\frac{\Delta_i^2 \log \log n}{n}\leq 1$. Our estimator is motivated by the classic Wyner-Ziv problem, and hence, we refer to it as
 the {\em Wyner-Ziv estimator}. The details of the algorithm are given in Section~\ref{s:sRMQ}.
 
 
Our protocol uses the same (randomized) $r$-bit quantizer for
each client's data and simply uses the sample mean of the quantized vectors as the estimate for $\bar{x}$. 
Furthermore, the common quantizer used by the clients is efficient and has nearly linear time-complexity of $O(d \log d)$. As was the case for RATQ,
our proposed quantizer first applies a random rotation  to the input vectors $x_i$ 
at client $i$
and the side information vector $y_i$ at the server. This ensures that the $\Delta_i$ upper bound on the $\ell_2$ distance of $x_i$ and $y_i$ is converted to roughly a $\Delta_i/\sqrt{d}$ upper bound on the $\ell_{\infty}$ distance between $x_i$ and $y_i$. This then enables us to use efficient one-dimensional  quantizers for each coordinate of the $x_i$, which can now operate with the knowledge that the server knows a $y_i$
with each coordinate within roughly ${\Delta_i}/{\sqrt{d}}$ of $x_i$'s coordinates.

Moreover, we show that this protocol $\pi^*_{\tt k}$ has optimal (worst-case) MSE up to an $O(\log \log n)$ factor. That is, we show that for any other
$r$-bit SMP protocol $\pi$ for $r\leq d$, we can find $\mathbf{x}$ and $\mathbf{y}$ satisfying~\eqref{eq:delta_cond} such that
\begin{align*}
\mathcal{E}(\pi, \mathbf{x}, \mathbf{y})={\Omega}\left(\min_{ i \in \{1, \ldots, n\}} \Delta_i^2 \cdot \frac{d }{nr} \right).
\label{e:lower_bound}
\end{align*}

In the \unknown setting, we propose a protocol $\pi^*_{\tt u}$ which adapts to the unknown distance $\Delta_i$ between $x_i$ and $y_i$ and, remarkably,  provides MSE guarantees dependent on $\mathbf{\Delta}$. Specifically, 
for the small-precision setting of $r\leq d$,
the protocol satisfies
 \[
 \mathcal{E}(\pi^*_u, \mathbf{x}, \mathbf{y})={O}\left(\sum_{i=1}^{n}\frac{\Delta_i}{n}  \cdot \frac{d \ln^*d}{nr} \right),
 \] 
 for all $\mathbf{x}$ and $\mathbf{y}$ in the unit Euclidean ball
 $\B:= \{x \in \R^d: \norm{x}_2 \leq 1\}$ and  satisfying~\eqref{eq:delta_cond}.
 Thus,  we improve upon the  optimal estimator for the no side information counterpart \theerthaedit{}{\eqref{eq:opt}} in the regime $\sum_{i=1}^{n}\frac{\Delta_i \ln^* d}{ n} \leq 1.$
 Once again, the quantizer employed by the protocol is efficient and has nearly linear time-complexity of $O(d \log d)$. At the heart of our proposed quantizer is the technique of correlated sampling from~\cite{Holenstein07} which enables to derive a $\mathbf{\Delta}$ dependent MSE bound. 
 
Furthermore, both our quantizers can be extended to the large-precision regime of  $r>d$. The quantizer for the \known setting directly extends by using $r/d$ bits per dimension. The MSE of the SMP protocol using this quantizer for all the clients is only a factor of $\log n + r/d$ from the lower bound derived in \cite{davies2020distributed} for the large-precision regime. The quantizer for the \unknown setting can be extended by sending the ``type'' of the communication vector, 
 following an idea proposed in Chapter \ref{Ch:Limits} for SimQ$^+$.  The MSE of the SMP protocol using this quantizer for all the clients falls as $2^{-r/d\ln^*d}$ as opposed to $d/r$ 
that can be obtained using naive extensions of our quantizer.

\subsection{Prior work}
The version of the distributed mean estimation problem with no side information at the server has been extensively studied. For any protocol in this setting operating with a precision constraint of $r \leq d$  bits per client, using a strong data processing inequality from \cite{duchi2014optimality}, \cite{suresh2017distributed} shows a lower bound on MSE of  $\displaystyle{\Omega\left(\frac{d}{nr}\right),}$ when all $x_i$s lie in the Euclidean ball of radius one.  \cite{suresh2017distributed} propose a rotation based  uniform quantization scheme which matches this lower bound up to a factor of $\log \log d$ for any precision constraint $r$.

The \known setting described above was first considered in \cite{davies2020distributed}.   The scheme of~\cite{davies2020distributed}
 relies on 
lattice quantizers with information theoretically optimal covering radius. Explicit lattices to be used and computationally efficient
decoding is not provided. 

In contrast, we provide explicit computationally efficient protocols
for both small- and large-precision settings. Also, we
establish lower bounds showing the optimality of our quantizer upto a multiplicative factor of $\log \log n$
in the small-precision regime of $r \leq d$.  In comparison, the scheme of \cite{davies2020distributed} is off by a factor of $\frac{d}{r}$ from this lower bound. Thus, when $r \ll d$, 
our scheme performs significantly better than 
that in \cite{davies2020distributed}.
We remark that the \unknown setting, which is perhaps more important in certain applications where estimating the distance of
side information of each client is infeasible, has not been considered before. 


 
\subsection*{Organization}
We will review some preliminaries in the next section.
Our results for the small-precision regime  in \known setting are provided in Section~\ref{s:known}
and in the \unknown setting are provided in Section~\ref{s:unknown}. In Section \ref{s:hpr}, we extend our results to the large-precision regime. Finally, we close with all the proofs in Section \ref{s:proof}.

\section{Preliminaries and the structure of our protocols}\label{s:preliminaries}
While our lower bound for the \known setting holds for an arbitrary SMP protocol,
all the protocols we propose in this chapter, for the no side information setting, as well as the \known and the \unknown settings in the side information case, have a common structure.
We use $r$-bit quantizers to form estimates of $x_i$s at the server and then compute the sample mean of
the estimates of $x_i$s. To describe our protocols and facilitate our analysis, we begin by 
concretely defining the distributed quantizers needed for this problem. Further, we present a simple result relating
the performance of the resulting protocol to the parameters of the quantizer.

An $r$-bit quantizer  $Q$ for input vectors in $\X \subset \R^d$ 
and side information $\Y\subset \R^d$
consists of randomized mappings\footnote{We can use public randomness $U$
for randomizing.}
$(\Qenc, \Qdec)$ with the encoder mapping $\Qenc:\X \to\{0,1\}^{r}$  used by the client to quantize
and the decoder mapping $\Qdec: \{0,1\}^r \times \Y \to \X$ used by the server to aggregate quantized vectors. 
The overall quantizer $Q$ is given by the composition mapping $Q(x, y)=\Qdec(
(\Qenc(x), y)$. 

In our protocols, for input $\mathbf{x}$ and side information $\mathbf{y}$,
client $i$ uses the encoder $\Qenc_i$ for the $r$-bit quantizer $Q_i$ to 
send $\Qenc_i(x_i)$. The server uses $\Qenc_i(x_i)$ and $y_i$ to form the estimate 
$\hat{x}_i=Q_i(x_i,y_i)$
of $x_i$. We assume that the randomness used in quantizers $Q_i$ for different
$i$ is independent, whereby $\hat{x}_i$ are independent of each other for different
$i$. Then server finally forms the estimate of the sample mean as
\begin{align}
  \hat{\bar{x}}:=\frac 1 n\sum_{i =1}^n \hat{x}_i.   
  \label{e:estimate}
\end{align}

For any quantizer $Q$, the following two quantities
will determine its performance when used in our distributed mean estimation protocol:
\eq{
\alpha(Q; \Delta) &:= \sup_{x\in \X, y\in \Y : \norm{x-y}_2\leq \Delta}\E{\norm{Q(x, y)-x}_2^2},\\
\beta(Q; \Delta) &:= \sup_{x\in \X, y\in \Y : \norm{x-y}_2\leq \Delta}\norm{\E{Q(x, y)-x}}_2^2,
}
where\footnote{The $\alpha$ above differs from the $\alpha_2$ defined in the first part of the thesis; the former characterizes the worst-case MSE, while the latter describes the worst-case $L_2$ norm. However, $\beta$ is similar to $\beta_2$ defined in the first part of the thesis, as both characterize the worst-case bias of the quantizer.} the expectation is over the randomization of the quantizer. Note that 
$\alpha(Q; \Delta)$ can be interpreted as the worst-case MSE  and $\beta(Q, \Delta)$ the worst-case bias 
over $x\in \X$ and $y\in \Y$ such that $\norm{x-y}_2\leq \Delta$.

The result below will be very handy for our analysis. 
\begin{lem}\label{l:main}
For $\mathbf{x}\in \X^n$ and $\mathbf{y}\in \Y^n$ satisfying~\eqref{eq:delta_cond} and $r$-bit quantizers $Q_i$, $i \in [n]$,
using independent randomness for different $i\in[n]$,
the estimate $\hat{\bar{x}}$ in~\eqref{e:estimate} 
and the sample mean  $\bar{x}$ in~\eqref{e:sample_mean}
satisfy
\[
\E{ \|\hat{\bar{x}} - \bar{x}\|_2^2}
\leq \sum_{i=1}^n\frac{\alpha(Q_i; \Delta_i)}{n^2} +  \sum_{i=1}^n\frac{\beta(Q_i; \Delta_i)}{n}.
\]
\end{lem}

 \section{Distributed mean estimation with no side information}
 As stated previously, in the setting of distributed mean estimation with no side information we have $\Delta_i=1$ and $y_i=0,$ $\forall ~i ~\in [n].$ We will therefore state our results under these assumptions for this case.
 Our protocol $\pi_n^\ast$ uses subsampled RATQ as the quantizer for each client, which is described in Section \ref{s:LowPrec}, with parameters of the Quantizer set as in  \eqref{e:RATQ_RCS_params} and \eqref{e:RATQ_unit_levels}.
 \begin{thm}\label{t:DME_noside}
For $n\geq 2$, fixed $\mathbf{\Delta}=(\Delta_1,\ldots, \Delta_n),$ $d \geq r \geq 2\left(3 +\ceil{\log (1+\ln^*({d}/{3}) ) }\right),$ and $\mathbf{y}$, where $\Delta_i=1$ and $y_i=0,$ $\forall ~i~ \in [n],$ the protocol $\pi^*_n$ with parameters set as in \eqref{e:RATQ_RCS_params} and \eqref{e:RATQ_unit_levels} is an $r$-bit protcol which satisfies 
\[\mathcal{E}(\pi^*_n, \mathbf{x}, \mathbf{y}) \leq (6+2\ceil{\log (1+\ln^\ast(d/3))})
\left(\sum_{i \in [n]}\frac{1}{n} \cdot \frac{d }{nr} \right).
\]

\end{thm}
\noindent The proof is a direct extension of the analysis of RATQ presented in Chapter \ref{Ch:RATQ} and is deferred to Section \ref{s:DME_noside_proof}.

This matches the lower bound of $\Omega\left(\displaystyle{\frac{d}{nr}}\right)$, derived in \cite[Theorem 5]{suresh2017distributed}, upto a tight $\log \log^* (d)$.  To the best of our knowledge, for $r \ll d$, this is the tightest known upper bound for distributed mean estimation with no-side information.  Moreover, the protocol $\pi_n^\ast$ can be efficiently implemented as the encoding and decoding complexity of RATQ is $d\log d$. 



\section{Distributed mean estimation with \known}\label{s:known}
In this section, we present our Wyner-Ziv estimator for the \known setting. As described in Section~\ref{s:preliminaries}, we use the the same (randomized) 
quantizer across all the clients and form the estimate of sample mean as in~\eqref{e:estimate}. We only need to define the common quantizer 
used by all the clients, which we do in Section \ref{s:sRMQ}. In Sections \ref{s:MQ} and \ref{s:RMQ}, we provide the basic
building blocks of our final quantizer.
Further, in Section~\ref{s:LB}, we derive a lower bound for the worst-case MSE that  
establishes  the near-optimality of our protocol. 
 Throughout we restrict to the small-precision setting of $r\leq d$.
\subsection{Modulo Quantizer (MQ)}\label{s:MQ}
The first subroutine used by our larger quantizer is the {\em Modulo Quantizer } (MQ).
MQ is a one dimensional distributed quantizer that can be applied to the input $x\in\R$
with side information $y \in \R$. We give an input parameter $\Delta^\prime$
to MQ where $|x-y|\leq \Delta^\prime$. In addition to $\Delta^\prime$,
 MQ also has the resolution parameter $k$ 
and the lattice parameter $\eps$
 as inputs.

For an appropriate
$\eps$ to be specified later, 
we consider the lattice $\Z_{\eps}= \{ \eps z: z \in \Z\}$. 
For a given input $x$, the encoder $\Qenc_{\tt M}$ 
finds the closest points in $\Z_{\eps}$ larger and smaller than $x$.
Then, one of these points is sampled randomly to get an unbiased estimate of $x$. The sampled point will be of the form $\tilde{z} \eps$, where $\tilde{z}$ is in $\Z$.  We note that
the chosen point $\tilde{z}$ satisfies
\begin{align}
\eps\E{\tilde{z}} &= x\,\, \text{and}
\nonumber
\\
|x-\eps \tilde{z}|&< \eps, \quad \text{almost surely}.
    \label{e:tildez_close}
\end{align}
The encoder sends $w=\tilde{z} \mod k$ to the decoder, which requires $\log k$ bits.

Upon receiving this $w$, the decoder $\Qdec$ looks at the set $\Z_{w,\eps}=\{(zk+w )\cdot \eps: z \in \Z\}$ and decodes the point closest to $y$, which we denote by $Q_{\tt M}(x, y)$.  
Note
that declaring $y$ will already give a MSE of less than $\Delta$. 
A useful property of this decoder is that
its output is always within a bounded distance from $y$; namely,
since in Step 1 of Alg.~\ref{a:D_MCQ} we look for the closest 
point to $y$ in the lattice $Z_{w,\eps}:=\{(zk+w)\cdot \eps: z\in \Z\}$,
the output $Q_{{\tt M}}(x,y)$ satisfies 
\begin{align}
|Q_{{\tt M}}(x,y)-y|\leq k\eps, \quad \text{almost surely}.
    \label{e:boundedness_MQ_decoder}
\end{align}

We summarize MQ in Alg.~\ref{a:E_MCQ} and~\ref{a:D_MCQ}.
\begin{figure}[ht]
\centering
\begin{tikzpicture}[scale=1, every node/.style={scale=1}]
\node[draw,text width= 11 cm, text height=,] {%
\begin{varwidth}{\linewidth}
            
            \algrenewcommand\algorithmicindent{0.7em}
\begin{algorithmic}[1]
\vspace{-0.5cm}
\Require Input $ x\in \R$, Parameters 
{$k$, $\Delta^{\prime}$, and $\eps$}

\State Compute $z_u = \ceil{x/\eps}$,  $z_l = \floor{x/\eps}$

 \State  Generate
     $
      {\tilde{z}} =
\begin{cases}
z_u, \quad w.p. ~ x/\eps - z_l
 \\ z_l,  \quad w.p. ~  
  z_u-x/\eps
\end{cases}
$
   
 \State \textbf{Output:} $\Qenc_{\tt M}(x)=\tilde{z} \text{ mod } k$
\end{algorithmic}  
\end{varwidth}};
 \end{tikzpicture}
 
 \renewcommand{\figurename}{Algorithm}
 \caption{Encoder  $\Qenc_{\tt M}(x)$ of MQ}\label{a:E_MCQ}
 \end{figure}
 
 \begin{figure}[ht]
\centering
\begin{tikzpicture}[scale=1, every node/.style={scale=1}]
\node[draw, text width= 11 cm, text height=,] {%
\begin{varwidth}{\linewidth}
            
            \algrenewcommand\algorithmicindent{0.7em}
 \renewcommand{\thealgorithm}{}
\begin{algorithmic}[1]
  \Require Input $w \in \{0, \ldots, k-1\}$, $y \in \R$
   \State Compute $\hat{z}= \arg\min \{ |(zk +w)\cdot \eps -y| \colon z \in \Z\} $
    \State \textbf{Output:} $\Qdec_{\tt M}(w, y)=(\hat{z}k+w)\eps$\label{step:output_coordinate}
\end{algorithmic}
\end{varwidth}};
 \end{tikzpicture}
 
 \renewcommand{\figurename}{Algorithm}
 \caption{Decoder $\Qdec_{\tt M}(w, y)$ of MQ}\label{a:D_MCQ}
 \end{figure}




The result below provides performance guarantees for $Q_{\tt M}$. The key observation is that the output $Q_{\tt M}(x, y)$ of the quantizer equals $\tilde{z}\eps$ with $\tilde{z}$ found at the encoder, if
$\eps$ is set appropriately. 
\begin{lem}\label{t:MQ}
Consider the Modulo Quantizer $Q_{\tt M}$ described in 
Alg.~\ref{a:E_MCQ} and~\ref{a:D_MCQ} with parameter
$\eps$ set to satisfy 
\begin{align}
 k\eps\geq 2(\eps + \Delta^\prime).
 \label{e:parameter_condition}
\end{align}
Then, for every $x, y$ in $\R$ such that $|x-y| \leq \Delta^\prime$,
the output $Q_{\tt M}(x, y)$ of MQ satisfies
 \begin{align*}
\E{Q_{\tt M}(x, y)}&=x\,\,\text{ and }
\\
|Q_{\tt M}(x, y)-x|&\leq \eps,
\quad \text{almost surely.}
 \end{align*}
 In particular, we can set $\eps=2\Delta^\prime/(k-2)$, 
 to get 
 $|Q_{\tt M}(x, y)-x|\leq 2\Delta^\prime/(k-2)$.
Furthermore, the output of $Q_{\tt M}$ can be described in $\log k$ bits.
\end{lem}

We close with a remark that the modulo operation
used in our scheme is the simplest and easily implementable version of 
classic
coset codes obtained using nested lattices used in 
distributed quantization ($cf.$~\cite{Forney88, Zamir02, liu2016polar}) and was used in~\cite{davies2020distributed} as well.

\subsection{Rotated Modulo Quantizer (RMQ)}\label{s:RMQ} We now describe {\em{Rotated Modulo Quantizer} (RMQ)}. RMQ  and the subsequent quantizers in this section will be used to quantize input vector $x$ in $\R^d$ with side information $y$ in $\R^d$, where $\norm{x-y}_2\leq \Delta$. RMQ first preprocesses the input $x$ and side information $y$ by randomly rotating them and then simply applies MQ for each coordinate. For rotation, we multiply both $x$ and $y$ with a random matrix $R$, given in \eqref{e:R}, which is sampled using shared randomness between the encoder and decoder.
We formally describe 
  the quantizer in  Alg.~\ref{a:E_RMQ} and~\ref{a:D_RMQ}.
  
\begin{rem}\label{r:subg}
We remark that the vector $R\left(x-y\right)$ has zero mean subgaussian coordinates with a variance factor of $\Delta^2/d$. From Lemma \ref{l:standar_subg}, this implies that for all coordinates $i$ in $[d]$, we have \[P\left(|R\left(x-y\right)(i)| \geq \Delta^{\prime} \right) \leq 2e^{-\frac{{\Delta^{\prime}}^{2}d}{2\Delta^2}}.\] 
This
observation allows us to use $\Delta^\prime\approx \Delta/\sqrt{d}$
for MQ applied to each coordinate.
\end{rem}

\begin{figure}[ht]
\centering
\begin{tikzpicture}[scale=1, every node/.style={scale=1}]
\node[draw,text width= 14 cm, text height=,] {%
\begin{varwidth}{\linewidth}
\algrenewcommand\algorithmicindent{0.7em}
\begin{algorithmic}[1]
\Require Input $ x\in \R^d$,  Parameters $k$ and $\Delta^\prime$
\State Sample $R$ as in \eqref{e:R} using public randomness
  \State $x^\prime = Rx$
\State \textbf{Output:}  $\Qenc_{{\tt M},R}(x)=[\Qenc_{{\tt M}}(x^\prime(1)), \ldots ,\Qenc_{{\tt M}}(x^\prime(d)]^T$
using parameters $k$, $\eps$, and $\Delta^\prime$
for $\Qenc_{{\tt M}}$ of Alg.~\ref{a:E_MCQ}
\end{algorithmic}  
\end{varwidth}};
 \end{tikzpicture}
 \renewcommand{\figurename}{Algorithm}
 
 \caption{Encoder  $\Qenc_{{\tt M},R}(x)$ of RMQ}\label{a:E_RMQ}
 \end{figure}
 \begin{figure}[ht]
\centering
\begin{tikzpicture}[scale=1, every node/.style={scale=1}]
\node[draw, text width= 14 cm, text height=,] {%
\begin{varwidth}{\linewidth}
            \algrenewcommand\algorithmicindent{0.7em}
 \renewcommand{\thealgorithm}{}
\begin{algorithmic}[1]
  \Require Input $w \in \{0, \ldots, k-1\}^d$, $y \in \R^d$, 
 \Statex ${~~~~~~~~}$ Parameters $k$ and $\Delta^\prime$
 \State Get $R$ from public randomness.
   \State $y^\prime = Ry$ 
    \State \textbf{Output:} $\displaystyle{\Qdec_{{\tt M},R}(w, y)=R^{-1} \sum_{i \in [d]}\Qdec_{{\tt M}}(w(i), y^\prime(i))e_i}$  
    \Statex using parameters $k$, $\eps$, and $\Delta^\prime$
for $\Qdec_{{\tt M}}$ of Alg.~\ref{a:D_MCQ}, 

\end{algorithmic}
\end{varwidth}};
 \end{tikzpicture}
 \renewcommand{\figurename}{Algorithm}
 \caption{Decoder $\Qdec_{{\tt M},R}(w, y)$ of RMQ}\label{a:D_RMQ}
 \end{figure}

 \begin{lem}\label{t:RMQ}
Fix $\Delta \geq 0$. Let $Q_{{\tt M}, R}$ be 
RMQ described in 
Alg.~\ref{a:E_RMQ} and~\ref{a:D_RMQ}. Then,  
 for\footnote{In the proof, we provide
 a general bound which holds for all $k$.} $k \geq 4$, $\delta\in(0,\Delta)$, $\Delta^{\prime}=\sqrt{6(\Delta^2/d)\ln (\Delta/\delta)}$
 and the parameter $\eps$ of MQ set to $\eps=2\Delta^\prime/(k-2)$, we have
 for $\X=\Y=\R^d$ that
 \begin{align*}
 &\alpha(Q_{{\tt M}, R}; \Delta) \leq \frac{24\, \Delta^2 }{(k-2)^2} \ln \frac{\Delta}{\delta}   + 154\, \delta^2 
 \quad \text{and}
 \\
 & \beta(Q_{{\tt M}, R}; \Delta) \leq 154\, \delta^2.
 \end{align*}
Furthermore, the output of quantizer {$Q_{{\tt M}, R}$} can be described in $d \log k$ bits.
 \end{lem}

\begin{rem}
The choice of $\Delta^{\prime}$ in the first statement of the Lemma \ref{t:RMQ} is based on Remark \ref{r:subg}.
  We note that $\delta$ is a parameter to control the bias incurred by our quantizer. 
{
By setting $\Delta^\prime =\Delta$ we can get an unbiased quantizer, but it only recovers
the performance obtained by simply using MQ for each coordinate, an algorithm
considered in~\cite{davies2020distributed} as well.
}  
\end{rem}

\subsection{Subsampled RMQ: A Wyner-Ziv quantizer for $\R^d$}\label{s:sRMQ}
 Our final quantizer is a modification of RMQ of previous section 
 where we make the precision less than $r$ bits by randomly sampling a subset of coordinates.
 Specifically, note that $\Qenc_{{\tt M},R}(x)$ sends $d$ binary strings of $\log k$
 bits each. We reduce the resolution by sending only a random subset $S$ of
 these strings. This subset is sampled using shared randomness and is available to
 the decoder, too. Note that $\Qdec_{{\tt M},R}$ applies $\Qdec_{\tt M}$
 to these strings separately; now, we use $\Qdec_{\tt M}$ to decode the entries in $S$ alone.
We describe the overall quantizer in Alg.~\ref{a:E_RCS_RMQ} and~\ref{a:D_RCS_RMQ}. 

\begin{figure}[ht]
\centering
\begin{tikzpicture}[scale=1, every node/.style={scale=1}]
\node[draw,text width= 14 cm, text height=,] {%
\begin{varwidth}{\linewidth}
            
            \algrenewcommand\algorithmicindent{0.7em}
\begin{algorithmic}[1]
\vspace{-0.5cm}
\Require Input $x\in \R$, Parameters $k$, $\Delta^{\prime}$, and $\mu$
\State Sample $S\subset[d]$ u.a.r. from all subsets of $[d]$ of
cardinality $\mu d$ and sample $R$ as in~\eqref{e:R} using public randomness

  \State \textbf{Output:} $\Qenc_{\tt WZ}(x)=\{\Qenc_{{\tt M}}(Rx(i)): i \in S \}$
  using parameters $k$, $\eps$, and $\Delta^\prime$
for $\Qenc_{{\tt M}}$ of Alg.~\ref{a:E_MCQ}
\end{algorithmic}  
\end{varwidth}};
 \end{tikzpicture}
 
 \renewcommand{\figurename}{Algorithm}
 \caption{Encoder  $\Qenc_{\tt WZ}(x)$ of subsampled RMQ}\label{a:E_RCS_RMQ}
 \end{figure}
 
 \begin{figure}[ht]
\centering
\begin{tikzpicture}[scale=1, every node/.style={scale=1}]
\node[draw, text width= 14 cm, text height=,] {%
\begin{varwidth}{\linewidth}
            
            \algrenewcommand\algorithmicindent{0.7em}
 \renewcommand{\thealgorithm}{}
\begin{algorithmic}[1]
  \Require Input $w \in \{0, \ldots, k-1\}^{\mu d}$, $y \in \R$
   \State Get $S$ and $R$ from public randomness
   \State Compute $\tilde{x}=(\Qdec_{{\tt M}}( w(i), Ry(i)), i\in S)$
   using parameters $k$, $\eps$, and $\Delta^\prime$ for $\Qdec_{{\tt M}}$ of Alg.~\ref{a:D_MCQ} 
   \State $\hat{x}_R =  \frac{1}{\mu}\sum_{i\in S}\left(\tilde{x}(i)-Ry(i) \right)e_i +Ry  $
    \State \textbf{Output:} $\displaystyle{\Qdec_{\tt WZ}(w, y)=R^{-1} \hat{x}_R}$
 
\end{algorithmic}
\end{varwidth}};
 \end{tikzpicture}
 
 \renewcommand{\figurename}{Algorithm}
 \caption{Decoder $\Qdec_{\tt WZ}(w, y)$ of subsampled RMQ}\label{a:D_RCS_RMQ}
 \end{figure}
 \begin{rem}
We remark that, typically, when implementing random sampling, we set the unsampled components to $0$, as was the case in Chapter \ref{Ch:RATQ}. 
However, to get $\Delta$ dependent bounds on  MSE, we set the unsampled coordinates
to the corresponding coordinate of side information and center our estimate appropriately to only have small bias.
\end{rem}
The result below relates the performance of 
our final quantizer $Q_{{\tt WZ}}$ to that of $Q_{{\tt M}, R}$, which
was already analysed in
the previous section.
\begin{lem}\label{t:RCS_RAQ_alpha_beta}
Fix $\Delta  > 0$.
Let $Q_{{\tt WZ}}$ and $Q_{{\tt M}, R}$ be the quantizers described in 
Alg.~\ref{a:E_RCS_RMQ} and~\ref{a:D_RCS_RMQ} and Alg.~\ref{a:E_RMQ} and~\ref{a:D_RMQ}, respectively.
Then, for $\mu d \in [d]$,  we have for $\X=\Y=\R^d$  that
\eq{ &\alpha(Q_{{\tt WZ}}; \Delta)\leq
\frac{\alpha(Q_{{\tt M}, R}; \Delta)}{\mu} + \frac{\Delta^2 }{\mu}\quad \text{and}
\quad \\& \beta(Q_{{\tt WZ}}; \Delta)=\beta(Q_{{\tt M}, R}; \Delta).}
Furthermore, the output of quantizer $Q_{{\tt WZ}}$ can be described in $\mu d \log k$ bits.
\end{lem}

We are now equipped to prove our first main result. Our protocol $\pi^*_{\tt k}$ uses $Q_{\tt WZ}$ for each client as described
in Section~\ref{s:preliminaries} and forms the estimate $\hat{\bar{x}}$
as in~\eqref{e:estimate}. 
We set the parameters needed for $Q_{\tt WZ}$ in Alg.~\ref{a:E_RCS_RMQ} and~\ref{a:D_RCS_RMQ} as follows: For client $i$, we 
set the parameters of MQ as 
\begin{align}
\delta =\frac{\Delta_i}{\sqrt{n}},
\quad
\log k= \ceil{\log (2+ \sqrt{12\ln n})},
\quad 
\Delta^{\prime}= \sqrt{6(\Delta_i^2/d)\ln (\Delta_i/\delta)},
\quad 
\eps=2\Delta^\prime/(k-2),
\label{e:param_sRMQ1}
\end{align}
and set the parameter $\mu$ as
\begin{align}
 \mu d = \left\lfloor\frac{r}{\log k}\right\rfloor.   
 \label{e:param_sRMQ2}
\end{align}
We characterize the resulting error performance in the next result.
\begin{thm}\label{t:DME_known}
For a $n\geq 2$, a fixed $\mathbf{\Delta}=(\Delta_1, ...,\Delta_n)$,
and  $d \geq r \geq 2\ceil{\log (2+ \sqrt{12\ln n})}$,
the protocol $\pi^*_k$ with parameters as set in \eqref{e:param_sRMQ1}
and \eqref{e:param_sRMQ2} is an $r$-bit protocol which 
satisfies 
\[
\mathcal{E}(\pi^*_k, \mathbf{x}, \mathbf{y})\leq
(79\, \lceil\log(2+\sqrt{12\ln n})\rceil + 26)\,
\left(\sum_{i=1}^n\frac{\Delta_i^2}{n} \cdot \frac{d }{nr} \right), 
\]
for all $\mathbf{x},\mathbf{y}$ satisfying~\eqref{eq:delta_cond}.
\end{thm}
\begin{proof}
Denoting by $Q_{i}$ the quantizer $Q_{\tt WZ}$
with parameters set for user $i$,
by Lemmas~\ref{l:main} and~\ref{t:RCS_RAQ_alpha_beta}, we get
\begin{align*}
{\E{ \|\hat{\bar{x}} - \bar{x}\|_2^2}}
&\leq \sum_{i=1}^n\frac{\alpha(Q_i; \Delta_i)}{n^2} +  \sum_{i=1}^n\frac{\beta(Q_i; \Delta_i)}{n}
\\
&\leq 
\frac 1{\mu n^2}\sum_{i=1}^n(\alpha(Q_{{\tt M}, R,i}; \Delta_i)
+\Delta_i^2) + 
\sum_{i=1}^n\frac{\beta(Q_{{\tt M}, R,i}; \Delta_i)}{n},
\end{align*}
where $Q_{{\tt M}, R,i}$ denotes RMQ with parameters set for user $i$.
Further, since $k \geq 4$ holds when $n\geq 2$ for our choice of parameters,
by using Lemma~\ref{t:RMQ} and substituting $\delta^2=\Delta_i^2/n$, we get
\begin{align*}
    \alpha(Q_{{\tt M}, R,i}; \Delta_i)
    &\leq \frac{12 \Delta_i^2 \ln n}{(k-2)^2} + \frac{ 154\Delta_i^2}{n},
    \\
    \beta(Q_{{\tt M}, R,i}; \Delta_i)&\leq \frac{ 154\Delta_i^2}{n},
\end{align*}
which with the previous bound gives
\begin{align*}
{\E{ \|\hat{\bar{x}} - \bar{x}\|_2^2}}
&\leq 
\frac 1{\mu d}\left(
\frac{12 \ln n}{(k-2)^2} + \frac{154}{n}
+1+ 154\mu
\right)
\sum_{i=1}^n\frac{d\Delta_i^2}{n^2}
\\
&\leq 
\frac {79\lceil\log(2+\sqrt{12\ln n})\rceil + 26}{r}
\sum_{i=1}^n\frac{d\Delta_i^2}{n^2},
\end{align*}
where in the final bound
we used our choice of $k$, the assumption that $n\geq 2$ (which
implies that $d\geq r\geq 6$), 
and the fact that $\lceil r/\log k\rceil\geq
r/2$ if $r\geq 2\log k$.
\end{proof}

\begin{rem} \label{r:DMQ_lp}
We note that by using MQ for each coordinate without rotating
(or even with rotation using $R$ as above)
and with $\Delta^\prime=\Delta_i$ yields MSE less than
\[
{O}\left(\sum_{i=1}^n\frac{\Delta_i^2}{n} \cdot \frac{d \log d}{nr}\right),
\]
for $r\leq d$. Thus, our approach above allows us to remove the
$\log d$ factor at the cost of a (milder for large $d$) $\log \log n$ factor.

\end{rem}

Thus, as can be seen from the lower bound presented in Theorem  \ref{t:lb_k} below, our  Wyner-Ziv estimator $\pi^*_k$ is nearly optimal. Finally, $Q_{\tt WZ}$ can be efficiently implemented as both the encoding and decoding procedures have nearly-linear time complexity\footnote{ The most expensive operation at  both the encoder and decoder of this estimator is the Hadamard matrix multiplication operation, which requires $d \log d$ real operations.} of $O(d \log d)$.

\subsection{Lower bound}\label{s:LB}
We now prove a lower bound on the MSE incurred by any SMP protocol using $r$ bits per client. 
The proof relies on the strong data processing inequality in \cite{duchi2014optimality} and is similar in structure to the lower bound for distributed mean estimation without side-information in \cite{suresh2017distributed}.
\begin{thm}\label{t:lb_k} 
{
Fix $\mathbf{\Delta}=(\Delta_1, \ldots, \Delta_n)$.
There exists
a universal constant $c<1$ 
such that for any $r$-bit SMP protocol $\pi $,
with $r \leq  c d$,
there exists input $(\mathbf{x}, \mathbf{y})\in \R^{2d}$ 
satisfying \eqref{eq:delta_cond} and such that 
\[\mathcal{E}(\pi, \mathbf{x}, \mathbf{y}) \geq  c \min_{i \in [d]}\Delta_{i}^2\cdot \frac{d}{nr}.
\]
}
\end{thm}

%


  \section{Distributed mean estimation for \unknown}\label{s:unknown}
  Finally, we present our Wyner-Ziv estimator for the \unknown setting. 
  We first, in Section \ref{s:CSI}, describe the idea of correlated sampling from \cite{Holenstein07}, which will serve as an essential building block for all our quantizers in this section. We then build towards our final quantizer, described in \ref{s:sRDAQ}, by first describing its simpler versions in Section \ref{s:DAQ} and \ref{s:RDAQ}. 
Once again, we restrict to the small-precision setting of $r\leq d$.

\subsection{The correlated sampling idea}\label{s:CSI}
Suppose we have two numbers $x$ and $y$ lying in $[0,1]$.
A $1$-bit unbiased estimator for $x$ is the random variable $\indic{\{U \leq x\}},$ where $U$ is a uniform random variable in $[0, 1]$. The variance of such an estimator is $x-x^2$. We 
consider a variant of this estimator given by: 
\begin{align}
\hat{X}=\indic{\{U \leq x\}} - \indic{\{U \leq y\}} +y,
\label{e:correlated_primitive}
\end{align}
where, like before, $U$ is a uniform random variable in $[0, 1].$ Such an estimator still uses only $1$-bit of information related to $x$. 
It is easy to check that this estimator
unbiased estimator of $x$, namely $\E{\hat X}=x$.
The variance of this estimator is given by
\[
\mathtt{Var}(\hat X) = \E{(\hat X - x)^2} = |x-y|-(x-y)^2,
\]
which is lower than that of the former quantizer when $x$ is close to $y$.
We build-on this basic primitive to obtain a quantizer with MSE
bounded above by a $\mathbf{\Delta}$-dependent expression, without requiring  the knowledge of 
$\mathbf{\Delta}$.

\subsection{Distance Adaptive Quantizer (DAQ)} \label{s:DAQ}
DAQ and subsequent quantizers in this Section will be described for input $x$ and side information $y$ lying in $\R^d$.
The first component of our quantizer, DAQ, which uses~\eqref{e:correlated_primitive} 
and incorporates the correlated sampling idea discussed earlier.
Both the encoder and the decoder of DAQ use the same  $d$  uniform random variables $\{U(i)\}_{i=1}^{d}$ between $[-1, 1]$, which are generated using public randomness.
At the encoder, each coordinate of vector $x$ is encoded to the bit $\indic{\{U(i) \leq x(i)\}}$. At the decoder, using the bits received from the encoder, side information $y$, and the public randomness $\{U(i)\}_{i=1}^{d}$, we first compute bits $\indic{\{U(i) \leq y(i)\}}$ 
for each $i\in[d]$. Then, the estimate of $x$ is formed as  follows:
\[
Q_{\tt D}(x, y)  =  \sum_{i=1}^{d} \left( \indic{\{U(i) \leq x(i)\}} - \indic{\{U(i) \leq y(i)\}}\right)e_i  +y. 
\]
We formally describe the quantizer in Alg. \ref{a:E_DAQ} and \ref{a:D_DAQ}.

\begin{figure}[ht]

\centering
\begin{tikzpicture}[scale=1, every node/.style={scale=1}]
\node[draw,text width= 14 cm, text height=,] {%
\begin{varwidth}{\linewidth}
            
            \algrenewcommand\algorithmicindent{0.7em}
\begin{algorithmic}[1]

\Require Input $ x\in \R^d$
\State Sample  $U(i) \sim Unif[-1, 1], \forall i \in [d]$
 \State  
     $
      \tilde{x} = \sum_{i=1}^{d}\indic{\{U(i) \leq x(i)\}} \cdot e_i 
$
   
 \State \textbf{Output:} $\Qenc_{\tt D}(x)=\tilde{x}$, where $\tilde{x}$ is
 viewed as binary vector of length $d$
\end{algorithmic}  
\end{varwidth}};
 \end{tikzpicture}
 
 \renewcommand{\figurename}{Algorithm}
 \caption{Encoder  $\Qenc_{\tt D}(x)$ of DAQ}\label{a:E_DAQ}
 \end{figure}
 
 \begin{figure}[ht]
\centering
\begin{tikzpicture}[scale=1, every node/.style={scale=1}]
\node[draw, text width= 14 cm, text height=,] {%
\begin{varwidth}{\linewidth}
            
\algrenewcommand\algorithmicindent{0.7em}
 \renewcommand{\thealgorithm}{}
\begin{algorithmic}[1]
  \Require Input $w \in \{0,1\}^d$, $y \in \R^d$, 
 \State Get $U(i), \forall i \in [d],$ using public randomness
   \State Set $\tilde{y}= \sum_{i=1}^{d}\indic{\{U(i) \leq y(i)\}} \cdot e_i $
    \State \textbf{Output:} $\Qdec_{\tt D}(w, y)= 2(w -\tilde{y})+y$, where $w$ is viewed as a vector
    in $\R^d$
    \label{step:output_coordinate}
\end{algorithmic}
\end{varwidth}};
 \end{tikzpicture}
 
 \renewcommand{\figurename}{Algorithm}
 \caption{Decoder $\Qdec_{\tt D}(w, y)$ of DAQ}\label{a:D_DAQ}
   
 \end{figure}

\noindent The next result characterizes the performance for DAQ.
\begin{lem}\label{t:DAQ}
Let $Q_{\tt D}$ 
denote DAQ described in Algorithms \ref{a:E_DAQ} and \ref{a:D_DAQ}. Then, for $\X=\Y=\B$ and every $\Delta > 0$, we have
\[
\alpha(Q_{\tt D}; \Delta) \leq 2\Delta\sqrt{d}\,\,\text{ and }\,\,\beta(Q_{\tt D}; \Delta)= 0.
\]
Furthermore, the output of quantizer $Q_{\tt D}$ can be described in $ d $ bits.
\end{lem}




\subsection{Rotated Distance Adaptive Quantizer (RDAQ)} \label{s:RDAQ} 
Next, we proceed as for the \known setting and add
a preprocessing step of rotating $x$ and $y$
using random
matrix $R$ of~\eqref{e:R}, which is sampled using shared randomness.
We remark that here random rotation is used to exploit the subgaussianity of the rotated $x$ and $y$, whereas in RMQ of previous section it was used to exploit the subgaussianity of  $x-y$. That is, RMQ exploited the fact that each coordinate of the Rotated vector  $R(x-y)$ is much smaller compared to each of the coordinate of $(x-y),$ whereas RDAQ exploits the fact that coordinates of both the rotated vectors $Rx$ and $Ry$ are much smaller relative to coordinates of $x$ and $y$.
After this rotation step, we proceed with a quantizer similar to DAQ, but
we quantize each coordinate at multiple ``scales.'' 
We describe this step in detail below.


\paragraph{Using multiple scales.}
In DAQ,
 we considered each coordinate of the input vector $x$ to be anywhere between
 $[-1,1]$ and used one uniform random variable
 for each coordinate.  Now, 
 we will use $h$ independent uniform random variables
 for each coordinate,
 each corresponding to a different scale $[-M_j, M_j]$, $j\in \{0,1,2,\ldots, h-1\}$.
 For convenience, we abbreviate $[h]_0:=\{0,1,2,\ldots, h-1\}$.
 
 Specifically, let $U(i, j)$ be distributed uniformly over $[-M_j, M_j]$,
 independently for different  $i\in [d]$
  and different $j\in[h]_0$.
 The values $M_j$s correspond to different 
 scales and are set, along with $h$, as follows:
For all  $j \in[h]_0$,
\begin{align}
M_{j}^2:=  \frac{6}{d} \cdot e^{*j}, \quad
  \log h :=\ceil{\log(1+ \ln^*(d/6))},
  \label{e:levels}
  \end{align}
 where $e^{*j}$ denotes the $j$th iteration of $e$ given by
 $\displaystyle{e^{*0}:=1, \quad e^{*1}:=e, \quad e^{*j}:=e^{e^{*(j-1)}}}$.
  All the $dh$ uniform random variables are generated using public randomness and are available to both the encoder and the decoder.

The intervals $[-M_j, M_j]$ are designed to minimize the MSE of our quantizer 
by tuning its ``resolution'' to the ``scale'' of the input,
and while still ensuring unbiased estimates.  Observe that this is the second time we are using the general idea of using multiple intervals for quantizing randomly rotated vectors, with the first time being RATQ in Chapter \ref{Ch:RATQ}.

\paragraph{Multiscale DAQ.}
After rotation, we proceed as in DAQ, except that 
we use different scale $M_j$ for different coordinates. 
Ideally, for the $i$th coordinate,  
we would like to use $M_{z^*(i)}$, where $z^*(i)$ is the smallest index such that  both $Rx(i)$ and $Ry(i)$ lie in $[-M_{z^*(i)}, M_{z^*(i)}]$.
However, since $y$ is not available to the encoder,
we simply resort to 
sending the smallest value $z(i)$ which is the smallest index such that  $ Rx(i) 
\in [-M_{z(i)}, M_{z(i)}]$ and apply the encoder of DAQ 
$h$ times to compress
$x$ at all scales, $i.e.$, we send $h$ bits
 $(\indic{\{U(i, j) \leq Rx(i)\}}, j\in [h]_0)$.

Thus, the overall number of bits used by RDAQ's encoder is $d \cdot( h+\ceil{\log h})$.
At RDAQ's decoder, using $z(i)$,
we compute the smallest index $z^*(i)$ containing both  $Rx(i)$ and $Ry(i)$. In effect, the decoder emulates the decoder for DAQ applied
to $Ry$, but for scale $M_{z^*(i)}$.
The encoding and decoding algorithm of RDAQ are described in Alg. \ref{a:E_RDAQ}  and \ref{a:D_RDAQ}, respectively. 


\begin{figure}[ht]
\centering
\begin{tikzpicture}[scale=1, every node/.style={scale=1}]
\node[draw, text width= 14 cm, text height=,] {%
\begin{varwidth}{\linewidth}
            
            \algrenewcommand\algorithmicindent{0.7em}
 \renewcommand{\thealgorithm}{}
\begin{algorithmic}[1]
\Require Input $x \in \B$
\State Sample $U(i, j) \sim Unif[-M_j, M_j]$, $i \in [d], j\in[h]_0$, and sample $R$ as in\eqref{e:R} using public randomness.

\State $x_R=Rx$
\For{$i \in [d]$} 
  \Statex \hspace{1cm} $z(i)= \min \{j \in [h]_0: |x_R(i)| \leq M_j\}$
  \EndFor

\For{$j \in [h]_0$} 
\Statex \hspace{1cm} $\tilde{x}_j= \sum_{i=1}^{d} \indic{\{U(i,j)\leq x_R(i)\}}e_i$
\EndFor

   \State \textbf{Output:} 
   $\Qenc_{{\tt D}, R}(x) = \left( [\tilde{x}_0, \ldots, \tilde{x}_{h-1}], z\right)$, where we view $\tilde{x}_j$s as binary vectors
\end{algorithmic}
\end{varwidth}};
 \end{tikzpicture}
 
 \renewcommand{\figurename}{Algorithm}
 \caption{Encoder $\Qenc_{{\tt D}, R}(x)$ at for RDAQ}\label{a:E_RDAQ}
 \end{figure}

\begin{figure}[ht]
\centering
\vspace{-0.1cm}
\begin{tikzpicture}[scale=1, every node/.style={scale=1}]
\node[draw, text width= 14 cm, text height=,] {%
\begin{varwidth}{\linewidth}
            
            \algrenewcommand\algorithmicindent{0.7em}
 \renewcommand{\thealgorithm}{}
\begin{algorithmic}[1]
\Require   Input $ (w, z) \in \{0, 1\}^{d \times h} \times [h]_0^d$
and $y\in \B$

  \State\label{step:1} Get $U(i, j)$, $i \in [d]$, $j\in[h]_0$, and $R$ using public randomness.

  \State $y_R=Ry$
  \For{$i \in [d]$} 
  \Statex \hspace{1cm} $z^{\prime}(i)= \min \{j \in \{[h]_0\}: |y_R(i)| \leq M_j\}$
  \Statex \hspace{1cm} $z^*(i)= \max\{z(i), z^{\prime}(i)\}$
  
  \EndFor
  
 \State\label{step:4} $w^{\prime}= \sum_{i =1}^d 2M_{z^*(i)}\left(w(i, z^*(i) )  - \indic{\{U(i, z^*(i))\leq y_R\}} \right)$
\State $\hat{x}_R= w^{\prime}+Ry$
   \State \textbf{Output:}
   $\displaystyle{\Qdec_{{\tt D}, R}(w, y) =R^{-1} \hat{x}_R.}$
  
\end{algorithmic}
\end{varwidth}};
 \end{tikzpicture}
 
 \renewcommand{\figurename}{Algorithm}
 \caption{Decoder $\Qdec_{{\tt D}, R}(x)$ for RDAQ }\label{a:D_RDAQ}
 \end{figure}


Then, the quantized output $Q_{{\tt D}, R}$ corresponding to input vector $x$ and side-information $y$ is
\begin{align*}
    Q_{{\tt D}, R}(x, y) = R^{-1} &\Bigg[\sum_{i=1}^d 2M_{z^*(i)}\left( \indic{\{U(i, z^*(i))\leq Rx(i)\}}   -\indic{\{U(i, z^*(i))\leq Ry(i)\}}\right)  +Ry \Bigg].
\end{align*}
We remark that since rotated  coordinates $Rx(i)$ and $Ry(i)$ have subgaussian tails, with very high probability $M_{z^*(i)} $ will be much less than $1$, which  helps in reducing the overall MSE significantly.
The performance of the algorithm is characterized below.
\begin{lem}\label{t:RDAQ}
Let $Q_{{\tt D}, R}$ be RDAQ described in Alg. \ref{a:E_RDAQ} and \ref{a:D_RDAQ}. Then,  for $\X=\Y=\B$ 
 and every $\Delta > 0$,
we have 
\[\alpha(Q_{{\tt D}, R}; \Delta) \leq  16 \sqrt{3} \Delta \quad \text{and} \quad \beta(Q_{{\tt D}, R}; \Delta) =0.\]
Furthermore, the output of quantizer $Q$ can be described in $d( h + \log h)$ bits.
 \end{lem}

\subsection{Subsampled RDAQ: A universal Wyner-Ziv quantizer for unit Euclidean ball}\label{s:sRDAQ}
Finally, we bring down the precision of RDAQ to $r$, as before
for the \known setting,
by retaining the output of RDAQ for only coordinates $i\in S$,
where $S$ 
 is generated uniformly at random from all subsets of $[d]$
of cardinality $\mu d$ using public randomness.
Specifically, we execute 
Alg.~\ref{a:E_RDAQ} and~\ref{a:D_RDAQ} with 
$S$ replacing $[d]$ and multiplying $w^\prime$ in Step 4 of
Alg.~\ref{a:D_RDAQ} by normalization factor of $d/|S|$.
The output of the resulting encoder is given by
\begin{align}\label{e:esRDAQ}
\Qenc_{{\tt WZ}, u}(x)= \{\Qenc_{{\tt D},  R}(x)(i): i \in S \},
\end{align}
where  $ \Qenc_{{\tt D},  R}(x)(i) $ represents the encoded bits 
$([\tilde{x}_0(i), \ldots, \tilde{x}_{h-1}(i)  ], z(i) )$
for the $i${th} coordinate using RDAQ, and
the output of the resulting decoder is given by
\begin{align}
 Q_{{\tt WZ}, u}(x, y) &= R^{-1} \Bigg[   \frac{1} {\mu}\sum_{i \in S}  2M_{z^*(i)}\Big(\indic{\{U(i, z^*(i))\leq Rx(i)\}}
 -\indic{\{U(i, z^*(i))\leq Ry(i)\}}\Big)  +Ry\Bigg].
 \label{e:s_RDAQ}
 \end{align}

\begin{lem}\label{t:sRDAQ}
Let $Q_{{\tt WZ}, u}$ be the quantizers described in \eqref{e:esRDAQ} and
 \eqref{e:s_RDAQ} and $Q_{{\tt D}, R}$ be RDAQ described in
 Alg.  \ref{a:E_RDAQ} and \ref{a:D_RDAQ}.
 Then, for $\mu d \in [d]$, $\X=\Y=\B$, and every $\Delta > 0$, we have 
 \[
 \alpha(Q_{{\tt WZ}, u}; \Delta) \leq  \frac{\alpha(Q_{{\tt D}, R}; \Delta)}{\mu} \quad\text{and} \quad\beta(Q_{{\tt WZ}, u}; \Delta) =0.
 \]
Furthermore, the output of quantizer $Q_{{\tt WZ}, u}$ can be described in $\mu d( h + \log h)$ bits.
 \end{lem}
 
 We are now equipped to prove our second main result. Our protocol $\pi^*_{\tt u}$ uses $Q_{{\tt WZ}, u}$ for each client as described
in Section~\ref{s:preliminaries} and forms the estimate $\hat{\bar{x}}$
as in~\eqref{e:estimate}. 
Unlike for the \known setting, we now use the
same parameters for $Q_{{\tt WZ},u}$
for all clients, given by
\begin{align}\label{e:param_sRDAQ}
&\mu d =  \left\lfloor\frac{r}{h +\log h}\right\rfloor. 
\end{align}

\begin{thm}  \label{t:DME_unknown}
For {$d \geq  r \geq 2(h+\log h)$}
and $h$ given in~\eqref{e:levels},
the $r$-bit protocol $\pi^*_u$ with parameters as set in \eqref{e:param_sRDAQ}
satisfies 
\[\mathcal{E}(\pi^*_u, \mathbf{x}, \mathbf{y})\leq 
(128\sqrt{3}\, (1+\ln^*(d/6)))
\left(\sum_{i \in [n]}\frac{\Delta_i}{n} \cdot \frac{d }{nr} \right),\]
for all $\mathbf{x},\mathbf{y}$ satisfying~\eqref{eq:delta_cond}, for every $\mathbf{\Delta}=(\Delta_1, ...,\Delta_n)$.
\end{thm}
\begin{proof}
Denote by $\hat{\bar{x}}$ the output of the protocol. Then,
by Lemmas~\ref{l:main} and Lemma~\ref{t:sRDAQ}, we get
\begin{align*}
\E{ \|\hat{\bar{x}} - \bar{x}\|_2^2}
&\leq \frac 1 {n^2\mu} \sum_{i=1}^n\alpha(Q_{{\tt D},R}; \Delta_i)
 \\
 &\leq 
 \frac {16\sqrt{3}} {n^2\mu} \sum_{i=1}^n\Delta_i,
\end{align*}
where the previous inequality is by Lemma~\ref{t:RDAQ}. 
The 
proof is completed by using 
$
\mu\geq \frac{r}{2d(h +\log h)}\geq
\frac{r}{4dh}$,
which follows from
\eqref{e:param_sRDAQ} and the assumption that $r\geq 2(h +\log h)$.
\end{proof}
The Wyner-Ziv estimator $\pi^*_u$ is universal in $\mathbf{\Delta}$: it operates without the knowledge of the distance between the input and the side information and yet gets
MSE depending on $\mathbf{\Delta}$. Moreover, it can be efficiently implemented as both the encoding and the decoding procedures have nearly linear time complexity of $O(d \log d)$.

\section{The large-precision regime}\label{s:hpr}
\subsection{RMQ in the large-precision regime.}
For the \known setting, our quantizer RMQ described in Alg. 4 and 5
remains
valid even for $r>d$. We will assume $r=m d$ for integer $m \geq 2$. For each client $i$, we set
\begin{align}
\delta = \frac{\Delta_i}{n^{\frac 12}(2^{r/d}-2)}, \quad \log k = \frac{r}{d}, \quad 
\Delta^\prime &= \sqrt{6(\Delta_i^2/d)\ln \Delta_i/\delta}, \quad \eps = \frac{2\Delta^{\prime}}{k-2}.
 \label{e:param_high_precision_known}
 \end{align}

The performance of 
protocol $\pi^*_k$  using RMQ with parameters set as in \eqref{e:param_high_precision_known} for each client can be characterized as follows.
\begin{thm}
For a fixed $\mathbf{\Delta}=(\Delta_1, ...,\Delta_n)$
and  $r=md$ for integer $m\geq 2$,
the protocol $\pi^*_k$ with parameters set as in  \eqref{e:param_high_precision_known}
satisfies 
\[
\mathcal{E}(\pi^*_k, \mathbf{x}, \mathbf{y})=\left(
12 \ln n+ \frac{24r}{d}+ {154}/{n}+
166
\right)
\left(\sum_{i \in
[n]}\frac{\Delta_i^2}{n} \cdot \frac{1}{n(2^{r/d}-2)^2} \right), 
\]
for all $\mathbf{x},\mathbf{y}$ satisfying~\eqref{eq:delta_cond}.
\end{thm}
\begin{proof}
Denoting by $Q_{i}$ the quantizer $Q_{{\tt M},R}$
with parameters set for client $i$,
by Lemmas~\ref{l:main} and~\ref{t:RMQ}, we get
\begin{align*}
{\E{ \|\hat{\bar{x}} - \bar{x}\|_2^2}}
&\leq \sum_{i=1}^n\frac{\alpha(Q_i; \Delta_i)}{n^2} +  \sum_{i=1}^n\frac{\beta(Q_i; \Delta_i)}{n}
\end{align*}
Further, since $k\geq 4$ holds when $r\geq 2d$ for our choice of parameters,
by using Lemma~\ref{t:RMQ} and substituting $\delta^2=\Delta_i^2/n(2^{r/d}-2)^2$, we get
  \begin{align*}
\alpha(Q_{i}; \Delta_i)&\leq \frac{12 \Delta_i^2\ln (n(2^{r/d}-2)^2)}{{(2^{{r}/{d}}-2)}^2}
+ \frac{ 154\Delta_i^2}{n(2^{{r}/{d}}-2)^2},
\\
\beta(Q_{i}; \Delta_i)&\leq  \frac{154\Delta_i^2}{n(2^{{r}/{d}}-2)^2}.
\end{align*}

which with the previous bound gives
\begin{align*}
{\E{ \|\hat{\bar{x}} - \bar{x}\|_2^2}}
&\leq 
\left(
12 \ln n+\frac{24r}{d}+ \frac{154}{n}+
154
\right)
\sum_{i=1}^n\frac{\Delta_i^2}{n^2(2^{r/d}-2)^2},
\end{align*}
where use the inequality $ \ln x \leq x$, $\forall x \geq 0$, to bound $\ln (2^{r/d}-2)^2/(2^{r/d}-2)^2$ by $1$.

\end{proof}

\begin{rem}
 Similar to Remark \ref{r:DMQ_lp}, we note that using MQ for each coordinate without rotating (or even with rotation using $R$ as above) with $\Delta^{\prime}=\Delta_i$ yields MSE less than 
 \[ O\left( \sum_{i=1}^{n}\frac{\Delta_i^2}{n}\cdot \frac{d}{n2^{2r/d}}\right),\]
 for $r \geq d$. Thus our approach above allows us to remove the $d$ factor at the cost of a (milder for large $d$) $\log n +r/d$ factor.

\end{rem}

\subsection{Boosted RDAQ: RDAQ in the large-precision regime.}
Moving to the \unknown setting, we describe an update to RDAQ
described in Alg. 10 and 11 for the large-precision setting.
For brevity, we denote by $m:=r/d$  the number of bits per dimension.
A straight-forward scheme to make use of the
high precision is to independently implement the RDAQ quantizer
approximately $ \floor{m/\ln^*d}$ times and use the average of the
quantized estimates as the final estimate.
We will see that the MSE incurred by such an
estimator is $O(\Delta \ln^*d/m)$.
We will show that this naive
implementation can be significantly improved and an exponential decay
in MSE with respect to $m$ can be achieved.

We boost RDAQs performance as follows.
Simply speaking, instead of sending the bits produced by multiple
instances of the encoder of RDAQ, we send the ``type'' of each sequence.
A similar idea appeared in~\cite{mayekar2020limits} for the case without
any side information.
At the encoding stage of RDAG
given in Alg. 10 and 11, after random rotation and computing
$z$ in Steps $1$ to $3$ of Alg. 10, we repeat
Step $4$
$N$ times with independent randomness each time and store
only the total number of ones seen for each coordinate $i$ and scale $j$.
Specifically, let $U_t(i, j)$ be an independent uniform random variable in
$[-M_j,M_j]$, for all $i \in [d], j\in [h]_0$, and $t \in [N]$, which are generated using
public randomness between the encoder and the decoder. Using this
randomness, we compute  
$\tilde{x}_{j,t}= \sum_{i=1}^{d} \indic{\{U_t(i,j)\leq x_R(i)\}}e_i$
for all $j \in [h]_0$. Then, instead of storing  
$\tilde{x}_{j,t}$ for each $j$ and $t$, we store the sum
$\sum_{t=1}^n\tilde{x}_{j,t}$ for each $j\in [h]_0$.
Since each coordinate of the sum can be stored in $\log N$ bits,
the new encoder's output can be stored in  $d(h\log N +\log h)$.
Thus, we can implement this scheme by using $m=(h\log N +\log h)$
bits per dimension.

At the decoding stage, we rotate $y$ and compute $z^*$ in precisely
the same manner as done in Steps 1 to 3
of the decoding Alg. 11 of RDAQ. Then,
  using the encoded input received, the side-information $y$, the same
  random variables $U_t(i,j)$ and random matrix $R$ used by the
  encoder, the final estimate $Q(x)$ is 
  \begin{align}\label{e:boosted_RDAQ}
    &Q(x)=R^{-1} \left( \frac{1}{N}\cdot \sum_{i \in [d]}\sum_{t \in
  [N]} \left( B^t_{i, Rx} - B^t_{i, Ry} \right)  e_i +Ry\right), 
  \end{align}
  where $B^t_{i, v} = \indic{\{U_n(i,z^*(i) ) \leq v(i)\}}$ for $v$ in
  $\R^d$. 

The result below characterizes the performance of our quantizer Boosted RDAQ
$Q$.
  \begin{lem}\label{t:bRDAQ}
Let $Q$ be Boosted RDAQ described above.
Then,  we have for $\X=\Y=\R^d$ and every $\Delta > 0$,
we have 
\[
\alpha_u(Q; \Delta) \leq  \frac{ 16 \sqrt{3}\Delta}{N}\,\,
\text{ and }\,\,
    \beta_u(Q; \Delta) =0.
    \]
    Furthermore, the output of the quantizer can be described in
    $d(h \log N + \log h)$ bits. 
  \end{lem}
  Thus, when we have a total precision budget of $r=dm$ bits using the
    Boosted RDAQ algorithm with  number of repetitions $N=
    2^{\floor{(m -\log h)/h}}$, we get an exponential decay in MSE
    with respect to $m$. 

We consider the protocol $\pi^*_u$ that uses the $Q$ above for each
client with $M_j$ and $h$ set as in \eqref{e:levels}, $i.e.$, with
\begin{align}
N=2^{\floor{(m -\log h)/h}},\,\, M_j^2= \frac{6e^{*j}}{d}, j\in[h]_0, ~~ \log h =\lceil \log (1+\ln^*(d/6) )\rceil.
\label{e:param_high_precision_unknown}
\end{align}
Therefore, by the previous lemma and Lemma \ref{l:main}, we 
get the following result.

\begin{thm}
For $r=dm$ with integer $m\geq h + \log h$,
the protocol $\pi^*_u$ with parameters as set in \eqref{e:param_high_precision_unknown}
satisfies
\[
\mathcal{E}(\pi^*_u, \mathbf{x}, \mathbf{y})=\sum_{i \in [n]}\frac{\Delta_i}{n} \cdot \frac{64\sqrt{3}}{n2^{{r}/(d(2+2\ln^*(d/6)) )}}
,
\]
for all {$\mathbf{x},\mathbf{y}$} satisfying \eqref{eq:delta_cond},
for every $\mathbf{\Delta}=(\Delta_1, ...,\Delta_n)$.
\end{thm}
\begin{proof}
Denote by $\hat{\bar{x}}$ the output of the protocol. Then,
by Lemmas~\ref{l:main} and Lemma~\ref{t:bRDAQ}, we get
\begin{align*}
\E{ \|\hat{\bar{x}} - \bar{x}\|_2^2}
&\leq \frac 1 {n^2} \sum_{i=1}^n\alpha(Q; \Delta_i)
 \\
 &\leq 
 \frac {16\sqrt{3}} {n^2 N } \sum_{i=1}^n\Delta_i,
\end{align*}
where the previous inequality is by Lemma~\ref{t:bRDAQ}. 
The 
proof is completed by using 
\[
N\geq
\frac{2^{m/h}}{2^{1+(\log h)/h}}\geq \frac{2^{m/h}}{4} \geq \frac{2^{m/(2+2\ln^*(d/6))}}{4},\]
where the first inequality follows from using $\lfloor x\rfloor \geq x-1$ for the floor function in the value of $N$ in \eqref{e:param_high_precision_unknown}, the second follows from  the fact that $\log x \leq x, \forall x \geq 0$, and the third follows from $\lceil x\rceil \leq x+1$  for the ceil function in the value of $h$ in \eqref{e:param_high_precision_unknown}.
\end{proof}

\section{Proofs of results}\label{s:proof}
\subsection{Proof of Lemma \ref{l:main}}

For the estimator $\hat{\bar {x}}$ in~\eqref{e:estimate},
with $\hat{x}_i=Q_i(x_i, y_i)$, we have

\eq{
&\E{\left\|\frac{1}{n}\cdot\sum_{i \in [n]}Q_i(x_i,y_i)- \frac{1}{n}\cdot \sum_{i \in [n]} x_i \right\|_2^2}\\
&= \frac{1}{n^2} \cdot \sum_{i \in [n]} \E{\norm{Q_i(x_i,y_i)-x_i}_2^2} + \frac{1}{n^2} \cdot \sum_{i \neq j}  \E{\langle Q_i(x_i,y_i)-x_i, Q_j(x_j,y_j)-x_j\rangle}\\
&= \frac{1}{n^2} \cdot \sum_{i \in [n]} \E{\norm{Q_i(x_i,y_i)-x_i}_2^2} + \frac{1}{n^2} \cdot \sum_{i \neq j}  {\langle \E{Q_i(x_i,y_i)}-x_i, \E{Q_j(x_j,y_j)}-x_j\rangle}\\
&= \frac{1}{n^2} \cdot \sum_{i \in [n]} \E{\norm{Q_i(x_i,y_i)-x_i}_2^2}+ \left(\frac{1}{n} \cdot \sum_{i} \|\E{Q_i(x_i,y_i)}-x_i\|_2\right)^2
\\
&\hspace{2cm}
- \frac{1}{n^2} \cdot \sum_{i}\|\E{Q_i(x_i,y_i)}-x_i\|_2^2
\\
&\leq \frac{1}{n^2} \cdot \sum_{i \in [n]} \E{\norm{Q_i(x_i,y_i)-x_i}_2^2}+ \frac{(n-1)}{n^2} \cdot \sum_{i}\|\E{Q_i(x_i,y_i)}-x_i\|_2^2,
}
where the second identity uses the independence of $Q_i(x_i,y_i)$ for different $i$
and the final step uses Jensen's inequality.
The result follows by bound each term using 
the fact that $\mathbf{x}$ and $\mathbf{y}$ satisfy (2)
and the 
definitions of $\alpha(Q_i,\Delta_i)$ and $\beta(Q_i,\Delta_i)$, for $i\in[n]$.
\qed

\subsection{Proof of Theorem \ref{t:DME_noside}}\label{s:DME_noside_proof}
We will need the following Lemma to complete the proof of Theorem \ref{t:DME_noside}.
 \begin{lem}\label{l:RCS_RATQ}
For any $r \geq \Omega(\log \ln^* d)$ and for any $Y$ such that $\norm{Y_2} \leq 1,$ let $Q$ be the composition of RCS with RATQ. Then, $Q(Y)$ can be represented in $r$ bits, $\E{Q(Y)\mid Y}=Y,$ and \[\E{\norm{Q(Y)-Y}_2^2\mid Y} \leq \frac{d (3+\ceil{\log(1+ \ln^*d/3)})}{r}.\]
\end{lem}
 \begin{proof}
 By the
 description of RATQ we have that $Q_{{\tt at}, R} =R^{-1} Q_{{\tt
 at}, I}(RY)$, where $Q_{{\tt at}, I}$ is as defined
 in~\eqref{e:Q_at_I}. Thus, using the fact that $R$ is a unitary
 matrix
\[
\E{ \norm{Q_{{\tt at}, R}(Y) -Y }_2^2 } =\E{ \norm{Q_{{\tt at}, I}(RY) -RY }_2^2}.
\]
 When the parameters are set as in~\eqref{e:RATQ_unit_levels}, we get
\[
RY(j) \leq M_{h-1} \text{ a.s., } \forall j \in [d],
\] 
whereby
\[
\E{ \norm{Q_{{\tt at}, R}(Y) -Y }_2^2 } =\E{ \sum_{j \in [d]}(Q_{{\tt at}, I}(RY)(j) -RY(j) )^2 \indic{RY(j) \leq M_{h-1}}}.    
\]
The proof is completed by noting that $Y$ satisfies $\norm{Y}_2\leq 1$
a.s., setting $m=3/d$ and $m_0=(2/d)\ln s$, and applying
Lemma~\ref{l:useful}.   
\end{proof}

Combining the Lemma above with Lemma \ref{l:main} completes the proof of upper bound.
\qed
\subsection{Proof of Lemma \ref{t:MQ}}
As mentioned in~\eqref{e:tildez_close},
 the integer $\tilde{z}$ found in Alg.~\ref{a:E_MCQ}
satisfies 
$\E{\tilde{z}\eps}=x$ and $|x-\tilde{z}\eps|< \eps$.
Therefore, 
it suffices to
show that the output of the quantizer satisfies $Q_{\tt M}(x,y)=\tilde{z}\eps$.

To see that $Q_{\tt M}(x,y)=\tilde{z}\eps$, 
denote 
the lattice used in decoding Alg.~\ref{a:D_MCQ}
as $\Z_{w,\eps}:=\{(zk+w )\cdot \eps: z \in \Z\}$. 
The decoding algorithm finds the point in $\Z_{w,\eps}$
that is closest to $y$. Note that
$w=\tilde{z}\mod k$, whereby $\tilde{z}\eps$ is a point in this lattice.
Further, 
for any other point $\lambda\neq \tilde{z}\eps$ in the lattice,  we must have 
\[
|\lambda-\tilde{z}\eps|\geq k\eps,
\]
and so, by triangular inequality, 
that 
\[
|\lambda -y|\geq |\lambda-\tilde{z}\eps|-|\tilde{z}\eps- y| \geq k\eps 
-|\tilde{z}\eps- y|.
\]
Thus, $\tilde{z}\eps$ is closer to $y$ than $\lambda$
if 
\begin{align}
k\eps> 2|\tilde{z}\eps- y|.
\label{e:closest}
\end{align}
Next, by using~\eqref{e:tildez_close} once again, we have
\[
|\tilde{z}\eps- y|\leq |\tilde{z}\eps- x| +|x- y|  < \eps+\Delta^\prime,
\]
which by condition~\eqref{e:parameter_condition} in the lemma 
implies that~\eqref{e:closest} holds. 
It follows that $|\lambda -y|>|\tilde{z}\eps- y|$ for every 
$\lambda\in \Z_{w,\eps}$, which shows that  
$Q_{\tt M}(x,y)=\tilde{z}\eps$ and completes the proof.
\qed
\subsection{Proof of Lemma  \ref{t:RMQ}}
Recall from Remark~\ref{r:subg} that
for the random matrix $R$ given in~\eqref{e:R}, for every vector
$z\in \R^d$, the random variables $Rz(i)$, $i\in [d]$, are sub-Gaussian with variance 
parameter $\|z\|_2^2/d$. Furthermore, we need the following bound
for ``truncated moments'' of sub-Gaussian random variables.
\begin{lem}\label{l:variance_tail}
For a sub-Gaussian random $Z$ with variance factor $\sigma^2$
and every $t\geq 0$, we have 
\[
\E{Z^2\indic{\{|Z|>t\}}}\leq 
2(2\sigma^2+t^2)e^{-t^2/2\sigma^2}.
\]
\end{lem}
\begin{proof}
Note that for any nonnegative random variable $U$, 
it can be verified that
\[
\E{U\indic{\{U>x\}}}= xP(U>x) + \int_{x}^\infty P(U>u)\,du.
\]
Upon substituting $U=Z^2$ and $x=t^2$, along with
the fact that $Z$ is sub-Gaussian with variance parameter $\sigma^2$,
we get 
\begin{align*}
    \E{Z^2\indic{\{Z^2>t^2\}}}&= t^2P(Z^2>t^2) + 
\int_{t^2}^\infty P(Z^2>u)\,du
\\
&\leq 2t^2e^{-t^2/2\sigma^2}+ 
2\int_{t^2}^\infty e^{-u/2\sigma^2}\,du
\\
&\leq 2(t^2+2\sigma^2)e^{-t^2/2\sigma^2},
\end{align*}
which completes the proof.
\end{proof}

We now handle the MSE $\alpha(Q)$ and bias $\beta(Q)$ separately below.
 \paragraph{Bound for MSE $\alpha(Q)$:}
 Denote by 
 $Q_{{\tt M}, R}(x, y)$ the final quantized value
 of the quantizer RMQ. For convenience, we abbreviate 
 \[
 \hat{x}_R :=
 R\, Q_{{\tt M}, R}(x, y). 
 \]
 Observe that $\hat{x}_R =\sum_{i \in [d]}
 Q_{\tt M}(Rx(i), Ry(i))e_i$, where   $ Q_{\tt M}$ is the MQ of Alg.~\ref{a:E_MCQ} and~\ref{a:D_MCQ} with parameters $k\geq$ and $\Delta^{\prime}$ set as in the statement of the lemma.  
 Since $R$ is a unitary transform, we have
\begin{align}
\E{\norm{Q_{{\tt M},R}(x,y)-x}_2^2}
&=\E{\norm{\hat{x}_{R}-Rx}_2^2}
\nonumber
\\
&=\sum_{i=1}^{d} \E{(\hat{x}_R(i) -Rx(i))^2}
\nonumber
\\
&= \sum_{i=1}^{d} \E{(\hat{x}_R(i)-Rx(i))^2 \indic{\{|R\left(x-y\right)(i)| \leq \Delta^{\prime}\}}}
\nonumber
\\
&\hspace{2cm}+ \sum_{i=1}^{d} \E{(\hat{x}_R(i)-Rx(i))^2 \indic{\{|R\left(x-y\right)(i)| \geq \Delta^{\prime}\}}}
\label{e:error_split}
\end{align}
We consider each error term on the right-side above separately. 
We can view the first term as the error corresponding to
MQ, when the input lies in its ``acceptance range.'' 
Specifically, under the event $\{|R\left(x-y\right)(i)| \leq
   \Delta^{\prime}$\}, we get
   by Lemma~\ref{t:MQ}
   that
   \[
   |\hat{x}_R(i)
   -Rx(i)| \leq \eps=
   \frac{2\Delta^{\prime}}{k-2}, \quad \text{{almost surely}},
   \]
   whereby 
  \begin{align}
  \sum_{i=1}^{d} \E{(\hat{x}_R(i)-Rx(i))^2 \indic{|R\left(x-y\right)(i)| \leq \Delta^{\prime}}}\leq d\,\eps^2.
  \label{e:error_term1}
  \end{align}
The second term on the right-side of~\eqref{e:error_split} corresponds
to the error due to ``overflow'' and is handled using concentration
bounds for the rotated vectors. Specifically, 
 we get
\begin{align}
\lefteqn{\sum_{i=1}^{d} \E{(\hat{x}_R(i)-Rx(i))^2 \indic{\{|R\left(x-y\right)(i)| \geq \Delta^{\prime}\}}}}
\nonumber
\\
&\leq 
2\sum_{i=1}^{d} \left[\E{(\hat{x}_R(i)-Ry(i))^2 \indic{\{|R\left(x-y\right)(i)| \geq \Delta^{\prime}\}}}
+\E{(Rx(i)-Ry(i))^2 \indic{\{|R\left(x-y\right)(i)| \geq \Delta^{\prime}\}}}
\right]
\nonumber 
\\
&\leq 2k^2\eps^2
\sum_{i=1}^{d}P(|R\left(x-y\right)(i)| \geq \Delta^{\prime})
+
2\sum_{i=1}^{d} 
\E{(Rx(i)-Ry(i))^2 \indic{\{|R\left(x-y\right)(i)| \geq \Delta^{\prime}\}}}
\nonumber
\\
&\leq 4dk^2\eps^2e^{-{d{\Delta^\prime}^2}/{2\Delta^2}}
+
2\sum_{i=1}^{d} 
\E{(Rx(i)-Ry(i))^2 \indic{\{|R\left(x-y\right)(i)| \geq \Delta^{\prime}\}}}
\nonumber
\\
&\leq 4dk^2\eps^2 e^{-{d{\Delta^\prime}^2}/{2\Delta^2}}+
4(2\Delta^2+d\Delta^{\prime 2})e^{-\frac{d{\Delta^\prime}^2}{2\Delta^2}},
\label{e:error_term2}
\end{align}
where the second inequality follows upon noting that from the description decoder of MQ in
Alg.~\ref{a:D_MCQ} that 
$|\hat{x}_R(i)-Ry(i)|\leq \eps k$ almost surely for
each $i\in [d]$; the third inequality uses
the fact that $R(x-y)(i)$ is sub-Gaussian with variance parameter
$\|x-y\|_2^2/d\leq \Delta^2/d$; and fourth inequality
is by Lemma~\ref{l:variance_tail}.


Upon combining~\eqref{e:error_split},~\eqref{e:error_term1}, and~\eqref{e:error_term2}, and substituting
$\eps=2\Delta^{\prime}/(k-2)$ and 
${\Delta^\prime}^2=6(\Delta^2/d) \log \Delta/\delta$, 
we obtain
\begin{align}
\label{e:break}
    \E{\norm{Q_{{\tt M},R}(x,y)-x}_2^2}
    &\leq 
    d\,\eps^2+ 4dk^2\eps^2 e^{-\frac{d{\Delta^\prime}^2}{2\Delta^2}}+
4(2\Delta^2+d\Delta^{\prime 2})e^{-\frac{d{\Delta^\prime}^2}{2\Delta^2}}
\\
\nonumber
&= 24\, \frac{\Delta^2}{(k-2)^2}\ln \frac \Delta \delta 
+96\delta^2 \left(\frac{k}{k-2}\right)^2\cdot\frac{\ln (\Delta/\delta)}{(\Delta/\delta)}
+ 8\delta^2 \cdot \frac{1+3\ln (\Delta/\delta)}{(\Delta/\delta)}
\\
\nonumber
&\leq 24\,\frac{\Delta^2}{(k-2)^2}\ln \frac \Delta \delta + \left(\frac{96}{e} \left(\frac{k}{k-2}\right)^2+\frac{24}{e^{2/3}}\right)\cdot \delta^2,
\end{align}
where we used $(1+3\ln u)/u \leq 3/e^{2/3}$ 
and $(\ln u)/u\leq 1/e$
for every $u>0$. We conclude by noting that
for $k\geq 4$, 
\[
\left(\frac{96}{e}\left(\frac{k}{k-2}\right)^2+\frac{24}{e^{2/3}}\right)\leq 154.
\]

 \paragraph{Bias $\beta(Q)$:}
 The calculation for the bias is similar to that we 
 used to bound the second term on the right-side of
 \eqref{e:error_split}. Using the notation 
 $\hat{x}_R$ introduced above, 
 we have 
\eq{
\lefteqn{\norm{\E{Q_{{\tt M}, R}}-x}_2}
\\
&=\norm{\E{R^{-1}\left(\hat{x}_R-Rx\right)}}_2
\\
&= \norm{R\E{R^{-1}\left(\hat{x}_R-Rx\right)}}_2 \\ &=\norm{\E{RR^{-1}\left(\hat{x}_R-Rx\right)}}_2 \\ &=\norm{\E{\hat{x}_R-Rx}}_2,
}
where the second identity holds since $R$ is a unitary matrix.

Further, since $Q_{{\tt M}}(x,y)$ is an unbiased estimate
of $x$ when $|x-y|\leq \Delta^\prime$ (see Lemma~\ref{t:MQ}), 
  by~\eqref{e:error_term1} and~\eqref{e:error_term2} we obtain
\eq{
\norm{\E{\hat{x}_R-Rx}}_2^2 &\leq \sum_{i=1}^{d} \E{\left(\hat{x}_R(i)-Rx(i)\right)\indic{|R\left(x-y\right)_i| \geq \Delta^{\prime})} }^2\\
&\leq  \sum_{i=1}^{d}\E{\left(\hat{x}_R(i)-Rx(i)\right)^2\indic{|R\left(x-y\right)(i)| \geq \Delta^{\prime})} }\\
&\leq 154\, \delta^2,
}
which completes the proof.
\qed

\subsection{Proof of Lemma  \ref{t:RCS_RAQ_alpha_beta}}
 \paragraph{Mean Square Error $\alpha(Q_{S,R})$:}

From the description of Algorithms \ref{a:E_RCS_RMQ} and \ref{a:D_RCS_RMQ}, we know that the quantized output of subsampled RMQ $Q_{\tt WZ}$ for an input $x$ is
\eq{
&Q_{\tt WZ}(x) =   R^{-1}\hat{x}_R \text{, where}\\
&\hat{x}_R=\frac{1}{\mu}\sum_{i\in
  [d]}\left( Q_{\tt M}( Rx(i), Ry(i)) -Ry(i) \right) \indic{\{i \in S\}}\, e_i +Ry,}
  and $Q_{\tt M}( Rx(i), Ry(i)) $ denotes the quantized output of the modulo quantizer for an input $Rx(i)$ and side-information $Ry(i)$.
  Use the shorthand $Q(Rx(i))$ for $Q_{\tt M}( Rx(i), Ry(i))$, we have
  \eq{
&\E{ \norm{Q_{\tt WZ}(x) -x}_2^2}\\  &= \sum_{i \in [d]} \E{\left( \frac{1}{\mu}\left( Q(Rx(i)) -Ry(i) \right) \indic{\{i \in S\}}- (Rx(i)-Ry(i)) \right)^2 }\\
&= \sum_{i \in [d]} \E{ \frac{1}{\mu^2} Q(Rx(i)) -Rx(i)^2 \indic{\{i \in S\}}}
\\
&\hspace{2cm}+\sum_{i \in [d]}\E{\left(\frac{1}{\mu}\left(Rx(i) -Ry(i)\right)\indic{\{i \in S\}}   - (Rx(i)-Ry(i)) \right)^2 } \\
&= \sum_{i \in [d]}\frac{1}{\mu} \E{\left( Q(Rx(i)) -Rx(i) \right)^2  }
+ \sum_{i \in [d]}
\E{\left(Rx(i) -Ry(i)\right)^2}\cdot
\E{\left(\frac{1}{\mu}\indic{\{i \in S\}}   - 1 \right)^2 } 
\\
&= \sum_{i \in [d]}\frac{1}{\mu} \E{\left( Q(Rx(i)) -Rx(i) \right)^2  } + \sum_{i \in [d]}
\E{\left(Rx(i) -Ry(i)\right)^2}\cdot \frac{1-\mu}{\mu}\\
&\leq \frac{\alpha(Q_{{\tt M}, R})}{\mu}+ \frac{\Delta^2}{\mu},
  }
  where we used the independence of $S$ and $R$ in the third identity
  and used the fact that $R$ is unitary in the final step.
  
  \paragraph{ Bias $\beta(Q_{S, R})$:} 
 This follows upon noting that the conditional expectation (over $S$) of the output of
 subsampled RMQ given $R$ 
 is the vector $R^{-1} \sum_{i \in [d]}Q_{\tt M}(Rx(i), Ry(i))e_i$, which, in turn, is equivalent in distribution to the output of RMQ.
\qed

\subsection{Proof of Theorem \ref{t:lb_k}}
We denote $\Delta_{min} =  \min_{i \in [d]}\Delta_{i}$
and set $y_i$s to be $0$.
Let $x_1,...,x_n$ be an $iid$ sequence with common
distribution 
such that for all $j \in [d]$ we have 
\[
x_1(j) =\begin{cases}  \frac{\Delta_{min}}{\sqrt{d}}
\quad \text{w.p.} \frac{1+\alpha(j)\delta}{2}\\
- \frac{\Delta_{min}}{\sqrt{d}}
\quad \text{w.p.} \frac{1-\alpha(j)\delta}{2},
\end{cases}
\]
where $\alpha \in \{-1, 1\}^d$ is generated uniformly at random.
We have the following Lemma for such $x_i$s, which 
provides a 
lower bound for the MSE of any estimator of the mean of the distribution of $x_i$s.

\begin{lem}\label{l:lowbound}
{For $x_1, ..., x_n$ generated as above and}
 any estimator $\hat{\bar{x}}$ of the mean formed using only $r$-bit
quantized version of $x_i$s, we have\footnote{Note that the side information $y_i$s are all set to $0$.}
\[\E{ \left\|\hat{\bar{x}}- \frac{\delta \Delta_{min}}{\sqrt{d}} \alpha\right\|_2^2} \geq c^\prime \cdot   \frac{d\Delta_{min}^2}{nr},\]
where $c^\prime <1$ is a universal constant. 
\end{lem}
\noindent Proof of Lemma \ref{l:lowbound} follows from either \cite[Proposition 2]{duchi2014optimality} or \cite[Theorem 11]{acharya2020general}.

The proof of Theorem~\ref{t:lb_k} is completed by using this claim. Specifically, using $2a^2+2b^2 \geq (a+b)^2$, we have
\eq{
2\E{ \norm{\hat{\bar{x}}- \bar{x} }_2^2} + 2\E{\norm{ \bar{x}- \frac{\delta \Delta_{min}}{\sqrt{d}} \alpha}_2^2}  \geq \E{ \norm{\hat{\bar{x}}- \frac{\delta \Delta_{min}}{\sqrt{d}} \alpha}_2^2},
}
which, along with the observation that
\[\E{\norm{ \bar{x}- \frac{\delta \Delta_{min}}{\sqrt{d}} \alpha}_2^2} \leq \frac{\Delta_{min}^2}{n},
\]
gives
\eq{
\E{ \norm{\hat{\bar{x}}- \bar{x} }_2^2} \geq & \frac{c^\prime d\Delta^2_{min}}{2nr} -\frac{\Delta^2_{min}}{n}\\
\geq & \frac{c^\prime \Delta^2_{min}d}{4nr},
}
when $(d/r)\geq 4/c^\prime$. The proof is completed by setting
$c=c^\prime/4$.
\qed

\begin{rem}
Since the lower bound in \cite{acharya2020general} holds for sequentially interactive protocols, 
if we allow interactive protocols for mean estimation where client $i$ gets to see the messages transmitted by the clients
$j$ in $[i-1]$, 
and can design its quantizers based on these previous messages, even then the lower bound above will hold. 
\end{rem}

\subsection{Proof of Lemma  \ref{t:DAQ}}

We will prove a general result which will not only prove Lemma \ref{t:DAQ} but will also be useful in the proof of Lemma \ref{t:RDAQ}. Consider $x$ and $y$ in $\R^d$ such that each coordinate of both $x$ and $y$ lies in $[-M, M]$. Also, consider the following generalization of DAQ:
\[
Q_{\tt D}(x, y)  =  \sum_{i=1}^{d}2 M\left( \indic{\{U(i) \leq x(i)\}} - \indic{\{U(i) \leq y(i)\}}\right)e_i  +y, 
\]
where $\{U_i\}_{i \in [d]}$ are $iid$ uniform random variables in $[-M,M]$. We will show that 
\begin{align}\label{e:intermediate}
\E{Q_{\tt D}(x, y)}=x \quad \text{and} \quad  \E{\norm{Q_{\tt D}(x, y)-x}_2^2} \leq 2M \norm{x-y}_1,
\end{align}
which upon setting $M=1$ proves Lemma \ref{t:DAQ}.
 
Towards proving $\eqref{e:intermediate}$, note that
from the estimate formed by $Q_{\tt D}$, it is easy to see that $\E{Q_{\tt D}(x, y)}=x$.
The MSE can be bounded as follows:
\eq{\E{\norm{Q_{\tt D}(x, y)-x}_2^2}&=\sum_{i=1}^{d} \E{   (2M\left(
    \indic{\{U_i \leq x(i)\}}  -\indic{\{U_i \leq y(i)\}} \right)- \left(x(i) -y(i)\right))^2  }  
\\
&= \sum_{i=1}^{d}  4M^2 \frac{|x(i)-y(i)|}{2M}-\norm{x-y}_2^2 
\\
&=2 M\norm{x-y}_1 - \norm{x-y}_2^2,
}
where we used the observations that $ 2M \left( \indic{\{U_i \leq x(i)\}}  -\indic{\{U_i \leq y(i)\}} \right)$ is an unbiased estimate of $\left(x(i) -y(i)\right)$ and that  $\left( \indic{\{U_i \leq x(i)\}}  -\indic{\{U_i \leq y(i)\}} \right)^2$ equals one if and only if exactly one of the indicators is one, which in turn happens with probability $\frac{|x(i)-y(i)|}{2M}$.\qed.

\subsection{Proof of Lemma  \ref{t:RDAQ}}
\paragraph{Worst-case bias $\beta(Q_{{\tt D}, R} \Delta)$:}
Since the final interval $[-M_{h-1}, M_{h-1}]$ contains $[-1, 1]$, we can see that $\E{Q_{{\tt D}, R} (x, y)} = x$.

~\paragraph{Worst-case MSE $\alpha(Q_{{\tt D}, R} ; \Delta)$:}
We denote by  $B^{x}_{ij}$ and $B^{y}_{ij}$ the bits
\eq{
B^{x}_{ij} = \indic{\{U(i,j) \leq Rx(i)\}} ~\text{ and}~
B^{y}_{ij} = \indic{\{U(i, j) \leq Ry(i)\}}.
}
Then, the final quantized value of the quantizer RDAQ  can be expressed as
$Q_{{\tt D}, R}(X) =  R^{-1} \hat{x}_R$ 
where, with 
$z^*(i)$  denoting the smallest $M_j$ such that
the interval $[-M_j, M_j]$ contains $Rx(i)$ and $Ry(i)$ and $[h]_0=\{0, \ldots, h-1 \}$,
\begin{align*}
\nonumber \hat{x}_R :=\sum_{i \in \{1, \ldots, d\}} \left( \sum_{j \in [h]_0} 2M_j \cdot \left( B^{x}_{ij}  - B^{y}_{ij} \right)  +Ry(i) \right)\indic {\{z^*(i)=j\}}  e_i.
\end{align*}

Since $R$ is a unitary transform, 
we get

\eq{
\E{\norm{Q_{{\tt D}, R}(x)-x}_2^2}
&=
\E{\norm{RQ_{{\tt D}, R}(x)-Rx}_2^2}
\\
&=\E{\norm{\hat{x}_R-Rx}_2^2} 
\\
&=\sum_{i \in [d]} \E{ (\hat{x}_R(i) -Rx(i))^2  }\\
&=\sum_{i \in [d]} \E{ \left(  \sum_{j \in [h]_0}(2M_j \cdot \left( B^{x}_{ij}  - B^{y}_{ij} \right)  +Ry(i) -Rx(i))\indic {\{z^*(i)=j\}}\right)^2  }\\
&=\sum_{i \in [d]}\sum_{j \in [h]_0}   \E{ \left( 2M_j \left( B^{x}_{ij}  - B^{y}_{ij} \right) +Ry(i) -Rx(i) \right)^2  \indic {\{z^*(i)=j\}} ,}}
where the last identity uses $\indic{\{z^*(i)=j_1\}} \indic{\{z^*(i)=j_2\}}=0$ for all $j_1 \neq j_2, $ to cancel the cross-terms in the expansion of  $(\hat{x}_R(i) -Rx(i))^2$. Conditioning on $R$
and using the independence 
of $\indic{\{z^*(i)=j\}}$ from the randomness used in MQ, we get

\begin{align}\E{\norm{Q_{{\tt D}, R}(x)-x}_2^2}
\nonumber
&=\sum_{i \in [d]}\sum_{j \in [h]_0}   \E{ \E{\left( 2M_j \left( B^{x}_{ij}  - B^{y}_{ij} \right) +Ry(i) -Rx(i) \right)^2\mid R } \indic {\{z^*(i)=j\}} }\\
\nonumber
&\leq \sum_{i \in [d]}\sum_{j \in [h]_0}   \E{ 2M_j |Rx(i) -Ry(i)| \indic{\{z^*(i)=j\}} },\\
\nonumber
&\leq \sum_{i \in [d]} \E{ 2M_0 |Rx(i) -Ry(i)|\indic{\{z^*(i)=0\}} }\\& \nonumber ~~~+ \sum_{i \in [d]}\sum_{j \in [h-1]}   \E{ 2M_j |Rx(i) -Ry(i)|\indic{\{z^*(i)=j\}} },\\
\nonumber
&\leq \sum_{i \in [d]} \E{ 2M_0 |Rx(i) -Ry(i)| }\\& ~~~+ \sum_{i \in [d]}\sum_{j \in [h-1]}   \E{ 2M_j |Rx(i) -Ry(i)|\indic{\{z^*(i)=j\}} },
\label{e:overallterm}
\end{align}

where the  first inequality follows from \eqref{e:intermediate} in the proof of Lemma \ref{t:DAQ}.

Next, noting that 
\[
\indic{\{z^*(i)=j\}} \leq  \indic{\{|RX(i)| \geq
  M_{j-1}\}}+ \indic{\{|RY(i)| \geq M_{j-1}\}} \quad \text{ almost surely}, 
  \]
  an
application of the Cauchy-Schwarz inequality yields
\begin{align}
\nonumber &\E{ 2M_j |Rx(i) -Ry(i)|\indic{\{z^*(i)=j\}} }\\ 
\nonumber
&\leq 2M_j \E{(Rx(i)-Ry(i))^2}^{1/2} \E{(\indic{\{|RX(i)| \geq
  M_{j-1}\}}+ \indic{\{|RY(i)| \geq M_{j-1}\}})^2}^{1/2}\\
 \nonumber &\leq 2M_j \E{(Rx(i)-Ry(i))^2}^{1/2} \left(2P(|Rx(i) |\geq M_{j-1} )+ 2P(|Ry(i)| \geq M_{j-1} )\right)^{1/2}\\
  &\leq 2M_j \E{(Rx(i)-Ry(i))^2}^{1/2} \left(8 e^{\frac{-d M^2_{j-1}}{2}}\right)^{1/2},
  \label{e:term_2}
  \end{align}
where the second ineqaulity uses $(a+b)^2 \leq 2a^2+2b^2$ and the third uses subgaussianity of $Rx(i)$ and $Ry(i)$.

Substituting the upper bound in \eqref{e:term_2} for the second term  in the RHS of \eqref{e:overallterm} and using $\E{X} \leq \E{X^2}^{1/2}$ for the first term, we get
\eq{
\E{\norm{Q_{{\tt D}, R}(x)-x}_2^2}
&\leq \sum_{i \in [d]}  \E{  |Rx(i) -Ry(i)|^2}^{1/2} \left(2M_0+\sum_{j \in [h-1]} 2M_j \cdot \left(8e^{-\frac{dM_{j-1}^2}{2}}\right)^{1/2}\right)\\
&\leq  \sqrt{d \cdot \E{ \norm{Rx -Ry}_2^2}} \left(2M_0+\sum_{j \in [h-1]} 2M_j \cdot \left(8e^{-\frac{dM_{j-1}^2}{2}}\right)^{1/2}\right)\\
&= \sqrt{d \cdot \norm{x -y}_2^2} \left(2M_0+\sum_{j \in [h-1]} 2M_j \cdot \left(8e^{-\frac{dM_{j-1}^2}{2}}\right)^{1/2}\right)\\
&=\sqrt{d \cdot \norm{x -y}_2^2} \left(2\sqrt{\frac{6}{d}}+\sum_{j \in [h-1]} 2 \sqrt{\frac{6e^{*j}}{d}} \cdot \left(8e^{-1.5e^{*(j-1)}}\right)\right)\\
&= 8\sqrt{3} \cdot \sqrt{ \norm{x -y}_2^2} \left( 1 + \sum_{j\in[h-1]} e^{-0.5 e^*(j-1)} \right)\\
& \leq 16 \sqrt{3}\cdot \sqrt{ \norm{x -y}_2^2},
\\}
where the second inequality uses the fact that $\sum_{i} \norm{a}_1 \leq \sqrt{d}\norm{a}_2$, the first and second indentities follow from the fact that $R$ is unitary transform and substituting for $M_i$s, the final inequality follows from the bound of $1$ for $\sum_{j=1}^{\infty}e^{-0.5 e^*(j-1)}$, which, in turn, can seen as follows
\eq{
e^{-0.5 e^*(j-1)} &={e^{-0.5}}+  {e^{-0.5e}}+ {e^{-0.5e^e}}
+\sum_{j=3}^{\infty}e^{-0.5{e^{*(j)}}}
\\
&\leq {e^{-0.5}}+  {e^{-0.5e}}+ {e^{-0.5e^e}}
+\sum_{j=3}^{\infty}e^{-0.5{je^e}}
\\
&\leq {e^{-0.5}}+  {e^{-0.5e}}+ {e^{-0.5e^e}}+\frac1{e^{e^e}-1}
\\
&\leq 1.
}
\qed

\subsection{Proof of Lemma \ref{t:sRDAQ}}

\paragraph{Worst-case bias $\beta(Q_{{\tt WZ}, u}; \Delta)$:}
It is straightforward to see that $\E{Q_{{\tt WZ}, u}(x)} = x$.

\paragraph{Worst-case MSE $\alpha(Q_{{\tt WZ}, u}; \Delta)$:}

We denote by  $B^{x}_{ij}$ and $B^{y}_{ij}$ the bits
\eq{
B^{x}_{ij} = \indic{\{U(i,j) \leq Rx(i)\}} ~\text{ and}~
B^{y}_{ij} = \indic{\{U(i, j) \leq Ry(i)\}}.
}
Then,
the quantized output can be stated as follows:
{noting that $Q_{{\tt WZ}, u}(x) =  R^{-1} \hat{x}_R$
where,  with 
$z^*(i)$  denoting the smallest $M_j$ such that
the interval $[-M_j, M_j]$ contains $Rx(i)$ and $Ry(i)$,}
\[
\hat{x}_R :=\left(\sum_{i \in \{1, \ldots, d\}}\sum_{j \in \{0, \ldots, h-1\}} 2M_j \cdot \left( B^{x}_{ij}  - B^{y}_{ij} \right) \indic {\{z^*(i)=j\}} \indic { \{ i \in S\}}\cdot e_i +Ry\right),
\]

Since $R$ is a unitary transform, 
the mean square error between
$Q_{{\tt WZ}, u}(x)$ and $x$ 
can be bounded as
in the proof of Lemma~\ref{t:RDAQ} as follows:
\eq{
\E{\norm{Q_{{\tt WZ}, u}(x)-x}_2^2}
&=\E{\norm{\hat{x}_R-Rx}_2^2}
\\
&=\E{\norm{\hat{x}_R-Rx}_2^2} 
\\
&=\sum_{i \in [d]} \E{ \hat{x}_R(i) -Rx(i))^2  }\\
&=\sum_{i \in [d]}\sum_{j \in [h]}   \E{ \left( 2M_j \left( B^{x}_{ij}  - B^{y}_{ij} \right)\indic { \{ i \in S\}} +Ry(i) -Rx(i) \right)^2  \indic {\{z^*(i)=j\}} }\\
&=\sum_{i \in [d]}\sum_{j \in [h]}   \E{ \E{\left( 2M_j \left( B^{x}_{ij}  - B^{y}_{ij} \right)\indic { \{ i \in S\}} +Ry(i) -Rx(i) \right)^2\mid R } \indic {\{z^*(i)=j\}} }\\
&\leq \sum_{i \in [d]}\sum_{j \in [h]}   \E{ \frac{2M_j}{\mu} \cdot |Rx(i) -Ry(i)|\cdot \indic{\{z^*(i)=j\}} },
}
where the inequality follows from similar calculations in the proof of Lemma \ref{t:DAQ}.
The rest of the analysis proceeds as that in the proof of Lemma \ref{t:RDAQ}.
\qed

\subsection{Proof of Lemma \ref{t:bRDAQ}}
For $Q(x)$ as in \eqref{e:boosted_RDAQ}, we have \[Q(x) =\sum_{i=1}^{N} q_i/N,\] where $q_i$ for all $i \in \{1, \ldots N\}$ is an unbiased estimate of $x$ and equals in distribution the output of the RDAQ quantizer for an input $x$ and side information $y$.  Moreover, $q_i$s are mutually independent conditioned on $R$. Therefore,
\eq{
\E{\norm{Q(x) -x}_2^2 } &= \E{\norm{\sum_{i=1}^{N}\frac{q_i}{N}-x}_2^2}\\
&=\E{ \E{\norm{\sum_{i=1}^{N}\frac{q_i}{N}-x}_2^2|R}}\\
&=\E{\sum_{i=1}^{N}\frac{1}{N^2}\E{\norm{q_i-x}_2^2|R}}\\
&\leq 16 \sqrt{3}\, \frac{\Delta}{N},
}
where the third identity follows from the conditional independence of
$q_i$s after conditioning on $R$ and the fact that $q_i$ is an
unbiased estimate of $x$.
The final inequality
 follows from the fact that $q_i$ equals in distribution the output of the RDAQ quantizer and then using Lemma \ref{t:RDAQ}.
\qed

\section{Concluding Remarks}
 In this chapter, we saw that having access to side-information helps for the problem of communication-constrained distributed mean estimation.  Using this side-information allows us to break the lower bounds for this problem in the no-side information setting. We suspect that identifying side-information sources and then using them will improve the performance in communication-constrained distributed learning scenarios. 

Finally, our techniques could also be used to further exploit the correlation between client data, as was shown in \cite{liang2021improved}.  \cite{liang2021improved} built upon our work and showed that our proposed quantizer RMQ could also be used to exploit the correlation between client data. Specifically, \cite{liang2021improved} showed that when client data is ``close'', the bound in Theorem \ref{t:DME_known} can be further improved. The key idea was to use the quantized data from clients as side-information to decode other clients' data.

\chapter{Revisiting Gaussian Rate-Distortion}\label{Ch:GRD}
\section{Synopsis}
We consider the problem of Gaussian rate-distortion in both the no side-information and side-information case. In the no side-information case, as a by-product of the quantizers designed in Chapter \ref{Ch:RATQ}, we obtain an efficient quantizer for Gaussian vectors
 which attains a rate very close to the
Gaussian rate-distortion function and is, in fact, universal for subgaussian input vectors. In the setting where the decoder has access to some side-information, popularly known as the Wyner-Ziv problem, we leverage the quantizers developed in Chapter \ref{Ch:DME} and obtain an efficient scheme in this setting. Once again, our scheme is universal for subgaussian vectors.

The results presented in this chapter are from \cite{mayekar2020ratqj} and \cite{mayekarwyner}.

\section{Introduction}

We revisit the classic Gaussian rate-distortion problem. In the classic Gaussian rate-distortion
   we seek to quantize a random Gaussian vector to within a
  specified mean squared error while using as few bits per dimension as possible ($cf.$~\cite{gallager1968information, CovTho06}). 
Typical
  fixed-length schemes for this problem draw on its duality with the
  channel coding problem and modify channel codes to obtain coverings;
  see, for instance,~\cite{martinian2006low, SommerFederShalvi08, yan2013polar}. 
However,  these schemes may not
  be acceptable for two reasons: First, the resulting complexity is
  still too high for hardware implementation; and second, the
  resulting schemes are not universal and are tied to Gaussian
  distributions, specifically.
  
  We also revisit the Gaussian Wyner-Ziv problem ($cf.$~\cite{wyner1976rate, Oohama97}). Similar to the problem described above, in the Gaussian Wyner-Ziv problem we seek to quantize a random Gaussian vector to within a
  specified mean squared error while using as few bits per dimension. Except, in this case, the decoder has access to a correlated Gaussian vector.
Practical codes for this problem can be found in ~\cite{Zamir02, Pradhan03, KoradaUrbanke10, LingGaoBelfiore12, LiuL15a}. However, once again, these codes are computationally too expensive and the analysis is tied to the Gaussian distribution.

For the Gaussian rate-distortion problem, we evaluate the performance of a subroutine of RATQ, ATUQ, presented in Chapter \ref{Ch:RATQ}.
Similarly, for the Gaussian Wyner-Ziv problem, we use the quantizer developed in the \known setting in Chapter \ref{Ch:DME}.
Our schemes in both settings have almost constant computational complexity per dimension and require a minuscule excess rate over the optimal asymptotic rate. Moreover, unlike the classical schemes for these problems, we do not require the distribution to be exactly Gaussian, and subgaussianity suffices. 
  

 \subsection*{Organization}
In Section \ref{s:GRD},  we describe the Gaussian rate-distortion problem, our scheme for this problem which employs quantizers from Chapter \ref{Ch:RATQ}, and its performance. In Section \ref{s:GWZ}, we describe the Gaussian Wyner-Ziv problem, our scheme  for this problem which employs quantizers from Chapter \ref{Ch:DME}, and its performance.
\section{The Gaussian rate-distortion problem}\label{s:GRD}

Consider a random vector $X=[X(1),\cdots, X(d)]^T$ with 
$iid$ components $X(1), \cdots, X(d)$ generated from
a zero-mean Gaussian distribution with variance $\sigma^2$. For a pair $(R,D)$ of nonnegative
numbers
is an {\em achiveable} rate-distortion pair if we can find a quantizer $Q_d$
of precision $dR$ and with mean square error $\E{\norm{X -
    Q_d(X)}_2^2}\leq dD$.
For $D>0$, denote by $R(D)$ the infimum over all
$R$ such that $(R,D)$ constitute an achievable rate-distortion pair
for all $d$ sufficiently large. A well-known result in information theory
characterizes $R(D)$ as follows  ($cf.$~\cite{CovTho06}):
\eq{
R(D)=
 \begin{cases}
\frac{1}{2} \log \frac{\sigma^2}{D} \quad &\text{ if } D \leq
\sigma^2,  \\ 
0 &\text{ if } D > \sigma^2.    
\end{cases}}
The function $R(D)$ is called the {\em Gaussian rate-distortion
  function}.

Over the years, several constructions using error correcting codes and
lattices have evolved that attain the rate-distortion function,
asymptotically for large $d$. In this section, we show that a slight
variant of ATUQ, too, attains a rate very close to the Gaussian
rate-distortion function, when applied to Gaussian random vectors.

Specifically, consider the quantizer $Q_{{\tt at}, I}$ described
earlier in~\eqref{e:Q_at_I}.
Recall that  $Q_{{\tt at}, I}$ can be
described by algorithm \ref{a:E_RATQ} and \ref{a:D_RATQ} with random
matrix $R$ replaced with $I$. That is, we
divide the input vector in $\ceil{d/s}$ subvectors and employ ATUQ to quantize
them. In fact, we will apply this quantizer not only to a Gaussian
random vector, but any random vector with subgaussian
components; the components need not even be independent.
Thus, we show that our quantizer is almost optimal {\em
  universally} for all subgaussian random vectors. 

We set the parameters $m$, $m_0$, $h$, $s$, and $\log(k+1)$ of $Q_{at, I}$ as follows:
\begin{align}\label{e:Param}
\nonumber 
&m = 3v, \quad m_0= 2v \ln s,\\& \nonumber \log h = \ceil{\log\left(1+\ln^*\left(\frac{4\ln(8\sqrt{2}v/D)}{3}\right)\right)},\\& \nonumber
s = \min \{ \log h, d\}, \\&
\text{ and } \log (k+1) =
\ceil{\log\left(2+\sqrt{\frac{18v+6v\ln s}{D}}\right)}. 
\end{align}
 \begin{thm}\label{t:gauss_rd}
   Consider a random vector $X$ taking values in $\R^d$ and with 
   components $X_i$, $1\leq i \leq d$ such that each $X_i$ is a
   centered subgaussian random variables with a variance factor $v$. 
 Let $Q_d$ be the $d$-dimensional $Q_{at, I}$ with parameters as in
 \eqref{e:Param}. Then, for $d\geq \log h$ and $D<v/4$, $Q_d$ gets the mean square error less than $dD$ using
 rate $R$ satisfying
 \[
 R \leq \frac{1}{2}\log\frac{v}{D}+ O\left(\log\log\log
 \log^*\log\left (\frac{v}{D}\right)\right).
 \]
 \end{thm}
 \begin{proof}
 
We split the overall mean square error into two terms and 
derive upper bounds for each of them. Specifically, we have
  \eq{
  \frac{1}{d}\cdot \E{\norm{X_d-Q_d(X_d)}_2^2}&=\frac{1}{d}\cdot \E{\sum_{i \in [d]}(X_d(i)-Q_d(X_d)(i))^2  \indic{\{|X_d(i)|\leq M_{h-1}\}}}\\&\hspace{0.5cm}+\frac{1}{d}\cdot \E{\sum_{i \in [d]}(X_d(i)-Q_d(X_d)(i))^2  \indic{\{|X_d(i)| > M_{h-1}\}}}.
  }
The second term on the right-side above can be bounded as follows:
 \eq{\frac{1}{d}\cdot \E{\sum_{i \in [d]}(X_d(i)-Q_d(X_d)(i))^2  \indic{\{|X_d(i)| > M_{h-1}}\}}
& =  \frac{1}{d}\cdot \E{\sum_{i \in [d]}X_d(i)^2  \indic{\{|X_d(i)| > M_{h-1}\}}}\\
&\leq \E{X_d(1)^4}^{1/2} P\left(|X_d(1)| > M_{h-1}\right)^{1/2}\\
&\leq 4\sqrt{2} v e^{-\frac{M_{h-1}^2}{4v},}
} 
where the first inequality follows by the Cauchy-Schwarz inequality and the second follows by Lemma \ref{l:standar_subg}.
 Note that $M_{h-1}^2\geq m e^{*(h-1)}\geq 3 ve^{*\ln^*(4\ln(8\sqrt{2}v/D)/3)}=4v\ln(8\sqrt{2}v/D)$, which with the previous bound leads to
  \eq{\frac{1}{d}\cdot \E{\sum_{i \in [d]}(X_d(i)-Q_d(X_d)(i))^2  \indic{\{|X_d(i)| > M_{h-1}\}}}
& \leq \frac{D}{2}.}
Furthermore, by Lemma~\ref{l:subg_mse} we have
\eq{\frac{1}{d}\cdot \E{\sum_{i \in [d]}(X_d(i)-Q_d(X_d)(i))^2  \indic{\{|X_d(i)|\leq M_{h-1}\}}}\leq \frac{9v+3v\ln s}{(k-1)^2}
\leq\frac{D}{2}, }
where the last equality holds since
$k \geq 1+ \sqrt{\frac{18v+6v\ln s}{D}}.$

It remains to bound the rate. Note that the 
overall resolution used for the entire vector is
\eq{
  d \log (k+1)+\ceil{\frac{d}{s}}\log h &\leq  d \ceil{\log\left(2+\sqrt{\frac{18v+6v\ln s}{D}}\right)}+ d + \log h
}
Therefore, for $d\geq \log h$ and $D<v/4$, the proof is completed by bounding the rate $R$ as 
\eq{ R&\leq  \log\left(2+ \sqrt{\frac v D}\,\sqrt{{18+6\ln \log h}}\right)+ 3\\
&\leq \frac{1}{2} \log \frac{v}{D} +\log\left(1+ \sqrt{18+6\ln \ceil{\log\left(1+\ln^*\left(\frac{4\ln(8\sqrt{2}v/D)}{3}\right)\right)}}\right)+3\\
&\leq \frac{1}{2} \log \frac{v}{D}+ O\left(\log 
\log \log \log^*\log \frac{v}{D}\right).
}

 \end{proof}
 
 
 We
 remark that the additional term is a small constant for reasonable
 values of the parameters $v$ and $D$. Note that our proposed
 quantizer just uses uniform quantizers with different dynamic ranges,
 and yet is almost universally rate optimal.   

\section{The Gaussian Wyner-Ziv problem}\label{s:GWZ}
Consider the random vectors $X=[X(1),\cdots, X(d)]^T$ and $Y=[Y(1),\cdots, Y(d)]^T$,  where the coordinates $\{X(i), Y(i)\}_{i=1}^{d}$ form an $iid.$ sequence. Furthermore, for all $i \in [d]$, \newest{let}
\[
X(i)=Y(i)+Z(i),
\]
where $Y(i)$ and $Z(i)$ are independent and zero-mean Gaussian random variables with variances $\sigma_y^2$ and $\sigma_z^2$, respectively. The encoder has access $X$, \newest{which it quantizes and sends to the decoder}. The decoder, \newest{on the other hand}, has access to $Y$ (note that encoder does not have acess to $Y$) and can use it to decode $X$. A pair $(R, D)$ of non-negative numbers is an achievable rate-distortion pair if we can find a quantizer $Q_d$ of precision $dR$
and with mean square error $\E{\norm{Q_d(X, Y)-X}_2^2}\leq dD$.  For $D \geq 0$, denote by $R_{\tt WZ}(D)$ the infimum over all $R$ such that $(R, D)$ constitute an achievable rate-distortion pair for all $d$ sufficiently large. From\footnote{The model considered in \cite{wyner1976rate} and perhaps the more popular Wyner-Ziv model is $Y=X+Z$. Nevertheless, through MMSE rescaling this model can be converted to $X=Y^{\prime}+Z^{\prime}$ (see, for instance, \cite{liu2016polar}).} \cite{wyner1976rate}, $R_{\tt WZ}(D)$ can be characterized as follows:
\eq{
R_{\tt WZ}(D) = \begin{cases} \frac{1}{2}\log \frac{\sigma_z^2}{D} \quad &\text{if} \quad D\leq \sigma_z^2,\\
0 \quad &\text{if} \quad D > \sigma_z^2.
\end{cases}
}
Several constructions that involve computational heavy methods such as error correcting codes and lattice encoding attain the rate-distortion function, asymptotically for large $d$.
 In this section, we show that modulo quantizer with parameters set appropriately attains a rate very close to the rate-distortion function \newest{$R_{\tt WZ}(D)$}. Moreover, we will show that this rate can be  achieved for arbitrary $Y$ and $Z$, as long as $Z$ is a zero mean subgaussian random variable with variance factor $\sigma_z^2$. Our proposed quantizer $Q_d(X, Y)$ uses the modulo quantizer to quantize $X(i)$ with side information $Y(i)$ at the decoder and the parameter $k, \Delta^{\prime}$ set as follows:

  \begin{align}
 \nonumber
&\delta = \sqrt{D/308}, 
\quad
\log k= \ceil{\log \left(2+ (\sigma_z/\sqrt{D})4\sqrt{3\ln (2\sqrt{77}\sigma_z/\sqrt{D})} \right)}\\
&
\Delta^{\prime}= \sqrt{6(\sigma_z^2)\ln (\sigma_z/\delta)},
\quad 
\eps=2\Delta^\prime/(k-2),
\label{e:param_sRMQ1}
 \end{align}
 \begin{thm}\label{t:G_WZ}
 Consider random vectors $X, Y$ in $\R^d$, where for all coordinates $i \in \{1, \ldots, d\}$, we have 
 \[
 X(i)=Y(i)+Z(i),
 \] and $Z(i)$ is a centered subgaussian random variable with variance factor of $\sigma_z^2$, \newest{independent of $Y(i)$}. Let $Q_d(X, Y)$  be the quantizer described above. Then, for $D \leq \frac{\sigma^2}{308}$, we have MSE less than $dD$ using rate satisfying
 \[R \leq \frac{1}{2}\log \frac{\sigma_z^2}{D}+O\left(\log \log \frac{\sigma_z^2}{D} \right).\]
 \end{thm}
 \begin{proof}

 The proof of this Theorem is similar to that of Lemma \ref{t:RMQ}.
 We denote by $Q(X(i), Y(i))$ the output of the modulo quantizer with side information $Y(i)$ and parameters $k$, $\Delta^{\prime}$ set as in \eqref{e:param_sRMQ1}. Then, we have
 
\begin{align}
\nonumber
\E{\norm{Q_d(X, Y)-X}^2} &\leq \sum_{i=1}^{d} \E{(Q(X(i), Y(i)) -X(i))^2}\\
\nonumber
&\leq \sum_{i=1}^{d} \E{(Q(X(i), Y(i))-X(i))^2 \indic{\{|\left(X(i)-Y(i)\right)| \leq \Delta^{\prime}\}}}
\\& \hspace{2cm}+ \sum_{i=1}^{d} \E{(Q(X(i), Y(i))-X(i))^2 \indic{\{|\left(X(i)-Y(i)\right)| \geq \Delta^{\prime}\}}}.
\label{e:error_split_GWZ}
\end{align}
 
We bound the first term on the right-side in a similar manner as the bound in \eqref{e:error_term1}. Specifically,
under the event 
$\{|X(i)-Y(i)| \leq
   \Delta^{\prime}\}$, we get
   by Lemma~\ref{t:MQ}
   that
   \[
   |Y(i)
   -X(i)| \leq \eps=
   \frac{2\Delta^{\prime}}{k-2}, \quad \text{{almost surely}},
   \]
   whereby 
  \begin{align}
  \sum_{i=1}^{d} \E{(Y(i)-X(i))^2 \indic{\{|X(i)-Y(i)\}| \leq \Delta^{\prime}}}\leq d\,\eps^2.
  \label{e:error_term1_GWZ}
  \end{align}

For the second term in the RHS note that $X(i)-Y(i)$ is subgaussian with  variance factor $\sigma_z^2$. Therefore, by proceeding in a similar manner as the derivation of \eqref{e:error_term2} we get
\begin{align}\nonumber &\sum_{i=1}^{d} \E{(Q(X(i), Y(i))-X(i))^2 \indic{\{|X(i)-Y(i)| \geq \Delta^{\prime}\}}}\\
&\leq 
2\sum_{i=1}^{d} \left[\E{( Q(X(i), Y(i))- Y(i))^2 \indic{\{|X(i)-Y(i)| \geq \Delta^{\prime}\}}}
+\E{(Y(i)-X(i))^2 \indic{\{|X(i)-Y(i)| \geq \Delta^{\prime}\}}}
\right]
\nonumber 
\\
&\leq 2k^2\eps^2
\sum_{i=1}^{d}P(|X(i)-Y(i)| \geq \Delta^{\prime})
+
2\sum_{i=1}^{d} 
\E{(X(i)-Y(i))^2 \indic{\{|X(i)-Y(i)| \geq \Delta^{\prime}\}}}
\nonumber
\\
&\leq 4dk^2\eps^2e^{-{d{\Delta^\prime}^2}/{2 \sigma_z^2}}
+
2\sum_{i=1}^{d} 
\E{(X(i)- Y(i))^2 \indic{\{|X(i)-Y(i)| \geq \Delta^{\prime}\}}}
\nonumber
\\
&\leq 4dk^2\eps^2 e^{-{{{\Delta^\prime}}^2}/{2\sigma_z^2}}+
4(2\sigma_z^2+d\Delta^{\prime 2})e^{-\frac{{\Delta^{\prime}}^2}{2\sigma_z^2}},
\label{e:error_term2_GWZ}
\end{align}
where the second inequality follows upon noting from the description decoder of MQ in
Alg.~\ref{a:D_MCQ} that 
$|Q(X(i), Y(i))- Y(i)|\leq \eps k$
almost surely for
each $i\in [d]$; the third inequality uses
the fact that $X(i)-Y(i)$ is sub-Gaussian with variance parameter
$\sigma_z^2$; and the fourth inequality
is by Lemma~\ref{l:variance_tail}.

Upon bounding the two terms on the right-side of \eqref{e:error_split_GWZ} 
from above
using \eqref{e:error_term1_GWZ}, \eqref{e:error_term2_GWZ},  we obtain
\begin{align}
\nonumber
\E{\norm{Q_d(X, Y)-X}^2}\leq d\eps^2+ 4dk^2\eps^2 e^{-{{\Delta^\prime}^2}/{2\sigma_z^2}}+
4(2\sigma_z^2+d\Delta^{\prime 2})e^{-\frac{{{\Delta^{\prime}}^2}}{2\sigma_z^2}}.
\end{align}
Note that the RHS in the upper bound above is precisely the same as in \eqref{e:break} with $\sigma_z^2$ replacing $\Delta^2/d$.Therefore proceeding in the same manner as in \eqref{e:break}, we get
\[\E{\norm{Q_d(X, Y)-X}^2}\leq 24 \frac{\sigma_z^2}{(k-2)^2}\ln \frac{\sigma_z}{\delta} + 154 \delta^2.\]
Substituting the value of $k$ and $\delta$ completes the proof.

 \end{proof}
 
 \section{Concluding Remarks}
 The key difference between our proposed quantizers and those proposed in the literature is that we don't focus on precisely matching the rate-distortion function. This allows us to design computationally efficient quantizers, which still have a rate close to the rate-distortion function.



%

%


\part{Source Coding Schemes for Timeliness}


\chapter{Minimum Age Source Codes}

\section{Synopsis}
A transmitter observing a sequence of independent and identically
distributed random variables seeks to keep a receiver updated about
its latest observations.  The receiver need not be apprised about each
symbol seen by the transmitter, but needs to output a symbol at each
time instant $t$.  If at time $t$ the receiver outputs the symbol seen
by the transmitter at time $U(t)\leq t$, the age of information at the
receiver at time $t$ is $t-U(t)$.  We study the design of lossless
source codes that enable transmission with minimum average age at the
receiver.  We show that the asymptotic minimum average age can be
attained up to a constant gap by the Shannon codes for a tilted version of the original
pmf generating the symbols, which can be computed easily by solving an
optimization problem.  Furthermore, we exhibit an example with
alphabet $\X$ \newer{where  Shannon} codes for the original pmf incur an
asymptotic average age of a factor $O(\sqrt{\log |\X|})$ more
than that achieved by our codes.  Underlying our prescription for
optimal codes is a new variational formula for integer moments of
random variables, which may be of independent interest.  Also, we
discuss possible extensions of our formulation to randomized schemes
\newer{and to the erasure channel,} and include a treatment of the related problem of
source coding for minimum average queuing delay.

The results presented in this Chapter are from \cite{mayekar2018optimal} and \cite{mayekar2018optimalb} 

\section{Introduction}
Timeliness is emerging as an important requirement for communication
{in} cyber-physical systems (CPS). {Broadly}, it refers to the
requirement of having the latest information from the transmitter
available at the receiver in a timely fashion.  It is important to
distinguish the requirement of timeliness from that of low delay
transmission: The latter places a constraint on the delay in
transmission of each message, while timeliness is concerned about how
recent is the current information at the receiver.  
In particular, the instantaneous staleness at the receiver is low if a message is received with low delay. 
However, the instantaneous staleness increases linearly at the receiver until a subsequent message is decoded successfully.
{A heuristically appealing metric that can capture the notion of timeliness of
  information in a variety of applications, termed its {\em age}, was first used
  in~\cite{KaulYatesGruteser11} for a setting involving queuing and
  link layer delays and was analyzed systematically for a queuing
  model in the pioneering work~\cite{KaulYatesGruteser12}; see~\cite{yates2015lazy,
    he2018optimal, sun2017update, bhambay2017differential, 
    bacinoglu2017scheduling, kosta2017age} for a sampling of
  subsequent developments in problems related to minimum age
  scheduling.} In this paper, we initiate a systematic study of the
design of source codes with the goal of minimizing the age of the
information at the receiver.

As a motivating application, consider 
  remote sensor data monitoring where
  at each instant the sensor observes real-valued, time-series measurements. For concreteness, the reader may consider voltage and current data recording using \newer{intelligent electronic devices} in a power distribution network. 
  The sensor communicates to a center over a network to enable fault detection and
  fault analysis. On the one hand, the communication protocol and buffer constraints at the sensor limits the rate at which the sensor can send data packets to the center. On the other hand, it is not very important for the center to get all the packets from the sensor.  
  Rather the center wants timely updates about the sensor observations.
  In fact, when operating with cheap hardware with limited front-end buffers, it is common to have observation values in the buffer
  overwritten as new recordings are made even before the previous one waiting in the buffer has been picked-up for processing. Our work focuses on data compression for such applications where there is no direct cost of skipping packets and the interest is only in timely updates.   

\subsection{Main Contributions}
Specifically, we consider the problem of source coding where a
transmitter receives symbols generated from a known distribution and
seeks to communicate them to a receiver in a timely fashion.\footnote{
This assumption of known distribution is realized in practice by building a model for sensor data offline, before initiating the live monitoring process.}
\newer{To that
end, it encodes a symbol $x$ to $e(x)$ using a variable length
prefix-free code $e$. } The coded sequence is then transmitted over a
noiseless communication channel that sends one bit per unit time.  We
restrict our treatment to a simple class of deterministic\footnote{Our
  analysis of average age extends to randomized schemes as well; see
  Section~\ref{s:extensions}.} update schemes, termed {\it memoryless
  update schemes}, where the transmitter does not have have a buffer to store the symbols it has seen previously and simply sends the next observed symbol once the 
  channel is free.

Specifically, denoting the source alphabet by $\X$,
the transmitter observes a symbol $X_t \in \X$ at each discrete time $t$. 
At time $t=1$, 
the transmitter communicates the symbol $X_1 = x_1$ by encoding it to codeword $e(x_1)$ of length $\ell(x_1)$ bits. 
This transmission requires $\ell(x_1)$ channel uses and is received perfectly at the decoder at time $1+\ell(x_1)$.  
Since the channel remains busy sending $e(x_1)$ for time instants $1$ to $\ell(x_1)$, the transmitter cannot send any new symbols during this period. 
At time $t' = 1+\ell(x_1)$, the transmitter observes the symbol $X_{t'} = x_{t'}$.
\newer{Under a memoryless update scheme, }
the transmitter cannot store the symbols seen during the time interval $\{2, \dots, \ell(x)\}$ and communicates codeword $e(x_{t'})$ over the next $\ell(x_{t'})$ channel uses, starting from the time instant $t'=1+\ell(x_1)$.
The communication process continues repeatedly in this fashion.

We emphasize that under memoryless schemes, the source symbols
generated and observed by the transmitter while the channel is busy
sending a previous symbol are simply skipped. This skipping is only
allowed when the channel is busy, and not at the will of the encoder
when the channel is free (see Section~\ref{s:extensions} for
discussion on randomized schemes that allow the transmitted to skip
symbols even when channel is free). Furthermore, the encoder need not
indicate to the decoder that a symbol has been skipped using a special
symbol -- the decoder can ascertain this from the received
communication since the  
channel is noiseless  \newer{and compression is done using prefix-free codes}.  


On the receiver side, at each instance $t$ the decoder outputs a time
$U(t)$ and the symbol $X_{U(t)}$ seen by the transmitter at time
$U(t)$.  Thus, the {\em age of information} at the receiver at time $t$ is
given by $A(t) = t - U(t)$.  
  We note that age of information measures timeliness at the receiver.
  When the transmitter skips source symbols, $U(t)$ remains unchanged at the receiver and the age $A(t)$ increases. Therefore, the age metric implicitly penalizes for
  skipping symbols.

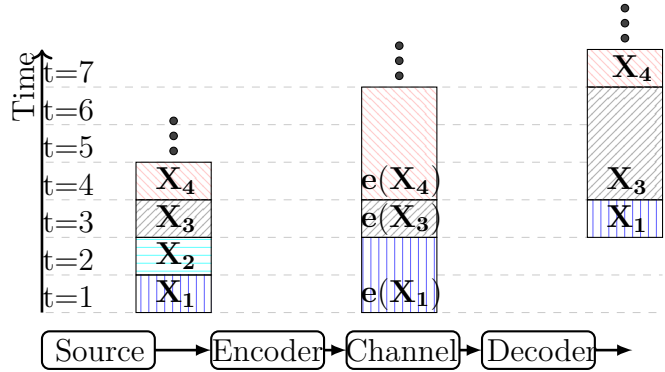
\begin{figure}[h]
\centering \input{Figures/figure-01.tex}
\caption{Illustration of a memoryless update scheme for the first 4 packets in the source-queue.}
\label{f:setup}
\end{figure}

We illustrate the setup in Figure~\ref{f:setup}. 
 In this example,
 the symbol $X_1$ generated at time $t=1$ is encoded to a two-bit codeword $e(X_1)$ and received at the decoder at time $t=3$ after two channel uses.
At time $t=2$, the transmitter skips symbol $X_2$  since the channel was busy sending $X_1$ when it arrived. Further, the decoder retains $U(t)=0$ since it has not received any symbol. 
 At time $t=3$, the decoder receives the codeword $e(X_1)$, updates $U(3) = 1$, and outputs the corresponding  symbol $X_1$. Thus, the age of information at the receiver at time $t = 3$ is $A(3) = 2$. Since the channel becomes available at time $t=3$, the transmitter encodes the symbol $X_3$ and transmits the one-bit codeword $e(X_3)$, which is received after a single channel-use at time $t=4$. 
At time $t=4$, the decoder outputs time $U(4) = 3$ with outputs the corresponding symbol $X_3$, 
and the age of information at the receiver is $A(4) = 1$. Once again, the channel becomes available at time $t=4$ for the transmitter. It encodes the current symbol $X_4$ into the codeword $e(X_4)$ of length $3$ bits and sends $e(X_4)$ over the channel; $e(X_4)$ is received at time $t=7$. 
The decoder retains the output $U(t) = 3$ and $X_{U(t)} = X_3$ for times $t \in \{4,5,6\}$. At time $t=7$, the decoder outputs time $U(7) = 4$ and the corresponding symbol $X_4$; the age of information at the receiver is $A(7) = 3$.

Our goal in this paper is to design prefix-free codes for which the
average age of the memoryless scheme above is minimized; namely codes
$e$ that minimize
\[
\bar{A}(e) = \lim_{T\to \infty} \frac 1T \sum_{t=1}^T A(t).
\]
This formulation is apt for the timely update problem where the
transmitter need not send each update and strives only to reduce the
average age of the information at the receiver.

Using a simple extension of the renewal reward
  theorem, we derive a closed form formula for the asymptotic average
age attained by a prefix-free code.  Interestingly, this formula is a
rational function of the first and the second moment of the random
codeword length.  Our main technical contribution in this paper is a
variational formula for the second moment of random variables that
enables an algorithm for finding the code that attains the minimum
asymptotic average age up to a constant gap.  The variational formula
is of independent interest and may be useful in other settings where
such cost functions arise; we point-out one such setting in
Section~\ref{s:extensions}. In fact, our prescribed prefix-free code
is a Shannon code\footnote{A Shannon
code for $P$ is a prefix-free code that assigns lengths $\ell_S(x) = \lceil -\log P(x)\rceil$ to
a symbol $x$ ($cf.$~\cite{CovTho06}).} for a tilted version of the original pmf. See \eqref{e:tilted_distribution} below for the \newer{description} of the tilted version; it can be computed by solving an optimization problem entailing entropy maximization.

The formula for average age that we derive yields an $O(\log |\X|)$
upper bound on the minimum average age, attained by a fixed length
code.  We show that the same upper bound of $O(\log|\X|)$ holds for
the average age of a Shannon code for the original distribution as
well. However, we exhibit an example where Shannon codes for the original
distribution have $\Omega(\log|\X|)$ age, while our aforementioned
proposed code 
yields an average age of $O(\sqrt{\log |\X|})$.

In addition to our basic formulation, we present a few extensions of
our formulations and other use cases for our proposed variational
formula. Specifically, while we restrict to deterministic schemes for
the most part, our analysis can be extended easily to analyze
randomized schemes where the encoder can choose to skip an available
transmission slot randomly. This idea of skipping transmission
  slots arises also in the recent work~\cite{sun2017update}, albeit in a slightly different context. We
  exhibit an example where a particular randomized scheme outperforms
  every deterministic scheme. However, our analysis is limited and does not
  completely clarify the role of randomization; for instance, it
  remains unclear for which distributions can randomized schemes
  strictly outperform deterministic ones.

In another direction, we consider the case where the transmission
channel is not error-free, but can erase each bit with a known
probability.  Furthermore, an ACK-NACK feedback indicating the success
of transmission is available.  Note that for the standard transmission
problem, the simple repeat-until-succeed scheme is optimal in this
setting.  Our analysis can be used to design the optimal source code
when we restrict our channel coding to this simple scheme.  However,
the optimality of the ensuing source-channel coding scheme remains
unclear.

Finally, we study the related problem of source coding for ensuring
minimum queuing delays.  This problem, introduced
in~\cite{Humblet78thesis}, is closely related to the minimum age
formulation of this paper.  Interestingly, our recipe for designing
update codes with minimum average age can be extended to this setting
as well.  However, here, too, our results are somewhat unsatisfactory:
Our approach only provides a solution to the real-relaxation of the
underlying integer-valued optimization problem and naive rounding-off
is far from optimal.  Nonetheless, we have included these extensions
in the current paper since they indicate the rich potential for our
proposed techniques and provide new formulations for future research.

\subsection{Prior Work}
The problem of designing update codes with low average age is related
to real-time source coding ($cf.$~\cite{Mahajan}) where we seek to
transmit a stream of data under strict delay bounds.  A related
formulation has emerged in the control over communication network
literature ($cf.$~\cite{TatikondaMitter04}) where an observation is
quantized and sent to an estimator/controller to enable control.
Here, too, the requirement is that of communication under bounded
delay.

An alternative formulation for minimum age source coding is considered
in the recent work~\cite{YatesDCC}. Unlike our formulation,
skipping of symbols is prohibited
in~\cite{YatesDCC}. Instead, the authors consider
fixed-to-variable length block codes and require that each coded
symbol be transmitted over a constant rate, noiseless bit-pipe. In
this setting, an exact expression for average age is not available,
and the authors take recourse to an approximation for average age. This
approximate average age is then optimized numerically over a set of
prefix-free codes using the algorithm in
\cite{larmore1989minimum}. The authors further reduce the computational complexity of this algorithm by using  the algorithm in \cite{baer2006source}.

A recent paper~\cite{ZhongYatesSoljanin18} extends this problem to include random arrival times of source symbols and applies the algorithm from~\cite{larmore1989minimum} for optimizing the cost function.
 Note that the cost function optimized
  in \cite{larmore1989minimum} is similar to the approximate average
  age of~\cite{YatesDCC , ZhongYatesSoljanin18}, but with one crucial difference:  While the former is monotonic in both first and second moments of
  random lengths, the latter is not. In absence of this monotonicity,
  the optimality of the solution produced by algorithm in
  \cite{larmore1989minimum} is not guaranteed for the cost functions in~\cite{YatesDCC, ZhongYatesSoljanin18}. In a related
work~\cite{zhong2017backlog}, the same authors point-out that the
average age can be further reduced by allowing the encoder to
dynamically control the block-length of the fixed-to-variable length
codes.

\newer{In contrast to~\cite{YatesDCC}, which is perhaps closest to our work, 
  we derive an exact expression for average age and rigorously establish the structural properties of the optimal solution to the relaxed problem.}
In fact, our proposed minimum average age problem differs from all these prior
formulations since we need not send the entire stream and are allowed
to skip some symbols. 
In our applications of interest, such as that of real-time sensor data monitoring outlined earlier, the allowed
communication rates are much lower than the rate at which data is
generated. Thus, there is no hope of transmitting all the data at
bounded delay, as mandated by the formulations available hitherto. 
Nonetheless, our setting is related closely to that
  in~\cite{YatesDCC} and provides a complementary
  formulation for age optimal source coding.
    We note that our focus is on settings where the alphabet size of the streaming symbols is large. In such settings, the average age for any memoryless update scheme would be much larger than a small constant. Therefore, it suffices to establish optimality up to small additive constants.


%
%

\subsection*{Some preliminaries} 
We recall the notions of Shannon lengths and Shannon codes, which will be used throughout. A source code is called {\em prefix-free} if no codeword is a prefix of another. 
\begin{defn}[Shannon lengths and Shannon codes for $P$]\label{d:shannon}
  For a pmf $P$ on an alphabet $\X$, the real-values
  $\ell(x)=-\log P(x), x \in \X$, are called the {\em Shannon lengths} for the pmf $P$. A prefix-free source code for $P$ with codeword lengths $\ell(x)=\ceil{-\log P(x)}, \ \forall x \in \X,$ is called a {\em Shannon code}\footnote{There can be different codes with codeword lengths required in our definition of a Shannon code. We simply refer to all of them as a Shannon code, since any of these can serve our purpose in this paper.} for the pmf $P$. 
\end{defn}

\subsection*{Organization}
The next section contains a formal description of our setting and a
formula for asymptotic average age of a code. Our main technical tool
is presented in Section~\ref{s:variational}, and we apply it to the
minimum average age code design problem in
Section~\ref{s:code_design}.  Numerical evaluations of our proposed
scheme for the family of Zipf distributions is presented in
Section~\ref{s:age_numerical}. Section~\ref{s:extensions} contains a
discussion on extensions to randomized schemes and erasure channel,
along with a treatment of source codes for minimum average waiting
time.  We provide all the proofs in the final section.

\section{Average age for memoryless update schemes}
\label{s:timely_updates} 

Consider a discrete-time system in which at every time instant $t$, a
transmitter observes a symbol $X_t$ generated from a finite alphabet
$\X$ with pmf $P$.  We assume that the sequence $\{X_t\}_{t=1}^\infty$
is {independent and identically distributed} (iid).  The transmitter
has a noiseless communication channel at its disposal over which it
can transmit one bit per unit time. A {\em memoryless update scheme}
consists of a prefix-free code, represented by its encoder $e:\X\to
\{0,1\}^*$, and a decoder which at each time instant $t$ declares a
time index $U(t)\leq t$ and an estimate $\hat{X}_{U(t)}$ for the
symbol $X_{U(t)}$ that was observed by the encoder at time $U(t)$. We
focus on error-free schemes and require $\hat{X}_{U(t)}$ to equal
$X_{U(t)}$ with probability $1$.

\begin{figure}
\centering
\input{Figures/figure-02a.tex}
\input{Figures/figure-02b.tex}
sh\caption{A sample path of $U(t)$, $A(t)$ corresponding to Figure \ref{f:setup} starting with $U(1)=0$.}
\label{f:Age_sample_path}
\end{figure}
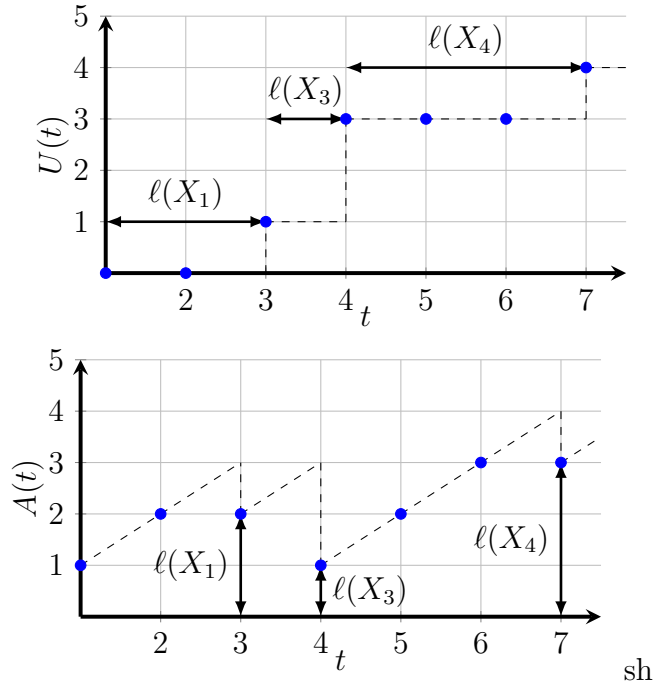

In a memoryless update scheme, once the encoder starts communicating a
symbol $x$, encoded as $e(x)$, it only picks up the next symbol once
all the bits in $e(x)$ have been transmitted successfully to the
receiver.  The time index $U(t)$ is updated to a new value only upon
receiving all the encoded bits for the current symbol. That is, if the
transmission of a symbol is completed at time $t-1$, \newer{the encoder will
start transmitting $e(X_t)$ in the next instant.} Moreover, if the
  final bit of $e(X_t)$ is received at time $t^\prime$, $U(t^\prime)$
  is updated to $t$. A typical sample path for $U(t)$ is given in
Figure~\ref{f:Age_sample_path}.
The age $A(t)$ of the symbol available at the receiver at time $t$ is
given by
\begin{align}
A(t)= t-U(t).  \nonumber
\end{align}
A more general treatment can allow errors in estimates of $X_{U(t)}$ as
well as encoders with memory, but we limit ourselves to the
simple error-free and memoryless setting in this paper.

We are interested in designing prefix-free codes $e$ that minimize the
average age for the memoryless update scheme described above.
\begin{defn}
The {\it average age} for a prefix-free code $e$, denoted
$\bar{A}(e)$, is given by
\begin{align}
\bar{A}(e) = \limsup_{T\to \infty}\frac 1 T \sum_{t=1}^T {(t-U(t))}.
\nonumber
\end{align}
\end{defn}
\newer{We remark that  $\bar{A}(e)$ can be viewed as the average area under the curve of $A(t)$ (w.r.t. $t$). }
Note that $\bar{A}(e)$ is random variable, nevertheless we will prove
that this random variable is a constant almost surely.  
For any symbol $x \in \X$, we denote the length of the codeword $e(x)$ by $\ell(x)$. 
Let $X \in \X$ be a random symbol with pmf $P$ over the alphabet $\X$, 
then the length of the random codeword $e(X)$ is denoted by 
\[L= \ell(X).\] The result below
  uses a simple extension of the classical renewal reward theorem \newer{($cf.$ \cite{ross1996stochastic})} to
  provide a closed form expression for $\bar{A}(e)$ in terms of the
  first and the second moments of $L$.
\begin{thm}\label{t:average_age}
Consider a random variable $X$ with pmf $P$ on $\X$. For a prefix-free
code $e$, the average age $\bar{A}(e)$ is given by
\begin{align}
\bar{A}(e) = \E {L}+ \frac{\E{L^2}}{2\E {L}} - \frac 12 \quad
a.s. \quad.
\label{e:cost_theta}
\end{align}
\end{thm}
\noindent The proof is deferred to Section \ref{p:avg_age}.

Denoting by $\bar{A}^*$ the minimum average age over all prefix-free
codes $e$, as a corollary of the characterization above, we can obtain
the following bounds for $\bar{A}^*$.
\begin{cor}\label{c:age_bounds}
For any pmf $P$ over $\X$, the optimal average age $\bar{A}^*$ is
bounded as
\begin{align}
\frac 32 H(P) -\frac 12 \leq \bar{A}^* \leq \frac 32\log|\X| + 1.
\nonumber
\end{align} 
\end{cor}
The proof of lower bound simply uses Jensen's inequality $\E{L^2} \geq \E{L}^2$ and the fact
that $\E{L} \geq H(P)$ for a prefix free code; the upper bound is
obtained by using codewords of constant length $\lceil
\log|\X|\rceil$.

Note that the lengths $\ell(x)$ are required to be nonnegative
integers. However, for any set of real-valued lengths $\ell(x)\geq 0$,
we can obtain integer-valued lengths by using the rounded-off values
$\lceil \ell(x)\rceil$. Unlike the average length cost, the average
age cost function identified in \eqref{e:cost_theta} is not an
increasing function of the lengths. Nevertheless, by
\eqref{e:cost_theta}, the average age $\bar{A}(e)$ achieved when we
use the rounded-off values can be bounded as follows: Denoting
  $\bar{L}:=\lceil\ell(X)\rceil$, we have
\begin{align}
\E{\bar{L}} + \frac{\E{\bar{L}^2}}{2\E {\bar{L}}} - \frac 12 & \leq
\E{L+1} + \frac{\E{(L+1)^2}}{2\E {L}} -\frac{1}{2} \nonumber \\ \nonumber & \leq \E{L} +
\frac{\E{L^2}}{2\E {L}}+ \frac{2\E{L}}{2\E {L}}\\& \hspace{2cm}+ \frac{1}{2\E {L}}
+\frac 12 \nonumber \\ & \leq \E {L} + \frac{\E{L^2}}{2\E {L}} +2.
\label{e:ceil_loss}
\end{align}

Accordingly, in our treatment below we shall ignore the integer
constraints and allow nonnegative real-valued length assignments.

Returning now to the bound of Corollary~\ref{c:age_bounds}, the upper
and lower bounds differ only by a constant $1.5$ when $P$ is uniform. In view of the foregoing discussion, Shannon codes for
a uniform distribution attain the minimum average age up to a constant gap. The next result
gives an upper bound on average age for Shannon codes for an arbitrary
$P$ on $\X$.
\begin{lem}
Given a pmf $P$ on $\X$, a Shannon code $e$ for $P$ has average age $\overline{A}(e)$ at most $O(\log |\X|)$.
\end{lem}
\begin{proof}
\newer{Let $\ell(X)$ denote the lengths of Shannon code corresponding to $P$ (see Definition \ref{d:shannon}).} We establish the claim using the standard bound
  $H(P^\prime) \leq \log |\X|$ for an appropriately chosen pmf
  $P^\prime$ on $\X$. Specifically, for the tilting of $P$ given by
$P^\prime(x)\propto \ell(x)P(x)$, we get

 \begin{align*}
\log |\X| & \geq \sum_{x \in \X} \frac{ P(x)\ell(x)}{\E{\ell(X)}}
\log \frac{\E{\ell(X)}}{P(x)\ell(x)} 
\\ &= \sum_{x \in \X}
\frac{ P(x)\ell(x)(-\log P(x))}{\E{\ell(X)}} \\& \hspace{1cm} - \sum_{x \in \X}
\frac{P(x)\ell(x)}{\E{\ell(X)}}\log\frac{\ell(x)}{\E{\ell(X)}}
\\ &\geq \sum_{x \in \X} \frac{ P(x)\ell(x)(-\log P(x))}{\E{\ell(X)}}
\\ & \hspace{1cm} - \sum_{x \in \X: \ell(x) \geq
  \E{\ell(X)}}\frac{P(x)\ell(x)}{\E{\ell(X)}}\log\frac{\ell(x)}{\E{\ell(X)}}.
 \end{align*}
Using $-\log P(x) \geq \ell(x)-1$ and $\ln x \leq \frac{x^2-1}{2x}$
for $x\geq 1$, we obtain
 \begin{align*}
\log |\X| & \geq \frac{\E{\ell^2(X)}}{\E{\ell(X)}} - 1 \\&\hspace{0.5cm}- \frac
1{2\ln 2}\cdot \sum_{x \in \X: \ell(x) \geq \E{\ell(X)}} P(x)
\left(\frac{\ell^2(x)}{\E{\ell(X)}^2} - 1\right) 
\\
& \geq \frac{\E{\ell^2(X)}}{\E{\ell(X)}} - 1 \\&\hspace{1cm} - \frac
1{2\ln 2}\cdot \sum_{x \in \X: \ell(x) \geq \E{\ell(X)}} P(x)\cdot
\frac{\ell^2(x)}{\E{\ell(X)}^2}
\\ & \geq
\frac{\E{\ell^2(X)}}{\E{\ell(X)}} - 1 - \frac 1{2\ln 2}\cdot
\sum_{x \in \X}\frac{ P(x)\ell^2(x)}{\E{\ell(X)}^2}
\\& 
\geq \frac{\E{\ell^2(X)}}{\E{\ell(X)}} - 1 - \frac 1{2\ln 2}\cdot
\sum_{x \in \X}\frac{ P(x)\ell^2(x)}{\E{\ell(X)}}
 \\ &\geq
\left(1- \frac{1}{2\ln
  2}\right)\cdot\frac{\E{\ell^2(X)}}{\E{\ell(X)}} -1, 
 \end{align*}
where the second-last inequality follows from the fact that $\E{\ell^2(X)} \geq \E{\ell(X)}$, which in turn follows from the fact that $\ell(X) \geq 1$. The proof is
completed by rearranging the terms.
\end{proof}

It is of interest to examine if, in general, a Shannon code for $P$
itself has average age close to $\bar{A}^*$, as was the case for the
uniform distribution. In fact, it is not the case. Below we exhibit a
pmf $P$ where the average age of a Shannon code for $P$ is
$\Omega(\log |\X|)$, namely the previous bound is tight, and yet a
Shannon code for another distribution (when evaluated for $P$) has an
average age of only $O(\sqrt{\log |\X|})$.
\begin{example}\label{ex:1}
Consider $\X = \{0,...,2^n\}$ and a pmf $P$ on $\X$ given by \eq{
  P(x)=\begin{cases} 1-\frac 1n, \quad &x =0 \\ \frac 1{n2^n}, \quad
  &x \in \{1, \dots ,2^n\}.
\end{cases}
}Using \eqref{e:cost_theta}, the average age $\bar{A}(e_P)$ for a
Shannon code for $P$ can be seen to satisfy $\bar{A}(e_P) \approx (n +
2\log n)/2$. On the other hand, if we instead use a Shannon code
for the pmf $Q$ given by \eq{ Q(x)=
\begin{cases}
 \frac {1}{2^{\sqrt{n}}}, \quad &x =0 \\ \frac
       {1-2^{-\sqrt{n}}}{2^{n}}, \quad &x \in \{1, \dots ,2^n\},
\end{cases}
}we get $\E{L} \approx \sqrt{n}$ and $E{L^2} \approx 2n$, whereby
$\bar{A}(e_{Q}) \approx 2 \sqrt{n}$, just $O(\sqrt{\log
  |\X|})$.\qed
\end{example}
Thus, one needs to look beyond the standard { Shannon
  codes for $P$} to find codes with minimum average
age. Interestingly, we show that \newer{Shannon codes for a
  tilted version of $P$} attain the optimal asymptotic average age (up
to the constant loss of at most $2.5$ bits incurred by rounding-off
lengths to integers). In particular, for the example above, our
proposed optimal codes will have an average age of only $O(\sqrt{\log
  |\X|})$ in comparison to $\Omega(\log |\X|)$ of Shannon codes for
$P$.

\mnote{At HT: I  modified the comment above. Note that the optimal titling is not Q but something else.}
A key technical tool in design of our codes is a variational formula
that will allow us to linearize the cost function in
\eqref{e:cost_theta}, thereby rendering Shannon codes for a tilted
distribution optimal. We present this in the next section.
\section{A variational formula for  $p$-norm} \label{s:variational}
The expression for average age identified in
Theorem~\ref{t:average_age} involves the second moment of the random
codeword length $L$. This is in contrast to the traditional variable
length source coding problem where the goal is to minimize the average
codeword length $\E L$. For this standard cost, Shannon codes which
assign a codeword of length $\lceil-\log P(x)\rceil$ to the symbol $x$
come within $1$-bit of the optimal cost (see, for instance,
\cite{CovTho06}).  A variant of this standard problem was studied in
\cite{Campbell}, where the goal was to minimize the log-moment
generating function $\log \E {\exp(\lambda L)}$. A different approach
for solving this problem is given in \cite{SundaresanHanewal} where
the {\it Gibbs variational principle} is used to linearize the
nonlinear cost function $\log \E {\exp(\lambda L)}$. The next result
provides the necessary variational formula to extend the
aforementioned approach to another nonlinear function, namely
${\|L\|_p := }(\E {L^p})^{\frac 1p}$ for $p> 1$.

We believe that our result is of independent interest, and present it in a general 
  form that applies to general distributions (and not just the discrete random variables considered  in this paper). To state the general result, we recall a basic notation from probability theory. For two probability measures $P$ and $Q$ on the same probability space such that $Q$ is absolutely continuous with respect to $P$, denoted $Q\ll P$, denote by $\frac{d Q}{d P}$ the Radon-Nikodym derivative of $Q$ with respect to $P$. Note that $\frac{d Q}{dP}$, too, is a random variable measurable with respect to the underlying sigma-algebra.
  A reader not familiar with these notions can see a standard textbook on probability theory for definitions. For the discrete case, $Q\ll P$ corresponds to the condition\footnote{ \newer{${\tt supp}(P)$ denotes the support of distribution $P$ over an alphabet $\X$, $i.e.$,
${\tt supp}(P):=\{x \in \X: P(x)>0\}$.
      .}} ${\tt supp}(Q)\subset {\tt supp}(P)$ and $\frac{d Q}{d P}$ equals the ratio of the pmfs of the distributions $Q$ and $P$.

  Note that expectations are always taken with respect to the reference measure.  In particular, the expectations without any subscript in  Theorem~\ref{t:variational_formula} below and its proof denote the expectation with respect to $P$, which is the reference measure in this case. The expectation in Remark \ref{r:P(X=0)=0} denotes the expectation  with respect to $R$. 

\begin{thm}\label{t:variational_formula}
For a real-valued random variable $X$ with distribution $P$ and $p\geq 1$ such
that $\|X\|_p < \infty$, we have
\begin{equation}
\| X\|_p = \max_{Q\ll P} \E{ \left(\frac{dQ}{dP}\right)^{\frac 1
    {p^\prime}} |X|}, \nonumber
\end{equation}
where $p^\prime = p/(p-1)$ is the H\"older conjugate of $p$.
\end{thm}
\begin{proof}
For $Q\ll P$ and $0<\alpha\neq 1$, let $D_\alpha(P,Q)$ denote the
R\'enyi divergence of order $\alpha$ between distributions $Q$ and $P$
(see~\cite{Ren61}), defined by
\begin{equation*}
D_\alpha(P,Q) := \frac 1{\alpha-1}\log \E
{\left(\frac{dQ}{dP}\right)^\alpha}.
\end{equation*}
It is well-known that $D_\alpha(P,Q)\geq 0$ with equality if and only if
$P=Q$. Consider the probability measure $P_p\ll P$ defined by 
\begin{equation*}
\frac{d P_p}{dP} :=\frac 1{\|X\|_p^p}\cdot |X|^p.
\end{equation*}
Then, for $\alpha = 1/p^\prime$,

\begin{align*}
0&\leq D_\alpha(P_p,Q)
= \frac 1{\alpha-1} \log\E{\left(\frac{dQ}{dP}\right)^\alpha
  \left(\frac{dP_p}{dP}\right)^{1-\alpha}} \\ &=-p
\log\E{\left(\frac{dQ}{dP}\right)^\alpha |X|}+p \log \|X\|_p,
\end{align*}
where the previous equality holds since $p(1-\alpha)=1$.  Thus, for
every $Q\ll P$,
\begin{equation*}
 \E{\left(\frac{dQ}{dP}\right)^\alpha |X|}\leq \|X\|_p,
\end{equation*}
with equality if and only if $P_p=Q$.
\end{proof}

\begin{rem}\label{r:P(X=0)=0}
 The given definition of R\'enyi divergence restricts
    Theorem \ref{t:variational_formula} to the case $P(X=0)=0$. To
    remove this restriction, the following general definition of
    R\'enyi divergence with respect to a common measure can be
    used: For all $Q, P \ll R$, define
\[
    D_{\alpha}(P,Q):= \frac 1{\alpha-1}
    \log\E{\left(\frac{dQ}{dR}\right)^\alpha
      \left(\frac{dP}{dR}\right)^{1-\alpha}} .
\]
    The proof then follows by using the positivity of
    $D_{\alpha}(P_p,Q)$, then by proceeding in the same manner as
    the previous proof.
  \end{rem}
Returning to the problem at hand, \newer{we apply the variational formula above to the $L_2$ norm of a discrete random variable. We highlight this special case separately below.}

\begin{cor}\label{c:2variational_formula}
For a discrete random variable $X$ with a pmf $P$ such
that $ \|X\|_2 < \infty$, we have
\[
{
\|X\|_2=\max_{{\tt supp}(Q)\subset {\tt supp}(P)}\sum_{x\in \X}\sqrt{Q(x)P(x)}x},
\]
where ${\tt supp}(P)$ denotes the support-set of the distribution $P$. 
\end{cor}

\section{Prefix-free codes with minimum average age}
\label{s:code_design} 

We now present a recipe for designing prefix-free codes with minimum
average age.  By Theorem~\ref{t:average_age}, we seek prefix-free
codes that minimize the cost
\begin{equation}
\E L +\frac{\E {L^2}}{2\E L},
\label{e:cost_simple}
\end{equation}
where $L= \ell(X)$ for $X$ with pmf $P$.  Recall that a prefix-free code
with lengths $\{\ell(x) {\in \N}, x\in \X\}$ exists if and only if lengths
satisfy Kraft's inequality ($cf.$~\cite{CovTho06}), $i.e.$, if and only if
\begin{equation}
\sum_{x\in \X}2^{-\ell(x)}\leq 1.
\label{e:Kraft_cond}
\end{equation}
Following the discussion leading to \eqref{e:ceil_loss}, we relax the
{integral} constraints for $\ell(x)$ and search over all real-valued
$\ell(x)\geq 0$ satisfying \eqref{e:Kraft_cond}. Specifically, we solve the relaxed optimization problem
\begin{equation}\label{e:relaxed}
\min_{\ell \in \Lambda} \E{L}+\frac{\E{L^2}}{2\E{L}},
\end{equation}
where \[\Lambda =\big\{\ell \in \R^{|\X|}: \sum_{x \in \X} 2^{-\ell(x)} \leq1, ~\ell(x) \geq 0 ~\forall x \in \X\big\}.\] As noticed
in~\eqref{e:ceil_loss}, this can incur a loss of only a constant.  A
key challenge in minimizing \eqref{e:cost_simple} is that it is
nonlinear.  We linearize this cost as follows:
\begin{enumerate}
\item \newer{Note first the identity below, which is obtained by maximizing the expression on the right-side:}
\begin{align}\label{e:varZ}
&\E L +\frac{\E {L^2}}{2\E L} = \max_{z\geq 0} \left(1 - \frac
  {z^2}{2}\right) \E L + z \|L\|_2.
\end{align}

\item Then, Corollary~\ref{c:2variational_formula} yields
\[
  {\|L\|_2=\max_{Q\ll
      P}\sum_{x\in \X}\sqrt{Q(x)P(x)}\ell(x)},
  \]
  which further leads to 
\begin{align*}
  \lefteqn{\E L +\frac{\E  {L^2}}{2\E L}}
\\
  &= \max_{z\geq 0} \left(1 -\frac {z^2}{2}\right) \E L + z\max_{Q\ll
  P}\sum_{x\in \X}\sqrt{Q(x)P(x)}\ell(x)
\\
&=\max_{z\geq
    0}\max_{Q\ll P} \sum_{x\in \X}g_{z,Q,P}(x)\ell(x),
\end{align*}
where
\begin{align}
\label{e:g_def}
g_{z,Q,P}(x) &:= \left(1- \frac{z^2}2\right)P(x) +z\sqrt{Q(x)P(x)}.
\end{align}
\end{enumerate}
As remarked earlier, as the source distribution $P$ is discrete, the constraint $Q \ll P$ simplifies to ${\tt supp}(Q)\subset {\tt supp}(P)$.
Thus, our goal is to identify the minimizer $\ell^*$ that achieves
\begin{align}
\Delta^*(P) = \min_{\ell \in \Lambda}\max_{z\geq 0}\max_{Q\ll P}
\sum_{x\in \X}g_{z,Q,P}(x)\ell(x).
\label{e:minmax_cost}
\end{align}

The result below captures our main observation and facilitates the
computation of optimal lengths attaining the minmax cost $\Delta^*(P)$.  
\begin{thm}[Structure of optimal codes]\label{t:update_optimal}
The optimal minmax cost $\Delta^*(P)$ in \eqref{e:minmax_cost}
satisfies
\begin{align}
\Delta^*(P) &= \max_{z\geq 0}\max_{Q\ll P} \min_{\ell \in \Lambda}
\sum_{x\in \X}g_{z,Q,P}(x)\ell(x) \nonumber \\ &=\max_{\substack{z \geq 0,Q\ll P, \\ (z,Q) \in \mathcal{G}}} \sum_{x\in \X}g_{z,Q,P}(x)
\log\frac{\sum_{x^\prime\in \X}g_{z,Q,P}(x^\prime)}{g_{z,Q,P}(x)},
\label{e:maxmin_cost}
\end{align}
where \eq{ \mathcal{G}:=\{z \geq 0 ,Q \in \R^{|\X|}:g_{z,Q,P}(x)\geq0
  \quad \forall x \in \X\}.  }  Furthermore, if $(z^*, Q^*)$ is the
maximizer of the right-side of \eqref{e:maxmin_cost}, then the minmax
cost \eqref{e:minmax_cost} is achieved uniquely by {the Shannon
  lengths}\footnote{{Recall that Shannon lengths for the pmf $P$ on $\X$ are given by $\ell(x)=-\log P(x) $, $x \in \X$, and are not necessarily integers.}} for the pmf
$P^*$ on $\X$ given by
\begin{equation}\label{e:tilted_distribution}
P^*(x) = \frac {g_{z^*,Q^*,P}(x)}{\sum_{x^\prime\in
    \X}g_{z^*,Q^*,P}(x^\prime)}.
\end{equation}

\end{thm}
 Thus, our prescription for design of source codes is simple: Use
  a Shannon code for $P^*$ instead of $P$. To compute $P^*$, we need
  to solve the optimization problem in \eqref{e:maxmin_cost}. Note that is unclear a priori that the minimum average age for the problem in $\eqref{e:relaxed}$ would correspond to Shannon lengths for some pmf since our cost function is
  not monotonic in expected length, whereby the optimal solution may
  not satisfy Kraft's inequality with equality. Nonetheless, we show
  that the Shannon lengths $-\log P^*(x)$ are optimal for the relaxed problem given by \eqref{e:relaxed}.

\newer{We note that our formal result above only provides a structural result for the optimal solution. But we believe that this structural result leads to a recipe to design practical algorithms for finding the optimal solution; we describe this recipe below.}
  Specifically, note that the resulting optimization problem for finding $P^*$
is one of entropy maximization for which several heuristic recipes are
available. Furthermore, we note the following structural
simplification for the optimal solution which shows that if
$P(x)=P(y)$, then $P^*(x)=P^*(y)$ must hold as well; the proof is
relegated to the Appendix.  Thus, the dimension of the optimization
problem~\eqref{e:maxmin_cost} can be reduced from $|\X|+1$ to $M_P+1$,
where $M_P$ denotes the number of distinct elements in the probability
multiset $\{P(x): x \in \X\}$. Let $A_1 \cdots A_{M_P}$ denote the
partition of $\X$ such that
\[
P(x)=P(y) \quad \forall x,y \in A_i, \quad \forall i \in [M_P].
\]

\begin{lem}\label{c:dim_red}
Suppose that $Q^*$ is an optimal $Q$ for \eqref{e:maxmin_cost}. Then,
$Q^*$ must satisfy
\begin{align}\label{e:property}
Q^*(x)=Q^*(y) \quad \forall x,y \in A_i, \quad \forall i \in [M_P].
\end{align}
\end{lem}

In proving Lemma~\ref{c:dim_red}, we use the fact that the cost
function in \eqref{e:maxmin_cost} is concave in $Q$ for each fixed $z$
and is concave in $z$ for each fixed $Q$ (see Lemma
  \ref{c:concavity}). However, it may not be jointly concave in
$(z,Q)$. Nevertheless, we apply standard numerical packages to
optimize it in the next section to quantify the performance of our
proposed codes and compare it with Shannon codes for the original
distribution $P$.  
\section{Numerical results for Zipf distribution}\label{s:age_numerical}

We program all our optimization problems
  in \textit{AMPL} \cite{AMPL} and solve it using
  \textit{SNOPT}~\cite{SNOPT} and \textit{CONOPT}~\cite{CONOPT}
  solvers.
Specifically, for the pmfs $P$ we consider in this section, we solve the optimization problem given by  \eqref{e:maxmin_cost}  to find the corresponding optimal $(z^*, Q^*)$.  In order to check if we have indeed found the optimal $(z^*, Q^*)$, we once again use Theorem \ref{t:update_optimal}.
  In particular, it follows from Theorem \ref{t:update_optimal} that the necessary and sufficient condition for a particular $(z, Q)$ to be the optimal solution is that   the value of the maximization problem \eqref{e:maxmin_cost} at $(z, Q)$  equals
  \[\E{-\log P^{\prime}(X)}  +\frac{\E {(\log P^{\prime}(X))^2}}{2\E{-\log P^{\prime}(X)}},
  \]
  where
  \[
  P^{\prime}(X)=\frac {g_{z,Q,P}(x)}{\sum_{x^\prime\in
      \X}g_{z,Q,P}(x^\prime)};
  \]
in all our numerical evaluations, the solution found by the solver satisfies this condition, which establishes its optimality.

We now illustrate our recipe for construction of prefix-free codes
that yield minimum average age for memoryless update schemes when $P$
is a Zipf distribution. 
 Specifically, we illustrate our qualitative results using
the ${\tt Zipf}(s,N)$ distribution with alphabet $\X = \{1, \cdots, N\}$
and given by \eq{ P(i) = \frac{i ^{-s}}{\sum_{j=1}^N j^{-s}}, \quad
  1\leq i \leq N.  }

\begin{figure}[th]
\centering \input{Figures/figure-03.tex}
\caption{ Comparison of proposed codes and Shannon codes for ${\tt
    Zipf}(s,256)$ with varying $s$.  The average age is computed using
  real-valued lengths as well as lengths rounded-off to integer
  values.} 

\label{f:zipf_comparison}
\end{figure}
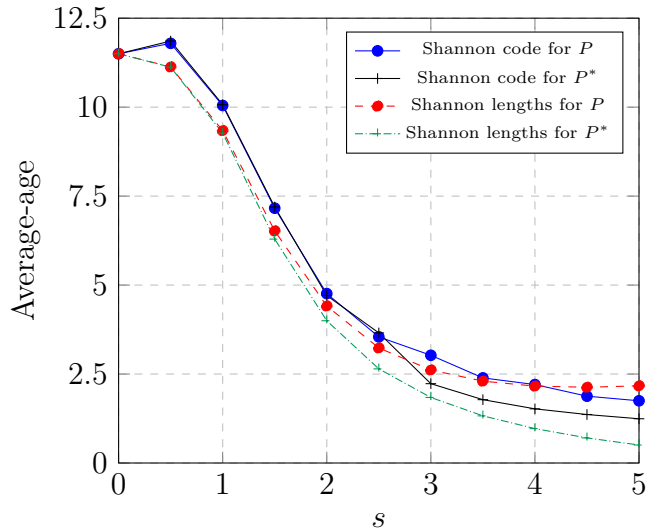

Heuristically, the  average age formula~\eqref{e:cost_theta} suggests that the differences between the performances of a code under average codeword length cost and the average age cost will be the most for ``peaky distribution,'' namely for distributions with heavy elements. The parameter $s$ of the Zipf distribution allows us to vary from a uniform distribution to a ``peaky distribution,'' making this family apt for our numerical study. Indeed, our numerical results confirm that our proposed scheme outperforms a Shannon code for $P$ when the parameter $s$ is high; see Figure~\ref{f:zipf_comparison}. When we
round-off real lengths to integers, the gains are subsided but still
exist. Further, when the parameter $s$ is close to $0$, { Shannon codes} for $P$ are close to
optimal.\newer{ With increase in $s$, the gain of our proposed schemes over Shannon codes starts becoming more prominent}. \newer{As an aside, Figure \ref{f:zipf_comparison} also provides an illustration of the non-monotonic nature of the average age function with respect to code lengths.}

\begin{figure}[ht]
\centering \input{Figures/figure-04.tex}
\caption{The pmf for $P^*$ and $P$ for ${\tt Zipf}(1,8)$.}
\label{f:P_star} 
\end{figure}
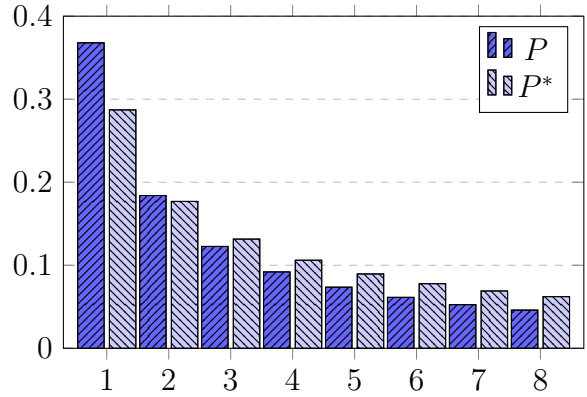

\begin{figure}[h]
\centering \input{Figures/figure-05.tex}
\caption{ Average age and average length for our update codes as a
  function of $H(P)$ for ${\tt Zipf}(s,256)$ with $s$ varying from $0$
  to $5$ at step sizes of $0.5$.  }
\label{f:ent}
\end{figure}
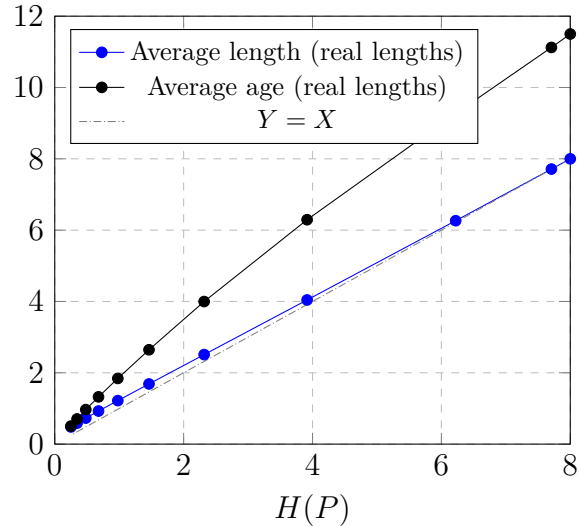

The distribution $P^*$ we use to construct our codes seems to be a
flattened version of the original Zipf distribution; we illustrate the
two distributions for ${\tt Zipf}(1,8)$ in Figure~\ref{f:P_star}.
As we see in Figure~\ref{f:P_star}, $P^*$ and $P$ are very close in
this case.  Indeed, we illustrate in Figure~\ref{f:ent} that the
average length $\E{L}$ when Shannon lengths $-\log P(x)$ are used and
when $-\log P^*(x)$ are used are very close\footnote{The difference of these two average lengths (averaged
  w.r.t. $P$) is given by the Kullback-Leibler divergence $D(P\|P^*)$;
  see~\cite{CovTho06}. }.  In
Figure~\ref{f:ent}, we note the dependence of average age on the
entropy of the underlying distribution $P$. As expected, average age
increases as $H(P)$ increases.

Thus, while Example~\ref{ex:1} illustrated high gains of the proposed
code over {Shannon codes} for $P$, for the specific case of Zipf
distributions the gains may not be large.  Characterizing this gain
for any given distribution is a direction for future research.

\section{Extensions}
\label{s:extensions}
\subsection{Randomization for Timely Updates}
\label{s:random_timely_updates}
We have restricted our treatment to deterministic {memoryless} update
schemes.  A natural extension to randomized memoryless schemes would
entail allowing the encoder to make a randomized decision to skip
transmission of a symbol even when the channel is free (we can
allocate a special symbol $\emptyset$ to signify no transmission to
the receiver).  Specifically, assume that we transmit the symbol
$\emptyset$ using a codeword of length $\ell(\emptyset)$ when we
choose not to transmit the observed symbol $x\in \X$. Denoting by
$\theta(x)$ the probability with which the encoder will transmit the
symbol $x$, the average age $\bar{A}(e,\theta)$ for the randomized scheme is given by
\begin{align}
\bar{A}(e,\theta)= \frac{\E {L(\theta)}}{\E{\theta(X)}}+
\frac{\E{L(\theta)^2}}{2\E {L(\theta)}} - \frac 12,
\label{e:randomized_avg_age}
\end{align}
where the random variable $L(\theta)$ is defined as follows:
\begin{align}\label{e:L(theta)} L(\theta) := \begin{cases}
      \ell(x),\quad w.p\quad P(x)\theta(x)\\ \ell(\emptyset),\quad
      w.p\quad 1-\E{\theta(X)}.
\end{cases}
\end{align}
Note that the expression in~\eqref{e:randomized_avg_age} is a slight generalization of  Theorem~\ref{t:average_age} and is derived in Section~\ref{p:avg_age}.

\begin{example}
Consider $\X = \{1,...,64\}$ and the following pmf; \eq{P(x)=
 \begin{cases} 1/4, \quad & x \in
   \{1,\dots,3\},\\ 1/244, \quad & x \in \{4, \dots ,64\}. \end{cases}
} Since $H(P) = 3.483$, Corollary~\ref{c:age_bounds} yields that the
average age of the deterministic memoryless update scheme is bounded
below by $4.724$. Next, consider a randomized update scheme with
$\theta(x)=1$ for $x\in \{1,2,3\}$ and $0$ otherwise.
For this choice, the effective pmf $P_\theta$ is uniformly distributed
over the symbols $\{1,2,3\}\cup\{\phi\}$. Thus, the optimal length
assignment for this case assigns $\ell(x)=2$ to all the symbols and
the average age equals $3.17$, which is less than the lower bound of
$4.724$ for the deterministic scheme.
\end{example}
The idea of skipping available transmission opportunities, i.e., not
transmitting even when the channel is free, to
  minimize average age appears in the recent work~\cite{sun2017update}
  as well, albeit in a slightly different setting.  Heuristically,
the randomization scheme above operates as we expect -- it ignores the
rare symbols which will require longer codeword lengths. In practice,
however, these rare symbols might be the ones we are interested
in. But keep in mind that our prescribed solution only promises to
minimize the average age and does not pay heed to any other
consideration. Furthermore, for a given randomization vector
  $\theta$, we can establish a result similar to
  Theorem~\ref{t:update_optimal}. This will lead to the design of almost
  optimal source codes for a given randomization vector $\theta$.
However, the joint optimality over the class of randomized schemes and
source coding schemes is still unclear.

 In a more comprehensive
treatment, one can study the design of update codes with other
constraints imposed. We foresee the use of
Corollary~\ref{c:2variational_formula} in these more general settings as
well.  
In another direction, we can consider the extension of our results to
the case when the transmission channel is an erasure channel with
probability of erasure $\epsilon$. If we assume the availability of perfect
feedback, a natural model for the link or higher layer in a network,
and restrict to simple repetition schemes where the transmitter keeps
on transmitting the coded symbol until it is received, our formula for
average age extends with (roughly) an additional multiplicative factor
of $1/(1-\epsilon)$.  Formally the average age over an erasure channel with
$\epsilon$ probability of erasure; a source code $e$, along with a
randomization vector $\theta$ and a repetition channel-coding scheme
yields the following average age
\begin{align*}
\bar{A}_\epsilon(e,\theta) = \frac 1 {1-\epsilon}\cdot \bar{A}(e,\theta) +
\frac{\epsilon}{2(1-\epsilon)}.
\end{align*} However, the optimality of repetition scheme is unclear, and 
the general problem constitutes a new formulation in joint-source
channel coding which is of interest for future research.

\subsection{Source Coding for Minimum Queuing Delay}
\label{S:Min_Del}
\begin{figure}[h]
\centering \input{Figures/figure-06.tex}
\caption{ Figure describes a typical sample-path for transmission of
  encoded symbols over a FCFS queuing system. Symbol $X_1$ arrives at
  some time instant $1$, it is encoded and transmitted over the
  channel. \newer{Recall that unlike the slotted setup of Figure~\ref{f:setup},
    the setup here is that of continuous time with Poisson arrivals.} 
  It is decoded at time instant $4$.  Symbol $X_2$ arrives
  in between time instants $2$ and $3$, and is placed in the queue, as
  the channel is busy transmitting $X_1$.  As soon as the channel
  becomes free at time instant $4$, an encoded version of $X_2$ is
  transmitted over it.
  Symbol $X_3$ arrives when the channel is free
  and is transmitted immediately.}
\label{f:min_avg_age}
\end{figure}
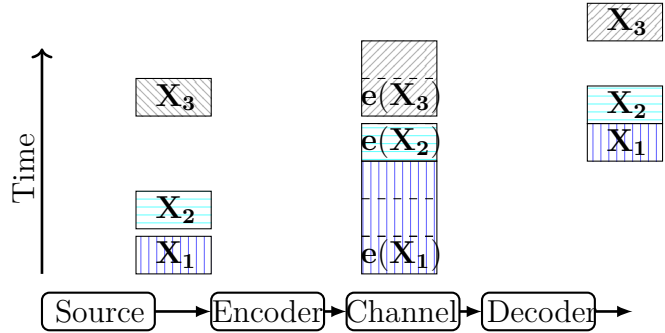

Next, we \newer{point out} a use case for Corollary~\ref{c:2variational_formula}
in a minimum queuing delay problem introduced
in~\cite{Humblet78thesis}.  The setting is closely related to our
minimum average age update formulation with two differences: First,
the arrival process of source symbols is a Poisson process of rate
$\lambda$; and second, the encoder is not allowed to skip source
symbols.  Instead, each symbol is encoded and scheduled for
transmission in a first-come-first-serve (FCFS) queue.  Our goal is to
design a source code that minimizes the average queuing delay
encountered by the source sequence.  Formally, the symbols
$\{X_n\}_{n=1}^\infty$ are generated iid from a finite alphabet $\X$,
using a common pmf $P$. Every incoming symbol $x$ is encoded as $e(x)$
using a prefix-free code specified by the encoder mapping $e:\X
\to \{0,1\}^*$, and the bit string $e(x)$ is placed in a queue. The
queue schedules bits for transmission using a FCFS policy. Each bit in
the queue is transmitted over a noiseless communication channel.
Denote by $A_n$ the time of successful arrival of the $n$th symbol. Also, denote by $D_n$ the time instant of successful reception of the $n$th
symbol $X_n$.  {That is, }$D_n$ is the instant at which the last bit
of $e(X_n)$ is received\footnote{\newer{Note both $A_n$ and $D_n$ may not be integer valued, unlike the age setup.}}. The delay for the $n$th symbol is given by
$D_n - A_n$; see Figure~\ref{f:min_avg_age} for an illustration.

Thus, if $\ell(x)$ is the length of the encoded symbol $e(x)$ in bits,
then the number of channel uses to transmit this symbol is $\ell(x)$,
whereby the service time of the $n^{th}$ arriving symbol is given by
$S_n = \ell(X_n)$. Since $\{X_n\}_{n=1}^\infty$ is iid and the encoder
mapping $e$ is fixed, the sequence $(S_n)_{n \in \N}$, too, is iid
with common mean $\E{L}$. Therefore, the resulting queue is an M/G/1
queuing system with Poisson arrivals of rate $\lambda$ and iid service
times $(S_n)_{n \in \N}$.  Note that this queue will be stable only if
$\lambda\E{S_n} = \lambda\E{L}< 1$.

We are interested in designing prefix-free  codes $e$ that
minimize the average waiting time defined as follows:
\begin{defn}
The {\it average {waiting time}} $D(e)$ of a source code $e$ is given
by
\begin{align*}
D(e) &:= \limsup_{N\to \infty}\frac 1 N \sum_{n=1}^N \E{D_n-A_n},
\end{align*}
where the expectation is over source symbol realizations
$\{X_n\}_{n=1}^\infty$ and arrival instants $\{A_n\}_{n \in \N}$.
\end{defn}
We seek prefix-free codes $e$ with the least possible average
waiting time $D(e)$.  In fact, a closed-form expression for $D(e)$ was
obtained in~\cite{Humblet78thesis}.  For clarity of exposition, we
denote the load for the queuing system above for a fixed $\lambda$ by
$\rho(L) {:=} \lambda\E{L}$.  Since $\rho(L) < 1$ for the queue to be
stable, the average codeword length $\E L$ must be strictly less than
a threshold $L_{\tt th} {:=} \frac{\E{L}}{\rho(L)} =
\frac{1}{\lambda}$ for the queue to be stable.
\begin{thm}[\cite{Humblet78thesis}]\label{t:average_delay}
Consider a random variable $X$ with pmf $P$ and a source code $e$
which assigns a bit sequence of length $\ell(x)$ to $x\in \X$.  Let
$L$ denote the random variable $\ell(X)$.  Then, the average waiting
time $D(e)$ for $e$ is given by
\begin{align}
D(e) &= \begin{cases} \frac{ \E{L^2}}{2(L_{\tt th} - \E{L})} +
  \E{L} , &\E{L} < L_{\tt th},\\ \infty, &\E{L} \geq L_{\tt
    th}.
\end{cases}
%
\label{e:delay_formula}
\end{align}
\end{thm}
Thus, the problem of designing source codes with minimum average
waiting time reduces to that of designing a prefix-free code that
minimizes the cost in \eqref{e:delay_formula}. This problem was first considered in \cite{Humblet78thesis}.  In fact, it was noted
in~\cite[Chapter 1, Section 3]{Humblet78thesis} that codes which
minimize the first moment are robust for \eqref{e:delay_formula}.
We will justify this empirical observation in Corollary
  \ref{c:KL-bound}.  However, optimal codes can differ from Shannon
codes for $P$.  Indeed, an algorithm for finding the optimal length
assignments $\ell(x)$, $x\in \X$, for a prefix-free code that
minimizes $\bar{D}(e)$ was presented in \cite{larmore1989minimum} and
the optimal code can be seen to outperform Shannon codes for $P$.
While this algorithm has complexity that is polynomial in the alphabet
size, it is computationally expensive for large alphabet sizes -- the
case of interest for our problem.

Interestingly, the cost function in \eqref{e:delay_formula} resembles
closely the expression we obtained for asymptotic average age and our
recipe used to design minimum average age codes can be applied to
design minimum average delay codes as well. The underlying
optimization problem can be solved numerically rather quickly, much
faster than the optimization in \cite{larmore1989minimum}.  However,
as before, our procedure can only handle the real-relaxation of the
underlying optimization problem, and unlike the previous case, naive
rounding-off to integer lengths yields a sub-optimal solution when
$(1-\rho(L))$ is small. Nonetheless, the minimum average waiting time
computed using our recipe serves as an easily computable lower bound
for the optimal $D(e)$. In fact, we observe in our numerical
simulations that the resulting lower bound is rather close to the
optimal cost obtained using~\cite{larmore1989minimum}.

Now, we describe the modification of our recipe to design codes with
$\E{L} < L_{\tt th}$ that minimize the cost
\begin{align}
\norm{L}_1 +\frac{\norm{L}_2^2 }{2(L_{\tt th}-\norm{L}_1)},
\label{e:qcost_simple}
\end{align}
where $L= \ell(X)$ for $X$ with pmf $P$. As before, we first obtain a
variational form of \eqref{e:qcost_simple} which entails a linear
function of lengths. Specifically, we have the following steps.
\begin{enumerate}
\item First, we obtain a polynomial form from the rational function:
  \eq{
&\frac{\norm{L}_2^2}{2(L_{\tt th} - \norm{L}_1)} = \max_{z \geq 0} z
    \norm{L}_2 - \frac{z^2}{2} (L_{\tt th} - \norm{L}_1).
}

\item 
Then, Corollary~\ref{c:2variational_formula} yields that the cost
in~\eqref{e:qcost_simple} equals
\begin{align*}
&\max_{z\geq 0}\max_{Q\ll P} \sum_{x\in \X}g_{z,Q,P}(x)\ell(x) -
  \frac{z^2}{2} L_{\tt th}
\end{align*}
where the $g_{z,Q,P}(x)$ is defined as
\begin{align*}
g_{z,Q,P}(x) &:= \left(1+ \frac{z^2}2\right)P(x)+z\sqrt{Q(x)P(x)}.
\end{align*}
\end{enumerate}
Thus, our goal reduces to identifying the minimizer $\ell^* \in
\Lambda$ that achieves
\begin{align}
 \Delta^*(P) = \min_{\substack{\ell\in \Lambda,\\ \E{L}<L_{\tt th}}
 }\max_{z\geq 0}\max_{Q\ll P} \sum_{x\in \X}g_{z,Q,P}(x)\ell(x) -
 \frac{z^2}{2} L_{\tt th}.\label{q:minmax_cost}
\end{align}
The result below is the counterpart of Theorem~\ref{t:update_optimal}
for minimum delay source codes and is proved in Section~\ref{p:3}.
\begin{thm}\label{t:qmaintheorem} 
Under the condition
\begin{equation}\label{eq:a_qmaintheorem}
H(X) + \log(1 + 1/\sqrt{2})< L_{\tt th},
\end{equation}
the optimal minmax cost $\Delta^*(P)$ in \eqref{q:minmax_cost}
satisfies
\begin{align}
\Delta^*(P) &= \max_{z\geq 0}\max_{Q\ll P} \min_{\substack{\ell\in
    \Lambda,\\ \E{L}<L_{\tt th}} } \sum_{x\in \X}g_{z,Q,P}(x)\ell(x)-
\frac{z^2}{2} L_{\tt th} \nonumber \\ &=\max_{z\geq 0}\max_{Q\ll P}
\sum_{x\in \X}g_{z,Q,P}(x)\log \frac{\sum_{x^\prime\in
    \X}g_{z,Q,P}(x^\prime)}{g_{z,Q,P}(x)} \nonumber \\ & \hspace{5cm} - \frac{z^2}{2} L_{\tt th}.
\label{q:maxmin_cost}
\end{align}
Furthermore, if $(z^*, Q^*)$ is the maximizer of the right-side of
\eqref{q:maxmin_cost}, then the minmax cost \eqref{q:minmax_cost} is
achieved uniquely by Shannon lengths for pmf
$P^*$ on $\X$ given by
\[
P^*(x) = \frac {g_{z^,Q^*,P}(x)}{\sum_{x^\prime\in
    \X}g_{z^*,Q^*,P}(x^\prime)}.
\]
\end{thm}
We remark that  $\eqref{e:delay_formula}$ implies that $H(X) < L_{\tt th}$ is essential for the existence of a prefix free source coding scheme with finite average delay. Thus, the condition $H(X)+ \log(1  + 1/\sqrt{2}) < L_{\tt th}$ is a mild one.

\begin{figure}[ht]
\centering
\begin{subfigure}[b]{\columnwidth}
\centering \input{Figures/figure-07a.tex}
\caption{Comparison of proposed codes with
  Larmore's Algorithms \cite{larmore1989minimum} for the distribution
  $P(1)=0.5$, and $P(i)=\frac{0.5}{ 255} \quad \forall i \in
  \{2,\cdots 256\}$.}
\label{f:MinDelay_comparisona}
\end{subfigure}
\hspace{0.1cm}

\begin{subfigure}[b]{\columnwidth}
\centering \input{Figures/figure-07b.tex}
\caption{Comparison of proposed codes with 
  Larmore's Algorithms \cite{larmore1989minimum} for the distribution
  $P(1)=0.6$, and $P(i)=\frac{0.4}{ 255} \quad \forall i \in
  \{2,\cdots 256\}$.}
\label{f:MinDelay_comparisonb}
\end{subfigure}
\caption{Comparison of proposed codes with  Larmore's Algorithms}
\end{figure}
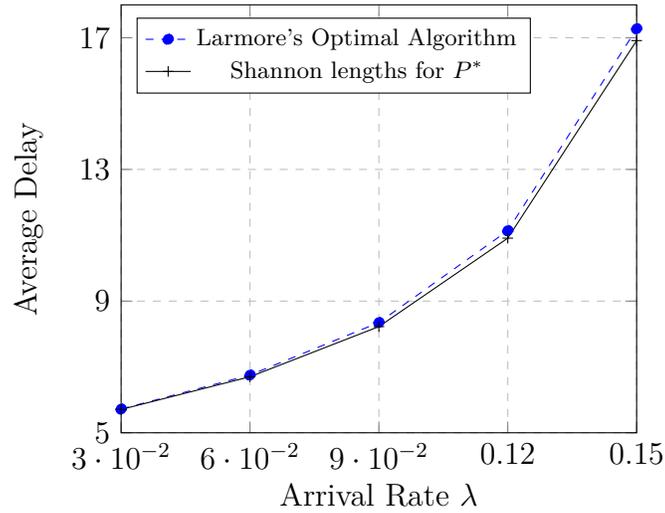
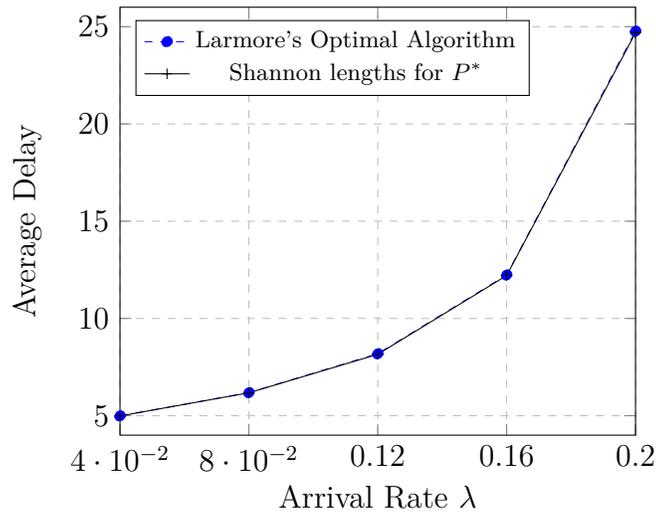

Thus, as before, the optimal codeword lengths for the relaxed problem (allowing real-valued lengths) correspond, once again, to Shannon lengths for a titled distribution
$P^*$.  As remarked earlier, the performance of the optimal source
code is known to be not too far from the Shannon code for $P$. This
observation can be justified by the following simple corollary of
Theorem~\ref{t:qmaintheorem}.
\begin{cor} \label{c:KL-bound}
The KL-Divergence between $P$, $P^*$ is bounded as \eq{D(P||P^*) \leq
  \log\left(1+\frac{1}{\sqrt{2}}\right) .}
  \label{q:corr}
\end{cor}
\begin{proof}
The proof follows from \eqref{e:q_corr}, which is in turn derived in
the proof of theorem \ref{t:qmaintheorem} in section \ref{p:3}.
\end{proof}
Thus, the average length for Shannon codes and our codes do not differ
by more than $\log(1+1/{\sqrt{2}})$ ($cf.$~\cite{CovTho06}).
Indeed, we note in
  Figures~\ref{f:MinDelay_comparisona},~\ref{f:MinDelay_comparisonb}
  via numerical simulations that the optimal cost
  in~\eqref{q:maxmin_cost} is very close to the performance of optimal
  codes designed using~\cite{larmore1989minimum}. \newer{This suggests that 
    possibly there is 
an appropriate rounding-off procedure for real-valued lengths that can
yield integer lengths with close to optimal performance}; devising such a
rounding-off procedure is an interesting research direction for the
future. We close this section by noting that analogous versions of Lemma~\ref{c:dim_red}
and Lemma~\ref{c:concavity} in the Appendix can be obtained for optimization problem \eqref{q:maxmin_cost}.

\section{Proofs}
\subsection{Proof of Theorem~\ref{t:average_age}}\label{p:avg_age}
We establish the expression for average age given in
\eqref{e:randomized_avg_age} for the more general class of randomized
schemes; Theorem~\ref{t:average_age} will follow upon setting
$\theta(x)=1$, for all $x \in \mathcal{X}$. Recall that the symbol $\emptyset$ is available only in the extended model in Section~\ref{s:extensions}, and not in the original model discussed in rest of the paper.
Note that the formula for
average age given in Theorem~\ref{t:average_age} is similar in form to
the expressions for average age derived in other settings;
see~\cite{KaulYatesGruteser12} for an example.

We will first set up some notation.
Let $S_0 := 0$ and \eq{ &S_k := \inf\{t > S_{k-1} :
  U(t)>U(t-1)\},~k \in \N.  }
\newer{Namely, $S_k$ is the time at which the decoder updates
its estimate for the symbol for the $k$th time.
Recall that $U(t)$ is incremented only
on successful reception at the receiver and is strictly increasing in $t$.
For brevity, we introduce the
notation $Y_k := S_k - S_{k-1}$
for the time between the $(k-1)$th and the $k$th information update at the decoder.
Further, denote by $Z_k:=S_{k}-U(S_k)$ the age at time $S_k$, which is simply the time taken for
  the successful reception of the symbol\footnote{This must be a symbol in $\X$ and not $\emptyset$ by the definition of $S_k$.} $x\in \X$ transmitted at time $U(S_k)$.}
Also, denote by $R_k$ \newer{the sum of instantaneous age  between $S_{k-1}$ and $S_k$} (the $k$th reward), namely
\eq{
R_k:= \sum_{t=S_{k-1}+1}^{S_{k}}(t-U(t)).
}
\mnote{"For Brevity," This  sentence is confusing too me. I preferred the preivious iteration of this sentence.}
\tnote{Is it okay now?}

Heuristically, our proof can be understood as follows. We note that the asymptotic average age is
roughly
\[
\frac{\sum_{k=1}^\infty R_k}{\lim_{k\to \infty} S_k}.
\]
It is easy to see that $\{Y_k\}_{k=1}^{\infty}$ is an iid sequence. Thus, if
$\{R_k\}_{k=1}^\infty$, too, was an iid sequence, we would obtain the
asymptotic average age to be $\E{R_1}/{\E{Y_1}}$ by the standard
Renewal Reward Theorem~\cite{ross1996stochastic}.  Unfortunately, this
is not the case.  But it turns out that the dependence in sequence
$\{R_k\}$ is only between consecutive terms.  Therefore, we can obtain
the same conclusion as above by dividing the sum $\sum_{k=1}^\infty
R_k$ into the sum of odd terms and even terms, each of which is in
turn a sum of iid random variables.

We will now proceed to prove that dependence in $R_k$ is between consecutive terms.  
Since $U(t)$ remains $U(S_{k-1})$ for all $t<S_k$, we get for $k\geq
1$ that
\begin{align}\label{e:Rk}
\nonumber
R_k &= \frac{(S_{k}-S_{k-1} -1)(S_{k}-S_{k-1})}{2} \\ \nonumber&\hspace{1cm} +(S_{k}-S_{k-1} 
  -1)\cdot(S_{k-1}-U(S_{k-1})) \\& \hspace{4.5cm} +S_{k}-U(S_{k}) \nonumber
\\
& =\frac 12 Y_k^2+Y_{k}\left(Z_{k-1} -  \frac 12 \right)+Z_k - Z_{k-1},
\end{align}
with $Z_0$ set to $0$.

Note that since the source sequence $\{X_n\}$ is iid and the randomization $\theta$ is stationary, the sequences $Y_k$ and $Z_k$
are iid, too. 
Therefore, the $(R_{2n})_{n \in \N}$ and $(R_{2n+1})_{n \in \N}$ are both\footnote{The initial term $R_1$ has a different distribution since $Z_0=0$.} 
iid sequences with $\E{R_{2n}} = \E{R_{2n+1}} = \E{R_2}$ for all $n$.   

Using this observation, we can obtain the following expression for the average age:
\begin{equation}
\bar{A}(e,\theta)=\frac{\E{R_2}}{\E{Y_1}}.
\label{eq:avgage_RY}
\end{equation}
Before we prove \eqref{eq:avgage_RY}, which is the main ingredient of
our proof, we evaluate the expression on the right-side. 

For $\E{Y_1}$, note that $Y_1$ gets incremented by $\ell(\emptyset)$ each
time $\emptyset$ is sent, and gets incremented finally by $\ell(x)$
once a symbol $x\in\X$ is sent. Thus, $Y_1$ takes the value
$\ell(x)+r\ell(\emptyset)$ with probability
$(1-\E{\theta(X)})^r\theta(x)P(x)$.
Denoting $\N_0=\N\cup \{0\}$, we get
\eq{
\E{Y_1}&= \sum_{x \in \X}\sum_{r \in \N_0} (\ell(x)+r\ell(\phi))P(x)\theta(x) (1-\E{\theta(X)})^r\\
&=\sum_{x \in \X} \sum_{r \in \N_0} \ell(x)P(x)\theta(x)(1-\E{\theta(X)})^r  \\&\hspace{1.3cm}+  
\sum_{x \in \X}\sum_{r \in \N_0}  r \ell(\phi)P(x)\theta(x) (1-\E{\theta(X)})^r\\
&=\frac{\sum_{x \in \X}\ell(x)P(x)\theta(x)}{\E{\theta(X)}}+\frac{\ell(\phi)(1-\E{\theta(X)})}{\E{\theta(X)}}\\
&=\frac{\E{L(\theta)}}{\E{\theta(X)}}.
}

For $\E{R_2}$, it follows from \eqref{e:Rk} that
\[
\E{R_2} = \frac 12 \E{Y_2^2} + \E{Y_2Z_1} -\frac12 \E{Y_2},
\]
since $\E{Z_2} = \E{Z_1}$. Also, note that $Z_1$ only depends on the symbol $x\in \X$ received at time $S_1$ which in turn can depend only on the symbols $X_n$ for $n \leq S_1-1$. 
On the other hand, $Y_2=S_2-S_1$ depends on symbols $X_n$ for $n\geq S_1$ and the outputs of the independent coin tosses corresponding to randomization $\theta$. 
Therefore, $Z_1$
is independent of $Y_2$, whereby
\[
\E{R_2} = \frac 12 \E{Y_2^2} + \E{Y_2}\left(\E{Z_1} -\frac12\right).
\]
Next, note that $Z_1$ takes the value $\ell(x)$, $x\in \X$, when the
symbol received at $S_1$ is $x$. This latter event happens with
probability 
\[
\sum_{r=0}^\infty (1- \E{\theta(X)})^r\theta(x)P(x) = \frac{\theta(x)P(x)}{\E{\theta(X)}},
\]
and so, by the definition of $L(\theta)$ in  \eqref{e:L(theta)},
\begin{align*}
\E{Z_1}&= \frac{\sum_x \ell(x)\theta(x)P(x)}{\E{\theta(X)}}
\\
&=\frac{\E{L(\theta)}}{\E{\theta(X)}}  - \frac{\ell(\emptyset)(1-\E{\theta(X)})}{\E{\theta(X)}}.
\end{align*}
Then by denoting $p_\emptyset=1-\E{\theta(X)}$, the second moment $\E{Y_1^2}$ can be computed by observing the following recursion:
\begin{align*}
&\E{Y_1^2} 
\\
&=\sum_{x \in \mathcal{X}}\sum_{r \in
  \N_0}(\ell(x)+r\ell(\emptyset))^2P(x)\theta(x) p_{\emptyset}^r 
\\ 
\nonumber 
&=\sum_{x \in \mathcal{X}}\ell(x)^2P(x)\theta(x)\\ \nonumber & \hspace{1cm} +p_{\emptyset}\sum_{x \in \mathcal{X}} \sum_{r \in \N} (\ell(x)+r\ell(\emptyset))^2P(x)\theta(x) p_{\emptyset}^{r-1}
\\ 
\nonumber 
&=\sum_{x \in \mathcal{X}}\ell(x)^2P(x)\theta(x)\\ \nonumber & \hspace{1cm}+p_{\emptyset}\sum_{x \in \mathcal{X}} \sum_{r \in \N} \big(\ell(x)+(r-1)\ell(\emptyset)\big)^2P(x)\theta(x) p_{\emptyset}^{r-1}
\\
&\hspace{1cm}+2\ell(\emptyset)p_{\emptyset}\sum_{x \in \mathcal{X}} \sum_{r \in \N}
 \big(\ell(x)+(r-1)\ell(\emptyset)\big)P(x)\theta(x) p_{\emptyset}^{r-1}
\\ 
&\hspace{1cm}+p_{\emptyset}\sum_{x \in \mathcal{X}} \sum_{r \in \N} 
\ell(\emptyset)^2P(x)\theta(x) p_{\emptyset}^{r-1}
\\
\nonumber &= \sum_{x \in \mathcal{X}}\ell(x)^2P(x)\theta(x)\\ \nonumber &\hspace{1cm}  +p_{\emptyset} \E{Y_1^2}
+2\ell(\emptyset)(1-\E{\theta(X)} )\E{Y_1}+\ell(\emptyset)^2p_{\emptyset},
\end{align*}
which upon rearrangement yields
\[
\E{Y_1^2} = \frac{\E{L(\theta)^2}}{\E{\theta(X)}} 
+2\E{Y_1}\cdot \frac{\ell(\emptyset)p_{\emptyset}}{\E{\theta(X)}}.
\]
Upon combining the relations derived above, we get
\[
\frac{\E{R_2}}{\E{Y_1}}= \frac{\E{L(\theta)^2}}{2\E{L(\theta)}}
+\frac{\E{L(\theta)}}{\E{\theta(X)}}-\frac 12,
\]
which with \eqref{eq:avgage_RY} completes the proof. 

It remains to establish \eqref{eq:avgage_RY}. The proof is a simple
extension of the renewal reward theorem to our sequence of rewards
$R_n$ in which adjacent terms\newer{ may be} dependent. We include it here for
completeness. Note that $(Y_n)_{n \in
  \N}$ is a sequence of non-negative iid random variables with mean
$\E{Y_1}$, and  $S_n=\sum_{k =1}^{n}Y_k$ for all $n\in \N$. The sequence
 $\{S_n\}$ serves as a sequence of renewal times  and $R_n$ denotes the reward accumulated in the $n$th
renewal interval (though not in the
standard iid sense). Define
$N(t)$ to be the number of receptions up to time $t>0$, $i.e.$,
\eq{
N(t)=\sup{\{n: S_n\leq t\}},
} 
and $R(t)$ to be the cumulative reward accumulated till time $t$,
$i.e.$, 
\eq{
R(t)=\sum_{k=1}^{N(t)}R_k.
}
With this notation, we have
\begin{align}
\frac{R(t)}{t}&=\frac{\sum_{k=1}^{N(t)}R_k }{t}
\\
&=\frac{\sum_{k=1}^{N(t)}R_k }{N(t)}.\frac{N(t)}{t}.
\label{e:avgage_factors}
\end{align}
Note that 
\eq{
\frac{\sum_{k=1}^{\floor{\frac{N(t)}{2}}}\sum_{i \in \{0,1\}
  }R_{2k+i}}{N(t)} & \leq \frac{\sum_{k=2}^{N(t)}R_k }{N(t)} \\& \leq
\frac{\sum_{k=1}^{\ceil{\frac{N(t)}{2}}}\sum_{i \in \{0,1\}
  }R_{2k+i}}{N(t)}.
}
We now analyze the two bounds in the previous set of inequalities. 
Since $\E{Y_1}$ is finite, we get (see \cite{ross1996stochastic} for a
proof) 
\begin{align}
\lim_{t \to \infty}\frac{N(t)}{t}\to \frac{1}{\E{Y_1}}
\quad a.s.,
\label{e:Nt_limit}
\end{align}
which also shows that $N(t) \to \infty \quad a.s.$ as $t
\to \infty$. 
Therefore, for $i\in\{0,1\}$,
\eq{
\frac{\sum_{k=1}^{\ceil{\frac{N(t)}{2}}}R_{2k+i}}{N(t)}
=\frac{\sum_{k=1}^{\ceil{\frac{N(t)}{2}}}R_{2k+i}}{\ceil{\frac{N(t)}{2}}}\cdot\frac{\ceil{\frac{N(t)}{2}}}{N(t)}  . 
}
Since $(R_{2k+i})_{k \in \N}$ is iid and $N(t) \to \infty
\quad a.s.$ as $t \to \infty$, strong law of large
numbers yields 
\eq{
\lim_{t \to 
    \infty}\frac{\sum_{k=1}^{\ceil{\frac{N(t)}{2}}}R_{2k+i}}{\ceil{\frac{N(t)}{2}}}= 
  \E{R_2} \quad a.s.  \quad \forall i \in \{0,1\},
}  
which further gives
\eq{
\lim_{t \to
    \infty}\frac{\sum_{k=1}^{\ceil{\frac{N(t)}{2}}}\sum_{i \in \{0,1\}
    }R_{2k+i}}{N(t)} =\E{R_2} \quad a.s.  .
}  
Similarly,
\eq{
\lim_{t \to
    \infty}\frac{\sum_{k=1}^{\floor{\frac{N(t)}{2}}}\sum_{i \in
      \{0,1\} }R_{2k+i}}{N(t)} =\E{R_2} \quad a.s.  
.}  
Combining the observations above, we get 
\eq{
\lim_{t \to \infty}\frac{\sum_{k=1}^{N(t)}R_k }{N(t)}
  =\E{R_2} \quad a.s.  
,}
which together with \eqref{e:avgage_factors} and \eqref{e:Nt_limit} yields \eqref{eq:avgage_RY}.

\subsection{Proof of Theorem~\ref{t:update_optimal}}
Our proof is based on noticing that the minmax cost $\Delta^*(P)$ in
\eqref{e:minmax_cost}  
involves weighted average length with weights $g_{z, Q, P}(x)$. In
fact, we will see below that there is no loss in restricting to
nonnegative weights, whereby our cost has a form of average length
with respect to a distribution that depends on $(z,Q)$. 
The broad idea of the proof is to establish that a optimal code corresponding
to the {\it least favorable} choice of $(z,Q)$ is minmax
optimal. However, the proof is technical since our cost function may not satisfy the assumptions in a standard
saddle-point theorem.

A simpler form of the minmax cost $\Delta^*(P)$ from \eqref{e:varZ} is given by
\begin{align}\label{eq:Delta_simple}
{\Delta^*(P)=\min_{\ell \in \Lambda}\max_{z \geq 0} f(\ell,z),} 
\end{align}
 where 
\begin{align}\label{e:f}
f(\ell,z):=-z^2 \frac{\E{L}}{2}+z\sqrt{\E{L^2}}+\E{L}.
\end{align}
We seek to apply the following version of  Sion's minmax theorem to the function $f$.
\begin{thm}[Sion's Minmax Theorem~\cite{sion1958general}]\label{t:Sion's t}
Let $\X$ be convex space and $\Y$ be a convex, compact space. Let $h$ be a function on $\X\times\Y$ which is convex on $\X$ for every fixed $y$ in $\Y$
and concave on $\Y$ for every fixed $x$ in $\X$. Then, 
$$\inf_{x \in \X} \sup_{y \in \Y} h(x,y)= \sup_{y \in \Y} \inf_{x \in \X} h(x,y).$$
\end{thm}
Indeed, the following lemma shows that our function $f$ satisfies the convexity requirements of Sion's minmax theorem. 
\begin{lem}\label{l:conv_conc_f}
 $f(\ell,z)$ is convex in $\ell$ for every fixed $z \geq 0$
  and concave in $z$ for a fixed $\ell \in \Lambda$.
\end{lem}
\begin{proof}
To show that $f(\ell,z)$ is a convex function of $\ell$ for every
fixed $z \geq 0$, it suffices to show that $\sqrt{\E{L^2}}$ is
convex in $L=\ell(X)$. To that end, let $L_1=\ell_1(X)$ and
$L_2=\ell_2(X)$, for some $\ell_1$ and $\ell_2$ in $\lambda$. For all
$\lambda \in [0,1]$,
\[
\sqrt{ \E{\left(\lambda L_1 + (1-\lambda)L_2 \right)^2}} \leq  \lambda \sqrt{\E{ L_1^2 }} + (1-\lambda) \sqrt{\E{ L_2^2 }},
\]
where the inequality is by  Minkowski inequality for $\norm{L}_2$.

The \newer{concavity} in $z$ can be seen easily by noticing that \newer{$\frac{\partial ^2f(\ell, z)}{\partial  z^2} \leq 0$} for all $\ell$ in $\lambda$.
\end{proof}

However, our underlying domains of optimization are not compact. Our proof below circumvents this difficulty by showing that we may replace one of the domains by a compact set. For ease of reading, we divide the proof into 3 steps; we begin by
summarize the flow at a high-level.  The first step is to show that this minmax cost remains unchanged
when the domain of $z$ is restricted to a bounded interval $[0,K]$ for
a sufficiently large $K$. 
This will allow us to interchange $\min_{l \in \Lambda}$ and $\max_{z\in [0,K]}$ in the second step by using Theorem~\ref{t:Sion's t} to obtain
\begin{align}
\Delta^*(P)=\max_{z \in [0,K]} \min_{\ell \in \Lambda}f(\ell,z). 
\label{e:restricted_range_z}
\end{align}
Furthermore, we then use Corollary~\ref{c:2variational_formula} to linearize the cost. But this brings in the maximization over an additional parameter $Q$, 
which we again interchange with the minimum over $\ell$ using Sion's minmax theorem (Theorem~\ref{t:Sion's t}). Note that the required convexity of the cost function is easy to see; we note it in the following lemma.

\begin{lem}\label{l:conv_conv_g}
For every fixed $z \geq 0$, $\sum_{x \in \X}g_{z,Q,P}(x)\ell(x)$ is convex in $\ell$ for a fixed $Q \ll P$
  and concave in $Q$ for a fixed $\ell \in \Lambda$.
 \end{lem}
 \begin{proof}
For every fixed $z \geq 0$, the cost function $\sum_{x \in \X}g_{z,Q,P}(x)\ell(x)$ is linear, and thereby convex, 
in $\ell$ for a fixed $Q$ . For concavity in $Q$, note that 
for a fixed $\ell \in \Lambda$, the function $\sqrt{Q(x)}$ is a concave function of $Q(x)$, for all $x$ in $\X$. 
\end{proof}
Thus, we obtain
\[
\Delta^*(P)=\max_{z \in [0, K], Q\ll P}\min_{\ell \in \Lambda}\sum_{x\in \X}g_{z,Q,P}(x)\ell(x).
\] 
In the final step, we will establish that the optimal code
for linear cost with weights corresponding to the least favorable $(z, Q)$ is minmax optimal.
We now present each step in detail. 

\paragraph*{Step 1} We begin by noting that there is no loss in
restricting to codes with\footnote{For simplicity, we assume that
  $\log \X$ is an integer.} $\E{L}\leq \log{|\mathcal{X}|}$. Indeed, 
note that for $\E{L}>\log{|\mathcal{X}|}$ the average age is bounded as
\begin{align}
\E{L}+\frac{\E{L^2}}{2\E{L}} \geq \frac{3}{2}\E{L}> \frac 3 2\log{|\mathcal{X}|},
\label{e:age_bound_large}
\end{align}
where we have used Jensen's inequality. 
On the other hand, a
fixed-length code with $\ell(x) = \log |\X|$ attains 
\begin{align}
\E{L}+\frac{\E{L^2}}{2\E{L}} =\frac 32 \log{|\mathcal{X}|},
\label{e:fixed-length}
\end{align}
which gives the desired form
\begin{align}
\nonumber
\Delta^*(P) & = \min_{\ell \in \Lambda, \E{L}\leq
  \log{\mathcal{X}}}\E{L}+\frac{\E{L^2}}{2\E{L}}\\ & = \min_{\ell \in \Lambda, \E{L}\leq \log{\mathcal{X}} }\max_{z \in \R} f(\ell,z),
\label{e:restrict_min}
\end{align}
\newer{where $f(\ell,z)$ is defined in \eqref{e:f}}.
Also, for a fixed $\ell$ in $\Lambda$ the function $f(\ell,z)$ attains its maximum at 
$z^*(\ell)$ given by
\[
z^*(\ell):= \frac{\sqrt{\E{L^2}}}{\E{L}}.
\]
For $\E{L}\leq \log |\X|$, the maximizer $z^*(\ell)$ is bounded
as\footnote{We assume without loss of generality that $P(x)>0$ for every $x\in \X$.}
\begin{align*}
z^*(\ell) &\leq \frac{\sqrt{\E{L^2}}}{H(X)} \\ &=
\frac{\sqrt{\sum_{x}P(x)\ell(x)^2}}{H(X)} \\ &\leq
\frac{\E{L}}{H(X)}\sqrt{\max_{x\in \X}\frac 1{P(x)}} \\ &\leq
\frac{\log |\X|}{H(X)}\sqrt{\frac 1{\min_{x\in \X}P(x)}},
\end{align*}
where the first inequality uses $\E{L}\geq H(X)$, which holds for
every prefix-free code, and the second holds since $\|a\|_2 \leq
\|a\|_1$ for any sequence $a= (a_1, ..., a_n)$. 
Denoting
\[
K := \frac{\log |\X|}{H(X)}\sqrt{\frac 1{\min_{x\in \X}P(x)}},
\]
\eqref{e:restrict_min} yields \eq{ \Delta^*(P)&=\min_{\ell \in
    \Lambda, \E{L}\leq \log{|\X| }}\max_{z \in [0, K]} f(\ell, z).  }
Next, we show that the minmax cost above remains unchanged when we
drop the constraint $\E{L}\leq \log |\X|$ in the outer
minimum,
 which will complete the first step
of the proof and establish \eqref{e:restricted_range_z}.
Indeed, since by \eqref{e:fixed-length} the minimum over $\ell\in
\Lambda$ such that $\E{L}\leq \log|\X|$ is at most $(3/2)\log |\X|$, it suffices to show that 
\begin{align}
\min_{\ell \in \Lambda, \E{L} > \log |\X|}\max_{z \in [0, K]} f(\ell,
z) > \frac 32 \log |\X|.
\label{e:lb_larger_EL}
\end{align}
We divide the proof of this fact into two cases. 
First consider the case when $\ell$ in $\Lambda$ is such that  $\E{L}> \log |\X|$ and $K \geq
z^*(\ell)$. Then,  $\max_{z\in [0,K]}f(\ell,z)$ equals $\max_{z \geq 0}f(\ell,z)$, which is bounded below by $(3/2)\log |\X|$ using
\eqref{e:age_bound_large} and the definition of $f(\ell,z)$.
For the second case when $\E{L}> \log |\X|$ and $K < z^*(\ell)$, we
have
\eq{
\max_{z \in[0,K]}f(\ell,z)
&=-K^2\frac{\E{L}}{2}+K\sqrt{\E{L^2}}+\E{L}
\\
&>K^2\frac{\E{L}}{2}  +\E{L}
\\
&>\frac 32\cdot \E{L} 
\\
&>\frac{3}{2}\cdot\log|\X|,} 
where the first inequality uses $K< z^*(\ell)=\sqrt{\E{L^2}}/{\E{L}}$
and the second holds since $K\geq 1$ from its definition. Therefore, we have established~\eqref{e:lb_larger_EL}, and so we have
\eq{
\Delta^*(P)
=\min_{\ell \in \Lambda, \E{L}\leq \log{|\X| }}\max_{z \in [0, K]} f(\ell, z)
=\min_{\ell \in \Lambda}\max_{z \in [0, K]} f(\ell, z).
}

\paragraph*{Step 2} 
By lemma \ref{l:conv_conc_f} , $f(\ell,z)$ is convex in $\ell$ for every fixed $z \geq 0$
  and concave in $z$ for a fixed $\ell \in \Lambda$, $z$ takes values in a convex compact set $[0,K]$, and 
the set  $\{\ell: \ell \in \Lambda\}$ is convex, we get from Sion's minmax theorem (Theorem~\ref{t:Sion's t}) that 
\eq{
\Delta^*(P)
=\min_{\ell \in
      \Lambda}\max_{z \in [0, K]} f(\ell,z)= \max_{z \in [0,
        K]}\min_{\ell \in \Lambda}f(\ell,z).
} 
Using Corollary~\ref{c:2variational_formula}, we have
 \eq{\|L\|_2 =\max_{Q\ll P}\sum_{x\in \mathcal{X}}Q(x)^{\frac{1}{2}}
    P(x)^{\frac{1}{2}}\ell(x),
}
which by the definition of $f$ in  \eqref{e:f} further yields
\begin{align}\label{eq:Delta_3}
{f(\ell,z)=\max_{Q\ll P}\sum_{x\in \X}g_{z,Q,P}(x)\ell(x),}
\end{align}
where
\begin{align*}
g_{z,Q,P}(x) = \left(1- \frac{z^2}2\right)P(x) +z\sqrt{Q(x)P(x)}.
\end{align*}
We have obtained 
 \begin{align}
 \Delta^*(P)=\max_{z \in [0, K]}\min_{\ell
   \in \Lambda}\max_{Q\ll P}\sum_{x\in \X}g_{z,Q,P}(x)\ell(x).
 \end{align}
From Lemma~\ref{l:conv_conv_g}, $\sum_{x \in \X}g_{z,Q,P}(x)\ell(x)$ is convex in $\ell$ , for all $Q \ll P$,
  and concave in $Q$, for a fixed $\ell \in \Lambda$.
 Furthermore, since the set $\{Q:Q\ll P\}$ is convex compact for a pmf $P$ on finite alphabet, using Sion's minmax
theorem (Theorem~\ref{t:Sion's t}) once again, we get 
\begin{align}
& \Delta^*(P)=\max_{z \in [0, K]}\max_{Q\ll P}\min_{\ell \in
    \Lambda}\sum_{x\in \X}g_{z,Q,P}(x)\ell(x),
\label{eq:Delta_4}
\end{align}
which completes our second step.

 \paragraph*{Step 3} By \eqref{eq:Delta_4}, we get
\begin{align}
& \Delta^*(P)\leq\max_{z \geq 0}\max_{Q\ll P}\min_{\ell \in
    \Lambda}\sum_{x\in \X}g_{z,Q,P}(x)\ell(x).
\nonumber
\end{align}
On the other hand, by \eqref{eq:Delta_simple} and \eqref{eq:Delta_3}  we have 
\begin{align} \nonumber
\Delta^*(P)& =\min_{\ell \in
    \Lambda}\max_{z \geq 0}\max_{Q\ll P}\sum_{x\in \X}g_{z,Q,P}(x)\ell(x)\\& \geq\max_{z \geq 0}\max_{Q\ll P}\min_{\ell \in
    \Lambda}\sum_{x\in \X}g_{z,Q,P}(x)\ell(x),
\nonumber
\end{align}
whereby
\begin{align}
\nonumber
 \Delta^*(P) &=\min_{\ell \in
    \Lambda}\max_{z \geq 0}\max_{Q\ll P}\sum_{x\in \X}g_{z,Q,P}(x)\ell(x) \\ &=\max_{z \geq 0}\max_{Q\ll P}\min_{\ell \in
    \Lambda}\sum_{x\in \X}g_{z,Q,P}(x)\ell(x),
\label{e:max_min}
\end{align}
which proves the first part of theorem \ref{t:update_optimal}.

Next, we claim that in the maxmin formula above, the maximum is attained by a $(z,Q)$ for which $g_{z,Q,P}(x)$ is non-negative for every $x$. Indeed, if for some $z,Q$ there exists an $x^\prime$ in $\X$ such that $g_{z,Q,P}(x^\prime)$ is negative, then the cost $\sum_{x\in \X}g_{z,Q,P}(x)\ell(x)$ is minimized by any $\ell$ such that $\ell(x^\prime)=\infty$ and the minimum value is $-\infty$. Such $z,Q$ clearly can't be the optimizer of the maxmin problem, 
since for $z=0$, we have $g_{z,Q,P} \ge 0$, which in turn leads to $\min_{\ell \in
    \Lambda}\sum_{x\in \X}g_{z,Q,P}(x)\ell(x) \geq 0$.

Finally, consider  $(z, Q)$ such that $g_{z,Q,P}(x)\geq 0$ for all $x \in \X$. For such a $(z, Q)$, we seek to identify the minimized $\ell$ below:
\begin{align}\label{eq:Delta_5}
&\min_{\ell \in \Lambda}\sum_{x\in \X}g_{z,Q,P}(x)\ell(x) \nonumber \\&=\sum_{x'\in
  \X}g_{z,Q,P}(x')\min_{\ell \in \Lambda} \sum_{x\in
  \X}\frac{g_{z,Q,P}(x)}{\sum_{x'\in \X}g_{z,Q,P}(x')}\ell(x).
\end{align}
Thus, our optimization problem reduces to the standard problem of designing minimum average length prefix-free codes for the pmf 
\[
P_{z,Q}(x) = \frac{g_{z,Q,P}(x)}{\sum_{x'\in \X}g_{z,Q,P}(x')}.
\]
By Shannon's source coding theorem for variable length codes, the minimum is achieved by 
\[
\ell^*_{z,Q}(x):=\log\newer{ \frac{\sum_{x^{\prime}\in
      \X}g_{z,Q,P}(x)}{g_{z,Q,P}(x)}.}
\] 
Furthermore, $\ell^*_{z,Q}$ is the unique minimizer in $\Lambda$. 

Consider now a maximizer $(z^*, Q^*)$ of the maxmin problem in \eqref{e:max_min}, and let $\ell^o = \ell^*_{z^*, Q^*}$. 
Then, by Lemma~\ref{l:unique_spl} in the appendix,$(\ell^o, (z^*, Q^*))$ is a saddle-point for the minmax problem in \eqref{e:max_min}. 
  Moreover, $\ell^o$ is the unique minmax optimal solution.

\subsection{Proof of Theorem~\ref{t:qmaintheorem}}\label{p:3}
Denoting 
\begin{align}\label{e:f2_def}
f( \ell,z)= -z^2 \frac{(L_{\tt
      th}-\E{L})}{2}+z\sqrt{\E{L^2}}+\E{L},
\end{align}     
the optimal cost $\Delta^*(P)$ can be written as 
\begin{align*}
\Delta^*(P)&= \inf_{\ell \in \Lambda , \E{L} < L_{\tt
    th}}\frac{\E{L^2}}{2(L_{\tt th}-\E{L})}+\E{L}\\&
= \min_{\ell \in \Lambda, \E{L} < L_{\tt
    th}}\max_{z \geq 0} f(\ell,z).
\end{align*}
This form is similar to the one we had in Theorem~~\ref{t:update_optimal}. But the proof there does not extend to the case at  hand. 
Specifically, note that for each $\ell$, $f(\ell,z)$ attains its maximum value for $z^*(\ell)=\frac{\sqrt{\E{L^2}}}{(L_{\tt th}-\E{L})}$ which, unlike the quantity that we obtained in the proof of Theorem~\ref{t:update_optimal}, is unbounded over the set of $\ell\in \Lambda$ such that $\E{L}\leq L_{\tt th}$. However, under the additional assumption $H(X)+\log (1+1/\sqrt{2})< L_{\tt th}$, we can provide a simpler alternative proof. 
We rely on the following lemma.
\begin{lem}\label{l:minmax_conditions}
Consider a function $h:\X\times \Y \to \R$ such that the set $\X$ is compact convex, the set $\Y$ is convex, $h(x, y)$ is a convex function of $x$ for every fixed $y$ and a concave function of $y$ for every fixed $x$. Suppose additionally that there exist a convex subset $\X_0$ of $\X$ and a compact convex subset $\Y_0$ of $\Y$ such that
\begin{enumerate}
\item for every for every $x\in \X_0$, an optimizer $y^*(x)\in \arg\max_{y\in \Y}h(x,y)$ belongs to $\Y_0$; and 
\item for every $y\in \Y_0$, an optimizer $x^*(y)\in \arg\min_{x\in \X} h(x,y)$ belongs to $\X_0$.
\end{enumerate}
Then, 
\[
\min_{x\in \X}\max_{y\in \Y}h(x,y)= \max_{y\in \Y}\min_{x\in \X}h(x,y).
\]
\end{lem}
\begin{proof}
Note that since for $x$ in $\X_0$, the $y$ that maximizes $h(x,y)$ over $\Y$ is in $\Y_0$, we get 
\[
\min_{x\in \X}\max_{y\in \Y}h(x,y)\leq
\min_{x\in \X_0}\max_{y\in \Y}h(x,y)
=\min_{x\in \X_0}\max_{y\in \Y_0}h(x,y).
\]
Further, by Sion's minmax theorem (Theorem~\ref{t:Sion's t}), the right-side equals
$\max_{y\in \Y_0}\min_{x\in \X_0}h(x,y)$. But by our second assumption, the restriction $x\in \X_0$ can be dropped, and we have
\[
\max_{y\in \Y_0}\min_{x\in \X_0}h(x,y)=
\max_{y\in \Y_0}\min_{x\in \X}h(x,y)\leq \max_{y\in \Y}\min_{x\in \X}h(x,y).
\]
Thus, we have shown $\min_{x\in \X}\max_{y\in \Y}h(x,y)\leq \max_{y\in \Y}\min_{x\in \X}h(x,y)$, which completes the proof since the inequality in the other direction holds as well.
\end{proof}
For our minmax cost, we will verify that both the conditions of the lemma above hold under the assumption $H(X)+\log (1+1/\sqrt{2}) <L_{\tt th}$. Indeed, first note that for any fixed $\ell\in \Lambda$ with $\E{L}\leq H(X)+\log(1+1/\sqrt{2})$, the maximizer $z$ of $f(\ell,z)$  given by $\sqrt{\E{L^2}}/(L_{\tt th} - \E{L})$ satisfies
\begin{align*}
&\frac{\sqrt{\E{L^2}}}{L_{\tt th} - \E{L}}\\&\leq 
\sqrt{\frac 1{\min_x P(x)}}\cdot \frac{\E{L}}{L_{\tt th} - \E{L}}
\\
&\leq \sqrt{\frac 1{\min_x P(x)}}\cdot \frac{H(X)+\log (1+1/\sqrt{2})}{
L_{\tt th} - H(X)-\log (1+1/\sqrt{2})}.
\end{align*}
Denote the right-side above by $K$ and $L_{\tt th}' = H(X)+\log (1+1/\sqrt{2})$. 
Therefore, with the set $\{\ell\in \Lambda, \E{L}\leq L_{\tt th}'\}$ in the role of $\X_0$ in Lemma~\ref{l:minmax_conditions}, 
the set $[0,K]$ can play the role of $\Y_0$.

To apply Lemma~\ref{l:minmax_conditions}, we require two
conditions to hold: first, that 
$f(l,z)$ is a convex function of $\ell$ for every fixed $z$ and a concave function of $z$ for every fixed $\ell$, second, that
for every $z\in [0,K]$, the minimizing $\ell$ 
satisfies $\E{L}\leq L_{\tt th} ^ \prime$.The first easily follows from \eqref{e:f2_def}. The proof of this fact is exactly the same as Lemma~\ref{l:conv_conc_f}. However, while the second condition can be
shown to be true, the proof of
this fact is almost the same as the proof of our theorem.  For
simplicity of presentation, we instead present an alternative proof of
the theorem that uses a slight extension of the lemma above.
Note that from our foregoing discussion and following the proof of the
lemma, 
we already have obtained
\[
\Delta^*(P)\leq \max_{z \in[0,K]}\min_{\ell \in \Lambda, \E{L} \leq {L^{\prime}}_{\tt th}}f(\ell,z).
\]
By using Corollary~\ref{c:2variational_formula}  and using Sion's minmax theorem once again, we get
\begin{align*}
  \lefteqn{\Delta^*(P)}
\\
  &
  \leq \max_{z \in [0, K]}\max_{Q\ll P}\min_{\ell \in \Lambda, \E{L} \leq {L^{\prime}}_{\tt th}}\sum_{x\in \X}  g_{z,Q,P}(x)\ell(x)- \frac{z^2}{2} L_{\tt th},
\end{align*}
where 
\begin{align*}
g_{z,Q,P}(x) := \left(1+ \frac{z^2}2\right)P(x)+z\sqrt{Q(x)P(x)}.
\end{align*}    
In the preceding argument, we can use Sion's minmax theorem as the following two conditions hold. First, for every fixed $z \geq 0$, the function $\sum_{x \in \X}  g_{z,Q,P}(x)\ell(x)- \frac{z^2}{2} L_{\tt th}$ is concave in $Q$ for a fixed $\ell \in \Lambda$ and convex in $\ell$ for a fixed $Q \ll P$. Second, the sets $\{Q: Q\ll P\}$ and $\{\ell \in \Lambda: \E{L} \leq {L^{\prime}}_{\tt th}\}$ are compact and convex. Proof of the first is exactly the same as  that of \ref{l:conv_conv_g}. Second is true as we have restricted to a finite alphabet $\X$.
Thus, we can proceed as in the proof of the lemma, but  we need to show now that for every $z\in [0,K]$ and $Q\ll P$, the optimal $\ell^*(z,Q)$ satisfies $\E{L^*}\leq L_{\tt th}^{\prime}$
. 
Indeed, consider the following optimization problem for a fixed  $z$, $Q$:
\begin{align*}
\min_{\ell \in \Lambda}& \sum_{x\in \X}g_{z,Q,P}(x)\ell(x) \\&=\left(\sum_{x'\in
  \X}\newer{g_{z,Q,P}(x^{\prime})}\right) \min_{\ell \in \Lambda} \sum_{x\in
  \X}\frac{g_{z,Q,P}(x)}{\sum_{x'\in \X}g_{z,Q,P}(x')}\ell(x).
\end{align*}
Since $\frac{g_{z,Q,P}(x)}{\sum_{x'\in \X}g_{z,Q,P}(x')}$ are nonnegative and add to $1$,
in the optimization problem above, we are minimizing the expected prefix
free lengths for a finite alphabet for a particular distribution. Thus,
by Shannon's Source Coding Theorem, the optimal $\ell^*_{z,Q}$ is given by
\newer{
\begin{align*}
\ell^*_{z,Q}(x):=\log \frac{\sum_{x^{\prime}\in\X}g_{z,Q,P}(x^{\prime})}
{g_{z,Q,P}(x)};
\end{align*}}
in fact, this optimizer is unique. But then for  every $x$ in $\X$, \newer{
\eq{
&\ell^*_{z,Q}(x)\\&=\log \frac{\sum_{x^{\prime}\in \X}g_{z,Q,P}(x^{\prime})}{g_{z,Q,P(x)}}\\ 
 &=\log\frac{\sum_{x^{\prime} \in \X}\left(1+\frac{z^2}2\right)P(x)+\sum_{x\in \X}z\sqrt{Q(x)P(x)}}{\left(1+\frac{z^2}2\right)P(x)+z\sqrt{Q(x)P(x)}}
  \\
& \leq \log
  \frac{1}{P(x)} \\ & ~ +\log\left(\frac{\left(1+ \frac{z^2}2\right)}{\left(1+
    \frac{z^2}2\right)+z\sqrt{\frac{Q(x)}{P(x)}}}
  +\frac{z}{\left(1+
    \frac{z^2}2\right)+z\sqrt{\frac{Q(x)}{P(x)}}}\right)
 \\
& \leq \log
  \frac{1}{P(x)}+\log\left(\frac{\left(1+ \frac{z^2}2\right)}{\left(1+
    \frac{z^2}2\right)}
  +\frac{z}{\left(1+
    \frac{z^2}2\right)}\right)        
     \\
    &\leq \log
  \frac{1}{P(x)}+\log\left(1+\frac{1}{\sqrt{2}}\right),
   }}
where the first inequality is by the Cauchy-Schwarz inequality, the second inequality follows upon noting that $\frac{Q(x)}{P(x)}$ is nonnegative, and the last inequality follows from the fact that $z^2/2+1 \geq\sqrt{2}z$ \newer{(which holds with equality at $z=\sqrt{2}$)}. 
Thus as a consequence of this inequality the expected code length of
such a code is upper bounded as
follows, 
\begin{align}\label{e:q_corr}
\E{L^*_{z,Q}}\leq 
    H(x)+\log\left(1+\frac{1}{\sqrt{2}}\right),
\end{align}
which in the manner of Lemma~\ref{l:minmax_conditions} gives
\begin{align}
\nonumber \Delta^*(P) = \max_{z \geq 0}\max_{Q\ll P}\min_{\ell \in
  \Lambda, \E{L} \leq {L}_{\tt th}}\sum_{x\in \X} g_{z,Q,P}(x)\ell(x)-
\frac{z^2}{2} L_{\tt th}.
\end{align}
Finally, it remains to establish that $\ell^*_{z^*,Q^*}$ is the unique minmax optimal solution. This can be shown in exactly the same manner as it was shown for Theorem~\ref{t:update_optimal} in the previous section; we skip the details.\qed



\subsection{A saddle-point lemma}
The following simple result is needed to establish the minmax
optimality of our scheme. The first part of the result claims that any
pair of minmax optimal $x$ and maxmin optimal $y$ forms a saddle
point, a well-known fact. The second part claims that if the minimizer
for the maxmin optimal $y$ is unique, then it must also be minmax
optimal and thereby constitute a saddle-point with $y$.

\begin{lem}\label{l:unique_spl} 
Consider the minmax problem $\displaystyle{\min_{x\in \X}\max_{y\in \Y}h(x,y)}$, and
assume that
\[
\min_{x\in \X}\max_{y\in \Y}h(x,y)=\max_{y\in \Y}\min_{x\in \X}h(x,y).
\]
Then, for every pair $(x^*, y^*)$ such that $\displaystyle{x^*\in\arg\min_{x \in
  \mathcal{X}}\max_{y \in \mathcal{Y}}h(x,y)}$ and $\displaystyle{y^*\in\arg\max_{y
  \in \mathcal{Y}}\min_{x \in \mathcal{X}}h(x,y)}$  
constitutes a saddle-point. Furthermore, if the minimizer $x^o(y^*)$
of $\min_{x \in \mathcal{X}}h(x,y^*)$ is unique, then $x^*=x^o(y^*)$ 
is the unique minmax optimal solution. 
\end{lem}
\begin{proof}
Since minmax and maxmin costs are assumed to be equal, by the definition of $x^*$ and
$y^*$, we have
\begin{align}
h(x,y^*) &\geq \max_{y^\prime \in \mathcal{Y}}\min_{x^\prime \in \mathcal{X}}h(x^\prime,y^\prime)
\nonumber
\\
&=\min_{x^\prime \in \mathcal{X}}\max_{y^\prime \in \mathcal{Y}}h(x^\prime,y^\prime)\geq
h(x^*,y),
\label{e:minmax_exchange}
\end{align}
for all $x$ in $\mathcal{X}$ and $y$ in $\mathcal{Y}$.
Upon substituting $ x^*$ for $x $ and $y^*$ for $y$, we get that $x^*$
   is a minimizer of $h(x,y^*)$ and $y^*$ a maximizer of $h(x^*,
y)$. Therefore, $(x^*, y^*)$ forms a saddle-point and $h(x^*,
y^*)=\min_{x\in\X}\max_{y\in\Y}h(x,y)$.

Turning now to the second part, suppose that $x^\prime$, too, is minmax optimal. 
Then, using \eqref{e:minmax_exchange} with $x=x^\prime$ and $y=y^*$, we get that $x^\prime$ must be a minimizer
of $h(x,y^*)$ as well. 
But since this minimizer is unique,  $x^\prime$ must coincide with $x^o$.

\end{proof}
\section*{Proof of Lemma~\ref{c:dim_red}}
Denoting
\[
c_P(z,Q):=\sum_{x\in \X}g_{z,Q,P}(x) \log\frac{\sum_{x^\prime\in
    \X}g_{z,Q,P}(x^\prime)}{g_{z,Q,P}(x)},
\]
 we begin by observing the
concavity of $c_P(z,Q)$.  Recall the notations $\G=\{z \geq 0, Q \in \R^{|\X|}:g_{z,Q,P}(x)\geq0 \quad \forall x \in \X\}$
and $g_{z,Q,P}(x)  = (1-z^2/2)P(x) + z\sqrt{Q(x)P(x)}$.

\begin{lem}\label{c:concavity}
The function $c_P(z,Q) $ is concave in $Q$ for each fixed $z$
and is concave in $z$ for each fixed $Q$, over the set $\G$.
\end{lem}
\begin{proof}
For the first part, \eqref{eq:Delta_5} yields that for every $(z,Q)\in
\mathcal{G}$,
\eq{
\sum_{x\in \X}g_{z,Q,P}(x) &\log\frac{\sum_{x^\prime\in
    \X}g_{z,Q,P}(x^\prime)}{g_{z,Q,P}(x)}\\ &=\min_{\ell \in
  \Lambda}\sum_{x\in \X}g_{z,Q,P}(x)\ell(x).
}
Also, for every fixed $z$, the function $g_{z,Q,P}(x)$ is concave in $Q$, and thereby $\sum_{x\in \X}g_{z,Q,P}(x)\ell(x)$, is concave in
 $Q$. Thus, since the minimum of concave functions is concave,
$c_P(z,Q)$ is concave in $Q$ for a fixed $z$. Similarly, we can show
concavity in $z$ for a fixed $Q$ since $g_{z,Q,P}(x)$ is concave in
$z$, too, for every fixed $Q$.  
\end{proof}

We now complete the proof of Lemma~\ref{c:dim_red}.
 We will show that for any $(z,Q)$ which is feasible for optimization
 problem \eqref{e:maxmin_cost}, we can find a feasible
 $(z,Q^\prime)$ with $Q^\prime$ satisfying \eqref{e:property}, and
$$ c_P(z,Q) \leq c_P(z,Q^\prime). $$

Indeed, consider
  $Q^\prime(x):=Q(A_i)/|A_i|$ for all $x \in
  \X$.  The remainder of the proof is divided into two parts, the
first proving the feasibility of $Q^\prime$ and the second 
 proving $ c_P(z,Q) \leq c_P(z,Q^\prime). $
\paragraph{Feasibility of $(z,Q^\prime)$}  

From the feasibility of
  $(z,Q)$, for all symbols $x$ in $A_i$ and for all $i$ in
$[M_P]$, $g_{z,Q,P}(x) \geq 0$, whereby 
\begin{align*}
\sum_{x \in A_i}
    g_{z,Q,P}(x) 
&=\sum_{x \in A_i}
 \left(1- \frac{{z}^2}2\right)P(x)
  \\&\hspace{3cm}  +{z}\sum_{x \in A_i}\sqrt{Q(x)P(x)} 
\\
&= \left(1- \frac{{z}^2}2\right)P(A_i)
    +{z}\sum_{x \in A_i}\sqrt{Q(x)P(x)} 
\\
&\geq \left(1-\frac{{z}^2}2\right)P(A_i)
+{z}\sqrt{Q^\prime(A_i)P(A_i)} 
\\
&= |A_i| g_{z,Q^{\prime},P}(x) 
\\
&\geq 0,
\end{align*}
where the first inequality is by Cauchy-Schwarz inequality, the positivity
 of $z$, and the assumption that $P(x) = P(A_i)/|A_i|$ for every $x$ in
 $A_i$, and the final identity uses definition of $Q^\prime$. This
 proves the feasibility of $(z, Q^\prime)$ for the optimization
 problem \eqref{e:maxmin_cost}.

\paragraph{Proof of optimality}  
Denoting by $\Pi(A_1)$ the set of all permutations of the elements of
$A_1$, let $Q^{\pi}$ be the distribution given by
\[ 
Q^\pi(x)=\begin{cases}
&Q(\pi(x)),\quad \forall x\in A_1
\\
&Q(x), \quad \text{otherwise}.
\end{cases}
\]
Then, the distribution $\overline{Q} = (1/|\Pi(A_1)|)\cdot\sum_{\pi\in \Pi(A_1)}Q^\pi$ satisfies
\[ 
\overline{Q}(x)=\begin{cases}
&\frac 1 {|A_1|} \cdot Q(A_1),\quad \forall x\in A_1
\\
&Q(x), \quad \text{otherwise}.
\end{cases}
\]
Since by Lemma \ref{c:concavity} $c_P(z,Q)$ is concave in $Q$ for every fixed $z$, we get
\[
c_P(z,\overline{Q}) \geq \frac{1}{|\Pi(A_1)|}\cdot\sum_{\pi \in\Pi(A_1)}
c_P(z,Q^\pi).
\]
Furthermore, note that $g_{z,Q^\pi, P}(x)= g_{z, Q, P}(\pi(x))$ 
since
$P(x)=P(A_1)/|A_1|$ for every $x$ in $A_1$, and thereby 
$c_P(z, Q^\pi) = c_P(z, Q)$ for every $\pi\in \Pi(A_1)$ . Therefore,
combining the observations above, we obtain $c_P(z,\overline{Q}) \geq
c_P(z, Q)$.

Repeating this argument by iteratively using permutations of $A_i$ for $i\geq 2$, we obtain the required inequality 
\[
c_P(z, Q') \geq c_P(z,Q).
\]

\qed

\section{Concluding Remarks}
In this chapter, we studied the source coding problem where the goal is to minimize
\[\E{L}+\frac{\E{L^2}}{2\E{L}}\]
subject to the constraint that the code lengths satisfy Kraft's inequality. We saw that this problem differs from the standard source coding problem where the goal is to minimize $\E{L},$ and the classic source-coding solutions such as deploying Shannon codes may be suboptimal. Our main result was a structural result showing that the optimal code lengths for the relaxed version of the problem are Shannon lengths for tilting of the original distribution. Our recipe to prove this result was to linearize the cost function in terms of length by first expressing as the optimal value of a quadratic maximization problem over a new variable. Then, we use a variational formula for the $L_2$ norm of a random variable to linearize the cost. We believe that our approach can be used to prove similar structural results for other source coding problems, thereby gaining computational insights into solving them, as we saw with the application of our recipe for the problem of designing source codes with minimum delay.



\bibliography{tit2018,references} \bibliographystyle{IEEEtranS}

\end{document}

%% file: preamble.tex
\usepackage{varwidth}
\usepackage[utf8]{inputenc} 
\usepackage[T1]{fontenc} 
\usepackage{hyperref}
\usepackage{url} 
\usepackage{booktabs} 
\usepackage{amsfonts} 
\usepackage{nicefrac}
\usepackage{microtype} 
\usepackage[noend]{algpseudocode}
\usepackage{soul}
\usepackage[T1]{fontenc}
\usepackage{lmodern}
\usepackage{appendix}
\usepackage{graphicx}
\usepackage[usenames,dvipsnames,table]{xcolor}
\usepackage{amsmath}
\usepackage{algorithm}
\usepackage[noend]{algpseudocode}
\algsetblock[Name]{Encoder}{Stop}{3}{1cm}
\algsetblock[Name]{Decoder}{Stop}{3}{1cm}
\algsetblock[Name]{SharedRandomness}{Stop}{3}{1cm}
\usepackage{mleftright}
\newcommand{\priv}{\varepsilon}
\usepackage[colorinlistoftodos,textsize=scriptsize]{todonotes}
\usepackage[mathscr]{eucal}

\usepackage[noadjust]{cite}
\usepackage{subcaption}
\usepackage{%
    amssymb,%
    amsthm,
    bbm,%
    etoolbox,%
    mathtools,%
    stmaryrd%
}

\usepackage{amsfonts}
\usepackage{amssymb}
\usepackage{enumitem}
\newlist{steps}{enumerate}{1}
\setlist[steps, 1]{label = \quad \quad \quad Step \arabic*:}
\usepackage{tikz,pgf,pgfplots,circuitikz}
\usepgflibrary{shapes}
\usetikzlibrary{%
    arrows,%
    decorations.text,%
    positioning,%
    scopes,%
    shapes,
    patterns%
}	

\theoremstyle{plain} \newtheorem{thm}{Theorem}[section]
\newtheorem{lem}[thm]{Lemma} 
\newtheorem{cor}[thm]{Corollary}

\theoremstyle{definition} \newtheorem{defn}[thm]{Definition}
 
\newtheorem{assum}[thm]{Assumptions} 
\newtheorem{example}[thm]{Example}

\theoremstyle{remark} \newtheorem{rem}{Remark}

\definecolor{lightgray}{gray}{0.9}

\newcommand{\ep}{\mathcal{E}}
\newcommand{\eqdef}{:=}
\newcommand{\eq}[1]{\begin{align*}#1\end{align*}}
\newcommand{\ceil}[1]{\left\lceil#1\right\rceil}%
\newcommand{\norm}[1]{\|#1\|}%

 \newcommand{\R}{\mathbb{R}}
\newcommand{\N}{\mathbb{N}}
\newcommand{\E}[1]{\mathbb{E}\left[#1\right]}
\newcommand{\EE}[2]{\mathbb{E}_{#1}\left[#2\right]}
\newcommand{\Z}{\mathbb{Z}} \newcommand{\B}{\mathscr{B}}
 
 \newcommand{\G}{\mathcal{G}} 
\newcommand{\V}{\mathcal{V}}
\newcommand{\X}{\mathcal{X}}
\newcommand{\Y}{\mathcal{Y}}
\newcommand{\W}{\mathcal{W}}
\newcommand{\oO}{\mathcal{O}}

\newcommand{\cX}{\mathcal{X}}
\newcommand{\cY}{\mathcal{Y}}

\newcommand{\eps}{\varepsilon}
\newcommand{\EEC}[2]{\EE{}{ #1 \;\middle\vert\; #2 }}
\newcommand{\uRoman}[1]{\uppercase\expandafter{\romannumeral#1}}

\DeclarePairedDelimiter\floor{\lfloor}{\rfloor}

\newcommand{\bPr}[1]{\Pr\mleft(#1\mright)}

\usepackage{subcaption}
\DeclareCaptionSubType*{algorithm}

\DeclareCaptionLabelFormat{alglabel}{Alg.~#2}

\newcommand{\Gconvex}{\mathcal{G_{\tt c}}}
\newcommand{\Gstrongconvex}{\mathcal{G_{\tt sc}}}
\newcommand{\p}{\mathbf{p}}

  \newcommand{\newer}[1]{{{#1}}}
  \newcommand{\newest}[1]{{{#1}}}
    \newcommand{\newerest}[1]{{{#1}}}

  \newcommand{\mnote}[1]{}
  \newcommand{\tnote}[1]{}
  \newcommand{\cnote}[1]{}
  \newcommand{\anote}[1]{}

\newcommand{\ignore}[1]{\leavevmode\unskip} 

\newcommand{\Qenc}{Q^{{\tt e}}}
\newcommand{\Qdec}{Q^{{\tt d}}}
\newcommand{\known}{known $\mathbf{\Delta}$ }
\newcommand{\unknown}{unknown $\mathbf{\Delta}$ }

\newcommand{\indic}[1]{\mathbbm{1}_{#1}}

\DeclareMathOperator{\sign}{sign}

\newcommand{\dtv}{\operatorname{d}_{\rm TV}}
\newcommand{\totalvardistrestr}[3][]{{\dtv^{#1}\mleft({#2, #3}\mright)}}
\newcommand{\totalvardist}[2]{\totalvardistrestr[]{#1}{#2}}
\newcommand{\kldiv}[2]{{\operatorname{D}\mleft({#1 \| #2}\mright)}}
\newcommand{\mutualinfo}[2]{ I\mleft(#1 \land #2\mright) }
\newcommand{\condmutualinfo}[3]{ I\mleft(#1 \land #2\mid #3\mright) }

\ifnum\withcolors=1
  \newcommand{\new}[1]{{\color{red} {#1}}} 
\else
  \newcommand{\new}[1]{{{#1}}}
\fi
\newcommand{\iid}{$iid$}
\newcommand\theerthaedit[2]{{\color{blue}{\st{#1}}}{\color{red}{#2}}}


%% file: notation.tex
%
%
 \begin{enumerate}

   \item Sets
   \begin{enumerate}
      \item   $\R^d$  is the set of $d$ dimensional vectors, where each coordinate can take any value on the real line.
    
      \item $\Z$ is the set of integers.
      \item $\N$ is the set of  positive integers.
      \item $[n] := \{1, \ldots, n\}$  is the set of number from $1$ to $n.$
      \item $|\X|$ is the cardinality of a discrete set $\X.$
      \item  $\{e_1, \ldots, e_d\}$ is the Euclidean basis of $\R^d,$ where $e_i$ is a $d$-dimensional vectors with $i$ coordinate equal to $1$ and rest of the coordinates equal to $0.$
        \end{enumerate}
      
 \item Random Variables and Events
    \begin{enumerate}
    \item 
      pmf is Probability mass function.
      \item
      pdf is Probability density function.
      \item 
      $iid$ is Independent and identically distributed.
      \item 
      $P(A)$ is Probability of event $A$.
      \item
      $\mathbb{E}[Z]$ is Expectation of the random variable $Z$.
\end{enumerate}

\item Norms
      \begin{enumerate} 
      \item $\norm{x}_{p} :=\sum_{i \in [d]} (|x(i)|^p)^{1/p}$ is the $\ell_p$-norm of $x \in \R^d$ .
      
      \item  $\norm{X}_p =\E{|X|^p}^{1/p}$ is the $L_p$ norm of a random variable $X$.
      \item Throughout the paper, $q$  denotes the H\"{o}lder
conjugate of $p$ (that is, $\frac{1}{p}+\frac{1}{q}=1$).
\end{enumerate}

\item Logarithms
\begin{enumerate}
\item 
The logarithm to the base 2 is denoted by $\log a$ and the logarithm to the base $e$ is denoted by $\ln a$. All the information theoretic measures considered in this paper -- such as Entropy, R\'enyi divergence,  Kullback-Leibler divergence, and Mutual Information -- are defined with logarithm to the base 2.
\item The iterated logarithms $\log^*(a)$  and $\ln^*(a)$ are defined  as the number of times $\log$ and $\ln$ must be iteratively
 applied to $a$ before the result is at most $1$.
\end{enumerate}

\item Maximum and minimum
\begin{enumerate}
\item We write  $a \vee
b$ and $a \wedge b$ for $\max\{a,b\}$ and $\min\{a, b\}$, respectively.
\end{enumerate}
 
   \end{enumerate}

%% file: Figures/Figure_SGD.tex
\begin{tikzpicture}[scale=0.6, transform shape,
    pre/.style={=stealth',semithick},
    post/.style={->,shorten >=0.5pt,>=stealth',semithick},
dimarrow/.style={->, >=latex, line width=2pt}
    ]
\clip (-0.5,6) rectangle  (19,-0.5) ;


\draw[black,thick, fill = tacream ] (12.5, 5.5) rectangle( 18, 0.5);
\node[align=center] at (15.2, 4) {\color{black}{\LARGE{Algorithm} \LARGE{$\pi$}}};

\draw[black,thick, fill = cyan!30! ] (0, 5.5) rectangle(6, 0.5);
\node[align=center] at (3,3.5) {\LARGE{First Order}};
\node[align=center] at (3,2.5) {{\LARGE{ Oracle} \huge{O}}};

\node[align=center] at (15, 2) {\color{black}{{\LARGE{Update $x_t$}}}};
\draw[dimarrow, orange] (12.45,4.5) to (6.05,4.5);
\node[align=center] at (9,4.75) 
{\LARGE{$x_t$}};

\draw[dimarrow, orange] (6.05,1.5) to (12.45, 1.5);
\node[align=center] at (9,1) 
{\LARGE{$\hat{g}(x_t)$}};
\end{tikzpicture}

%% file: Figures/Figure_SGD3.tex
\begin{tikzpicture}[scale=0.6, transform shape,
    pre/.style={=stealth',semithick},
    post/.style={->,shorten >=0.5pt,>=stealth',semithick},
dimarrow/.style={->, >=latex, line width=2pt},
normline/.style={-, line width=2pt}
    ]
\clip (-0.5,6) rectangle  (19,-0.6) ;

\draw[black,thick, fill = tacream ] (12.5, 5.5) rectangle( 18, 0.5);
\node[align=center] at (15.2, 4) {\color{black}{\LARGE{Algorithm} \huge{$\pi$}}};

\node[align=center] at (15, 2) {\color{black}{{\LARGE{Update $x_t$}}}};
\draw[black,thick, fill = cyan!30! ] (0, 5.5) rectangle(6, 0.5);
\node[align=center] at (3,3.5) {\LARGE{First Order}};
\node[align=center] at (3,2.5) {{\LARGE{ Oracle} \huge{O}}};

\draw[dimarrow, orange] (12.45,4.5) to (6.05,4.5);
\node[align=center] at (7.5,4.9) 
{\LARGE{$x_t$}};

\draw[dimarrow, orange] (9,4.5) to (9,2.5);
\node[align=center] at (9.7,3.5) 
{\LARGE{$W_t$}};
\draw[normline, orange] (6.05,1.5) to (8,1.5);
\node[align=center] at (7,0.9) 
{\LARGE{$\hat{g}(x_t)$}};

\draw[dimarrow, orange] (10,1.5) to (12.45,1.5);
\draw[black, thick, fill = tagreen  ] (8, 2.5) rectangle(10.5, 0.5);
\node[align=center] at (11.25, 0.9) 
{\LARGE{$Y_t $}};
\node[align=center] at (14.5, -0.25) 
{\LARGE{
$Y_t \mid \hat{g}(x_t) \sim W_t(\cdot \mid \hat{g}(x_t))$}};
\node[align=center] at (9.25, 1.5) 
{\huge{$W_t$}};
\node[align=center] at (8.75, 1.5) 
{};

\end{tikzpicture}

%% file: Figures/Figure_SGD5.tex
\begin{tikzpicture}[scale=0.6, transform shape,
    pre/.style={=stealth',semithick},
    post/.style={->,shorten >=0.5pt,>=stealth',semithick},
dimarrow/.style={->, >=latex, line width=2pt},
normline/.style={-, line width=2pt}
    ]
\clip (-0.5,6) rectangle  (19,-0.6) ;

\draw[black,thick, fill = tacream ] (12.5, 5.5) rectangle( 18, 0.5);
\node[align=center] at (15.2, 4) {\color{black}{\LARGE{Algorithm} \huge{$\pi$}}};

\node[align=center] at (15, 2) {\color{black}{{\LARGE{Update $x_t$}}}};
\draw[black,thick, fill = cyan!30! ] (0, 5.5) rectangle(6, 0.5);
\node[align=center] at (3,3.5) {\LARGE{First Order}};
\node[align=center] at (3,2.5) {{\LARGE{ Oracle} \huge{O}}};

\draw[dimarrow, orange] (12.45,4.5) to (6.05,4.5);
\node[align=center] at (7.5,4.9) 
{\LARGE{$x_t$}};

\draw[normline, orange] (6.05,1.5) to (8,1.5);
\node[align=center] at (7,0.9) 
{\LARGE{$\hat{g}(x_t)$}};

\draw[dimarrow, orange] (10,1.5) to (12.45,1.5);
\draw[black, thick, fill = tagreen  ] (8, 2.5) rectangle(10.5, 0.5);
\node[align=center] at (11.25, 0.9) 
{\LARGE{$Y_t $}};
\node[align=center] at (14.5, -0.25) 
{\LARGE{
$Y_t \mid \hat{g}(x_t) \sim W_t(\cdot \mid \hat{g}(x_t))$}};
\node[align=center] at (9.25, 1.5) 
{\huge{$W_t$}};
\node[align=center] at (8.75, 1.5) 
{};

\end{tikzpicture}

%% file: Figures/figure-01.tex
\begin{tikzpicture}
[scale=1, transform shape,
    pre/.style={=stealth',semithick},
    post/.style={->,shorten >=1pt,>=stealth',semithick},
dimarrow/.style={->, >=latex, line width=1pt}
    ]
\draw[<-,line width=1pt] (0.25,3.5) to (0.25,0);
\node[align=center, rotate=90] at (0.0,3.0) {Time}; 

\draw [dashed, color=black!20] (0.3, 0.0) to (8.5,0.0);
\node[align=center] at (0.6,0.2) {t=1}; 
\draw [dashed, color=black!20] (0.3, 0.5) to (4.5, 0.5);
\node[align=center] at (0.6,0.7) {t=2}; 
 \draw [dashed, color=black!20] (5.5, 0.5) to (8.5,0.5);
 \draw [dashed, color=black!20] (0.3, 2*0.5) to (8.5,2*0.5);
\node[align=center] at (0.6,0.7+0.5) {t=3}; 
\draw [dashed, color=black!20] (0.3, 3*0.5) to (8.5,3*0.5);
\node[align=center] at (0.6,0.7+2*0.5) {t=4}; 
\draw [dashed, color=black!20] (0.3, 4*0.5) to (4.5, 4*0.5);
\node[align=center] at (0.6,0.7+3*0.5) {t=5};
\draw [dashed, color=black!20] (5.5, 4*0.5) to(7.5,4*0.5);
\draw [dashed, color=black!20] (0.3, 5*0.5) to (4.5, 5*0.5);
\node[align=center] at (0.6,0.7+4*0.5) {t=6}; 
\draw [dashed, color=black!20] (5.5, 5*0.5) to(7.5,5*0.5);
\draw [dashed, color=black!20] (0.3, 6*0.5) to (8.5,6*0.5);
\node[align=center] at (0.6,0.7+5*0.5) {t=7};

\draw[thick, rounded corners=1.mm] (0.25, -0.25) rectangle (1.75, -.75);  
\node[align=center] at (1.02,-0.5) {Source};  
\draw[dimarrow] (1.75,-0.5) to (2.5,-0.5);
\draw[thick, rounded corners=1.mm] (2.5, -0.25) rectangle (4, -.75);
\node[align=center] at (3.25,-0.5) {Encoder};
\draw[dimarrow] (4,-0.5) to (4.3,-0.5);
\draw[thick, rounded corners=1.mm] (4.3,-0.25) rectangle (5.8, -0.75);
\node[align=center] at (5.1,-0.5) {Channel };
\draw[dimarrow] (5.8,-0.5) to (6.1,-0.5);
\draw[thick, rounded corners=1.mm] (6.1, -0.250) rectangle (7.6, -0.75);
\node[align=center] at (6.9,-0.5) {Decoder};
\draw[dimarrow] (7.6,-0.5) to (8.1,-0.5);
 
\draw[pattern=vertical lines, pattern color=blue!60] (1.5,0) rectangle (2.5,0.5); 
\node[align=center] at (2.1,0.25) {$\bold{X_1}$ };

\draw[pattern=horizontal lines, pattern color=cyan!60] (1.5,0.5) rectangle (2.5,1); 
\node[align=center] at (2.1,0.75) {$\bold{X_2}$ };

\draw[pattern=north east lines, pattern color=gray!60] (1.5,1) rectangle (2.5,1.5); 
\node[align=center] at (2.1,1.25) {$\bold{X_3}$ };

\draw[pattern=north west lines, pattern color=red!20] (1.5,1.5) rectangle (2.5,2); 
\node[align=center] at (2.1,1.75) {$\bold{X_4}$ };


\draw[ color=black, fill = black!75] (2,2.15) circle (0.05cm); 
\draw[ color=black, fill = black!75] (2,2.35) circle (0.05cm); 
\draw[ color=black, fill = black!75] (2,2.55) circle (0.05cm); 

\draw[pattern=vertical lines, pattern color=blue!60] (4.5,0) rectangle (5.5,1); 

\node[align=center] at (5.1,0.25) {$\bold{e(X_1)}$ };

\draw[pattern=north east lines, pattern color=gray!60] (4.5,1) rectangle (5.5,1.5); 
\node[align=center] at (5.1,1.25) {$\bold{e(X_3)}$ };

\draw[pattern=north west lines, pattern color=red!20] (4.5,1.5) rectangle (5.5,3); 
\node[align=center] at (5.1,1.75) {$\bold{e(X_4)}$ };

\draw[ color=black, fill = black!75] (5,3.15) circle (0.05cm); 
\draw[ color=black, fill = black!75] (5,3.35) circle (0.05cm); 
\draw[ color=black, fill = black!75] (5,3.55) circle (0.05cm);


\draw[pattern= vertical lines, pattern color=blue!60] (7.5,1) rectangle (8.5,1.5); 
\node[align=center] at (8.1,1.25) {$\bold{X_1}$ };

\draw[pattern=north east lines, pattern color=gray!60] (7.5,1.5) rectangle (8.5,3); 
\node[align=center] at (8.1,1.75) {$\bold{X_3}$ };

\draw[pattern=north west lines, pattern color=red!20] (7.5,1.5+3*0.5) rectangle (8.5,2+3*0.5); 
\node[align=center] at (8.1, 1.75+3*0.5)  {$\bold{X_4}$};
\draw[ color=black, fill = black!75] (8,3.15+0.5) circle (0.05cm); 
\draw[ color=black, fill = black!75] (8,3.35+0.5) circle (0.05cm); 
\draw[ color=black, fill = black!75] (8,3.55+0.5) circle (0.05cm);
\end{tikzpicture}

%% file: Figures/figure-02a.tex
\begin{tikzpicture}[
soldot/.style={color=blue, only marks, mark=*}, 
dimarrow/.style={<->, >=latex, line width=1pt}
]

\begin{axis}[
width=8.5cm,height=5cm,
    axis line style={line width=1.5pt},
	xmajorgrids=true,
    ymajorgrids=true,
    grid style=solid,
	axis x line=middle, axis y line=middle,	
	y label style={at={(axis description cs:-0.05,.5)},rotate=90,anchor=south},
	x label style={at={(axis description cs:.5,-.15)},anchor=south},
    xmin=1, xmax=7.5,
    ymin=0, ymax=5,
    xtick={1,2,...,7},
    ytick={0,1,2,...,5},
	x label style={at={(axis description cs:.5,-.25)},anchor=south},
	xlabel={$t$},
    ylabel={$U(t)$}
]

\foreach \x in {1,2}
\addplot[soldot] coordinates{(\x,0)};
\foreach \x in {3}
\addplot[soldot] coordinates{(\x,1)};
\foreach \x in {4}
\addplot[soldot] coordinates{(\x,3)};
\foreach \x in {5}
\addplot[soldot] coordinates{(\x,3)};
\foreach \x in {6}
\addplot[soldot] coordinates{(\x,3)};
\foreach \x in {7}
\addplot[soldot] coordinates{(\x,4)};

\draw[dimarrow] (axis cs:1,1)  -- (axis cs:3,1)node[midway,above] {$\ell(X_1)$} ;
\draw[dashed] (axis cs:3,0)  -- (axis cs:3,1) ;

\draw[dimarrow] (axis cs:3,3)  -- (axis cs:4,3)node[midway,above] {$\ell(X_3)$} ;
\draw[dashed] (axis cs:3,1)  -- (axis cs:4,1) ;
\draw[dashed] (axis cs:4,1)  -- (axis cs:4,3) ;

\draw[dashed] (axis cs:4,3)  -- (axis cs:7,3) ;
\draw[dimarrow] (axis cs:4,4)  -- (axis cs:7,4)node[midway,above] {$\ell(X_4)$} ;
\draw[dashed] (axis cs:7,3)  -- (axis cs:7,4) ;
\draw[dashed] (axis cs:7,4)  -- (axis cs:7.5,4) ;
\end{axis}
\end{tikzpicture} 

%% file: Figures/figure-02b.tex
\begin{tikzpicture}[
soldot/.style={color=blue, only marks, mark=*}, 
dimarrow/.style={<->, >=latex, line width=1pt}
]

\begin{axis}[
width=8.5cm,height=5cm,
    axis line style={line width=1.5pt},
	xmajorgrids=true,
    ymajorgrids=true,
    grid style=solid,
	axis x line=middle, axis y line=middle,	
	y label style={at={(axis description cs:-0.05,.5)},rotate=90,anchor=south},
	x label style={at={(axis description cs:.5,-.15)},anchor=south},
    xmin=1, xmax=7.5,
    ymin=0, ymax=5,
    xtick={1,2,...,7},
    ytick={0,1,2,...,5},
	x label style={at={(axis description cs:.5,-.25)},anchor=south},
	xlabel={$t$},
    ylabel={$A(t)$}
]

%
\foreach \x in {1,2}
\addplot[soldot] coordinates{(\x,\x)};
\foreach \x in {3}
\addplot[soldot] coordinates{(\x,2)};
\foreach \x in {4}
\addplot[soldot] coordinates{(\x,1)};
\foreach \x in {5}
\addplot[soldot] coordinates{(\x,2)};
\foreach \x in {6}
\addplot[soldot] coordinates{(\x,3)};
\foreach \x in {7}
\addplot[soldot] coordinates{(\x,3)};

\draw[dashed] (axis cs:1,1)  -- (axis cs:2,2) ;
\draw[dashed] (axis cs:2,2)  -- (axis cs:3,3) ;
\draw[dashed] (axis cs:3,2)  -- (axis cs:3,3) ;
\draw[dimarrow] (axis cs:3,0)  -- (axis cs:3,2) node[midway,left] {$\ell(X_1)$};
\draw[dashed] (axis cs:3,2)  -- (axis cs:4,3) ;
\draw[dashed] (axis cs:4,3)  -- (axis cs:4,1) ;
\draw[dimarrow] (axis cs:4,0)  -- (axis cs:4,1) node[midway,right] {$\ell(X_3)$};
\draw[dashed] (axis cs:4,1)  -- (axis cs:6,3) ;
\draw[dashed] (axis cs:6,3)  -- (axis cs:7,4) ;
\draw[dashed]  (axis cs:7,4) -- (axis cs:7,3);
\draw[dimarrow] (axis cs:7,3)  -- (axis cs:7,0) node[midway,left] {$\ell(X_4)$};
\draw[dashed]  (axis cs:7,3) -- (axis cs:7.5,3.5);
\end{axis}
\end{tikzpicture} 

%% file: Figures/figure-03.tex
\begin{tikzpicture}[thick,scale=1, every node/.style={transform shape}]
\begin{axis}[ width=8.5cm,height=7.5cm,
    xlabel={$s$},
    ylabel={Average-age},
    xmin=0, xmax=5,
    ymin=0, ymax=12.5,
    xtick={0,1, 2,3, 4, 5},
    ytick={0,2.5,5,7.5,10,12.5},
    legend style={font=\fontsize{6.5}{8}\selectfont},
    legend pos=north east,
    xmajorgrids=true,
    ymajorgrids=true,
    grid style=dashed,
]

\addplot[color=blue , solid, mark=*, mark options={blue}]table [x=SV, y=PShannonAge, col sep=comma] {Figures/AOIZ_check.csv}; 
\addplot+[color=black ,solid, mark=+, mark options={black}] table [x=SV, y=PAvgAge, col sep=comma] {Figures/AOIZ_check.csv}; 
\addplot+[color=red ,dashed, mark=*, mark options={red}]table [x=SV, y=UBAvgAge, col sep=comma] {Figures/AOIZ_check.csv}; 
\addplot+[color=ForestGreen ,densely dashdotted, mark=+, mark options={ForestGreen}] table [x=SV, y=AvgAge, col sep=comma] {Figures/AOIZ_check.csv}; 
        \legend{Shannon code for $P$,Shannon code for $P^*$,
Shannon lengths for $P$,  Shannon lengths for $P^*$,}

\end{axis}
\end{tikzpicture}

%% file: Figures/figure-04.tex
\begin{tikzpicture}[thick,scale=1, every node/.style={transform shape}]
    \begin{axis}[
    width=8.5cm,height=6cm,
            ybar,            
            legend pos=north east,
            xtick={1,2,3,4,5,6,7,8},
            ymin=0,ymax=0.4,  
            ymajorgrids=true,
                grid style=dashed,        
        ]
        
             \addplot[
    black,
    fill=blue!60,
    postaction={
        pattern=north east lines
    }
] table [x=I, y=p, col sep=comma] {Figures/AOIZ_TD_S1_N8.csv};
             
             \addplot[
    black,
    fill=blue!20,
    postaction={
        pattern=north west lines
    }
]  table [x=I, y=tilteddistribution, col sep=comma] {Figures/AOIZ_TD_S1_N8.csv}; 
        \legend{$P$,$P^*$}
    \end{axis}
\end{tikzpicture}

%% file: Figures/figure-05.tex
\begin{tikzpicture}[thick,scale=1, every node/.style={transform shape}]
\begin{axis}[
    xlabel={$H(P)$},
    ylabel={},
    xmin=0, xmax=8,
    ymin=0, ymax=12,
    xtick={0,  2,  4, 6,8},
    ytick={0,2,4,6,8,10,12},
    legend style={font=\footnotesize},
    legend pos=north west,
    xmajorgrids=true,
    ymajorgrids=true,
    grid style=dashed,
]
 
\addplot[color=blue, mark=*] table [x=Entropy, y=CodeLength, col sep=comma] {Figures/AOIZ_check.csv};
\addplot[color=black ,solid, mark=*] table [x=Entropy, y=AvgAge, col sep=comma] {Figures/AOIZ_check.csv}; 
 \addplot +[color=gray ,densely dashdotted, mark= ]table [x=Entropy, y=Entropy, col sep=comma] {Figures/AOIZ_check.csv}; 
\legend{Average length (real lengths), Average age (real lengths), $Y=X$} 
\end{axis}
\end{tikzpicture}

%% file: Figures/figure-06.tex
\begin{tikzpicture}
[scale=1, transform shape,
    pre/.style={=stealth',semithick},
    post/.style={->,shorten >=1pt,>=stealth',semithick},
dimarrow/.style={->, >=latex, line width=1pt}
    ]
\draw[<-,line width=1pt] (0.25,3) to (0.25,0);
\node[align=center, rotate=90] at (0.0,1.4) {Time}; 
\draw[thick, rounded corners=1.mm] (0.25, -0.25) rectangle (1.75, -.75);  
\node[align=center] at (1.02,-0.5) {Source};  
\draw[dimarrow] (1.75,-0.5) to (2.5,-0.5);
\draw[thick, rounded corners=1.mm] (2.5, -0.25) rectangle (4, -.75);
\node[align=center] at (3.25,-0.5) {Encoder};
\draw[dimarrow] (4,-0.5) to (4.3,-0.5);
\draw[thick, rounded corners=1.mm] (4.3,-0.25) rectangle (5.8, -0.75);
\node[align=center] at (5.1,-0.5) {Channel };
\draw[dimarrow] (5.8,-0.5) to (6.1,-0.5);
\draw[thick, rounded corners=1.mm] (6.1, -0.250) rectangle (7.6, -0.75);
\node[align=center] at (6.9,-0.5) {Decoder};
\draw[dimarrow] (7.6,-0.5) to (8.1,-0.5);
 
\draw[pattern=vertical lines, pattern color=blue!60] (1.5,0) rectangle (2.5,0.5); 
\node[align=center] at (2.1,0.25) {$\bold{X_1}$ };

\draw[pattern=horizontal lines, pattern color=cyan!60] (1.5,0.6) rectangle (2.5,1.1); 
\node[align=center] at (2.1,0.85) {$\bold{X_2}$ };


\draw[pattern=north west lines, pattern color=gray!60] (1.5,2.1) rectangle (2.5,2.6); 
\node[align=center] at (2.1,2.35) {$\bold{X_3}$ };


\draw[pattern=vertical lines, pattern color=blue!60] (4.5,0) rectangle (5.5,1.5); 
\node[align=center] at (5.1,0.25) {$\bold{e(X_1)}$ };
\draw [dashed] (4.5,0.5) to (5.5,0.5);
\draw [dashed] (4.5,1) to (5.5,1);

\draw[pattern=horizontal lines, pattern color=cyan!60] (4.5,1.5) rectangle (5.5,2); 
\node[align=center] at (5.1,1.75) {$\bold{e(X_2)}$ };

\draw[pattern=north east lines, pattern color=gray!60] (4.5,2.1) rectangle (5.5,3.1); 
\node[align=center] at (5.1,2.35) {$\bold{e(X_3)}$ };


\draw [dashed] (4.5,0.5) to (5.5,0.5);
\draw [dashed] (4.5,2) to (5.5,2);
\draw [dashed] (4.5,2.6) to (5.5,2.6);


\draw[pattern= vertical lines, pattern color=blue!60] (7.5,1.5) rectangle (8.5,2); 
\node[align=center] at (8.1,1.75) {$\bold{X_1}$ };

\draw[pattern=horizontal lines, pattern color=cyan!60] (7.5,2) rectangle (8.5,2.5); 
\node[align=center] at (8.1,2.25) {$\bold{X_2}$ };

\draw[pattern=north east lines, pattern color=gray!60] (7.5,3.1) rectangle (8.5,3.6); 
\node[align=center] at (8.1,3.35) {$\bold{X_3}$ };
\end{tikzpicture}

%% file: Figures/figure-07a.tex
\begin{tikzpicture}[thick,scale=1, every node/.style={transform shape}]
\begin{axis}[
    xlabel={Arrival Rate $\lambda$},
    ylabel={Average Delay},
    xmin=0.03, xmax=.15,
    ymin=5, ymax=18,
    xtick={ .03,  .06, .09,.12, .15},
    ytick={5,9,13,17},
    legend style={font=\fontsize{9}{10}\selectfont},
    legend pos=north west,
    xmajorgrids=true,
    ymajorgrids=true,
    grid style=dashed,
]

\addplot+[color=blue, dashed, mark=*, mark options={blue}]table [x=ArrivalRate, y=OptimalDelay(LarmoresAlgo), col sep=comma] {Figures/Comprsn_Shan_MinDel_Codes.csv}; 
\addplot+[color=black, solid, mark=+] table [x=ArrivalRate, y=maxminLowerBound, col sep=comma] {Figures/Comprsn_Shan_MinDel_Codes.csv}; 
        \legend{
Larmore's Optimal Algorithm,  Shannon lengths for $P^*$,}
\end{axis}
\end{tikzpicture}

%% file: Figures/figure-07b.tex
\begin{tikzpicture}[thick,scale=1, every node/.style={transform shape}]
\begin{axis}[
    xlabel={Arrival Rate $\lambda$},
    ylabel={Average Delay},
    xmin=0.04, xmax=.20,
    ymin=4, ymax=26,
    xtick={ .04,  .08, .12,.16, .20},
    ytick={5,10,15,20,25},
    legend style={font=\fontsize{9}{10}\selectfont},
    legend pos=north west,
    xmajorgrids=true,
    ymajorgrids=true,
    grid style=dashed,
]

\addplot+[color=blue ,dashed,  mark=*, mark options={blue}]table [x=ArrivalRate, y=OptimalDelay(LarmoresAlgo), col sep=comma] {Figures/Comprsn_Shan_MinDel_Codes_1.csv}; 
\addplot+[color=black, solid, mark=+] table [x=ArrivalRate, y=maxminLowerBound, col sep=comma] {Figures/Comprsn_Shan_MinDel_Codes_1.csv}; 
        \legend{ 
Larmore's Optimal Algorithm,  Shannon lengths for $P^*$,}
                                                                                                                                                                                                                                                                                                                                                                                                                                                                                                                                                                                                                                                                                                                                                                                                                                                                                                                                                                                                                                                                                                                                                                                                                                                                                                                                                                                                                                                                                                                                                                                                                                                                                                                                                                                                                                                                                                                                                                                                                                                                                                                                                                                                                                                                                                                                                                                                                                            \end{axis}
\end{tikzpicture}